\documentclass[10pt,rmp,aps,floatfix,showpacs,twocolumn]{revtex4-1}
\usepackage{graphicx}
\usepackage{amsmath, amsfonts, amssymb}
\usepackage{bm}
\usepackage{mathrsfs}
\usepackage[breaklinks=true]{hyperref}
\hypersetup{
colorlinks=true,
linkcolor=black,
linktoc=black,
citecolor=black,
filecolor=black,
menucolor=black,
pagecolor=black,
urlcolor=black
}
\usepackage{breakurl}

\begin{document}

\title{Extremely high-intensity laser interactions with fundamental quantum systems}

\author{A. Di Piazza}
\email{dipiazza@mpi-hd.mpg.de}
\author{C. M\"uller}
\email{carsten.mueller@tp1.uni-duesseldorf.de. New address: Institut f\"{u}r Theoretische Physik I, Heinrich-Heine-Universit\"{a}t D\"{u}sseldorf, Universit\"{a}tsstra\ss e 1, 40225 D\"{u}sseldorf, Germany}
\author{K. Z. Hatsagortsyan}
\email{k.hatsagortsyan@mpi-hd.mpg.de}
\author{C. H. Keitel}
\email{keitel@mpi-hd.mpg.de}
\affiliation{Max-Planck-Institut f\"ur Kernphysik, Saupfercheckweg 1, 69117 Heidelberg, Germany}

\date{\today}

\begin{abstract}
The field of laser-matter interaction traditionally deals with the response of atoms, molecules and plasmas to an external light wave. However, the recent sustained technological progress is opening up the possibility of employing intense laser radiation to trigger or substantially influence physical processes beyond atomic-physics energy scales. Available optical laser intensities exceeding $10^{22}\;\text{W/cm$^2$}$ can push the fundamental light-electron interaction to the extreme limit where radiation-reaction effects dominate the electron dynamics, can shed light on the structure of the quantum vacuum, and can trigger the creation of particles like electrons, muons and pions and their corresponding antiparticles. Also, novel sources of intense coherent high-energy photons and laser-based particle colliders can pave the way to nuclear quantum optics and may even allow for potential discovery of new particles beyond the Standard Model. These are the main topics of the present article, which is devoted to a review of recent investigations on high-energy processes within the realm of relativistic quantum dynamics, quantum electrodynamics, nuclear and particle physics, occurring in extremely intense laser fields.
\end{abstract}


\maketitle
\tableofcontents

\section{Introduction}

The first realization of the laser in 1960 \cite{Maiman_1960} is one of the most important technological breakthroughs. Nowadays lasers are indispensable tools for investigating physical processes in different areas ranging from atomic and plasma physics to nuclear and high-energy physics. This has been possible mainly due to the continuous progress made along two specific directions: decrease of the laser pulse duration and increase of the laser peak intensity \cite{Mourou_2011}. On the one hand, multiterawatt laser systems with a pulse duration in the femtosecond time scale are readily available nowadays and different laboratories have succeeded in the generation of single attosecond pulses. Physics at the attosecond time scale has been the subject of the recent review \onlinecite{Krausz_2009}. In this review it has been pointed out how pulses in the attosecond domain allow for the detailed investigation of the electron motion in atoms and during molecular reactions. The production of ultrashort pulses is strongly connected with the increase of the laser peak intensity. This is not only because temporal compression evidently implies an increase in intensity at a given laser energy, but also because higher intensities allow, in general, for controlling faster physical processes, which in turn can be exploited for generating correspondingly shorter light pulses.

Not long after the invention of the laser, available intensities were already sufficiently high to trigger nonlinear optical effects like second harmonic generation. It is, however, only after the experimental implementation of the Chirped Pulse Amplification (CPA) technique \cite{Strickland_1985} that it has been possible to reach the intensity threshold of $10^{14}\text{-} 10^{15}\;\text{W/cm$^2$}$ corresponding to electric field amplitudes of the same order as the Coulomb field in atoms. At such intensities the interplay between the laser and the atomic field significantly alters the electron's dynamics in atoms and molecules and this can be exploited, for example, for generating high-frequency radiation in the extreme-ultraviolet (XUV) and soft-x-ray regions (high-order harmonic generation or HHG) \cite{Protopapas_1997,Agostini_2004}. HHG as well as atomic processes in intense laser fields have been recently reviewed in \onlinecite{Winterfeldt_2008,Teubner_2009,Fennel_2010}, with specific emphasis on the control of high-harmonic spectra by spatio-temporal shaping of the driving pulse \cite{Winterfeldt_2008}, on harmonic generation in laser-plasma interaction \cite{Teubner_2009} and on the dynamics of clusters in strong laser fields \cite{Fennel_2010}. 

By increasing the optical laser intensity to the order of $10^{17}\text{-} 10^{18}\;\text{W/cm$^2$}$, another physically important regime in laser-matter interaction is entered: the relativistic regime. In such intense electromagnetic fields an electron reaches relativistic velocities already within one laser period, the magnetic component of the Lorentz force becomes of the same order of magnitude of the electric one, and the electron's motion becomes highly nonlinear as a function of the laser's electromagnetic field. Although the increasing influence of the magnetic force causes a suppression of atomic HHG in the relativistic domain, the highly nonlinear motion of the electrons in such strong laser fields is at the origin of numerous new effects as relativistic self-focusing in plasma and laser wakefield acceleration \cite{Mulser_b_2010}. In the recent reviews \onlinecite{Salamin_2006,Mourou_2006,Ehlotzky_2009}, different processes occurring at relativistic laser intensities are discussed. In particular, in \onlinecite{Ehlotzky_2009}, QED processes like Compton, Mott and M\o ller scattering in a strong laser field are covered, in \onlinecite{Mourou_2006}, technical aspects and new possibilities of the CPA techniques are reviewed together with relativistic effects in laser-plasma interaction as, for example, self-induced transparency and wakefield generation, while in \onlinecite{Salamin_2006}, spin-effects as well as relativistic multiphoton and tunneling recollision dynamics in laser-atom interactions are reviewed. Also in the same year another review was published on nonlinear collective photon interactions, including vacuum-polarization effects in a plasma \cite{Marklund_2006}. Whereas, the physics of plasma-based laser electron accelerators is the main subject covered in \onlinecite{Esarey_2009,Malka_2011}, with a special focus on the different phases involved (electron injection and trapping, and pulse propagation) and on the role of plasma instabilities in the acceleration process. Finally, in \onlinecite{Ruffini_2010_b} different processes related to electron-positron ($e^+\text{-}e^-$) pair production are reviewed with special emphasis on those occurring in the presence of highly-charged ions and in astrophysical environments.

In the present article we address physical processes that mainly occur at optical laser intensities mostly larger than $10^{21}\;\text{W/cm$^2$}$, i.e., well exceeding the relativistic threshold. After reporting on the latest technological progress in optical and x-ray laser technology (Sec. \ref{NRS}), we review some basic results on the classical and quantum dynamics of an electron in a laser field (Sec. \ref{FED}). Then, we bridge to lower-intensity physics by reviewing more recent advances in relativistic ionization and HHG in atomic gases (Sec. \ref{RD}). The main subject of the review, i.e., the response of fundamental systems like electrons, photons and even the vacuum to ultra-intense radiation fields is covered in Secs. \ref{TCS}-\ref{MPP}. As will be seen, such high laser intensities represent a unique tool to investigate fundamental processes like multiphoton Compton scattering (Sec. \ref{TCS}), to clarify conceptual issues like radiation reaction in classical and quantum electrodynamics (Sec. \ref{RR}) and to investigate the structure of the quantum vacuum (Sec. \ref{VPEs}). Also, other fundamental quantum-relativistic phenomena like the transformation of pure light into massive particles as electrons, muons and pions (and their corresponding antiparticles) can become feasible and can even limit the attainability of arbitrarily high laser intensities (Secs. \ref{PP}-\ref{MPP}). Finally, we will also review recent suggestions on employing novel high-frequency lasers and laser-accelerated particle beams to directly trigger nuclear and high-energy processes (Secs. \ref{NP}-\ref{Part_Phys}). The main conclusions of the article will be presented in Sec. \ref{Conlc}.

Units with $\hbar=c=1$ and the space-time metric $\eta^{\mu\nu}=\text{diag}(+1,-1,-1,-1)$ are employed throughout this review.

\section{Novel radiation sources}
\label{NRS}

In this section we review the latest technical and experimental progress in laser technology. We will discuss optical and x-ray laser systems separately. The latter are especially useful for $e^+\text{-}e^-$ pair production, for direct laser-nucleus interaction, as well as as probes for vacuum-polarization effects (see in particular Secs. \ref{VPEs}, \ref{PP} and \ref{NP}). For overviews of feasible accelerators also of relevance for the present review see, e.g., \onlinecite{Wilson_b_2001,Esarey_2009,Malka_2011,PDG_2010} and the relevant original literature as quoted in the respective sections.

\subsection{Strong optical laser sources}
\label{L_O}

As has been mentioned in the Introduction, since the invention of the CPA technique \cite{Strickland_1985} laser peak intensities have been boosted by several orders of magnitude. Another amplification technique called Optical Parametric Chirped Pulse Amplification (OPCPA), based on the nonlinear interaction among laser beams in crystals, was suggested almost at the same time as the CPA and proved to be promising as well \cite{Piskarskas_1986}. As a result of the increase in available laser intensities, exciting perspectives have been envisaged in different fields spanning from atomic to plasma and even nuclear and high-energy physics \cite{Gerstner_2007,Feder_2010,Mourou_2010,Tajima_2010}.

The group of G. Mourou at the University of Michigan (Michigan, USA) holds the record so far for the highest laser intensity ever achieved of $2\times 10^{22}\;\text{W/cm$^2$}$ \cite{Yanovsky_2008}, while no experiments have been performed at this intensity yet. This record intensity has been reached when the HERCULES laser was upgraded to become a $300\;\text{TW}$ Ti:Sa
system, amplified via CPA and capable of a repetition rate of $0.1\;\text{Hz}$. The 4-grating compressor allowed for a pulse duration of about $30\;\text{fs}$ and adaptive optics together with a $f/1$ parabola enabled to focus the beam down to a diameter of about $1.3\;\text{$\mu$m}$. This experimental achievement on the laser intensity pushed the capabilities of a multiterawatt laser almost to the limit. 

The $1\text{-PW}$ threshold has been already reached and even exceeded in various laboratories. For example, the Texas Petawatt Laser (TPL) at the University of Texas at Austin (Texas, USA) has exceeded the Petawatt threshold thanks to the OPCPA technique, by compressing an energy of $186\;\text{J}$ in a pulse lasting only $167\;\text{fs}$ \cite{TPL}. The TPL has been employed for investigating laser-plasma interactions at extreme conditions, particularly relevant for astrophysics. Also, the two laser systems Vulcan \cite{Vulcan} and Astra Gemini \cite{AG} at the Central Laser Facility (CLF) in the United Kingdom provide powers of the order of $1\;\text{PW}$. The Vulcan facility can deliver an energy of $500\;\text{J}$ in a pulse lasting $500\;\text{fs}$. It is a Nd:YAG laser system amplified via CPA and can provide intensities up to $10^{21}\;\text{W/cm$^2$}$. Whereas, the Astra Gemini laser consists of two independent Ti:Sa laser beams of $0.5\;\text{PW}$ each (energy of $15\;\text{J}$ and a pulse duration of $30\;\text{fs}$), with a maximum focused intensity of $10^{22}\;\text{W/cm$^2$}$. The particular layout of the Astra Gemini laser renders this system especially versatile for unique applications in strong-field physics, where two ultrastrong beams are required. Two laser systems are likely to be updated to the Petawatt level in Germany. The first one is the Petawatt High-Energy Laser for heavy Ion eXperiments (PHELIX) Nd:YAG laser at the Gesellschaft f\"{u}r Schwerionenforschung (GSI) in Darmstadt, capable now of delivering an energy of $120\;\text{J}$ in about $500\;\text{fs}$ \cite{PHELIX}. The 1-PW threshold should be reached by increasing the pulse energy to $500\;\text{J}$. Combined with the highly-charged ion beams at GSI, the PHELIX facility can be attractive for experimental investigations in strong-field quantum electrodynamics (QED). The second system to be updated to $1\;\text{PW}$ is the Petawatt Optical Laser Amplifier for Radiation Intensive Experiments (POLARIS) laser in Jena \cite{Hein_2010}. At the moment, a power of about $100\;\text{TW}$ (energy of $10\;\text{J}$ for a pulse duration of $100\;\text{fs}$) has been reached and the goal of $1\;\text{PW}$ power should be achieved by compressing $120\;\text{J}$ in about $120\;\text{fs}$. The Scottish Centre for the Application of Plasma-based Accelerators (SCAPA) research center is one of the main initiatives within the Scottish Universities Physics Alliance (SUPA) project dedicated to the high-power laser interaction with plasmas. A laser system will be developed, which will generate pulses of 5-7 J energy and of 25-30 fs duration at a repetition rate of 5 Hz, corresponding to a peak power of 200-250 TW, with potential for future upgrades to the petawatt level \cite{SCAPA_2012}.

The 1-PW threshold has been also exceeded in Ti:Sa laser systems like those described in \onlinecite{Sung_2010} (energy of $34\;\text{J}$ for a pulse duration of $30\;\text{fs}$) and in \onlinecite{Wang_2011} (energy of $32.3\;\text{J}$ for a pulse duration of $27.9\;\text{fs}$) and constructed at the Advanced Photonics Research Institute (APRI) at Gwangju (Republic of Korea) and at the Beijing National Laboratory for Condensed Matter Physics in Beijing (China), respectively. The BELLA (Berkeley Lab Laser Accelerator) is a Ti:Sa laser system under construction at the Lawrence Berkeley National Laboratory (LBNL) at Berkeley (California, USA) which will also reach the 1-PW threshold by compressing $40\;\text{J}$ in $40\;\text{fs}$ at a repetition rate of $1\;\text{Hz}$ \cite{BELLA}.

All the above systems require laser energies larger than $10\;\text{J}$ and this limits the repetition rate of existing petawatt lasers in the best situation to about $0.1\;\text{Hz}$ \cite{Sung_2010}. The Ti:Sa Petawatt Field Synthesizer (PFS) system under development in Garching (Germany) aims to be the first high-repetition rate petawatt laser system with an envisaged repetition rate of $10\;\text{Hz}$ \cite{PFS}. By adopting the OPCPA technique the PFS should reach the petawatt level by compressing an energy of about $5\;\text{J}$ in $5\;\text{fs}$ \cite{Major_2010}. For a recent review on petawatt-class laser systems, see \onlinecite{Korzhimanov_2011}. 

Finally, we also want to mention other high-power lasers, mainly devoted to fast ignition and characterized by relatively long pulses of the order of $1\;\text{ps}\text{-} 1\;\text{ns}$. Among others we mention the OMEGA EP system at Rochester (New York, USA) (energy of $1\;\text{kJ}$ for a pulse duration of $1\;\text{ps}$) \cite{OMEGA_EP} and the National Ignition Facility (NIF) at the Lawrence Livermore National Laboratory (LLNL) at Livermore (California, USA) (energy of $2\;\text{MJ}$ distributed in 192 beams with a pulse duration of about $3\text{-}10\;\text{ns}$) \cite{NIF}. Another high-power laser facility is the PETawatt Aquitaine Laser (PETAL) in Le Barp close to Bordeaux (France), which is a multi-petawatt laser, generating pulses with energy up to 3.5 kJ and with a duration of 0.5 to 5 ps \cite{PETAL}. The PETAL facility is planned to be coupled to the Laser M\'{e}gaJoule (LMJ) under construction in Bordeaux (France). In the LMJ a total energy of $1.8\;\text{MJ}$ is distributed in a series of 240 laser beamlines, collected into eight groups of 30 beams with a pulse duration of $0.2\text{-} 25\;\text{ns}$ \cite{LMJ_2011}.

\subsubsection{Next-generation $10\;\text{PW}$ optical laser systems}

The possibility of building a $10\;\text{PW}$ laser system is under consideration in various laboratories. At the CLF in the United Kingdom the $10\;\text{PW}$ upgrade of the Vulcan laser has already started \cite{Vulcan_10PW}. The new laser will provide beams with an energy of $300\;\text{J}$ in only $30\;\text{fs}$ via the OPCPA. A $10\;\text{PW}$ laser system is in principle capable of unprecedented intensities larger than $10^{23}\;\text{W/cm$^2$}$ if the beam is focused to about $1\;\text{$\mu$m}$. The front-end stage of the Vulcan $10\;\text{PW}$ is already completed and it delivers pulses with about $1\;\text{J}$ of energy at a central wavelength of $0.9\;\text{$\mu$m}$, with sufficient bandwidth to support a pulse with duration less than 30 fs.

Another $10\;\text{PW}$ laser project is the ILE APOLLON to be realized at the Institut de Lumi\'{e}re Extreme (ILE) in France \cite{Chambaret_2009}. The laser pulses are expected to deliver an energy of $150\;\text{J}$ in $15\;\text{fs}$ at the last stage of amplification after the front end (energy of $100\;\text{mJ}$ in less than $10\;\text{fs}$), with a repetition rate of one shot per minute. Laser intensities of the order of $10^{24}\;\text{W/cm$^2$}$ are envisaged at the ILE APOLLON system, entering the so-called ultrarelativistic regime, where also ions (rest energy of the order of $1\;\text{GeV}$) become relativistic within one laser period of such an intense laser field.

We mention here also the PEtawatt pARametric Laser (PEARL-10) project at the Institute of Applied Physics of the Russian Academy of Sciences in Nizhny Novgorod (Russia), which is an upgrade of the present 0.56-PW laser employing the OPCPA technique, to $10\;\text{PW}$ ($200\;\text{J}$ of energy compressed in $20\;\text{fs}$) \cite{Korzhimanov_2011}.

\subsubsection{Multi-Petawatt and Exawatt optical laser systems}

\begin{figure}
\begin{center}
\includegraphics[width=0.8\linewidth]{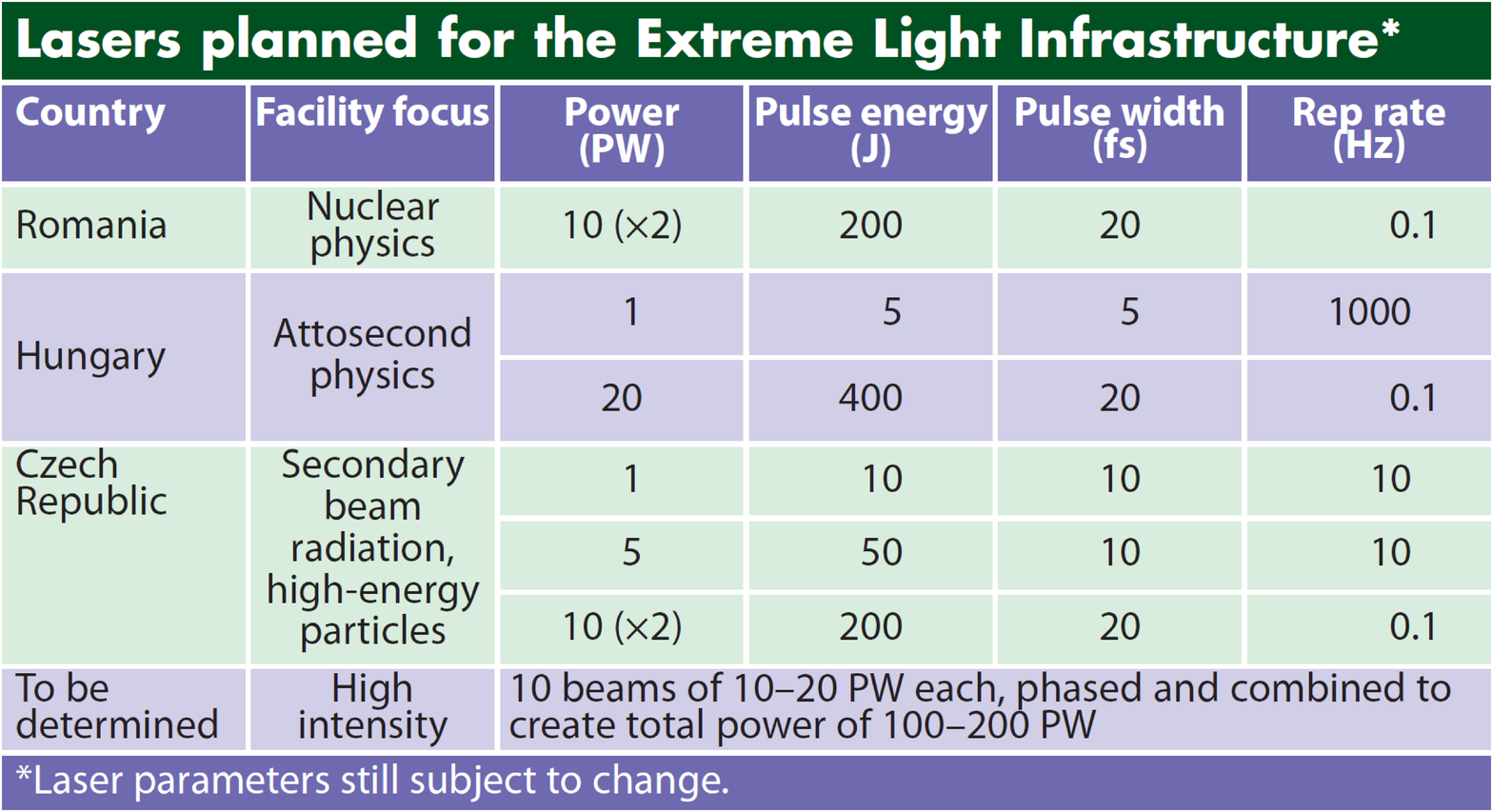}
\end{center}
\caption{(Color online) Summary of the four pillars of ELI. A power value of $10(\times2)\;\text{PW}$ indicates the availability of two laser systems each with $10\;\text{PW}$ power. Reprinted with permission from \onlinecite{Feder_2010}. Copyright 2010, American Institute of Physics.}
\label{Megalaser}
\end{figure}

The Extreme Light Infrastructure (ELI) \cite{ELI} (see Fig. \ref{Megalaser}), the Exawatt
Center for Extreme Light Studies (XCELS) \cite{XCELS} and the High Power laser Energy Research (HiPER) facility at the CLF in the United Kingdom \cite{HiPER} are envisaged laser systems with a power exceeding the $100\;\text{PW}$ level. 

ELI is a large-scale laser facility consisting of four ``pillars'' (see Fig. \ref{Megalaser}): one devoted to nuclear physics, one to attosecond physics, one to secondary beams (photon beams, ultrarelativistic electron and ion beams) and one to high-intensity physics.
This last one is of relevance here and it is supposed to comprise ten beams each with a power of $10\text{-} 20\;\text{PW}$ that, when combined in phase, should deliver a single beam of about $100\text{-} 200\;\text{PW}$ at a repetition rate of one shot per minute. The relatively high repetition rate is obtained by compressing in each beam alone $0.3\text{-} 0.4\;\text{kJ}$ of energy in a pulse of $15\;\text{fs}$. We mention that one of the aims of the ILE APOLLON system is to provide a prototype of the $10\text{-} 20\;\text{PW}$ beams, that will be then employed at ELI. In the high-field pillar of ELI ultrahigh intensities exceeding $10^{25}\;\text{W/cm$^2$}$ are envisaged, which are well above the ultrarelativistic regime. At such intensities, it will be possible to test different aspects of fundamental physics for the first time.

The XCELS infrastructure is planned to be built in Nizhny Novgorod (Russia) and it will consist of 12 beams each with energy of $300\text{-}400\;\text{J}$ and with duration of $20\text{-}30\;\text{fs}$. The pulse resulting from the superposition of these beams is expected to have a power of $200\;\text{PW}$, a pulse duration of about $25\;\text{fs}$, and divergence less than 3 diffraction limits (at a central wavelength of $0.91\;\text{$\mu$m}$). Apart from aiming to overcome the $100\;\text{PW}$ threshold, the main priorities of XCELS are the creation of sources of attosecond and subattosecond, X-ray and $\gamma$-ray pulses, the development of laser based electron and ion accelerators with electron and ion energies exceeding 100 GeV and up to 10 GeV, respectively, the realization in laboratory of astrophysical and early-cosmological conditions and the investigation of the structure of the quantum vacuum.

The main goal of the other large-scale facility HiPER is the first demonstration of laser-driven fusion, or fast ignition. To this end HiPER will deliver: 1) an energy of about $200\;\text{kJ}$ distributed in 40 beams with a pulse duration of several nanoseconds and a photon energy of $3\;\text{eV}$ in the compression side; 2) an energy of about $70\;\text{kJ}$ distributed in 24 beams with a pulse duration of $15\;\text{ps}$ and a photon energy of $2\;\text{eV}$ in the ignition side. Employing HiPER for high-intensity physics would imply a feasible reconfiguration of the ignition side to deliver $10\;\text{kJ}$ in only $10\;\text{fs}$ via the OPCPA technique. This would render HiPER a laser facility with Exawatt ($10^{18}\;\text{W}$) power and with a potential intensity of $10^{26}\;\text{W/cm$^2$}$.

Finally, we briefly mention the GEKKO EXA facility conceptually under design in Osaka (Japan) \cite{Gekko_EXA}. This facility is expected to deliver pulses of $2\;\text{kJ}$ energy and of $10\;\text{fs}$ duration corresponding to $200\;\text{PW}$ and with an intensity up to $10^{25}\;\text{W/cm$^2$}$.

\subsection{Brilliant x-ray laser sources}
\label{x_ray}
Strong optical laser systems are sources of coherent radiation at wavelengths of the order of $1\;\text{$\mu$m}$, corresponding to photon energies of the order of $1\;\text{eV}$. Considerable efforts have been devoted in the past few years to develop coherent radiation sources at photon energies larger than $100\;\text{eV}$. The discovery of the Self-Amplified Spontaneous Emission (SASE) regime \cite{Bonifacio_1984} has opened the possibility of employing Free Electron Lasers (FELs) to generate coherent light at such short wavelengths. In a FEL relativistic bunches of electrons pass through a spatially-periodic magnetic field (undulator) and emit high-energy photons. In the SASE regime the interaction of the electron bunch with its own electromagnetic field ``structures'' the bunch itself into slices (microbunches) each one emitting coherently even at wavelengths below $1\;\text{nm}$ (FELs at such small wavelengths are dubbed X-ray Free Electron Lasers (XFELs)).

\begin{figure}
\begin{center}
\includegraphics[width=0.8\linewidth]{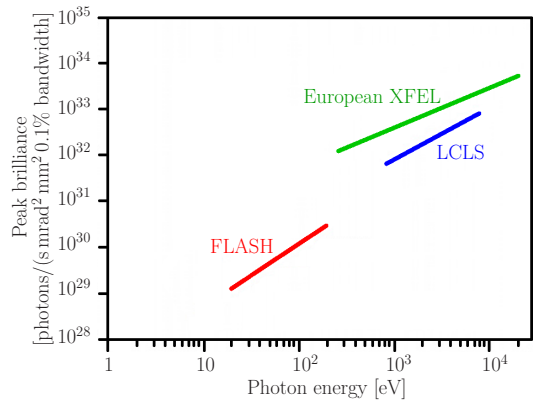}
\end{center}
\caption{(Color online) Comparison among the peak brilliances of the three facilities FLASH, LCLS and European XFEL as a function of the laser photon energy. An envisaged peak brilliance of $5\times 10^{33}$\;\text{photons/(s\,mrad$^2$\,mm$^2$\,0.1\% bandwidth)} at a photon energy of $12.4\;\text{keV}$ for the SACLA facility is reported in \onlinecite{XFEL}. See also \onlinecite{XFEL}.}
\label{XFEL}
\end{figure}

The Free-Electron Laser in Hamburg (FLASH) facility at the Deutsches Elektronen-SYnchrotron (DESY) in Hamburg (Germany) \cite{FLASH} is one of the most brilliant operating FEL's. It delivers short pulses (duration of about $10\text{-} 100\;\text{fs}$) of coherent radiation in the extreme ultraviolet-soft x-ray regime (fundamental wavelength from $60\;\text{nm}$ down to $6.5\;\text{nm}$ corresponding to photon energies from $21$ to $190\;\text{eV}$) at a repetition rate of $100\;\text{kHz}$. The intense electron beams available at FLASH (total charge of $0.5\text{-} 1\;\text{nC}$ at an energy of $1\;\text{GeV}$) allow for peak brilliance of the photon beam of about $10^{29}\text{-} 10^{30}\;\text{photons/(s\,mrad$^2$\,mm$^2$\,0.1\% bandwidth)}$ (see Fig. \ref{XFEL}), exceeding the peak brilliance of conventional synchrotron light sources by several orders of magnitude.
The Linac Coherent Light Source (LCLS) at Stanford (California, USA) uses the electron beams generated by the Stanford Linear ACcelerator (SLAC) at the National Accelerator Laboratory to generate flashes of coherent x-ray radiation of unprecedented brilliance \cite{LCLS} (see also \onlinecite{Emma_2010}). Since the electron beam energy can be varied from $4.5\;\text{GeV}$ to $14.4\;\text{GeV}$, accordingly the wavelength of LCLS can be tuned from $1.5\;\text{nm}$ to $0.15\;\text{nm}$ (corresponding to photon energies from $0.8\;\text{keV}$ to $8\;\text{keV}$). The peak brilliance of the LCLS is of the order of $10^{32}\text{-} 10^{33}$\;\text{photons/(s\,mrad$^2$\,mm$^2$\,0.1\% bandwidth)} (see Fig. \ref{XFEL}), the pulse duration is typically of about $40\text{-}80\;\text{fs}$ up to $500\;\text{fs}$, which can be decreased to $10\;\text{fs}$ in the low-charge electron beam mode, and the repetition rate is $120\;\text{Hz}$.

Another XFEL in operation is the SPring-8 Angstrom Compact free electron LAser (SACLA) at the RIKEN Harima Institute in Japan \cite{SACLA}. The electron accelerator, based on a conducting C-band high-gradient radiofrequency acceleration system, and the short-period undulator allow for a relatively compact facility of around 700 m in length (compared, for example, with the about 2 km of the LCLS). SACLA employs the 8 GeV electron beam of the Super Photon Ring - 8 GeV (SPring-8) accelerator \cite{SPring8} and has generated x-ray beams with $0.08\;\text{nm}$ wavelength (corresponding to a photon energy of $15.5\;\text{keV}$) at a repetition rate of $60\;\text{Hz}$ (see also the caption of Fig. \ref{XFEL}).

The European XFEL is under development at DESY in Hamburg (Germany) \cite{XFEL}. It is expected to deliver x-ray pulses with a peak brilliance up to about $5\times 10^{33}\;\text{photons/(s\,mrad$^2$\,mm$^2$\;0.1\% bandwidth)}$ at the unprecedented repetition rate of $27\;\text{kHz}$. The electron accelerator provides an electron beam with maximal energy of $17.5\;\text{GeV}$ able to generate laser pulses with a central wavelength of $0.05\;\text{nm}$, which corresponds to a photon energy of $24.8\;\text{keV}$ and with a pulse duration of $100\;\text{fs}$. Moreover, the European XFEL will be a versatile machine consisting of three photon beamlines: the SASE-1 and SASE-2, with linearly polarized photons with energy in the range $3.1\text{-} 24.8\;\text{keV}$ and the SASE-3, with linearly or circularly polarized photons of energy in the range $0.26\text{-} 3.1\;\text{keV}$.

We also mention that coherent attosecond pulses of XUV radiation (photon energy of the order of 100 eV) have been generated employing HHG in a gaseous medium \cite{Agostini_2004}. This technique allows for the production of beams with central photon energy up to several keVs \cite{Sansone_2006,Popmintchev_2009}, though with intensities several orders smaller than XFELs. Less stable sources of coherent soft x-rays are the so-called x-ray lasers, which are based on the amplification of spontaneous emission by multiply ionized atoms in dense plasmas created by intense laser pulses \cite{Zeitoun_2004,Wang_2008,Suckewer_2009}.

\section{Free electron dynamics in a laser field}\label{FED}

In this Section we review, for the benefit of the reader, some important basic results on the dynamics of a free electron in a laser field (see also the review \onlinecite{Eberly_1969}) and link them to recent investigations on the subject. Results in the realm of classical and quantum electrodynamics are considered separately. Radiation-reaction and electron self-interaction effects are not included here and their discussion is developed in Sec. \ref{RR}.

\subsection{Classical dynamics}
 \label{FED_C}
The motion of a charged particle in a laser field is usually associated to an oscillation along the laser polarization direction. This is pertinent to the non-relativistic regime, while the charge dynamics in the relativistic domain is enriched by new features like the drift along the laser propagation direction and other non-dipole effects (like the well-known figure-8 trajectory), as well as by the sharpening of the trajectory at those instants where the velocity along the polarization direction reverses. As a consequence, laser-driven relativistic free electrons also emit high harmonics of the laser frequency (see Sec. \ref{TCS}).

The classical motion of an electron (electric charge $e<0$ and mass $m$) in an arbitrary external electromagnetic field $F^{\mu\nu}(x)$ is determined by the Lorentz equation $mdu^{\mu}/ds=eF^{\mu\nu}u_{\nu}$, where $u^{\mu}=dx^{\mu}/ds$ is the electron four-velocity and $s$ its proper time \cite{Landau_b_2_1975}. If the external field is a plane wave, the field tensor $F^{\mu\nu}(x)$ depends only on the dimensional phase $\phi=(n_0x)$, where $n_0^{\mu}=(1,\bm{n}_0)$, with $\bm{n}_0$ being the unit vector along the propagation direction of the wave. In this case, for an arbitrary four-vector $v^{\mu}=(v^0,\bm{v})$ it is convenient to introduce the notation $v_{\parallel}=\bm{n}_0\cdot\bm{v}$, $\bm{v}_{\perp}=\bm{v}-v_{\parallel}\bm{n}_0$ and $v_-=(n_0v)=v^0-v_{\parallel}$. The four-vector potential of the wave can be chosen in the Lorentz gauge as $A^{\mu}(\phi)=(0,\bm{A}(\phi))$, with $A_-(\phi)=-A_{\parallel}(\phi)=0$. We indicate as $p^{\mu}=(\varepsilon,\bm{p})=mu^{\mu}$ the (kinetic) four-momentum of the electron. Since a plane-wave field depends only on $\phi$, the canonical momenta $\bm{p}_{\perp}(\phi)+e\bm{A}(\phi)$ and $p_-(\phi)$ are conserved as they are the conjugated momenta to the cyclic coordinates $\bm{x}_{\perp}$ and $t+x_{\parallel}$, respectively. For $p^{\mu}(\phi_0)=p_0^{\mu}=(\varepsilon_0,\bm{p}_0)=m\gamma_0(1,\bm{\beta}_0)$ being the initial condition for the electron's four-momentum at a given phase $\phi_0$, the above-mentioned conservation laws already allow to write the electron's four-momentum at an arbitrary phase $\phi$ as \cite{Landau_b_2_1975}
\begin{align}
\label{Free_Sol_0}
\begin{split}
\varepsilon(\phi)=&\varepsilon_0-e\frac{\bm{p}_{0,\perp}\cdot[\bm{A}(\phi)-\bm{A}(\phi_0)]}{p_{0,-}}\\
&+\frac{e^2}{2}\frac{[\bm{A}(\phi)-\bm{A}(\phi_0)]^2}{p_{0,-}},
\end{split}\\
\label{Free_Sol_1_2}
\bm{p}_{\perp}(\phi)=&\bm{p}_{0,\perp}-e[\bm{A}(\phi)-\bm{A}(\phi_0)],\\
\label{Free_Sol_3}
\begin{split}
p_{\parallel}(\phi)=&p_{0,\parallel}-e\frac{\bm{p}_{0,\perp}\cdot[\bm{A}(\phi)-\bm{A}(\phi_0)]}{p_{0,-}}\\
&+\frac{e^2}{2}\frac{[\bm{A}(\phi)-\bm{A}(\phi_0)]^2}{p_{0,-}},
\end{split}
\end{align}
where the on-shell condition $\varepsilon(\phi)+p_{\parallel}(\phi)=[\bm{p}^2_{\perp}(\phi)+m^2]/p_{0,-}$ was employed. For the paradigmatic case of a linearly polarized monochromatic plane wave, it is $A^{\mu}(\phi)=A_0^{\mu}\cos(\omega_0\phi)$, with $A_0^{\mu}=(0,E_0\bm{u}/\omega_0)$, where  $E_0$ is the electric field amplitude, $\omega_0$ the angular frequency and $\bm{u}$ the polarization direction (perpendicular to $\bm{n}_0$). 

The above analytical solution indicates that even if an electron is initially at rest, it becomes relativistic within one laser period $T_0=2\pi/\omega_0$ if the parameter
\begin{equation}
\label{xi_0}
\xi_0=\frac{|e|\sqrt{-A_0^2}}{m}=\frac{|e|E_0}{m\omega_0}
\end{equation}
is of the order of or larger than unity. In the relativistic regime the magnetic component of the Lorentz force, which depends on the electron's velocity, becomes comparable to the electric one and the electron's dynamics becomes highly nonlinear in the laser-field amplitude. Thus, the parameter $\xi_0$ is known as classical nonlinearity parameter. An heuristic interpretation of the parameter $\xi_0$ is as the work performed by the laser field on the electron in one laser wavelength $\lambda_0=T_0$ in units of the electron mass, which clearly explains why relativistic effects become important at $\xi_0\gtrsim 1$. Alternatively, Eqs. \eqref{Free_Sol_0}-\eqref{Free_Sol_3} indicate that the figure-8 trajectory has a longitudinal (transverse) extension of the order of $\lambda_0\xi_0^2$ ($\lambda_0\xi_0$), implying that the electron trajectory deviates from the unidirectional oscillating one and becomes nonlinear in the field amplitude at $\xi_0\gtrsim 1$. Note that numerically $\xi_0=6.0\sqrt{I_0[10^{20}\;\text{W/cm$^2$}]}\lambda_0[\mu\text{m}]=7.5\sqrt{I_0[10^{20}\;\text{W/cm$^2$}]}/\omega_0[\text{eV}]$, where $I_0=E_0^2/4\pi$ is the wave's peak intensity, and that $\xi_0$ is gauge- and Lorentz-invariant: the gauge invariance has to be intended with respect to gauge transformations which do not alter the dependence of the four-vector potential on $\phi$ (see \onlinecite{Heinzl_2009} for a thorough analysis of this issue). The solution in Eqs. \eqref{Free_Sol_0}-\eqref{Free_Sol_3} also indicates that in the ultrarelativistic regime at $\xi_0\gg 1$, the electron acquires a drift momentum along the propagation direction of the laser field which is proportional to $\xi_0^2$, in contrast to the transverse momentum which is proportional to $\xi_0$. In the case of an electron initially at rest, for example, the momentum $\bm{p}(\infty)$ of the electron after the laser pulse has been switched off ($\bm{A}(\infty)=\bm{0}$) has the components $\bm{p}_{\perp}(\infty)=e\bm{A}(\phi_0)$ and $p_{\parallel}(\infty)=e^2\bm{A}^2(\phi_0)/2m$.

Realistic laser pulses, as those produced in laboratories, have a more complicated structure than a plane wave, essentially because they are spatially focused on the transverse planes and the area of the focusing spot changes along the laser's propagation axis. Generally speaking, if the radius of the minimal focusing area (spot radius) is much larger than the central wavelength of the laser pulse, then the pulse can be reasonably approximated by a plane wave. A Gaussian beam in the paraxial approximation offers a more accurate analytical description of a realistic laser pulse, which shows a Gaussian profile in the transverse planes \cite{Salamin_2002a}. The dynamics of an electron in such a field cannot be derived analytically and a numerical solution of the Lorentz equation is required \cite{Salamin_2002b}. 

\subsection{Quantum dynamics}
\label{FED_Q}
In the realm of relativistic quantum mechanics, i.e., when $e^+\text{-}e^-$ pair production is negligible (see also Sec. \ref{PP}) and the single-particle quantum theory is applicable, the dynamics of an electron in an external electromagnetic field with four-vector potential $A^{\mu}(x)$ is described by the Dirac equation 
\begin{equation}
\label{Dir_Eq}
\{\gamma^{\mu}[i\partial_{\mu}-eA_{\mu}(x)]-m\}\Psi=0,
\end{equation}
where $\gamma^{\mu}$ are the Dirac matrices and where $\Psi(x)$ is the four-component electron bi-spinor \cite{Landau_b_4_1982}. Analogously to the classical case, if the external field is a plane wave, the Dirac equation can be solved exactly. If $p_0^{\mu}=(\varepsilon_0,\bm{p}_0)$ and $\sigma_0/2=\pm 1/2$ are the electron's four-momentum and spin at $\phi\to -\infty$ and if $A^{\mu}(-\infty)=0$, the positive-energy ($\varepsilon_0>0$) solution $\Psi_{p_0,\sigma_0}(x)$ of Eq. \eqref{Dir_Eq} reads \cite{Volkov_1935,Landau_b_4_1982}
\begin{equation}
\label{V_S}
\Psi_{p_0,\sigma_0}(x)=\left[1+\frac{e}{2p_{0,-}}\hat{n}_0\hat{A}(\phi)\right]\frac{u_{p_0,\sigma_0}}{\sqrt{2V\varepsilon_0}}e^{iS_{p_0}},
\end{equation}
where in general $\hat{v}=\gamma^{\mu}v_{\mu}$ for a generic four-vector $v^{\mu}$, where $u_{p_0,\sigma_0}$ is a positive-energy free bi-spinor \cite{Landau_b_4_1982}, $V$ is the quantization volume, and where
\begin{equation}
S_{p_0}(x)=-(p_0x)-\int_{-\infty}^{\phi}d\phi'\left[\frac{e(p_0A(\phi'))}{p_{0,-}}-\frac{e^2A^2(\phi')}{2p_{0,-}}\right]
\end{equation}
is the classical action of an electron in the plane wave \cite{Landau_b_2_1975}. The above electron states are known as positive-energy Volkov states. The negative-energy states $\Psi_{-p_0,-\sigma_0}(x)$ can be formally obtained by the replacements $p_0^{\mu}\rightarrow-p_0^{\mu}$ and $\sigma_0\rightarrow -\sigma_0$ in Eq. \eqref{V_S} except for the energy in the square root (the resulting bi-spinor $u_{-p_0,-\sigma_0}$ is the corresponding negative-energy free bi-spinor \footnote{We point out that the discussed Volkov states $\Psi_{\pm p_0,\pm\sigma_0}(x)$ are the so-called Volkov in-states, as they transform into free-states in the limit $t\to-\infty$ \cite{Fradkin_b_1991}. Volkov out-states, which transform into free-states in the limit $t\to\infty$, can be derived analogously and differ from the Volkov in-states only by an inconsequential constant phase factor (recall that $\bm{A}(\infty)=\bm{0}$).} \cite{Landau_b_4_1982}). Although it has been shown long ago that positive- and negative-energy Volkov states form a complete set of orthogonal states on the hypersurfaces $\phi=\text{const}$ \cite{Ritus_1985}, the corresponding property on the hypersurfaces $t=\text{const}$ is not straightforward and it has been proved only recently (see \onlinecite{Ritus_1985,Zakowicz_2005}, and \onlinecite{Boca_2010} for a proof of the orthogonality and of the completeness of the Volkov states, respectively). 

\begin{figure}
\begin{center}
\includegraphics[width=0.8\linewidth]{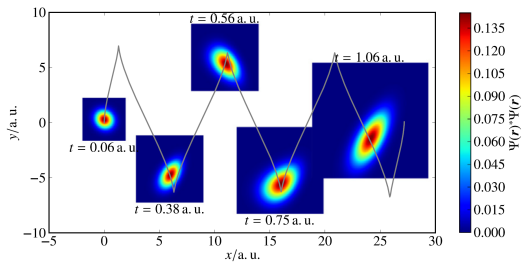}
\end{center}
\caption{(Color) Free wave packet evolution in a plane wave field. The solid gray line indicates the center of mass trajectory, coinciding essentially with the classical trajectory, and the laser pulse travels from left to right. The blue regions indicate the copropagating self-adaptive numerical grid. Time and space coordinates are given in ``atomic units'', with $1\;\text{a.u.}=24\;\text{as}$ and $1\;\text{a.u.}=0.05\;\text{nm}$, respectively. From \onlinecite{Bauke_2011b}.}
\label{El_dyn}
\end{figure}

Since the Volkov states form a basis of the space of the solutions of Dirac equation in a plane wave, they can be employed to build electron wave packets and study their evolution. A pedagogical example of laser-induced Dirac dynamics is displayed in Fig. \ref{El_dyn} for a plane wave with peak intensity of $6.3\times 10^{23}\;\text{W/cm$^2$}$ and central wavelength of $2\;\text{nm}$.
The figure shows the drift of the wave packet in the propagation direction of the wave, its spreading and its shearing due to non-dipole effects. In \onlinecite{Fillion-Gourdeau_2012} an alternative method of solving the time-dependent Dirac equation in coordinate space is presented, which explicitly avoids the fermion doubling, i.e., the appearance of unphysical modes when the Dirac equation is discretized.

As in the classical case, we shortly mention here the paradigmatic case of a monochromatic, linearly polarized plane-wave field $A^{\mu}(\phi)=A_0^{\mu}\cos(\omega_0\phi)$. In this case the action $S_{p_0}(x)$ can be written in the form $S_{p_0}(x)=-(q_0x)+\text{``oscillating terms''}$, with \cite{Ritus_1985}
\begin{equation}
\label{q}
q_0^{\mu}=p_0^{\mu}+\frac{m^2\xi_0^2}{4p_{0,-}}n_0^{\mu}.
\end{equation}
The four-vector $q_0^{\mu}$ plays the role of an ``effective'' four-momentum of the electron in the laser field and it is indicated as electron ``quasimomentum''. The corresponding electron ``mass'' $\sqrt{q_0^2}=m^*=m\sqrt{1+\xi_0^2/2}$ is known as electron's dressed mass. The results for the quasimomentum $q_0^{\mu}$ and the dressed mass $m^*$ in the case of a circularly polarized laser field with the same amplitude and frequency is obtained from the above ones with the replacement $\xi_0^2\rightarrow 2\xi_0^2$. The quasimomentum coincides classically with the average momentum of the electron in the plane wave. Correspondingly, the mass dressing depends only on the classical nonlinearity parameter $\xi_0$ and it is an effect of the quivering motion of the electron in the monochromatic wave (see also the recent review \onlinecite{Ehlotzky_2009}). As we will see in Sec. \ref{TCS_General}, it is important that conservation laws in QED processes in the presence of a monochromatic plane-wave field involve the quasimomentum $q_0^{\mu}$ for the incoming electrons rather than the four-momentum $p_0^{\mu}$. The question of the electron dressed mass in pulsed laser fields has been investigated in \onlinecite{Heinzl_2010b,Mackenroth_2011}. 

In the realm of QED the parameter $\xi_0$ can also be heuristically interpreted as the work performed by the laser field on the electron in the typical QED length $\lambda_C=1/m\approx 3.9\times 10^{-11}\;\text{cm}$ (Compton wavelength) in units of the laser photon energy $\omega_0$ (see Eq. \eqref{xi_0}). This qualitatively explains why multiphoton effects in a laser field become important at $\xi_0\gtrsim 1$, such that the laser field has to be taken into account exactly in the calculations \cite{Ritus_1985}. In the framework of QED this is achieved by working in the so-called Furry picture \cite{Furry_1951}, where the $e^+\text{-}e^-$ field $\Psi(x)$ is quantized in the presence of the plane-wave field. This amounts essentially in employing the Volkov (dressed) states and the corresponding Volkov (dressed) propagators \cite{Ritus_1985} instead of free particle states and free propagators to compute the amplitudes of QED processes. In the Furry picture the effects of the plane wave are accounted for exactly and only the interaction between the $e^+\text{-}e^-$ field $\Psi(x)$ and the radiation field $\mathcal{F}^{\mu\nu}(x)\equiv\partial^{\mu}\mathcal{A}^{\nu}(x)-\partial^{\nu}\mathcal{A}^{\mu}(x)$ is accounted for by means of perturbation theory. The complete evolution of the system ``$e^+\text{-}e^-$ field+radiation field'' is obtained by means of the $S$-matrix
\begin{equation}
\label{S}
S=\mathcal{T}\left[\text{exp}\left(-ie\int d^4x\bar{\Psi}\gamma^{\mu}\Psi\mathcal{A}_{\mu}\right)\right],
\end{equation}
where $\mathcal{T}$ is the time-ordering operator and $\bar{\Psi}(x)=\Psi^{\dag}(x)\gamma^0$. For an initial state containing only a single electron with four-momentum $p_0^{\mu}$, the quantitative description of the interaction between the electron, the laser field and the radiation field involves in particular the gauge- and Lorentz-invariant quantum parameter
\begin{equation}
\label{chi_0}
\chi_0=\frac{|e|\sqrt{-(F_{0,\mu\nu}p_0^{\nu})^2}}{m^3}=\frac{p_{0,-}}{m}\frac{E_0}{F_{\text{cr}}},
\end{equation}
where $F_{\text{cr}}=m^2/|e|=1.3\times 10^{16}\;\text{V/cm}=4.4\times 10^{14}\;\text{G}$ being the critical electromagnetic field of QED \cite{Ritus_1985}. The definition of $F_{\text{cr}}$ indicates that a constant and uniform electric field with strength of the order of $F_{\text{cr}}$, provides an $e^+\text{-}e^-$ pair with an energy of the order of its rest energy $2m$ in a distance of the order of the Compton wavelength $\lambda_C$, implying the instability of the vacuum under $e^+\text{-}e^-$ pair creation in the presence of such a strong field \cite{Sauter_1931,Heisenberg_1936,Schwinger_1951}. In Eq. \eqref{chi_0} we considered the case of a linearly polarized plane wave of the form $A^{\mu}(\phi)=A_0^{\mu}\psi(\phi)$, with $\psi(\phi)$ being an arbitrary function with $\max|d\psi(\phi)/d\phi|=1$ and we introduced the tensor amplitude $F_0^{\mu\nu}=k_0^{\mu}A_0^{\nu}-k_0^{\nu}A_0^{\mu}$, with $k_0^{\mu}=\omega_0n_0^{\mu}$ . For an ultrarelativistic electron initially counterpropagating with respect to the plane wave it is $\chi_0=5.9\times 10^{-2}\varepsilon_0[\text{GeV}]\sqrt{I_0[10^{20}\;\text{W/cm$^2$}]}$. The parameter $\chi_0$ can be interpreted as the amplitude of the electric field of the plane wave in the initial rest-frame of the electron in units of the critical field of QED and it controls the magnitude of pure quantum effects like the photon recoil in multiphoton Compton scattering and spin effects. This is why it is known as  ``nonlinear quantum parameter''.

Since the probability $dP_e/dVdt$ per unit volume and unit time of a quantum process is a gauge- and Lorentz invariant quantity, for those processes in a plane-wave field involving an incoming electron, as, e.g., multiphoton Compton scattering, it can depend only on the two parameters $\xi_0$ and $\chi_0$ \cite{Ritus_1985}. For an electromagnetic field $F^{\mu\nu}(x)=(\bm{E}(x),\bm{B}(x))$ either constant or slowly-varying, the quantity $dP_e/dVdt$, calculated in the latter case in the leading order with respect to the fields' derivatives, can in principle also depend on the two field invariants
\begin{align}
\mathscr{F}(x)&=\frac{1}{4}F^{\mu\nu}(x)F_{\mu\nu}(x)=-\frac{1}{2}[E^2(x)-B^2(x)],\\
\mathscr{G}(x)&=\frac{1}{4}F^{\mu\nu}(x)\tilde{F}_{\mu\nu}(x)=-\bm{E}(x)\cdot\bm{B}(x)
\end{align}
which identically vanish for a plane wave. In the second equation $\tilde{F}_{\mu\nu}(x)=\epsilon_{\mu\nu\alpha\beta}F^{\alpha\beta}(x)/2$ is the dual field of $F^{\mu\nu}(x)$ and $\epsilon^{\mu\nu\alpha\beta}$ is the four-dimensional completely anti-symmetric tensor with $\epsilon^{0123}=+1$ (since $\mathscr{G}(x)$ is actually a pseudo-scalar function, the probability $dP_e/dVdt$ can only depend on $\mathscr{G}^2(x)$). Note, however, that if $|\mathscr{F}(x)|,|\mathscr{G}(x)|\ll \min(1,\chi^2(x))F_{\text{cr}}^2$, with $\chi(x)=|e|\sqrt{|(F_{\mu\nu}(x)p_0^{\nu})^2|}/m^3$, then the dependence of $dP_e/dVdt$ on $\mathscr{F}(x)$ and $\mathscr{G}(x)$ can be neglected. In this case the probability $dP_e/dVdt$ essentially coincides with the analogous quantity calculated for a constant crossed field $F_0^{\mu\nu}$, with the replacement $F_0^{\mu\nu}\rightarrow F^{\mu\nu}(x)$ \cite{Ritus_1985}. For a monochromatic plane wave with angular frequency $\omega_0$ this occurs if $\xi_0\gg 1$. As will be seen in Sec. \ref{TCS_General}, this condition corresponds, e.g., to the formation time of multiphoton Compton scattering ($\sim m/|e|E_0$) being much shorter than the laser period $T_0$.

As has been mentioned, the $S$-matrix in Eq. \eqref{S} describes all possible electrodynamical processes among electrons, positrons and photons. The above considerations can be easily adapted for discussing processes involving an initial positron. Whereas, the probability $dP_{\gamma}/dVdt$ of a quantum process in a plane-wave field involving an incoming photon, as, e.g., multiphoton $e^+\text{-}e^-$ pair production, depends on the parameters $\xi_0$ and
\begin{equation}
\label{kappa_0}
\varkappa_0=\frac{|e|\sqrt{-(F_{0,\mu\nu}k^{\nu})^2}}{m^3}=\frac{k_-}{m}\frac{E_0}{F_{\text{cr}}},
\end{equation}
where $k^{\mu}=(\omega,\bm{k})$ is the four-momentum of the incoming photon (see \onlinecite{Ritus_1985} and Secs. \ref{VPEs}-\ref{Cascade}). For a photon counterpropagating with respect to the plane wave it is $\varkappa_0=5.9\times 10^{-2}\omega[\text{GeV}]\sqrt{I_0[10^{20}\;\text{W/cm$^2$}]}$. In the case of multiphoton $e^+\text{-}e^-$ pair production, the parameter $\varkappa_0$ can be interpreted as the amplitude of the electric field of the plane wave in units of the critical field $F_{\text{cr}}$ in the center-of-mass system of the created electron and positron \cite{Ritus_1985}. The above remarks on processes occurring in a constant or slowly-varying background field $F^{\mu\nu}(x)$ and involving an incoming electron, also apply to the case of an incoming photon once one replaces $\chi(x)$ with $\varkappa(x)=|e|\sqrt{|(F_{\mu\nu}(x)k^{\nu})^2|}/m^3$.

\section{Relativistic atomic dynamics in strong laser fields}
\label{RD}

When super-intense infrared laser pulses, as those described in Sec. \ref{L_O}, impinge on an atom, the latter is immediately partly or fully ionized \cite{Protopapas_1997,Becker_2002,Keitel_2001}. The ejected electrons experience the typical ``zig-zag'' motion of a free electron in both laser polarization and propagation directions (see Eqs. \eqref{Free_Sol_1_2}-\eqref{Free_Sol_3} and Fig. \ref{El_dyn}) and will not, in general, return to the ionic core. With an enhanced binding force on the remaining electrons, the ionization dynamics becomes increasingly complex and may experience subtle relativistic and correlation effects. When the binding force of the ionic core and that of the applied laser field eventually become comparable, the electrons may in special cases return to and interact with the parent ion (rescattering \cite{Kuchiev_1987,Corkum_1993,Schafer_1993}). This interaction leads, for example, to the ejection of other electrons, to the absorption of energy in a scattering process or to the emission of high-harmonic photons in case of recombination.

\subsection{Ionization}
\label{RD_I}
Previously, atomic or molecular ionization was studied with laser pulses of intensity below $10^{16}$ W/cm$^2$, and the relativistic laser-matter interaction was dominated by the plasma community. The pioneering experiment reported in \onlinecite{Moore_1999} on the ionization behavior of atoms and ions in interaction with a laser with intensity of $3\times 10^{18}$ W/cm$^2$ has thus attracted considerable interest. The laser magnetic field component was shown to alter the direction of ionization characteristically (see Sec. \ref{FED_C}). This is because in the relativistic regime the ionized electron in a laser field acquires a large momentum along the laser propagation direction (see Eq. \eqref{Free_Sol_3}) and photoelectrons are emitted mostly in that direction within a characteristic opening angle $\theta$: $\tan\theta\sim p_{\perp}(\infty)/p_{\parallel}(\infty)\sim 2/\xi_0$ (see the discussion below Eq. \eqref{xi_0}). With highly-charged ions becoming more easily available in a wide range of charges, e.g., via super-strong laser fields or by passing the ion beams through metallic foils, relativistic laser-induced ionization has been further studied \cite{Chowdhury_2001,Dammasch_2001,Yamakawa_2003,Yamakawa_2004,Gubbini_2005,DiChiara_2008,Palaniyappan_2008,DiChiara_2010}.

In this situation rescattering is generally suppressed and multiple ionization of atoms and ions takes place mostly via direct ionization, especially including tunneling. On the theoretical side attention was then focused on the relativistic generalization \cite{Popov_1997,Milosevic_2002,Popov_2004,Popov_2006} of the so-called Perelomov-Popov-Terent'ev (PPT) theory or Ammosov-Delone-Krainov (ADK) model \cite{Perelomov_1967,Ammosov_1986}, which describes atomic ionization in the quasistatic tunneling regime. While the common intuitive interpretation of the laser induced tunneling fails in the relativistic regime \cite{Reiss_2008}, a revised picture has been proposed in \onlinecite{Klaiber_2012}. The Strong Field Approximation (SFA) \cite{Keldysh_1965,Faisal_1973,Reiss_1980}, which treats in a universal way both the multiphoton and the tunneling regimes of strong-field ionization, has also been extended to the relativistic regime \cite{Reiss_1990,Reiss_1990_JOSAB}. Both the PPT theory and the SFA assume that the direct ionization process occurs as a single-electron phenomenon and thus neglects atomic structure effects.
\begin{figure}
\begin{center}
\includegraphics[width=0.8\linewidth]{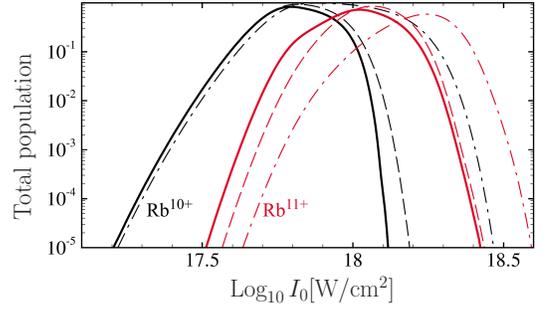}
\end{center}
\caption{(Color) The total Rb$^{10+}$ (black lines) and Rb$^{11+}$ (red lines) ion populations in a gaseous target as a function of the peak intensity of a linearly-polarized laser field with a wavelength of $0.8\;\text{$\mu$m}$ and with a pulse duration of $5\;\text{fs}$. The solid lines display two-electron inelastic tunneling, the  dashed lines one-electron inelastic tunneling and the dashed-dotted lines the results via the PPT theory. Adapted from \onlinecite{Zon_2009}.}
\label{Zon}
\end{figure}

When the tunneling process proceeds very fast, multi-electron correlation effects can occur due to the so-called shake-up processes. Thus, the detachment of one electron from the atom or ion via tunneling modifies the self-consistent potential sensed by the remaining electrons and may result, consequently, in the excitation of the atomic core (inelastic tunneling). A strong excitation may also trigger the simultaneous escape of several electrons from the bound state through the potential barrier (collective tunneling). These effects were known to occur also in the nonrelativistic regime \cite{Zon_1999,Zon_2000} but were concealed by competing rescattering effects. In \onlinecite{Zon_2009}, it was shown that, in the relativistic regime, the role of inelastic and collective tunneling can significantly increase and the relativistic PPT rate has been generalized in this respect. In a linearly polarized field, the rate $R_{\text{coll}}^{(N)}$ of inelastic collective tunneling of $N$ equivalent electrons from the outer shell of an ion is described by the following universal formula \cite{Zon_2009}:
\begin{equation}
\begin{split}
R_{\text{coll}}^{(N)}=&\frac{m\sqrt{6}}{\alpha^{3(N-1)}} C^{2N}_{\kappa l}\frac{M!(l+1/2)^N}{2^MN^{M+1}\sqrt{\pi}}\prod _{j=1}^N\frac{(l+m_j)!}{(m_j!)^2(l-m_j)!}\\
&\times \mathcal{I}^2\kappa^{3N-1}\left(\frac{2E_a}{E_0}\right)^{2(\nu-1)N-M+1/2}\text{e}^{-2NE_a/3E_0},
\end{split}
\end{equation}
where $\alpha=e^2\approx 1/137$ is the fine-structure constant, $m_1,\ldots,m_N$ are the magnetic quantum numbers of the bound electrons, $M=\sum_{j=1}^N m_j$, $l$ is the orbital quantum number of the electrons, $\kappa=\sqrt{2I_p^{(N)}/mN}$, $I_p^{(N)}=\sum_{j=1}^N(I_{p,j}^{(0)}-\Delta_j)$, $I_{p,j}^{(0)}$ is the $j$th ionization potential of a parent ion, $\Delta_j$ is the energy of the core excitation, $E_a=\kappa^3F_{\text{cr}}$ is the atomic field, $Z|e|$ is the charge of the residual ion, $\mathcal{I}$ is the adimensional overlap integral (see \onlinecite{Zon_2009} for its precise definition) and $C_{\kappa l}\approx(2/\nu)^{\nu}/\sqrt{2\pi\nu}$, with $\nu=Z\alpha/\kappa$. According to the calculations in \onlinecite{Zon_2009}, inelastic and collective tunneling effects contribute significantly to the relativistic ionization dynamics at intensities larger than $10^{18}$ W/cm$^2$, thus changing the ionization probability by more than one order of magnitude (see Fig. \ref{Zon}).

Spin effects of bound systems in strong laser fields were shown to moderately alter the quantum dynamics and its associated radiation via spin-orbit coupling in highly-charged ions already at an intensity of $\sim 10^{17}\;\text{W/cm$^2$}$ \cite{Hu_1999,Walser_2002}. More recently a nonperturbative relativistic SFA theory has been developed, describing circular dichroism and spin effects in the ionization of helium in an intense circularly polarized laser field \cite{Faisal_2011}. Here, two-photon ionization has been studied in the nonrelativistic intensity range $10^{13}\text{-} 10^{15}$ W/cm$^2$ with a photon energy of 45 eV, yielding small relative spin-induced corrections of the order of $10^{-3}$.

\begin{figure}
\begin{center}
\includegraphics[width=0.8\linewidth]{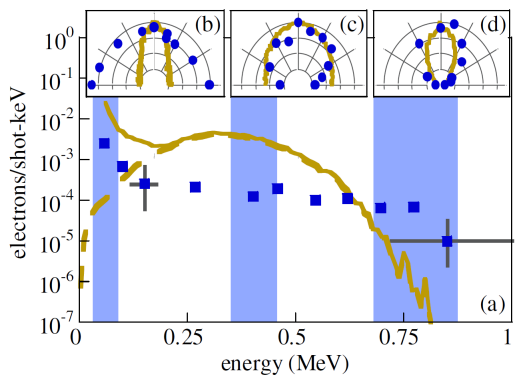}
\end{center}
\caption{(Color) (a) Experimental photoelectron spectra for argon at $I_0=1.2\times 10^{19}$ W/cm$^2$ and at an angle of $62^{\circ}$ from the laser propagation direction. Analytical results are shown for all photoelectrons (continuous line) and for the L-shell (dashed line). The angular distributions are at an electron energy of (b) 60 keV, (c) 400 keV, and (d) 770 keV. From \onlinecite{DiChiara_2008}.}
\label{walker}
\end{figure}

A series of experiments has been devoted to the measurement of atomic multi-electron effects in relativistically strong laser fields. In \onlinecite{DiChiara_2008}, the energy distribution of the ejected electrons and the angle-resolved photoelectron spectra for atomic photoionization of argon at $I_0\sim 10^{19}$ W/cm$^2$ have been investigated experimentally. Here, isolation of the single-atom response in the multicharged environment has been achieved by measuring photoelectron yields, energies, and angular distributions as functions of the sample density. Ionization of the entire valence shell along with several inner-shell electrons was shown at $I_0\sim 10^{17}\text{-}10^{19}$ W/cm$^2$. A typical spectrum in the case of linear polarization is displayed in Fig. \ref{walker}. 
An extended plateau-like structure appears in the spectrum due to the electrons originating from the L-shell and the longitudinal component of the focused laser field. A surprising feature is observed in the energy-resolved angular distribution. In contrast to the nonrelativistic case with increasing rescatterings and, thus, angular-distribution widths at high energies, here azimuthally isotropic angular distributions are observed at low energies ($\sim 60$ keV in Fig. \ref{walker}), which become narrower for high-energy photoelectrons. The authors attribute the anomalous broad angular distribution for low-energy electrons to electron-correlation effects. A similar experiment on the energy- and angle-resolved photoionization was later reported for xenon at a laser intensity of $10^{19}$ W/cm$^2$ \cite{DiChiara_2010}. For energies below 0.5 MeV, the yield and the angular distribution were shown not to be described by a  one-electron strong-field model, but rather involve most likely multielectron and high-energy atomic excitation processes. A further experiment on relativistic ionization of the methane molecule at $I_0\sim 10^{18}\text{-} 10^{19}$ W/cm$^2$ \cite{Palaniyappan_2008} indicated that molecular mechanisms of ionization play no role, and that C$^{5+}$ ions are produced at these intensities mostly via the cross-shell rescattering atomic ionization mechanism. All these experimental results still await an accurate theoretical description. 

On the computational side, various numerical methods have been developed to describe the laser-driven relativistic quantum dynamics in highly-charged ions. A Fast-Fourier-Transform split-operator code was implemented in \onlinecite{Mocken_2008} for solving the Dirac equation in 2+1 dimensions by employing adaptive grid and parallel computing algorithms. Another method has been developed in \onlinecite{Selsto_2009} to solve the 3D Dirac equation by expanding the angular part of the wave-function in spherical harmonics. The latter was applied to hydrogenlike ions in intense high-frequency laser pulses with emphasis on investigating the role of negative-energy states.
In \onlinecite{Bauke_2011}, the classical relativistic phase-space averaging method, generalized to arbitrary central potentials, and the enhanced time-dependent Dirac and Klein-Gordon numerical treatments are employed to investigate the relativistic ionization of highly-charged hydrogenlike ions in short intense laser pulses. For ionization dynamics beyond the tunneling regime, quantum mechanical and classical methods give similar results, for laser wavelengths from the near-infrared region to the soft x-ray regime. Furthermore a useful procedure has been developed, which employs the over-the-barrier ionization yields for highly-charged ions, to determine the peak laser field strength of short ultrastrong pulses in the range $I_0\sim 10^{18}\text{-} 10^{26}$ W/cm$^2$ \cite{Hetzheim_2009}.  In addition, in this article the ionization angle of the ejected electrons is investigated by the full quantum mechanical solution of the Dirac equation and the laser field strength is shown to be also linked to the electron emission angle.
The magnetic field-induced tilt in the lobes of the angular distributions of photoelectrons in laser-induced relativistic ionization has also been discussed in \onlinecite{Klaiber_2007}.

\begin{figure}
\begin{center}
\includegraphics[width=0.8\linewidth]{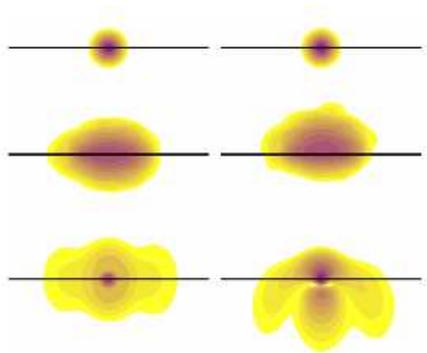}
\end{center}
\caption{(Color online)  Dipole (left) and nondipole (right) probability densities of the Kramers-Henneberger wave-function in the $x\text{-}z$ plane for a $x$-polarized, 10-cycle sin-like pulse propagating in the positive $z$ direction (upward), with $E_0=1.5\times 10^{11}\;\text{V/cm}$ and $\omega_0=54\;\text{eV}$. The snapshots are taken at $t=0$, $t=T_0/2$, and $t=1.8\,T_0$ from top to bottom. The length of the horizontal line corresponds to about $50\,a_B\approx 2.7\;\text{nm}$, with $a_B=\lambda_C/\alpha\approx 5.3\times 10^{-9}\;\text{cm}$ being the Bohr radius . Note that the scale is logarithmic with four contours per decade. From \onlinecite{Foerre_2006}.}
\label{Forre}
\end{figure}

There are also several new theoretical results for the ionic quantum dynamics in strong high-frequency laser fields, in the so-called stabilization regime, where the ionization rate decreases or remains constant also with increasing laser intensity. An unexpected nondipole effect has been reported in \onlinecite{Foerre_2006} via numerically solving the Schr\"odinger equation for a hydrogenic atom beyond the dipole approximation. For this purpose the Kramers-Henneberger transformation \cite{Kramers_1956,Henneberger_1968} has been employed, i.e., the transformation to the instantaneous rest-frame of a classical free electron in the laser field, and the terms $\sim \xi_0^2$ have been neglected in the Hamiltonian (the value of $\xi_0$ considered was approximately $0.14$). In Fig. \ref{Forre}, the resulting angular distribution of the ejected electrons in the nondipole regime of stabilization displays a third unexpected lobe anti-parallel to the laser propagation direction, together with the two expected lobes along the laser polarization direction. As a classical explanation, a drift along the laser propagation direction was identified for the bound electron wave packet in the nondipole case (see the middle panel of Fig. \ref{Forre}). Inside the laser field the electron has a velocity component along the positive $z$ axis but this velocity tends to zero at the end of the pulse. Thus, the electromagnetic forces alone do not change the electron momentum along the propagation direction at the end of the pulse. The net effect of the Coulomb forces on the electron wave packet is consequently a momentum component along the negative $z$ axis: the electron, which is most probably situated in the upper hemisphere over the pulse, undergoes a momentum kick in the negative $z$ direction each time it passes close to the nucleus. A similar effect has been reported for molecules \cite{Forre_2007}.

Radiative recombination, being the time-reversed process of photoionization, of a relativistic electron with a highly-charged ion in the presence of a very intense laser field has been considered in \onlinecite{Mueller_2009c}. It was shown that the strong coupling of the electron to the laser field may lead to a very broad energy spectrum of emitted recombination photons, with pronounced side wings, and to characteristic modifications of the photon angular distribution.

Specific features of nondipole quantum dynamics in strong and ultrashort laser pulses have also been investigated employing the so-called Magnus approximation \cite{Dimitrovski_2009}. The dominant nondipole effect is found to be a shift of the entire wave-function towards the propagation direction, inducing a substantial population transfer into states with similar geometry.

The recent experiment reported in \onlinecite{Smeenk_2011} addresses the question of how the photon momenta are shared  between the electron and ion during laser-induced multiphoton ionization. Theoretically, this problem requires a nondipole treatment, even in the nonrelativistic case, to take into account explicitly the laser photon momentum. Energy-conservation of $\ell$-photon ionization here means that $\ell\omega_0=I_p+U_p+K$, where $I_p$ is the ionization energy of the atom, $U_p=e^2E_0^2/4m\omega_0^2$ is the ponderomotive energy and $K$ is the electron's kinetic energy. The experimental results in \onlinecite{Smeenk_2011}, obtained using laser fields with wavelength of $0.8\;\text{$\mu$m}$ and $1.4\;\text{$\mu$m}$ in the intensity range of $10^{14}\text{-}10^{15}\;\text{W/cm$^2$}$ has show that the fraction of the momentum, corresponding to the number of observed photons needed to overcome the ionization energy $I_p$, is transferred to the created ion rather than to the photoelectron. The electron carries only the momentum corresponding to the kinetic energy $K$, while the ponderomotive energy and the corresponding portion of the momentum are transferred back to the laser field. This experiment shows that the tunneling concept for the ionization dynamics is only an approximation. In fact, the quasistatic tunneling provides no mechanism to transfer linear momentum to the ion, a conclusion that agrees with recent concerns in \onlinecite{Reiss_2008}.

\subsection{Recollisions and high-order harmonic generation}
\label{HHG}
Tunneling in the nonrelativistic regime is generally followed by recollisions with the parent ion along with various subsequent effects  \cite{Kuchiev_1987,Corkum_1993,Schafer_1993}. 
A characteristic feature of strong-field processes in the relativistic regime is the suppression of recollisions due to the magnetically induced relativistic drift of the ionized electron in the laser propagation direction (see Sec. \ref{FED_C}). Although relativistic effects become significant when the parameter $\xi_0$ exceeds unity, signatures of the drift in the laser propagation direction can be observed already in the weakly relativistic regime $\xi_0\lesssim 1$. The drift will have a significant impact on the electron's rescattering probability if, at the instant of recollision, the drift distance $d_{\parallel}$ in the laser propagation direction is larger than the electron's wave packet size $a_{wp,\parallel}$ in that direction \cite{Walker_2006}. The drift distance is approximately given by $d_{\parallel}\sim \lambda_0\xi_0^2/2$ (see, e.g., Eq. \eqref{Free_Sol_3}). Instead, the wave packet size $a_{wp,\parallel}$ can be estimated from $a_{wp,\parallel}\sim v_{\parallel} \Delta t$, where $v_{\parallel}$ is a typical electron velocity along the laser propagation direction and $\Delta t$ is the excursion time of the electron in the continuum. The velocity $v_{\parallel}$ can be related to the tunneling time $\tau_{\text{tun}}$ via the time-energy uncertainty: $mv_{\parallel}^2/2\sim 1/\tau_{\text{tun}}$. In turn, one can estimate the tunneling time $\tau_{\text{tun}}$ as $\tau_{\text{tun}}\sim l_{\text{tun}}/v_b$, where $l_{\text{tun}}\sim I_p/|e|E_0$ is the tunneling length and $v_b\sim \sqrt{2I_p/m}$ the velocity of the bound electron. In the above estimate, it was assumed that the work carried out by the laser field along the tunneling length equals $I_p$. Thus, at the rescattering moment $\Delta t\sim T_0$, the wave packet size $a_{wp,\parallel}$ is of the order of $\lambda_0\sqrt{|e|E_0/\sqrt{m^3I_p}}$ and the role of the drift can be characterized by means of the parameter $r=(d_{\parallel}/a_{wp,\parallel})^2$ as estimated by
\begin{equation}
r \sim  \xi_0^3\frac{\sqrt{2mI_p}}{16\omega_0}.
\label{drift}
\end{equation}
The condition $r\gtrsim 1$ determines the parameter region over which the signature of the drift becomes conspicuous.

As an alternative view on the relativistic drift, the ionized electron here misses the ionic core when it is ionized with zero momentum. Nevertheless, the recollision will occur if the electron is ionized with an appropriate initial momentum $p_d$ ($\sim m\xi_0^2/4$, see Eq. \eqref{Free_Sol_3}), opposite to the laser propagation direction. The probability $P_i(p_d)$ of this process is exponentially damped, though, due to the nonzero momentum $p_d$ (see, e.g., \onlinecite{Salamin_2006}):
\begin{equation}
P_i(p_d)\sim \exp \left[ -\frac{2}{3}\frac{(2mI_p)^{3/2}}{m|e|E_0}\left(1+\frac{p_d^2}{4mI_p}\right)\right]. 
\label{TI_pd}
\end{equation}
The drift term in the exponent proportional to $p_d^2$ will be important if $\sqrt{2mI_p}p_d^2/m|e|E_0\gtrsim 1$, which is equivalent to the condition $r\gtrsim 3$. At near-infrared wavelengths ($\omega_0\approx 1\;\text{eV}$) and for the ionization energy $I_p=13.6\;\text{eV}$ of atomic hydrogen, it becomes relevant at laser intensities $I_0$ approximately above $3\times10^{16}$ W/cm$^2$. Then, HHG and other recollision phenomena are suppressed.

The attainability of relativistic recollisions would, however, be very attractive for ultrahigh HHG \cite{Kohler_2012} as well as 
for the realization of laser controlled high-energy \cite{Hatsagortsyan_2006} and nuclear processes \cite{Milosevic_2004,Chelkowski_2004}.
Various methods for counteracting the relativistic drift have been proposed, such as by utilizing highly-charged ions \cite{Hu_2001,Hu_2002} which move relativistically against the laser propagation direction \cite{Mocken_2004,Chirila_2004}, by employing Positronium (Ps) atoms \cite{Henrich_2004}, or through preparing antisymmetric atomic \cite{Fischer_2007} and molecular \cite{Fischer_2006} orbitals. Here the impact of the drift of the ionized electron is reduced by the increase of the laser frequency in the system's center of mass, an equally strong drift via two constituents with equal mass or via appropriate initial momenta from antisymmetric orbitals, respectively. 

On the other hand, the laser field can also be modified to suppress the relativistic drift by employing tightly focused laser beams \cite{Lin_06}, two counter-propagating laser beams with linear polarization \cite{Keitel_1993,Kylstra_2000,Taranukhin_2000,Taranukhin_2001,Taranukhin_2002} or equal-handed circular polarization \cite{Milosevic_2004}. In the first two cases, the longitudinal component in the tightly focused laser beam may counteract the drift, or the Lorentz force may be eliminated in a small area near the antinodes of the resulting standing wave, respectively. In the third case involving circularly polarized light, the relativistic drift is eliminated because the electron velocity is oriented in the same direction as the magnetic field. This setup is well suited for imaging attosecond dynamics of nuclear processes but not for HHG because of the phase-matching problem \cite{Liu_2009}. In the weakly relativistic regime the Lorentz force may also be compensated by a second weak laser beam polarized along the direction of propagation of the strong beam \cite{Joachain_2002}. Furthermore, the relativistic drift can be significantly reduced by means of special tailoring of the driving laser pulse, which strongly reduces the time when the electron's motion is relativistic with respect to a sinusoidal laser pulse \cite{Klaiber_2006,Klaiber_2007}. Two consecutive laser pulses \cite{Verschl_2007a} or a single laser field assisted by a strong magnetic field can also be used to reverse the drift \cite{Verschl_2007b}. In addition two strong Attosecond Pulse Trains (APTs) \cite{Hatsagortsyan_2008} or an infrared laser pulse assisted by an APT \cite{Klaiber_2008} have been employed to enhance relativistic recollisions. In fact, due to the presence of the APT the ionization can be accomplished by one XUV photon absorption and the relatively large energy $\omega_X$ of the XUV-photon with $\omega_X=I_p+p_d^2/2m$ can compensate the subsequent momentum drift $p_d\sim m\xi_0^2/4$ in the infrared laser field.
\begin{figure}
\begin{center}
\includegraphics[width=0.8\linewidth]{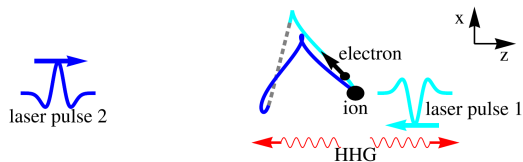}
\caption{(Color) The HHG setup with two counterpropagating APTs. After ionization by the laser pulse 1, the ejected electron is driven in the same pulse (light blue), propagates freely after the pulse 1 has left (gray dashed) and is driven back to the ion by the laser pulse 2 (dark blue). From \onlinecite{Kohler_2011}.} \label{setup}
\end{center}
\end{figure}

The main motivation for the realization of relativistic recollisions is 
the extension of HHG towards the hard x-ray regime with obvious benefits for time-resolved high-resolution imaging. In the past couple of decades, nonrelativistic atomic HHG \cite{Corkum_1993,Lewenstein_1994} has been developed as a reliable source of coherent XUV radiation and attosecond pulses \cite{Agostini_2004} opening the door for attosecond time-resolved spectroscopy \cite{Krausz_2009}. Nonrelativistic HHG in an atomic gas medium allows already to generate coherent x-ray photons up to keV energies \cite{Sansone_2006} and to produce XUV pulses shorter than 100 as \cite{Goulielmakis_2008}. The most favorable conversion efficiency for nonrelativistic keV harmonics is anticipated with mid-infrared driving laser fields \cite{Popmintchev_2009,Chen_2010}. However, progress in this field has slowed down, especially because of the inhibition, alluded to above, of recollisions due to optical driving-field intensity above $3\times10^{16}$ W/cm$^2$. This indicates the limit on the cut-off frequency $\omega_{c}$ of nonrelativistic HHG to $\omega_{c}\approx 3.17 U_p\sim 10$ keV.

Another factor hindering HHG at high intensities is the less favorable phase-matching. In strong laser fields, outer-shell electrons are rapidly ionized and produce a large free electron background causing a phase mismatch between the driving laser wave and the emitted x-rays. The feasibility of phase-matched relativistic HHG in a macroscopic ensemble was first investigated in \onlinecite{Kohler_2011}. Here, the driving fields are two counterpropagating APTs consisting of 100 as pulses with a peak intensity of the order of $10^{19}$ W/cm$^2$ (see Fig. \ref{setup}). The electron is driven to the continuum by the laser pulse 1 in Fig. \ref{setup}, followed by the usual relativistic drift. Thereafter, the laser pulse 2 overtakes the electron, reverses the drift and imposes the rescattering, yielding a much higher HHG signal than for a conventional laser field at the same cut-off energy. Here phase-matching can be fulfilled due to an additional intrinsic phase specific to this setup, depending on the time delay between the pulses and on the pulse intensity. The latter, being unique for this laser setup, mainly affects the electron excursion time and varies along the propagation direction. The phase-matching is achieved by modifying the laser intensity along the propagation direction and by balancing the phase slip due to dispersion with the indicated intrinsic phase. Note, however, that HHG in the relativistic regime has been observed experimentally rather efficiently in laser plasma interactions \cite{Dromey_2006}.

\section{Multiphoton Thomson and Compton scattering}
\label{TCS}
In this section we discuss one of the most fundamental processes in QED in a strong laser field: the emission of radiation by an accelerated electron. After reporting on recent theoretical investigations on this process, we discuss its possible applications for producing high-energy photon beams.

\subsection{Fundamental considerations}
\label{TCS_General}
When an electron is wiggled by an intense laser wave, it emits electromagnetic radiation. This process occurs with absorption of energy and momentum by the electron from the laser field and it is named as multiphoton Thomson scattering or multiphoton Compton scattering, depending on whether quantum effects, like photon recoil, are negligible or not. Multiphoton Thomson and Compton scattering in a strong laser field have been studied theoretically since a long time (see \onlinecite{Sarachik_1970,Salamin_1998,Sengupta_1949} for multiphoton Thomson scattering and \onlinecite{Goldman_1964,Nikishov_1964_a,Brown_1964} for multiphoton Compton scattering). The classical calculation of the emitted spectrum is based on the analytical solution in Eqs. \eqref{Free_Sol_0}-\eqref{Free_Sol_3} of the Lorentz equation in a plane wave and the substitution of the corresponding electron trajectory in the Li\'{e}nard-Wiechert fields \cite{Jackson_b_1975,Landau_b_2_1975}. Whereas, as we have discussed in Sec. \ref{FED_Q}, the quantum calculation of the amplitude of the process is performed in the Furry picture of QED. As a result, the total emission probability depends only on the two Lorentz- and gauge-invariant parameters $\xi_0$ (see Eq. \eqref{xi_0}) and $\chi_0$ (see Eq. \eqref{chi_0}). 

The parameter $\xi_0$ has already been discussed in Sec. \ref{FED_C}. In the contest of multiphoton Compton scattering this parameter controls in particular the effective order $\ell_{\text{eff}}$ of the emitted harmonics, which, for an ultrarelativistic electron, can be estimated in the following way. In order to effectively emit a frequency $\omega'$, the formation length $l_f$ of the process must not exceed the coherence length $l_{\text{coh}}$, because, otherwise, interference effects would hinder the emission. Since an electron with instantaneous velocity $\bm{\beta}$ and energy $\varepsilon=m\gamma=m/\sqrt{1-\beta^2}\gg m$ mainly emits along the direction of $\bm{\beta}$, within a cone with apex angle $\vartheta\sim 1/\gamma\ll 1$, the formation length $l_f$ can be estimated from $l_f\sim \varrho/\gamma$, with $\varrho$ being the instantaneous radius of curvature of the electron trajectory  \cite{Jackson_b_1975}. On the other hand, $l_{\text{coh}}=\pi/\omega'(1-\beta \cos \vartheta)\sim \gamma^2/\omega'$ \cite{Jackson_b_1975,Baier_b_1998}. By requiring that $l_f\lesssim l_{\text{coh}}$, we obtain the following estimate for the largest-emitted frequency (cut-off frequency) $\omega'_c$: $\omega'_c\sim \gamma^3/\varrho$. Now, in the average rest-frame of the electron, i.e., in the reference frame where the average electron velocity along the propagation direction of the laser vanishes (see Eq. \eqref{Free_Sol_3}), it is $\gamma^{\star}\sim \xi_0$ (corresponding to the energy $\varepsilon^{\star}\sim m\xi_0$) and $l_f^{\star}\sim \lambda_0^{\star}/\xi_0$, where the upper index $^{\star}$ indicates the variable in this frame. Consequently, $\omega_c^{\prime\,\star} \sim \xi_0^3 \omega^{\star}_0$ and the effective order of the emitted harmonics is $\ell_{\text{eff}}\sim \xi_0^3$ (note that $\ell_{\text{eff}}$ is a Lorentz scalar). As the order of the emitted harmonics corresponds quantum mechanically to the number of laser photons absorbed by the electron during the emission process, the parameter $\xi_0$ is also said to determine the ``multiphoton'' character of the process.

On the other hand, the nonlinear quantum parameter $\chi_0$ (see Eq. \eqref{chi_0}) in the contest of multiphoton Compton scattering controls the importance of quantum effects as the recoil of the emitted photon. In fact, we can estimate classically the importance of the emitted photon recoil from the ratio $\omega'_c/\varepsilon$ and our considerations above exactly indicate that $\omega'_c/\varepsilon\sim \xi_0^2 \omega^{\star}_0/m\sim \chi_0$. Thus, multiphoton Thomson scattering is characterized by the condition $\chi_0\ll 1$, while multiphoton Compton scattering by $\chi_0\gtrsim 1$. This result can also be obtained in the case of a monochromatic laser wave starting from the energy-momentum conservation relation
\begin{equation}
\label{Cons_law}
q_0^{\mu}+\ell k_0^{\mu}=q^{\prime\mu}+k^{\prime\mu}
\end{equation}
in the case in which $\ell$ laser photons are absorbed in the process \cite{Ritus_1985}. Here $q_0^{\mu}$ and $q^{\prime\mu}$ are the quasimomenta of the initial and final electron (see Eq. \eqref{q}) and $k^{\prime\mu}=\omega'n^{\prime\mu}$ is the four-momentum of the produced photon ($n^{\prime 2}=0$). From this expression it is easy to obtain the energy $\omega'$ of the emitted photon as
\begin{equation}
\label{omega_p}
\omega'=\frac{\ell\omega_0p_{0,-}}{(n'p_0)+\left(\ell\omega_0+\frac{m^2\xi_0^2}{4p_{0,-}}\right)n'_-}.
\end{equation}
By reminding that $\ell_{\text{eff}}\sim \xi_0^3$ and by estimating the typical emission angle of the photon \cite{Mackenroth_2011}, it is possible to show that $\omega'\sim \chi_0\varepsilon_0$ at $\xi_0\gg 1$. As it has also been throughroughly investigated analytically and numerically in \onlinecite{Boca_2011,Seipt_2011}, multiphoton Compton and Thomson spectra coincide in the limit $\chi_0\to 0$, although in \onlinecite{Seipt_2011} differences have been observed numerically in the detailed structure of the classical and quantum spectra also for $\chi_0\ll 1$. The most important difference between classical and quantum spectra is certainly the presence of a sharp cut-off in the latter as an effect of the photon recoil: the energy of the photon emitted in a plane wave is limited by the initial energy of the electron\footnote{In the case of a plane-wave background field, this limitation rather concerns the quantity $k'_-$ of the emitted photon, as $k'_-=p_{0,-}-p'_-<p_{0,-}$. However, for an ultrarelativistic electron with $p_{0,-}\gg m\xi_0$ and initially counterpropagating with respect to the laser field, it is $p_{0,-}\approx 2\varepsilon_0$, $k'_-\approx 2\omega'$ and $p'_-\approx 2\varepsilon'$ \cite{Baier_b_1998}.}. This does not occur classically, as there the frequency of the emitted radiation does not have the physical meaning of photon energy. The dependence of the energy cut-off on the laser intensity has been recently recognized as a possible experimental signature of multiphoton Compton scattering \cite{Harvey_2009}.

\begin{figure}
\begin{center}
\includegraphics[width=0.8\linewidth]{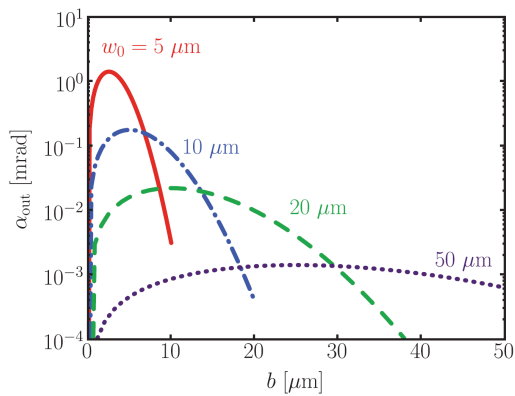}
\caption{(Color online) Deflection $\alpha_{\text{out}}$ of an electron initially counterpropagating with respect to a laser field with an energy of 3 J and a pulse duration of $20$ fs, as a function of the impact parameter $b$ for different laser waist radii $w_0$. From \onlinecite{Heinzl_2010}.
} \label{Heinzl}
\end{center}
\end{figure}

First calculations on multiphoton Thomson and Compton scattering have mainly focused on the easiest case of a monochromatic background plane wave, either with circular or linear polarization. The main results of these investigations, like the dependence of the emitted frequencies on the laser intensities have been recently reviewed in \onlinecite{Ehlotzky_2009}. The complete description of the multiphoton Compton scattering process with respect to the polarization properties of the incoming and the outgoing electrons, and of the emitted photon in a monochromatic laser wave has been presented in \onlinecite{Ivanov_2004}. Recently significant attention has been devoted to the investigation of multiphoton Thomson and Compton scattering in the presence of short and even ultrashort plane-wave pulses (we recall that such pulses have still an infinite extension in the directions perpendicular to the propagation direction). In \onlinecite{Boca_2009} multiphoton Compton scattering has been considered in the presence of a pulsed plane wave. The angular-resolved spectra are practically insensitive to the precise form of the laser pulse for $\omega_0\tau_0\ge 20$, with $\tau_0$ being the pulse duration. The main differences with respect to the monochromatic case are: 1) a broadening of the lines corresponding to the emitted frequencies; 2) the appearance of sub-peaks, which are due to interference effects in the emission at the beginning and at the end of the laser pulse. On the one hand, the continuous nature of the emission spectrum in a finite pulse in contrast to the discrete one in the monochromatic case has a clear mathematical counterpart. In both cases, in fact, the total transverse momenta $\bm{P}_{\perp}$ with respect to the laser propagation direction and the total quantity $P_-$ are conserved in the emission process (see Sec. \ref{FED}). However, in the monochromatic case the following additional conservation law holds (for a linearly polarized plane wave, see Eq. \eqref{Cons_law})
\begin{equation}
\varepsilon_0+p_{0,\parallel}+\frac{m^2\xi_0^2}{2p_{0,-}}+2\ell\omega_0=\omega'+k'_{\parallel}+\varepsilon'+p'_{\parallel}+\frac{m^2\xi_0^2}{2p'_-},
\end{equation}
so that the resulting four-dimensional energy-momentum conservation law allows only for the emission of the discrete frequencies in Eq. \eqref{omega_p}. On the other hand, the appearance of sub-peaks has been in particular investigated in \onlinecite{Heinzl_2010}, where it has been found that the number $N_{\text{s-p}}$ of sub-peaks within the first harmonic scales linearly with the pulse duration $\tau_0$ and with $\xi_0^2$: $N_{\text{s-p}}=0.24\,\xi_0^2\tau_0[\text{fs}]$. In this paper the effects of spatial focusing of the driving laser pulse are also discussed. The authors investigate in particular the dependence of the deflection angle $\alpha_{\text{out}}$ undergone by the electron after colliding head-on with a Gaussian focused beam as a function of the impact parameter $b$ (see Fig. \ref{Heinzl}).

By further decreasing the laser pulse duration, it has been argued that effects of the relative phase between the pulse profile and the carrier wave (the so-called Carrier Envelope Phase (CEP)) should become visible in multiphoton Thomson and Compton scattering. In \onlinecite{Boca_2009} the case of ultrashort pulses with $\omega_0\tau_0\gtrsim 4$ has been discussed and effects of the CEP on the harmonic yield in specific frequency ranges have been observed. In \onlinecite{Mackenroth_2010} the dependence of the angular distribution of the emitted radiation in multiphoton Thomson and Compton scattering on the CEP of few-cycles pulses has been exploited to propose a scheme to measure the CEP of ultrarelativistic laser pulses (intensities larger than $10^{20}\;\text{W/cm$^2$}$). The method is essentially based on the high directionality of the photon emission by an ultrarelativistic electron, because the trajectory of the electron, in turn, also depends on the laser's CEP. Accuracies in the measurement of the CEP of the order of a few degrees are theoretically envisaged. Multiphoton Compton scattering in one-cycle laser pulses has been considered in \onlinecite{Mackenroth_2011} and a substantial broadening of the emission lines with respect to the monochromatic case has been observed. The high-directionality of radiation emitted via multiphoton Thomson scattering has also been employed as a diagnostic tool in \onlinecite{Har-Shemesh_2012}, where a new rather precise method has been proposed to measure the peak intensity of strong laser fields (intensities between $10^{20}\;\text{W/cm$^2$}$ and $10^{23}\;\text{W/cm$^2$}$) from the angular aperture of the photon spectrum.

The study of multiphoton Thomson and Compton scattering in short laser pulses has also stimulated the investigation of scaling laws for the photon spectral density \cite{Heinzl_2010,Seipt_2011,Seipt_2011b,Boca_2011b}. For example, in \onlinecite{Heinzl_2010} a scaling law has been found for backscattered radiation in the case of head-on laser-electron collisions, which simplifies the averaging over the electron-beam phase space. A more general scaling law has been determined in \onlinecite{Seipt_2011b}, which relaxes the previous assumptions on head-on collision and on backscattered radiation employed in \onlinecite{Heinzl_2010}. Moreover, in \onlinecite{Seipt_2011} a simple relation is determined between the classical and the quantum spectral densities. Finally, in \onlinecite{Boca_2011b} it is found that in the ultrarelativistic case $\gamma_0\gg 1$ the angular distribution of the emitted radiation, integrated with respect to the photon energy, only depends on the ratio $\xi_0/\gamma_0$ and not on the independent values of $\xi_0$ and $\gamma_0$ (see also \onlinecite{Mackenroth_2010}).

In the above-mentioned publications the spectral properties of the emitted radiation in the classical and quantum regimes have been considered. In \onlinecite{Zhang_2008,Kim_2009}, instead, the temporal properties of the emitted radiation in multiphoton Thomson scattering have been investigated. In both papers the feasibility of generating single attosecond pulses is discussed.

Photoemission by a single-electron wave packet via Thomson scattering in a strong laser field has been discussed in \onlinecite{Peatross_2008}. It was shown that the partial emissions from the individual electron momentum components do not interfere when the driving field is a plane wave. In other words, the size of the electron wave packet, even when it spreads to the scale of the wavelength of the driving field, does not affect the Thomson emission.

Finally, we shortly mention that multiphoton effects in Thomson and Compton scattering have been measured in various laboratories. The second-harmonic radiation was first observed in the collision of a 1 keV electron beam with a Q-switched Nd:YAG laser, although the laser intensity was such that $\xi_0\approx 0.01$ \cite{Englert_1983}, and then in the interaction of a mode-locked Nd:YAG laser ($\xi_0=2$) with plasma electrons \cite{Chen_1998}. Multiphoton Thomson scattering of laser radiation in the x-ray domain has been reported in \onlinecite{Babzien_2006} (see \onlinecite{Pogorelsky_2000} for a similar proof-of-principle experiment). Single-shot measurements of the angular distribution of the second harmonic (photon energy $6.5\;\text{keV}$) at various laser polarizations have been carried out by employing a 60 MeV electron beam and a subterawatt CO$_2$ laser beam with $\xi_0=0.35$. In the prominent SLAC experiment \cite{Bula_1996} multiphoton Compton emission was detected for the first time. In this experiment an ultrarelativistic electron beam with energy of about $46.6\;\text{GeV}$ collided with a terawatt Nd:glass laser with an intensity of $10^{18}\;\text{W/cm$^2$}$ ($\xi_0\approx 0.8$ and $\chi_0\approx 0.3$) and four-photon Compton scattering has been observed indirectly via a nonlinear energy shift in the spectrum of the outcoming electrons.

\subsection{Thomson- and Compton-based sources of high-energy photon beams}
\label{TC_Sources}
The single-particle theoretical analysis presented above indicates that high-energy photons can be emitted via multiphoton Thomson and Compton scattering of an ultrarelativistic electron. For example, an electron with initial energy $\varepsilon_0\gg m$ colliding head-on with an optical laser field ($\omega_0\approx 1\;\text{eV}$) of moderate intensity ($\xi_0\lesssim 1$) is barely deflected by the laser field ($\varrho\sim \lambda_0\gamma_0/\xi_0\gg \lambda_0$) and potentially emits photons with energies $\omega[\text{keV}]\lesssim 3.8\times 10^{-3}\,\varepsilon_0^2[\text{MeV}]$. This feature has boosted the idea of so-called Thomson- and Compton-based sources of high-energy photons as a valid alternative to conventional synchrotron sources, the main advantages of the former being the compactness, the wide tunability, the shortness of the photon beams in the femtosecond scale and the potential for high brightness. Unlike the experiments on multiphoton Thomson and Compton scattering where laser systems with $\xi_0\gtrsim 1$ are generally employed, Thomson- and Compton-based photon sources preferably require lasers with $\xi_0\lesssim 1$, such that multiphoton effects are suppressed and shorter bandwidths of the photon beam are achieved. On the other hand, the electron beam quality is crucial for Thomson- and Compton-based radiation sources. In particular, the brightness of the photon beam scales inversely quadratically with the electron beam emittance, and linearly with the electron bunch current density.

Proof-of-principle experiments have demonstrated Thomson- and Compton-based photon sources by crossing a high-energy laser pulse with a picosecond relativistic electron beam from a conventional linear electron accelerator \cite{Ting_1995,Ting_1996,Schoenlein_1996,Leemans_1996,Pogorelsky_2000,Chouffani_2002,Sakai_2003}. We also mention the benchmark experiment carried out at LLNL, where photons with an energy of $78\;\text{keV}$ have been produced with a total flux of $1.3\times 10^6\;\text{photons/shot}$, by colliding an electron beam with an energy of $57\;\text{MeV}$ with a Ti:Sa laser beam with an intensity of about $10^{18}\;\text{W/cm$^2$}$ ($\xi_0\approx 0.5$) \cite{Gibson_2004}. 

Another achievement in the development of Thomson- and Compton-based photon sources has been the experimental realization of a compact all-optical setup, where the electrons are accelerated by an intense laser. In the first experiment with an all-optical setup \cite{Schwoerer_2006}, x-ray photons in the range of 0.4 keV to 2 keV have been generated. In this experiment the electron beam was produced by a high-intensity Ti:Sa laser beam ($I_0\approx 2\times 10^{19}\;\text{W/cm$^2$}$) focused into a pulsed helium gas jet. The characteristic feature of the all-optical setup is that the electron bunches and, consequently, the generated x-ray photon beams have an ultrashort duration ($\sim 100\;\text{fs}$) and a linear size of the order of $10\;\mu$m. Another advantage is that the electrons can be precisely synchronized with the driving laser field. In order to further improve the all-optical setup, design parameters for a proof-of-concept experiment have been analyzed in \onlinecite{Hartemann_2007}. For the calculation of the Compton scattering parameters, a 3D Compton scattering code has been used, which was extensively tested for Compton scattering experiments performed at LLNL \cite{Hartemann_2005,Brown_2004,Rosenzweig_2004,Hartemann_2004} (see \onlinecite{Sun_2011} for an alternative numerical simulation scheme). It is shown that x-ray fluxes exceeding $10^{21}$ s$^{-1}$ and a peak brightness larger than $10^{19}$ photons/(s\,mrad$^2$\,mm$^2$\,0.1\% bandwidth) can be achieved at photon energies of about 0.5 MeV. A few years later the Compton-based photon source MonoEnergetic Gamma-ray (MEGa-ray) has been designed at LLNL \cite{Gibson_2010}. Production of gamma-rays ranging from 75 keV to 0.9 MeV has been demonstrated with a peak spectral brightness of $1.5\times 10^{15}$ photons/(s\,mrad$^2$\,mm$^2$\,0.1\% bandwidth) and with a flux of $1.6\times 10^5$ photons/shot. An experimental setup for high-flux gamma-ray generation has been constructed in the Saga Light-Source facility in Tosu (Japan), by colliding a 1.4 GeV electron beam with a CO$_2$ laser (wavelength $10.6\;\text{$\mu$m}$) \cite{Kaneyasu_2011}. A flux of about $3.2\times 10^7\;\text{photons/s}$ gamma photons with energy larger than 0.5 MeV has been obtained.

\begin{figure}
\begin{center}
\includegraphics[width=0.8\linewidth]{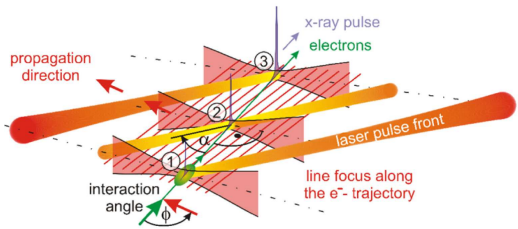}
\end{center}
\caption{(Color) Schematic setup of TWTS with the red lines indicating the laser focal lines. In the notation of \onlinecite{Debus_2010} $\phi$ is the angle between the initial electrons' velocity and the laser's wave vector and $\alpha$ is the angle between the laser pulse front and the laser propagation direction. Adapted from \onlinecite{Debus_2010}.}
\label{Debus1}
\end{figure}

In the basic setups of Thomson- and Compton-based photon sources the electrons experience the intense laser field for a time-interval much shorter than that needed to cross the whole laser beam, the former being of the order of the laser's Rayleigh length divided by the speed of light. Thus, the quest for a more intense laser pulse at a given power in order to increase the photon yield implies a tighter focusing and therefore a shorter effective interaction time, which in turn causes a broadening of the photon spectrum. In \onlinecite{Debus_2010} the Traveling-Wave Thomson Scattering (TWTS) setup is proposed, which allows the electrons to stay in the focal region of the laser beam during the whole crossing time (see Fig. \ref{Debus1}). This is achieved by employing cylindrical optics to focus the laser field only along one direction (red lines in Fig. \ref{Debus1}) and, depending on the angle between the initial electron velocity and the laser wave-vector, by tilting the laser pulse front. As a result, an interaction length $\sim 1\;\text{cm}\text{-}1\;\text{m}$ can be achieved and, correspondingly very large photon fluxes (up to $5\times 10^{10}$ photons/shot at 20 keV).
An alternative way of reaching longer effective laser-electron interaction times has been proposed in \onlinecite{Karagodsky_2010}, where a planar Bragg structure is employed to guide the laser pulse and realize Thomson/Compton scattering in a waveguide. In this way, the yield of x-rays can be enhanced by about two orders of magnitude with respect to the conventional free-space Gaussian-beam configuration at given electron beam and injected laser power in both configurations. However, there are two constraints specific to this setup. On the one hand, the electron beam has to have a small angular spread in order to be injected into the planar Bragg structure without causing wall damage. On the other hand, the laser field strength has to be such that $\xi_0 \lesssim 8\times 10^{-4}$ to avoid surface damage.

\begin{figure}
\begin{center}
\includegraphics[width=0.8\linewidth]{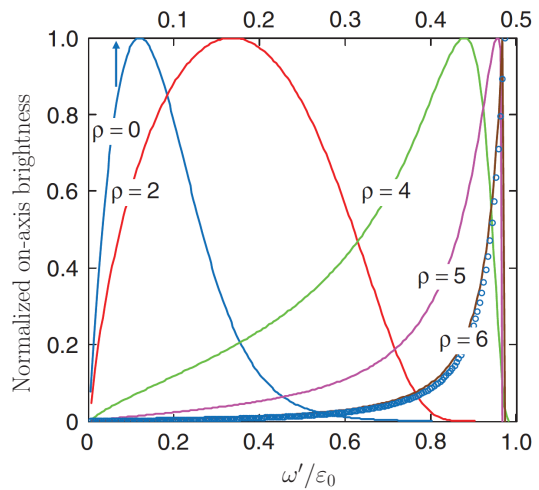}
\end{center}
\caption{(Color online) Normalized on-axis brightness for different values of the rapidity $\rho=\cosh^{-1}\gamma_0$ at a plasma temperature of 20 keV. See \onlinecite{Hartemann_2008} for the meaning of the blue circles at $\rho=6$. Adapted from \onlinecite{Hartemann_2008}.}
\label{Hartemann_pileup}
\end{figure}

Finally, in \onlinecite{Hartemann_2008} a setup has been proposed to obtain bright GeV gamma-rays via Compton scattering of electrons by a thermonuclear plasma. In fact, a thermonuclear deuterium-tritium plasma produces intense blackbody radiation with a temperature $\sim 20$ keV and a photon density $\sim 10^{26}$ cm$^{-3}$ \cite{Tabak_1994}. When a thermal photon with energy $\omega \sim 1$ keV counterpropagates with respect to a GeV electron ($\gamma_0\sim 10^3$), a Doppler-shifted high-energy photon $\omega' \sim \gamma_0^2 \omega\sim 1$ GeV can be emitted on axis, i.e., in the same direction of the incoming electron. Since $\omega'$ has to be smaller than the initial electron energy $\varepsilon_0$, a kinematical photon pileup is induced in the emitted photon spectrum at $\varepsilon_0$ \cite{Zeldovich_1969} (see Fig. \ref{Hartemann_pileup}). This results in a quasimonochromatic GeV gamma-ray beam with a peak brightness $\gtrsim 10^{30}$ photons/(s\,mrad$^2$\,mm$^2$\,0.1\% bandwidth), comparable with that of the FLASH (see Sec. \ref{x_ray}).

The unique features of Thomson- and Compton-based photon sources render them a powerful experimental device. For example, they can be employed for medical radio-isotope production and photo-fission, and for studying nuclear resonance fluorescence for \textit{in situ} isotope detection \cite{Albert_2010}. In this respect, such photon sources will represent the main experimental tool for nuclear-physics investigations at the Romanian pillar of ELI (see Fig. \ref{Megalaser}). Other possible applications, at photon energies beyond the MeV threshold, include the production of positron beams \cite{Omori_2003,Hugenschmidt_2012} as well as the investigations of high-energy processes occurring in gamma-gamma and gamma-lepton collisions \cite{Telnov_1990}.

\section{Radiation reaction}
\label{RR}
The issue of ``radiation reaction'' (RR) is one of the oldest and most fundamental problems in electrodynamics. Classically it corresponds to the determination of the equation of motion of a charged particle, an electron for definiteness, in a given electromagnetic field $F^{\mu\nu}(x)$. In fact, the Lorentz equation $mdu^{\mu}/ds=eF^{\mu\nu}u_{\nu}$ (see Sec. \ref{FED_C}) does not take into account that the electron, while being accelerated, emits electromagnetic radiation and loses energy and momentum in this way. The first attempt of taking into account the reaction of the radiation emitted by the electron on the motion of the electron itself (from here comes the expression ``radiation reaction''), was accomplished by H. A. Lorentz in the nonrelativistic regime \cite{Lorentz_b_1909}. Starting from the known Larmor formula $\mathcal{P}_L=(2/3)e^2\bm{a}^2$ for the power emitted by an electron with instantaneous acceleration $\bm{a}$, Lorentz argued that this energy-loss corresponds to a ``damping'' force $\bm{F}_R=(2/3)e^2d\bm{a}/dt$ acting on the electron. The expression of the damping force was generalized to the relativistic case by M. Abraham in the form \cite{Abraham_b_1905}
\begin{equation}
\label{F_R}
F^{\mu}_R=\frac{2}{3}e^2\left(\frac{d^2u^{\mu}}{ds^2}+\frac{du^{\nu}}{ds}\frac{du_{\nu}}{ds}u^{\mu}\right).
\end{equation}
In order to solve the problem of a radiating electron self-consistently, P. A. M. Dirac suggested in \onlinecite{Dirac_1938} to start from the coupled system of Maxwell and Lorentz equations
\begin{equation}
\label{M_L}
\left\{
\begin{aligned}
&\partial_{\mu}F^{\mu\nu}_T=4\pi j^{\nu}\\
&\partial_{\lambda}F_{T,\mu\nu}+\partial_{\mu}F_{T,\nu\lambda}+\partial_{\nu}F_{T,\lambda\mu}=0\\
&m_0\frac{du^{\mu}}{ds}=eF_T^{\mu\nu}u_{\nu},
\end{aligned}
\right.
\end{equation}
where $F_T^{\mu\nu}(x)=F^{\mu\nu}(x)+F_S^{\mu\nu}(x)$, with $F_S^{\mu\nu}(x)$ being the ``self'' electromagnetic field generated by the electron four-current $j^{\mu}(x)=e\int ds\delta(x-x(s))u^{\mu}$ and where the meaning of the symbol $m_0$ for the electron mass will be clarified below. In order to write an ``effective'' equation of motion for the electron which includes RR, one ``removes'' the degrees of freedom of the electromagnetic field \cite{Teitelboim_1971}. This is achieved in \onlinecite{Landau_b_2_1975} at the level of the Lagrangian of the system electron+electromagnetic field and an interesting connection of the RR problem with the derivation of the so-called Darwin Lagrangian is indicated. By working at the level of the equations of motion \eqref{M_L}, one first employs the Green's function method and formally determines the retarded solution $F_{T,\text{ret}}^{\mu\nu}(x)$ of the (inhomogeneous) Maxwell's equations: $F_{T,\text{ret}}^{\mu\nu}(x)=F^{\mu\nu}(x)+F_{S,\text{ret}}^{\mu\nu}(x)$ \cite{Teitelboim_1971}. Substitution of $F_{T,\text{ret}}^{\mu\nu}(x)$ in the Lorentz equation eliminates the electromagnetic field's degrees of freedom, but it is not straightforward because $F_{T,\text{ret}}^{\mu\nu}(x)$ has to be calculated at the electron's position, where the electron current diverges. This difficulty is circumvented by modeling the electron as a uniformly charged sphere of radius $a$ tending to zero. After performing the substitution, and by neglecting terms which vanish in the limit $a\to 0$, one obtains the equation $(m_0+\delta m)du^{\mu}/ds=eF^{\mu\nu}u_{\nu}+F_R^{\mu}$ with $\delta m=(4/3)e^2/a$ being formally diverging. However, it is important to note that the only diverging term in the limit $a\to 0$ is proportional to the electron four-acceleration. At this point a sort of ``classical renormalization principle'' is employed, saying that what one measures experimentally as the physical electron mass $m$ is the overall coefficient of the four-acceleration $du^{\mu}/ds$. Therefore, one sets $m=m_0+\delta m$ and obtains the so-called Lorentz-Abraham-Dirac (LAD) equation:
\begin{equation}
\label{LAD}
m\frac{du^{\mu}}{ds}=eF^{\mu\nu}u_{\nu}+\frac{2}{3}e^2\left(\frac{d^2u^{\mu}}{ds^2}+\frac{du^{\nu}}{ds}\frac{du_{\nu}}{ds}u^{\mu}\right).
\end{equation}
We point out that renormalization in quantum field theory is based on the fact that the bare quantities, like charge and mass, appear in the Lagrangian density of the theory, which is not an observable physical quantity. On the other hand, the bare electron mass $m_0$, which is formally negatively diverging for $m$ to be finite, appears here in the system of equations \eqref{M_L}, which should ``directly'' provide classical physical observables, like the electron trajectory. On the other hand, it is also known that the LAD equation is plagued with physical inconsistencies like, for example, the existence of the so-called ``runaway'' solutions with an exponentially-diverging electron acceleration, even in the absence of an external field (see the books \onlinecite{Hartemann_b_2001,Rohrlich_b_2007} for reviews on these issues).

It was first shown in \onlinecite{Landau_b_2_1975} that in the nonrelativistic limit the RR force given by Eq. \eqref{F_R} is much smaller than the Lorentz force, if the typical wavelength $\lambda$ and the typical field-amplitude $F$ of the external electromagnetic field fulfill the two conditions 
\begin{align}
\label{Cond_LL}
\lambda\gg \alpha\lambda_C, && F\ll \frac{F_{\text{cr}}}{\alpha},
\end{align}
where $F_{\text{cr}}$ is the critical electromagnetic field of QED (see Sec. \ref{FED_Q}). This allows for the reduction of order in the LAD equation, i.e., for the substitution of the electron acceleration in the RR force via the Lorentz force divided by the electron mass. In order to perform the analogous reduction of order in the relativistic case, the conditions \eqref{Cond_LL} have to be fulfilled in the instantaneous rest-frame of the electron \cite{Landau_b_2_1975}. The result is the so-called Landau-Lifshitz (LL) equation
\begin{equation}
\label{LL}
\begin{split}
m\frac{d u^{\mu}}{ds}=&eF^{\mu\nu}u_{\nu}+\frac{2}{3}e^2\left[\frac{e}{m}(\partial_{\alpha}F^{\mu\nu})u^{\alpha}u_{\nu}\right.\\
&\left.-\frac{e^2}{m^2}F^{\mu\nu}F_{\alpha\nu}u^{\alpha}+\frac{e^2}{m^2}(F^{\alpha\nu}u_{\nu})(F_{\alpha\lambda}u^{\lambda})u^{\mu}\right].
\end{split}
\end{equation}
The LL equation is not affected by the shortcomings of the LAD equation: for example, it is evident that if the external field vanishes, so does the electron acceleration. Most importantly, the conditions \eqref{Cond_LL} in the instantaneous rest-frame of the electron have always to be fulfilled in the realm of classical electrodynamics, i.e., if quantum effects are neglected. In order for this to be true, in fact, the two weaker conditions $\lambda\gg \lambda_C$ and $F\ll F_{\text{cr}}$ have to be fulfilled in the instantaneous rest-frame of the electron: the first guarantees that the electron's wave function is well localized and the second ensures that pure quantum effects, like photon recoil or spin effects are negligible \cite{Landau_b_4_1982,Baier_b_1998,Ritus_1985} (see also Sec. \ref{FED_Q}). This observation led F. Rohrlich to state recently that the LL equation is the ``physically correct'' classical relativistic equation of motion of a charged particle \cite{Rohrlich_2008}. Rohrlich's statement is also supported by the findings in \onlinecite{Spohn_2000}, where it is shown that the physical solutions of the LAD equation, i.e., those which are not runaway-like, are on the critical manifold of the LAD equation itself and are governed there exactly by the LL equation. On the other hand, since the LL equation is derived from the LAD equation, one may still doubt on its rigorous validity, due to the application in the latter equation of the ``suspicious'' classical mass-renormalization procedure.  However, this procedure is avoided in \onlinecite{Gralla_2009} by employing a more sophisticated zero-size limiting procedure, where also the charge and the mass of the particle are sent to zero but in such a way that their ratio remains constant. The authors conclude that at the leading-order level the LL equation represents the self-consistent perturbative equation of motion for a charge without electric and magnetic moment. The motion of a continuous charge distribution interacting with an external electromagnetic field is also investigated by a self-consistent model and at a more formal level in \onlinecite{Burton_2007}.

From the original derivation of the LL equation from the LAD equation in \onlinecite{Landau_b_2_1975}, it is expected that the two equations should predict the same electron trajectory, possibly with differences smaller than the quantum effects. This conclusion has been recently confirmed by analytical and numerical investigations in \onlinecite{Hadad_2010} for an external plane-wave field with linear and circular polarization and in \onlinecite{Bulanov_2011} for different time-dependent external electromagnetic field configurations. An effective numerical method to calculate the trajectory of an electron via the LL equation, which explicitly maintains the relativistic covariance and the mass-shell condition $u^2=1$, has been advanced in \onlinecite{Harvey_2011}. An alternative numerical method for determining the dynamics of an electron including RR effects has been proposed in \onlinecite{Mao_2010}.

We should emphasize that the LL equation is not the only equation which has been suggested to overcome the inconsistencies of the LAD equation. A list of alternative equations can be found in the recent review \cite{Hammond_2010_b} (see also \onlinecite{Seto_2011}). A phenomenological equation of motion, including RR and quantum effects related to photon recoil, has been suggested in \onlinecite{Sokolov_2009,Sokolov_2010} (see also Sec. \ref{QRR}). The authors write the differential variation of the electron momentum as due to two contributions: one arising from the external field and one corresponding to the recoil of the emitted photon. The resulting equation can be written as the system
\begin{equation}
\label{Sok_Eq}
\left\{
\begin{aligned}
&m\frac{dx^{\mu}}{d\tau}=p^{\mu}+\frac{2}{3}e^2\frac{\mathcal{I}_{\text{QED}}}{\mathcal{I}_L}\frac{eF^{\mu\nu}p_{\nu}}{m^2}\\
&\frac{dp^{\mu}}{d\tau}=eF^{\mu\nu}\frac{dx_{\nu}}{d\tau}-\mathcal{I}_{\text{QED}}\frac{p^{\mu}}{m},
\end{aligned}
\right.
\end{equation}
where $\tau$ is the time in the ``momentarily comoving Lorentz frame'' of the electron where the spatial components of $p^{\mu}$ instantaneously vanish, $\mathcal{I}_{\text{QED}}$ is the quantum radiation intensity \cite{Ritus_1985} and $\mathcal{I}_L=(2/3)\alpha\omega_0^2\xi_0^2$. The expression of $\mathcal{I}_{\text{QED}}$ in the case of a plane wave is employed, which is valid only for an ultrarelativistic electron in the presence of a slowly-varying and undercritical otherwise arbitrary external field (see Sec. \ref{FED_Q}).

It has also to be stressed that the original LAD equation is still the subject of extensive investigation (the first study of the LAD equation in a plane-wave field was performed in \onlinecite{Hartemann_1996}). In \onlinecite{Ferris_2011}, for example, the origin of the Schott term in the RR force, i.e., the term proportional to the derivative of the electron acceleration (see Eq. \eqref{F_R}), is thoroughly investigated and in \onlinecite{Kazinski_2011} the asymptotics of the physical solutions of the LAD equation at large proper times are obtained. Whereas, in \onlinecite{Noble_2011} a kinetic theory of RR is proposed, based on the LAD equation and applicable to study systems of many particles including RR (this last aspect is also considered in \onlinecite{Rohrlich_b_2007}). Enhancement of RR effects due to the coherent emission of radiation by a large number of charges is discussed in \onlinecite{Smorenburg_2010}. In this respect, we mention here that, in order to investigate strong laser-plasma interactions at intensities exceeding $10^{23}\;\text{W/cm$^2$}$, RR effects have been also implemented in Particle-In-Cell (PIC) codes \cite{Zhidkov_2002,Tamburini_2010,Tamburini_2011} by modifying the Vlasov equation for the electron distribution function according to the LL equation. Specifically, in \onlinecite{Zhidkov_2002} it is shown that in the collision of a laser beam with intensity $I_0=10^{23}\;\text{W/cm$^2$}$ with an overdense plasma slab, about $35\%$ of the absorbed laser energy is converted into radiation and that the effect of RR amounts to about $20\%$. One-dimensional \cite{Tamburini_2010} and three-dimensional \cite{Tamburini_2011} PIC simulations have shown that RR effects strongly depend on the polarization of the driving field: while for circular polarization they are negligible even at $I_0\sim 10^{23}\;\text{W/cm$^2$}$, at those intensities they are important for linear polarization. The simulations also show the beneficial effects of RR in reducing the energy spread of ion beams generated via laser-plasma interactions (see also Sec. \ref{LA}). A different beneficial effect of RR on ion acceleration has been found in \onlinecite{Chen_2011} for the case of a transparent plasma: RR strongly suppresses the backward motion of the electrons, cools them down and increases the number of ions to be bunched and accelerated. Finally, the system in Eq. \eqref{Sok_Eq} has been implemented in a 3D PIC code in \onlinecite{Sokolov_2009} showing that a laser pulse with intensity $10^{22}\;\text{W/cm$^2$}$ loses about $27\%$ of its energy in the collision with a plasma slab. The same system of equations has been employed to study the penetration of ultra-intense laser beams into a plasma in the hole-boring regime \cite{Naumova_2009} and to investigate the process of ponderomotive ion acceleration at ultrahigh laser intensities in overcritical bulk targets \cite{Schlegel_2009}.

\subsection{The classical radiation dominated regime}
As it was already observed in \onlinecite{Landau_b_2_1975}, the fact that the RR force in the LL equation has to be much smaller than the Lorentz force in the instantaneous rest-frame of the electron does not exclude that some components of the two forces can be of the same order of magnitude in the laboratory system. This occurs if the condition $\alpha\gamma^2 F/F_{\text{cr}}\sim 1$ is fulfilled at any instant, with $\gamma$ being the relativistic Lorentz factor of the electron and $F$ the amplitude of the external electromagnetic field. For an ultrarelativistic electron this condition can be fulfilled also in the realm of classical electrodynamics (quantum recoil effects are negligible if $\gamma F/F_{\text{cr}}\ll 1$) and it characterizes the so-called Classical Radiation-Dominated Regime (CRDR). The CRDR has been investigated in \onlinecite{Shen_1970} for a background constant and uniform magnetic field. In \onlinecite{Koga_2005} an equivalent definition of the CRDR in the presence of a background laser field has been formulated, as the regime where the average energy radiated by the electron in one laser period is comparable with the initial electron energy. By estimating the radiated power $\mathcal{P}_L$ from the relativistic Larmor formula $\mathcal{P}_L=-(2/3)e^2(du_{\mu}/ds)(du^{\mu}/ds)$ \cite{Jackson_b_1975} with $du^{\mu}/ds\to(e/m)F^{\mu\nu}u_{\nu}$, one obtains that $\mathcal{P}_L\sim \alpha\chi_0\xi_0\varepsilon_0$. Therefore, the conditions of being in the CRDR are
\begin{align}
\label{R_C}
R_C=\alpha\chi_0\xi_0\approx 1, && \chi_0\ll 1,
\end{align}
where, as has been seen in Sec. \ref{FED_Q}, the second condition ensures in particular that the quantum effects like photon recoil are negligible. The same condition $R_C\approx 1$ has been obtained in \onlinecite{Di_Piazza_2008_a} by exactly solving the LL equation \eqref{LL} for a general plane-wave background field. The analytical solution shows in fact that for an ultrarelativistic electron the main effect of RR is due to the last term in Eq. \eqref{LL}. As a consequence, while the quantity $u_-(\phi)$ is constant if the equation of motion is that due to Lorentz, it decreases here with respect to $\phi$ as $u_-(\phi)=u_{0,-}/h(\phi)$, where $u_0^{\mu}$ is the four-velocity at an initial $\phi_0$ and
\begin{equation}
\label{h}
h(\phi)=1+\frac{2}{3}\frac{R_C}{\omega_0}\int_{\phi_0}^{\phi}d\varphi \left(\frac{d\psi(\varphi)}{d\varphi}\right)^2,
\end{equation}
where the four-potential of the wave has been assumed to have the form $A^{\mu}(\phi)=A_0^{\mu}\psi(\phi)$ (see Sec. \ref{FED_Q}, below Eq. \ref{chi_0}). This effect has been recently suggested in \onlinecite{Harvey_2011b} as a possible signature to measure RR (see also \onlinecite{Lehmann_2011}). The two conditions in Eq. \eqref{R_C} are, in principle, compatible for sufficiently large values of $\xi_0$. For example, for an optical ($\omega_0=1\;\text{eV}$) laser field with an average intensity of $10^{24}\;\text{W/cm$^2$}$ and for an electron initially counterpropagating with respect to the laser field with an energy of $20\;\text{MeV}$, it is $\chi_0=0.16$ and $R_C=1.3$. This example shows that, in general, it is not experimentally easy to enter the CRDR at least with presently-available laser systems. In \onlinecite{Di_Piazza_2009} a different regime has been investigated, which is parametrically less demanding than the CRDR but in which the effects of RR are still large. In this regime the change in the longitudinal (with respect to the laser field propagation direction) momentum of the electron due to RR in one laser period is of the order of the electron's longitudinal momentum itself in the laser field. As a result, it is found that in the ultrarelativistic case and for a few-cycle pulse, if the conditions
\begin{equation}
\label{cond_domin2}
R_C\gtrsim \frac{4\gamma_0^2-\xi_0^2}{2\xi_0^2}>0
\end{equation}
are fulfilled, then the electron is reflected in the laser field only if RR is taken into account (see also \onlinecite{Harvey_2011c} for a recent investigation of the electron's dynamics in the two complementary regimes $2\gamma_0\lessgtr \xi_0$ including RR effects). This can have measurable effects if one exploits the high directionality of the radiation emitted by an ultrarelativistic electron (see Sec. \ref{TCS_General}). The results in \onlinecite{Di_Piazza_2009} show in fact that the apex angle of the angular distribution of the emitted radiation, with and without RR effects included, may differ by more than $10^{\circ}$ already at an average optical laser intensity of $5\times 10^{22}\;\text{W/cm$^2$}$ ($\xi_0\approx 150$) and at an initial electron energies of $40\;\text{MeV}$ ($2\gamma_0\approx 156$) for which $R_C\approx 0.08$. Small RR effects on photon spectra emitted by initially bound electrons had already been predicted via numerical integration of the LL equation in \onlinecite{Keitel_1998} well below the CRDR.

\subsection{Quantum radiation reaction}
\label{QRR}
The shortcomings of the classical approaches to the problem of RR suggest that it can be fully understood only at the quantum level. In the seminal paper \onlinecite{Moniz_1977} the origin of the classical inconsistencies, like the existence of runaway solutions of the LAD equation, were clarified in the nonrelativistic case. The authors first show that such inconsistencies are also absent in classical electrodynamics if one considers charge distributions with a typical radius larger than the classical electron radius $r_0=\alpha\lambda_C\approx 2.8\times 10^{-13}\;\text{cm}$. Going to the nonrelativistic quantum theory and by analyzing the Heisenberg equations of motion of the electron in an external time-dependent field, the authors conclude that the quantum theory of a pointlike particle does not admit any runaway solutions, provided that the external field varies slowly along a length of the order of $\lambda_C$ (this is an obvious assumption in the realm of nonrelativistic theory, as time-dependent fields with typical wavelengths of the order of $\lambda_C$ would in principle allow for $e^+\text{-}e^-$ pair production, see Sec. \ref{PP}). From this point of view a classical theory of RR has only physical meaning as the classical limit ($\hbar\to 0$) of the corresponding quantum theory and the authors indicate that the resulting equation of motion is the nonrelativistic LL equation with the bare mass $m_0$ (it is shown that the electrostatic self-energy of a point charge vanishes in nonrelativistic quantum electrodynamics). On the other hand, if one considers the quantum equations of motion of a charge distribution and performs the classical limit before the pointlike limit, then the classical equations of motion of the charge distribution are, of course, recovered and, once the point-like limit is then performed, runaway solutions appear again. The nonrelativistic form of the LL equation has been also recovered from quantum mechanics in \onlinecite{Krivitskii_1991} by including radiative corrections to the time-dependent electron momentum operator in the Heisenberg representation, and by calculating the time-derivative of the average momentum in a semiclassical state.

The situation in the relativistic theory is less straightforward because relativistic quantum electrodynamics, i.e., QED, is a field theory fundamentally different from classical electrodynamics. The first theory of relativistic quantum RR goes back to W. Heitler and his group \cite{Heitler_b_1984,Jauch_b_1976}. However, the evaluations of the QED amplitudes in Heitler's theory involve the solution of complicated integral equations and it has given a practical result only in the calculation of the total energy emitted by a nonrelativistic quantum oscillator, with and without RR. 

At first sight one would say that RR effects are automatically taken into account in QED, because the electromagnetic field is treated as a collection of photons that take away energy and momentum, when they are emitted by charged particles. However, photon recoil is always proportional to $\hbar$, making it a purely quantum quantity with no classical counterpart. Moreover, if one calculates the spectrum of multiphoton Compton scattering in an external plane-wave field, for example, and then performs the classical limit $\chi_0\to 0$, one obtains the corresponding multiphoton Thomson spectrum calculated via the Lorentz equation and not via the LAD or the LL equation (see also Sec. \ref{TCS_General}). Finally, it has also been seen that in classical electrodynamics the RR effects may not be a small perturbation on the Lorentz dynamics and they cannot be obtained as the result of a single limiting procedure. Otherwise they would always appear as a small correction. 

In order to understand what RR is in QED, it is more convenient to go back to Eq. \eqref{M_L} and to notice that the LAD, namely the LL, equation is equivalent to the coupled system of Maxwell and Lorentz equations. If one determines the trajectory of the electron via the LL equation and then calculates the total electromagnetic field $F_T^{\mu\nu}(x)$ via the Li\'{e}nard-Wiechert four-potential \cite{Landau_b_2_1975}, one has solved completely the classical problem of the radiating electron in the given electromagnetic field. As has been discussed in Sec. \ref{FED_Q}, the solution of the analogous problem in strong-field QED would correspond to completely determine the $S$-matrix in Eq. \eqref{S}, as well as the asymptotic state $|t\to+\infty\rangle$ for the given initial state $|t\to-\infty\rangle=|e^-\rangle$, which represents a single electron. The first-order term in the perturbative expansion of the $S$-matrix corresponds to the process of multiphoton Compton scattering and then, classically, to the Lorentz dynamics. Whereas, all high-order terms give rise to radiative corrections and to high-order coherent and incoherent (cascade) processes, and determine what we call ``quantum RR''. Here, by high-order coherent processes is meant those involving more than one basic QED process (photon emission by an electron/positron or $e^+\text{-}e^-$ photoproduction) but all occurring in the same formation region. Analogously, in higher-order incoherent or cascade processes each basic QED process occurs in a different formation region. Now, in the case of a background plane wave at $\xi_0\gg 1$ and $\chi_0\lesssim 1$, the quantum effects are certainly important but the radiative corrections and higher-order coherent processes scale with $\alpha$ and can be neglected \cite{Ritus_1972}. Also, if $\chi_0$ does not exceed unity then the photons emitted by the electron are mainly unable, by interacting again with the laser field, to create $e^+\text{-}e^-$ pairs, as the pair production probability is exponentially suppressed (see also Secs. \ref{PP} and \ref{Cascade}). Therefore, it can be concluded that at $\xi_0\gg 1$ and $\chi_0\lesssim 1$, RR in QED corresponds to the overall photon recoil experienced by the electron when it emits many photons consecutively and incoherently \cite{Di_Piazza_2010}. 

\begin{figure}
\begin{center}
\includegraphics[width=0.8\linewidth]{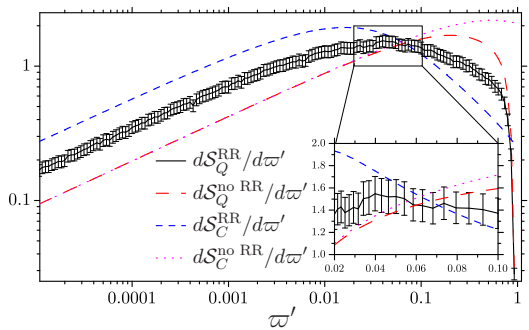}
\end{center}
\caption{(Color online) Quantum photon spectra as a function of $\varpi'=k'_-/p_{0,-}$ calculated with (solid, black line) and without (long dashed, red line) RR and the corresponding classical ones with (short dashed, blue line) and without (dotted, magenta line) RR. The error bars in the quantum spectrum with RR stem from numerical uncertainties in multidimensional integrations. The numerical parameters in our notation are: $\varepsilon_0=1\;\text{GeV}$, $\omega_0=1.55\;\text{eV}$ and $I_0=10^{23}\;\text{W/cm$^2$}$ ($R_Q=1.1$ and $\chi_0=1.8$). Adapted from \onlinecite{Di_Piazza_2010}. }
\label{QRR_Spectrum}
\end{figure}

A qualitative understanding of the above conclusion can be attained by assuming that $\chi_0\ll 1$ and by estimating the average number $N_{\gamma}$ of photons emitted by an electron in one laser period at $\xi_0\gg 1$. Since the probability of emitting one photon in a formation length is of the order of $\alpha$ and since one laser period contains about $\xi_0$ formation lengths \cite{Ritus_1985} then $N_{\gamma}\sim \alpha\xi_0$. Also, the typical energy $\omega'$ of a photon emitted by an electron is of the order $\omega'\sim \chi_0\varepsilon_0$, then the average energy $\mathcal{E}$ emitted by the electron is $\mathcal{E}\sim \alpha\xi_0\chi_0\varepsilon_0=R_C\varepsilon_0$. This estimate is in agreement with the classical result obtained from the LL equation. In other words, the classical limit of RR in this regime corresponds to the emission of a higher and higher number of photons all with an energy much smaller than the electron energy, in such a way that even though the recoil at each emission is almost negligible, the cumulative effect of all photon emissions may have a finite nonnegligible effect. Note that $\omega'$ and $N_{\gamma}$ are both pure quantum quantities and only their product $\mathcal{E}$ has a classical analogue in the limit $\chi_0\to 0$. These considerations have allowed for the introduction in \onlinecite{Di_Piazza_2010} of the Quantum Radiation-Dominated Regime (QRDR), which is characterized by multiple emission of photons already in one laser period. This regime is then characterized by the conditions
\begin{align}
\label{R_Q}
R_Q=\alpha\xi_0\approx 1, && \chi_0\gtrsim 1.
\end{align}
Quantum photon spectra have been calculated numerically in \onlinecite{Di_Piazza_2010} without RR, i.e., by including only the emission of one photon (four-momentum $k^{\prime\mu}$), and with RR, i.e., by including multiple-photon emissions (and by integrating with respect to all the four-momenta of the emitted photons except one indicated as $k^{\prime\mu}$). The results show that in the QRDR the effects of RR are essentially three (see Fig. \ref{QRR_Spectrum}): 1) increase of the photon yield at low photon energies; 2) decrease of the photon yield at high photon energies; 3) shift of the maximum of the photon spectrum towards low photon energies. 
Figure \ref{QRR_Spectrum} also shows that the classical treatment of RR (via the LL equation) artificially overestimates the above effects, the reason being that quantum corrections decrease the average energy emitted by the electron with respect to the classical value \cite{Ritus_1985}. However, at $\chi_0\ll 1$, i.e., when the recoil of each emitted photon is much smaller than the electron energy, then the quantum spectra converge into the corresponding classical ones. As mentioned in Sec. \ref{RR}, a semiclassical phenomenological approach to RR in the quantum regime has been proposed in \onlinecite{Sokolov_2009,Sokolov_2010b}.

Finally, the quantum modifications induced by the electron's self-field onto the Volkov states (see Eq. \eqref{V_S}) have been recently investigated in \onlinecite{Meuren_2011}. It is found that the classical expression of the electron quasimomentum $q_0^{\mu}$ in a linearly polarized plane wave (see Eq. \eqref{q}) admits a correction depending on the quantum parameter $\chi_0$ and also that self-field effects induce a peculiar dynamics of the electron spin.

\section{Vacuum-polarization effects}
\label{VPEs}
\begin{figure}
\begin{center}
\includegraphics[width=5cm]{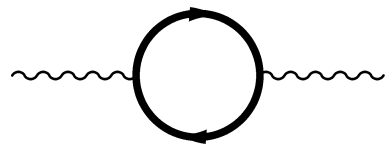}
\end{center}
\caption{Vacuum polarization diagram in an external background electromagnetic field. The thick electron lines indicate electron propagators calculated in the Furry picture, accounting exactly for the presence of the background field.}
\label{VP}
\end{figure}

QED predicts that photons interact with each other also in vacuum \cite{Landau_b_4_1982}. Effects arising from this purely quantum interaction are referred to as vacuum polarization effects. This is in contrast to classical electrodynamics where the linearity of Maxwell's equations in vacuum forbids self-interaction of the electromagnetic field in the vacuum itself. The possibility of photon-photon interaction in vacuum, within the framework of QED, can be understood qualitatively by observing that a photon may locally ``materialize'' into an $e^+\text{-}e^-$ pair which, in turn, interacts with other photons. For the same reason a background electromagnetic field can influence photon propagation (see Fig. \ref{VP} and \onlinecite{Landau_b_4_1982}).
In the latter case the extension $l_f$ of the region where this transformation occurs, i.e., its formation length, depends, in principle, on the structure of the background field \cite{Baier_2005}. However, in some cases it can be estimated qualitatively via the Heisenberg uncertainty principle from the typical momentum $p$ flowing in the $e^+\text{-}e^-$ loop in Fig. \ref{VP}. We consider, for example, a constant background electromagnetic field (or a slowly-varying one, at leading order in the space-time derivatives of the field itself). In this case, if the energy $\omega$ of the incoming photon (see Fig. \ref{VP}) is at most of the order of $m$, then the momentum $p$ flowing in the $e^+\text{-}e^-$ loop is of the order of $m$ and $l_f\sim 1/p\sim\lambda_C$. If $\omega\gg m$ the analysis is more complicated and the formation length strongly depends on the structure of the background field. 

From the theoretical point of view it is convenient to distinguish between low-energy vacuum-polarization effects if $\omega\ll m$ and high-energy ones if $\omega\gtrsim m$.

\subsection{Low-energy vacuum-polarization effects}
\label{Diff_Low}

\begin{figure}
\begin{center}
\includegraphics[width=0.8\linewidth]{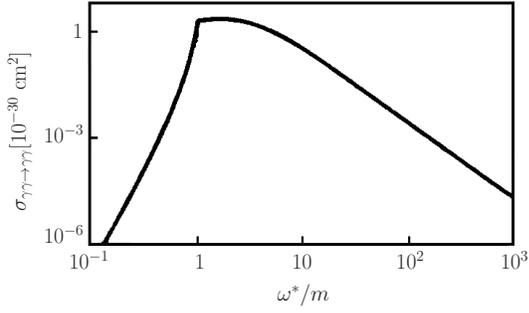}
\end{center}
\caption{Cross section of real photon-photon scattering as a function of the energy of the colliding photons in their center-of-momentum system in units of the electron mass. Adapted from \onlinecite{Landau_b_4_1982}. Copyright Elsevier (1982).}
\label{Gamma_Gamma}
\end{figure}

The scattering in vacuum of a real photon by another real photon is possibly the most fundamental vacuum-polarization process \cite{Landau_b_4_1982} and it has not yet been observed experimentally. The total cross section of the process depends only on the Lorentz-invariant parameter $\eta=(k_1k_2)/m^2$, with $k_1^{\mu}$ and $k_2^{\mu}$ being the four-momenta of the colliding photons or, equivalently, on the energy $\omega^*$ of the two colliding photons in their center-of-momentum system ($\eta=2\omega^{*\,2}/m^2$). This process has been investigated in \onlinecite{Euler_1936_a} in the low-energy limit $\eta\ll 1$ and then in \onlinecite{Akhiezer_1937} in the high-energy limit $\eta\gg 1$. The complete expression of the cross section $\sigma_{\gamma\gamma\to\gamma\gamma}$ was calculated in \onlinecite{Karplus_1951} and can also be found in \onlinecite{Landau_b_4_1982} (see also Fig. \ref{Gamma_Gamma}).
In the low-energy limit $\eta\ll 1$ the cross section $\sigma_{\gamma\gamma\to\gamma\gamma}$ is given by \cite{Landau_b_4_1982}
\begin{align}
\sigma_{\gamma\gamma\to\gamma\gamma}=\frac{973}{81000\pi}\alpha^4\lambda_C^2\eta^3 && \eta\ll 1
\end{align}
and in the high-energy one $\eta\gg 1$ by \cite{Landau_b_4_1982,Baier_1974}
\begin{align}
\begin{split}
\sigma_{\gamma\gamma\to\gamma\gamma}=&\frac{1}{\pi}\left[\frac{108}{5}+\frac{13}{2}\pi^2-8\pi^2\zeta(3)+\frac{148}{225}\pi^4\right.\\
&-24\zeta(5)\Big]\alpha^4\lambda_C^2\frac{1}{\eta},
\end{split}&& \eta\gg 1
\end{align}
where $\zeta(x)$ is the Riemann zeta function \cite{NIST_b_2010}. In terms of the center-of-momentum energy $\omega^*$, the cross section becomes $\sigma_{\gamma\gamma\to\gamma\gamma}[\text{cm$^2$}]=7.4\times 10^{-66}(\omega^*[\text{eV}])^6$ at $\omega^*\ll m$ and $\sigma_{\gamma\gamma\to\gamma\gamma}[\text{cm$^2$}]=5.4\times 10^{-36}/(\omega^*[\text{GeV}])^2$ at $\omega^*\gg m$. The steep dependence of $\sigma_{\gamma\gamma\to\gamma\gamma}$ on $\eta$ for $\eta\ll 1$ is the main reason why real photon-photon scattering has, so-far, eluded experimental observation (see the reviews \onlinecite{Salamin_2006,Marklund_2006} for experiments and experimental proposals until 2005 aiming to observe real photon-photon scattering in vacuum). 

However, various proposals have been put forward recently in order to observe this process by colliding strong laser beams which contain a large number of photons. A common theoretical starting point of all these proposals is the ``effective-Lagrangian'' approach \cite{Dittrich_b_1985,Dittrich_b_2000}. In this approach the interaction among photons in vacuum is described via an effective Lagrangian density of the electromagnetic field. By starting from the total Lagrangian density of the classical electromagnetic field and of the quantum $e^+\text{-}e^-$ Dirac field, one integrates out the degrees of freedom of the latter field and is left with a Lagrangian density depending only on the electromagnetic field. As has been seen above, the formation region of photon-photon interaction at low energies is of the order of $\lambda_C$, therefore if the classical electromagnetic field $F^{\mu\nu}(x)=(\bm{E}(x),\bm{B}(x))$ comprises only wavelengths much larger than $\lambda_C$, the interaction is approximately pointlike and the effective Lagrangian density is accordingly a local quantity. Also, since the effective Lagrangian density is a Lorentz invariant, it can depend only on the electromagnetic field invariants $\mathscr{F}(x)$ and $\mathscr{G}^2(x)$ already introduced in Sec. \ref{FED_Q}. The complete expression of the effective Lagrangian density was reported for the first time in \onlinecite{Heisenberg_1936,Weisskopf_1936} (see also \onlinecite{Schwinger_1951}) and it is known as the Euler-Heisenberg Lagrangian density. Here we are interested only in the experimentally-relevant low-intensity limit $|\mathscr{F}(x)|,|\mathscr{G}(x)|\ll F_{\text{cr}}^2$ and the leading-order Euler-Heisenberg Lagrangian density $\mathscr{L}_{EH}(x)$ reads \cite{Dittrich_b_1985,Dittrich_b_2000}
\begin{equation}
\label{L_EH}
\mathscr{L}_{EH}(x)=-\frac{1}{4\pi}\mathscr{F}(x)+\frac{\alpha}{360\pi^2}\frac{4\mathscr{F}^2(x)+7\mathscr{G}^2(x)}{F^2_{cr}}.
\end{equation}
Different experimental observables have been suggested to detect low-energy vacuum-polarization effects. Those will be reviewed next.

\subsubsection{Experimental suggestions for direct detection of photon-photon scattering}

\begin{figure}
\begin{center}
\includegraphics[width=0.8\linewidth]{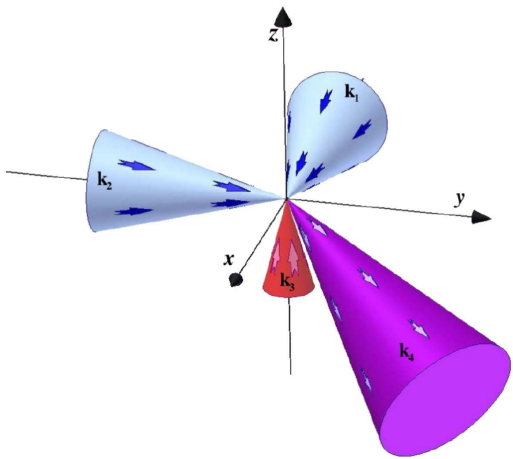}
\end{center}
\caption{(Color) Schematic ``three-dimensional'' setup for laser-assisted photon-photon scattering involving two incoming beams (in blue), an assisting one (in red) and a scattered one (in violet). From \onlinecite{Lundstroem_2006}.}
\label{Gamma_Gamma_Lundstroem}
\end{figure}

The most direct way to search for photon-photon scattering events in vacuum by means of laser fields is to let two laser beams collide and to look for scattered photons. However, by employing a third ``assisting'' laser beam, if one of the final photons is kinematically allowed to be emitted along this beam with the same frequency and polarization, then the number of photon-photon scattering events can be coherently enhanced \cite{Varfolomeev_1966}. In this laser-assisted setup the ``signal'' of photon-photon scattering is, of course, the remaining outgoing photon. In \onlinecite{Lundstroem_2006,Lundin_2007} an experiment has been suggested to observe laser-assisted photon-photon scattering with the Astra-Gemini laser system (see Sec. \ref{L_O}). The authors found a particular ``three-dimensional'' setup, which turns out to be especially favorable for the observation of the process (see Fig. \ref{Gamma_Gamma_Lundstroem}).
The number $N_{\gamma}$ of photons scattered in one shot for an optimal choice of the geometrical factors and of the polarization angles between the incoming and the assisting beams is found to be
\begin{equation}
N_{\gamma}\approx 0.25\frac{P_1[\text{PW}]P_2[\text{PW}]P_3[\text{PW}]}{(\lambda_4[\text{$\mu$m}])^3}.
\end{equation}
Here, $P_1$ and $P_2$ are the powers of the incoming beams, $P_3$ is the power of the assisting beam and $\lambda_4$ is the wavelength of the scattered wave to be measured. By plugging in the feasible values for Astra-Gemini $P_1=P_2=0.1\;\text{PW}$ and $P_3=0.5\;\text{PW}$, one obtains $N_{\gamma}\approx 0.07$, i.e., roughly one photon scattered every 15 shots (for the parameters of Astra Gemini the wavelength of both incoming beams is chosen as $0.4\;\text{$\mu$m}$, that of the assisting beam as $0.8\;\text{$\mu$m}$, so that the wavelength $\lambda_4=0.276\;\text{$\mu$m}$ of the scattered photon is different from those of the incoming and assisting beams).

\begin{figure}
\begin{center}
\includegraphics[width=0.8\linewidth]{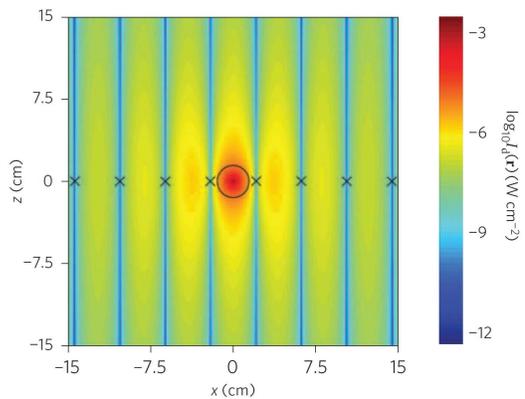}
\end{center}
\caption{(Color) Intensity $I_d$ of the vacuum-scattered wave for a probe beam propagating along the positive  $y$ direction and colliding with two strong beams aligned along the $x$ axis. The crosses correspond to coordinates $x_n$ according to the classical prediction $x_n=(n+1/2)\lambda_pd/D$, where $n$ in an integer number, $\lambda_p$ is the wavelength of the probe field, $d$ is the distance between the interaction region and the observation screen and $D$ is the distance between the centers of the two strong beams. The numerical values of the parameters can be found in \onlinecite{King_2010}. Adapted from \onlinecite{King_2010}.}
\label{DS}
\end{figure}

The quantum interaction among photons in vacuum has been exploited in \onlinecite{King_2010} to propose, for the first time, a double-slit setup comprised only of light (see also \onlinecite{Marklund_2010}). In this setup two strong parallel beams collide head-on with a counterpropagating probe pulse. The photons of the probe have the choice to interact either with one or with the other strong beam, and, when scattered, they are predicted to build an interference pattern with alternating minima and maxima typical of double-slit experiments (see Fig. \ref{DS}). Also, if one of the slits is closed, i.e., if the probe collides with only one strong beam, the interference fringes disappear. 
The key idea behind this setup is that the vacuum-scattered beam (intensity $I_d$), although propagating along the same direction as the probe, has a much wider angular distribution than the latter, offering the possibility of detecting vacuum-scattered photons outside the focus of the undiffracted probe beam. For a strong-field intensity of $I_0\approx 5\times 10^{24}\;\text{W/cm$^2$}$ that may be in the near future be available at ELI or at HiPER and for a probe beam with wavelength $\lambda_p=0.527\;\text{$\mu$m}$ and intensity $I_p=4\times 10^{16}\;\text{W/cm$^2$}$, it is predicted that about four photons per shot would contribute to build up the interference pattern in the observable region (it is the region outside the circle in Fig. \ref{DS}, where $I_d>100\,I_p$). The diffraction of a probe beam in vacuum by a single focused strong laser pulse is also investigated in \onlinecite{Tommasini_2010} in the case of almost counterpropagating beams. For optimal lasers parameters and at a strong-laser power of $100\;\text{PW}$ the diffracted vacuum signal is predicted to be measurable in a single shot. The effects on photon-photon scattering of the temporal profile of the laser pulses have been recently investigated in \onlinecite{King_2012}, showing a suppression of the number of vacuum scattered photons with respect to the infinite-pulse (monochromatic) case.

The concept of Bragg scattering has been exploited in \onlinecite{Kryuchkyan_2011} to observe the scattering of photons by a modulated electromagnetic-field structure in vacuum. If a probe wave passes through a series of parallel strong laser pulses and if the Bragg condition on the impinging angle is fulfilled, the number of diffracted photons can be strongly enhanced. At a fixed intensity for each strong beam the enhancement factor with respect to laser-assisted photon-photon scattering is equal to the number of beams in the periodic structure. However, in experiments usually the total energy of the laser beams is fixed and an enhancement by a factor of two is predicted. By considering $N_G$ equal Gaussian pulses propagating along the $x$ direction and their centers separated by a distance $D > 2w_z$ from each other, the resulting photon-photon scattering probability will be proportional to the phase-matching factor $\mathscr{P}$, with
\begin{equation}
\mathscr{P}=\frac{\sin^2(\delta k_z N_GD/2)}{\sin^2(\delta k_z D/2)},
\end{equation}
where the vector $\delta \bm{k}=\bm{k}_2-\bm{k}_1$ is the difference between the wave vectors of the reflected and incident waves. The Bragg condition is satisfied for $\delta k_z=2\pi l/D$, with $l$ being an integer. By employing ten optical laser beams with a wavelength of $1\;\text{$\mu$m}$ and each with an intensity of $2.3\times 10^{23}\;\text{W/cm$^2$}$, about five vacuum-scattered photons are predicted per shot. Finally, an enhancement of vacuum-polarization effects in laser-laser collision has been predicted in \onlinecite{Monden_2011} by employing strong laser beams with large angular aperture. For example, it is predicted that the number of vacuum-radiated photons will be enhanced by two orders of magnitude, if the angular aperture of the colliding beams is increased from $53^{\circ}$ to $103^{\circ}$.

Other experimental suggestions to measure photon-photon scattering in vacuum can be found in \onlinecite{Eriksson_2004,Tommasini_2008}.

\subsubsection{Polarimetry-based experimental suggestions}

The expression of the Euler-Heiseberg Lagrangian density in Eq. \eqref{L_EH}, suggests to interpret a region where only an electromagnetic field is present as a material medium characterized by a polarization $\bm{P}_{EH}=\partial \mathscr{L}_{EH}/\partial\bm{E}-\bm{E}/4\pi$ and a magnetization $\bm{M}_{EH}=\partial \mathscr{L}_{EH}/\partial\bm{B}+\bm{B}/4\pi$ \cite{Jackson_b_1975} given by
\begin{align}
\label{P_EH}
\bm{P}_{EH}&=\frac{\alpha}{180\pi^2 F_{\text{cr}}^2}\left[2(E^2-B^2)\bm{E}+7(\bm{E}\cdot\bm{B})\bm{B}\right],\\
\label{M_EH}
\bm{M}_{EH}&=\frac{\alpha}{180\pi^2 F_{\text{cr}}^2}\left[2(B^2-E^2)\bm{B}+7(\bm{E}\cdot\bm{B})\bm{E}\right].
\end{align}
Note that an arbitrary single plane wave cannot ``polarize'' the vacuum, as in this case $\bm{P}_{EH}$ and $\bm{M}_{EH}$ identically vanish. Equations \eqref{P_EH} and \eqref{M_EH} indicate that the presence of an electromagnetic field in the vacuum alters the vacuum's refractive index. The situation is even more complicated because of the vectorial nature of the background electromagnetic field which polarizes the vacuum and introduces a privileged direction in it. As a result, the vacuum's refractive index is altered in a way that depends, in general, on the mutual polarizations of the probe electromagnetic field and of the background field: the polarized vacuum behaves as a birefringent medium. For example, in the case of an arbitrary constant electromagnetic field $(\bm{E},\bm{B})$, the refractive indices $n_{EH,1/2}$ of a wave propagating along the direction $\bm{n}$ and polarized along one of the two independent directions $\bm{u}_1=\bm{\mathcal{E}}/|\bm{\mathcal{E}}|$ and $\bm{u}_2=\bm{\mathcal{B}}/|\bm{\mathcal{B}}|$, with $\bm{\mathcal{E}}=\bm{E}-(\bm{n}\cdot\bm{E})\bm{E}+\bm{n}\times\bm{B}$ and $\bm{\mathcal{B}}=\bm{B}-(\bm{n}\cdot\bm{B})\bm{B}-\bm{n}\times\bm{E}$ are given by \cite{Dittrich_b_2000}
\begin{align}
\label{n_EH_1}
n_{EH,1}&=1+\frac{4\alpha}{90\pi}\frac{(\bm{n}\times\bm{E})^2+(\bm{n}\times\bm{B})^2-2\bm{n}\cdot(\bm{E}\times\bm{B})}{F^2_{\text{cr}}},\\
\label{n_EH_2}
n_{EH,2}&=1+\frac{7\alpha}{90\pi}\frac{(\bm{n}\times\bm{E})^2+(\bm{n}\times\bm{B})^2-2\bm{n}\cdot(\bm{E}\times\bm{B})}{F^2_{\text{cr}}},
\end{align}
respectively.

The birefringence of the polarized vacuum is exploited in \onlinecite{Heinzl_2006} to show that if a linearly-polarized probe x-ray beam (wavelength $\lambda_p$) propagates along a strong optical standing wave, then it emerges from the interaction elliptically polarized with ellipticity $\epsilon$ given by
\begin{equation}
\epsilon=\frac{2\alpha}{15}\kappa\frac{l_{0,R}}{\lambda_p}\frac{I_0}{I_{\text{cr}}},
\end{equation}
where $\kappa\sim 1$ is a geometrical factor and $l_{0,R}$ is the Rayleigh length of the intense laser beam. If this beam is generated by a laser like ELI ($I_0\sim 10^{25}\;\text{W/cm$^2$}$), values of the ellipticities of the order of $10^{-7}$ are predicted at $\lambda_p=0.1\;\text{nm}$. Recent advances on x-ray polarimetry allow for measurement of ellipticities of the order of $10^{-9}$ at a wavelength of $0.2\;\text{nm}$ \cite{Marx_2011}. In \onlinecite{Ferrando_2007} a phase-shift has been found theoretically to be induced by vacuum-polarization effects when two laser beams cross in the vacuum, which is predicted to be measurable at laser intensities available at ELI or at HiPER.

When an electromagnetic wave with wavelength $\lambda$ impinges upon a material body, the features of the scattered radiation depend on the so-called diffraction parameter $\mathscr{D}=l_{\perp}^2/\lambda d$ \cite{Jackson_b_1975}. Here, $l_{\perp}$ is the spatial dimension of the body perpendicular to the propagation direction of the incident wave and $d$ is the distance of the screen, where the radiation is detected, from the interaction region. The near region, $\mathscr{D}\gg 1$, is known as the ``refractive-index limit'', because the effects of the presence of the body can be described as if the wave propagates through a medium with a given refractive index. However, if $\mathscr{D}\lesssim 1$ then the diffraction effects become important and description of the wave-body interaction only in terms of a refractive index is in general not possible. This aspect has been pointed out in \onlinecite{Di_Piazza_2006} within the context of light-light interaction in vacuum (see also \onlinecite{Di_Piazza_2007b}). Tight focusing required to reach high intensities usually renders the interaction region so small that diffraction effects may become substantial at typical experimental conditions. In some cases diffractive effects reduce by an order of magnitude the values of the ellipticity calculated via the refractive-index approach and also induce a rotation of the main axis of the polarization ellipse with respect to the initial polarization direction of the probe field \cite{Di_Piazza_2006}. Tight focusing of the strong polarizing beam requires quite a detailed mathematical description employing a realistic focused Gaussian beam, while a simpler description was employed for the usually weakly-focused probe beam. This prevented the applicability of the results in the so-called far region where $\mathscr{D}\ll 1$ and where the spatial spreading of the probe field is also important. This assumption has been recently removed in \onlinecite{King_2010_a}, where it was also pointed out that by considering the diffraction of a probe beam by two separated beams instead of that by a single standing wave, an increase in the ellipticity and in the rotation of polarization angle by a factor 1.5 is expected. 

A different method based on the phase-contrast Fourier imaging technique has been suggested in \onlinecite{Homma_2011} to detect vacuum birefringence. This technique provides a very sensitive tool to measure the absolute phase shift of a probe beam when it crosses an intense laser field. Numerical simulations demonstrate the feasibility of measuring vacuum birefringence also by employing an optical probe field and a 100-PW strong laser beam.

Photon ``acceleration'' in vacuum due to vacuum polarization has been studied in \onlinecite{Mendonca_2006}. This effect corresponds to a shift of the photon frequency when it passes through a strong electromagnetic wave. If $k^{\mu}_p$ is the four-momentum of a probe photon with energy $\omega_p$ when it enters a region where a strong laser beam is present, then, due to vacuum-polarization effects, $\omega_p$ becomes sensitive to the gradient of the intensity of the strong beam. As a result, a frequency up-shift (down-shift) is predicted at the rear (front) of the strong beam.

According to Eqs. \eqref{n_EH_1} and \eqref{n_EH_2} the phase velocity of light in vacuum is smaller than unity. This circumstance has been exploited in \onlinecite{Marklund_2005}, where Cherenkov radiation by ultrarelativistic particles moving with constant velocity in a photon gas has been predicted, if the speed of the particle exceeds the phase velocity of light. Finally, in \onlinecite{Zimmer_2012} an induced electric dipole moment of the neutron has been proposed as a signature of the polarization of the QED vacuum.

\subsubsection{Low-energy vacuum-polarization effects in a plasma}

Equations \eqref{n_EH_1} and \eqref{n_EH_2} indicate that vacuum-polarization effects elicited by a plane wave with intensity $I_0$ would alter the vacuum refractive index by an amount of the order of $(\alpha/45\pi)(I_0/I_{\text{cr}})$. It was first realized in \onlinecite{Di_Piazza_2007_a} that this aspect can be in principle significantly improved in a plasma. For the sake of simplicity the case of a cold plasma was considered and the vacuum-polarization effects were implemented in the inhomogeneous Maxwell's equations as an additional ``vacuum four-current'' \cite{Di_Piazza_2007_a}. Now, unlike in the vacuum, the field invariant $\mathscr{F}(x)$ for a single monochromatic circularly polarized plane wave does not vanish in a plasma. Thus, vacuum-polarization effects in a plasma already arise in the presence of a single traveling plane wave. In \onlinecite{Di_Piazza_2007_a} the vacuum-corrected refractive index $n$ of a two-fluid electron-ion plasma (ion mass, density and charge number given by $m_i$, $n_i$ and $Z$, respectively) in the presence of a circularly polarized plane wave has been found as
\begin{equation}
\label{n}
n=\sqrt{n_0^2+\frac{2\alpha}{45\pi}\frac{I_0}{I_{\text{cr}}}(1-n_0^2)^2},
\end{equation}
where
\begin{equation}
n_0=\sqrt{1-\frac{4\pi e^2n_i}{m\omega_0^2}\left(\frac{1}{\sqrt{1+\xi_0^2}}+\frac{Z}{\sqrt{(m_i/m)^2+Z^2\xi_0^2}}\right)},
\end{equation}
is the refractive index of the plasma without vacuum-polarization effects  \cite{Mulser_b_2010}. Equation \eqref{n} already indicates the possibility of enhancing the effects of vacuum polarization by working at laser frequencies $\omega_0$ such that $n_0\ll 1$, i.e., close to the effective plasma critical frequency. This region of parameters is in general complex to investigate, due to the arising of different instabilities. However, the idealized situation investigated in \onlinecite{Di_Piazza_2007_a} shows, at least in principle, the possibility of enhancing the vacuum-polarization effects by an order of magnitude at a given intensity $I_0$ with respect, for example, to the results in \onlinecite{Di_Piazza_2006}. The effects of the presence of an additional strong constant magnetic field have been analyzed in \onlinecite{Lundin_2007}. The general theory presented in this paper covers different waves propagating in a plasma as Alfv\'{e}n modes, whistler modes, and large-amplitude laser modes. We also mention the recent paper \onlinecite{Bu_2010}, in which the photon acceleration process has been investigated in a cold plasma and the reference \onlinecite{Brodin_2007}, where vacuum-induced photon splitting in a plasma is studied. Finally, we mention the possibility of testing nonlinear vacuum QED effects in waveguides. In \onlinecite{Brodin_2001} the generation of new modes in waveguides due to vacuum-polarization effects has been predicted. More recently signatures of nonlinear QED effects in the transmitted power along a waveguide have been analyzed in \onlinecite{Ferraro_2010}.

\subsection{High-energy vacuum-polarization effects}
\label{Diff_High}
Generally speaking the treatment of vacuum-polarization effects for an incoming photon with energy $\omega$ in the presence of a background electromagnetic field with a typical angular frequency $\omega_b$ cannot be performed in an effective Lagrangian approach if $\omega\omega_b/m^2\gtrsim 1$: the incoming photon ``sees'' the nonlocality of its interaction with the background electromagnetic field through its local ``transformation'' into an $e^+\text{-}e^-$ pair (see Fig. \ref{VP}). The technical difficulty in treating vacuum-polarization effects at high energies arises from the fact that the interaction between the virtual $e^+\text{-}e^-$ pair and the background field has to be accounted for exactly. This has been accomplished for the background field of a nucleus with charge number $Z$ such that $Z\alpha\sim 1$ and Delbr\"{u}ck scattering and photon splitting in such a field have also been observed experimentally (see the reviews \onlinecite{Milstein_1994,Lee_2003}). 

As we have recalled in Sec. \ref{FED_Q}, the Dirac equation in a background plane wave described by the four-vector potential $A^{\mu}(\phi)$ can be solved exactly and analytically. Accordingly, the exact electron (Volkov) propagator $G(x,y|A)$ in the same background field has also been determined (see, e.g., \onlinecite{Ritus_1985}). The so-called ``operator technique'', developed in \onlinecite{Baier_1976_a,Baier_1976_b}, turns out to be very convenient for investigating vacuum-polarization effects at high energies (the operator technique for a constant background field was developed in \onlinecite{Schwinger_1951,Baier_1975_a,Baier_1975_b}). In this technique a generic electron state $\Psi(x)$ in a plane-wave field and the propagator $G(x,y|A)$ are intended as the configuration representation of an abstract state $|\Psi\rangle$ and of an operator $G(A)$ such that $\Psi(x)=\langle x|\Psi\rangle$ and $G(x,y|A)=\langle x|G(A)|y\rangle$. In particular, since the propagator $G(x,y|A)$ is the solution of the equation $\{\gamma^{\mu}[i\partial_{\mu}-eA_{\mu}(\phi)]-m\}G(x,y|A)=\delta(x-y)$, then the abstract operator $G(A)$ is simply
\begin{equation}
G(A)=\frac{1}{\gamma^{\mu}[P_{\mu}-eA_{\mu}(\phi)]-m},
\end{equation}
with $P^{\mu}$ being the four-momentum operator. Evaluation via the operator technique of the matrix element corresponding to a generic vacuum-polarization process is then carried out by manipulating abstract operators, which is easier than by working with the corresponding quantities in configuration space.

In \onlinecite{Di_Piazza_2007} the operator technique has been employed to calculate the rate of photon splitting in a strong laser field for an incoming photon with four-momentum $k^{\mu}$. This was the first investigation of a QED process involving three Volkov propagators. The calculated rate is valid for an arbitrary plane-wave field, provided that radiative corrections can be neglected, i.e., at $\alpha\varkappa_0^{2/3}\ll 1$, with $\varkappa_0=(k_-/m)(E_0/F_{\text{cr}})$, in the most unfavorable regime $\xi_0,\varkappa_0\gg 1$ \cite{Ritus_1972}. In fact, as we have mentioned in Sec. \ref{FED_Q}, QED processes involving the collision of a photon and an intense plane wave are controlled by the two Lorentz- and gauge-invariant parameters $\xi_0$ and $\varkappa_0$. In \onlinecite{Di_Piazza_2007} it turned out to be more convenient to perform a parametric study of the photon-splitting rate by varying the two parameters $\xi_0$ and $\eta_0=\omega_0k_-/m^2$ (note that $\varkappa_0=\eta_0\xi_0$). By employing the Furry theorem \cite{Landau_b_4_1982}, it is shown that photon splitting in a laser field only occurs with absorption of an odd number of laser photons. In particular, if the strong field is circularly polarized and if it counterpropagates with respect to the incoming photon, conservation of the total angular momentum along the propagation direction of the beams implies that for $\eta_0\ll 1$ photon splitting can occur only via absorption of one or of three laser photons. 

In \onlinecite{Di_Piazza_2008} a physical scenario has been advanced in which nonperturbative vacuum-polarization effects can be in principle observed (here ``nonperturbative'' means ``high-order'' in the quantum nonlinearity parameter, see below). In this scenario a high-energy proton collides head-on with a strong laser field. The quantum interaction of the Coulomb field of the proton with the laser field allows for a merging of laser photons into a single high-energy photon. The use of a proton, instead of an electron, for example, is required in order to suppress the background process of multiphoton Thomson/Compton scattering, where again many photons of the laser can be directly absorbed by the proton and converted into a single high-energy photon (see Sec. \ref{TCS_General}). In fact, the kinematics of the two processes, vacuum-mediated laser photon merging and multiphoton Thomson/Compton scattering is the same, except that in the treatment in \onlinecite{Di_Piazza_2008} only an even number of laser photons can merge in the vacuum-mediated process. The probability of $\ell$-photon Thomson/Compton scattering of a particle with charge $Q$ and mass $M$ depends on the parameter $\xi_{0,c}=|Q|E_0/M\omega_0$ and scales as $\xi_{0,c}^{2\ell}$ at $\xi_{0,c}\ll 1$. Thus, the use of a ``heavy'' particle like a proton (mass $m_p=1.8\times 10^3\,m=938\;\text{MeV}$) is essential to suppress this background process. Note that in order to have $\xi_{0,p}=|e|E_0/m_p\omega_0\approx 1$ for a proton, a laser field with intensity of the order of $10^{24}\;\text{W/cm$^2$}$ is required. It is found that for an optical background field such that $\xi_0\gg 1$, the amplitude of the $(2\ell)$-photon merging process depends only on the nonlinear quantum parameter 
\begin{equation}
\chi^{(2\ell)}_{0,p}=\frac{E_0}{F_{\text{cr}}}\frac{2\ell(1+v_p)\omega_0}{m}\frac{1-\cos\vartheta}{1+v_p\cos\vartheta},
\end{equation}
where $v_p$ is the proton velocity and $\vartheta$ is the angle between the direction of the emitted photon and the propagation direction of the plane wave. By colliding a proton beam, of energy available at the Large Hadron Collider (LHC) of the order of $7\;\text{TeV}$ \cite{LHC} with an optical laser beam of intensity of $3\times 10^{22}\;\text{W/cm$^2$}$, it is demonstrated that the vacuum-mediated merging of two laser photons and the analogous two-photon Thomson/Compton scattering have comparable rates, implying that the inclusive signal should be twice the one expected without vacuum-polarization effects. By accounting for the details of the proton beams available at LHC and of the laser system PFS (see Sec. \ref{L_O}), about 670 two-photon merging events and about 5 four-photon merging events are expected per hour. In the discussed setup most of the photons are emitted almost in the direction of the proton velocity ($\vartheta\approx \pi$) in which $\chi^{(2)}_{0,p}\sim 1$. It is also indicated in \onlinecite{Di_Piazza_2008} that the use of the perturbative expression of the laser-photon merging rate at leading order in $\chi^{(2)}_{0,p}\ll 1$ would lead to an error of about $30\%$. Other setups for observing vacuum-polarization effects in laser-proton collisions have been discussed in \onlinecite{Di_Piazza_2008d}, involving, for example, XFEL or single intense XUV pulses. Finally, the process of Delbr\"{u}ck scattering in a combined Coulomb and laser field has been studied in \onlinecite{Di_Piazza_2008c}. Here an incoming photon is scattered by the Coulomb field of a nucleus and by a strong laser field. While the presence of the laser field is taken into account exactly in the calculations, only leading-order effects in the nuclear parameter $Z\alpha$ are accounted for. Analogously to \onlinecite{Di_Piazza_2008}, it is found that high-order nonlinear corrections in the parameter $\varkappa_0$ to the cross section of the process already become important at $\varkappa_0\approx 0.2$. For example, these corrections amount to about $50\%$ at $\varkappa_0=0.35$.

\section{Electron-positron pair production}
\label{PP}
\begin{figure}
   \centering
   \includegraphics[width=0.8\linewidth]{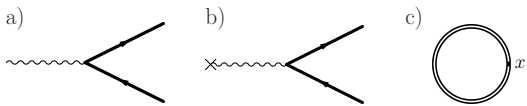}
   \caption{Feynman diagrams corresponding to processes (i) (part a)), (ii) (part b)) and (iii) (part c)), respectively. The thick continuous lines in parts a) and b) indicate Volkov positive- and negative-energy states. The crossed vertex in part b) stands for the Coulomb electromagnetic field. The diagram in part c) is related to the vacuum current $j_{\text{vac}}^{\mu}(x)$, that one has to determine in order to calculate the $e^+\text{-}e^-$ pair yield \cite{Dittrich_b_2000}. The double line indicates the electron propagator calculated in the Furry picture including exactly the background standing wave.}
   \label{PP_13}
\end{figure}

One of the most important predictions of QED has been the possibility of transforming light into matter \cite{Dirac_1928}. If two photons with four-momenta $k_1^{\mu}$ and $k_2^{\mu}$ collide at an angle such that the parameter $\eta=(k_1k_2)/m^2$ exceeds two, the creation of an $e^+\text{-}e^-$ pair becomes kinematically allowed (Breit-Wheeler $e^+\text{-}e^-$ pair production \cite{Breit_1934}). Shortly after the realization of the laser in 1960, theoreticians started to study possibilities for the creation of $e^+\text{-}e^-$ pairs from vacuum by very strong laser fields \cite{Reiss_1962, Nikishov_1964_a, Yakovlev_1966}. Because of constraints from energy-momentum conservation, a single plane-wave laser field cannot create pairs from vacuum, no matter how intense it is. For a single plane wave, in fact, all the photons propagate along the same direction and the parameter $\eta$ vanishes identically for any pair of photons in the plane wave. Thus, an additional source of energy is therefore required to trigger the process of pair production in a plane wave. There are essentially three different possibilities (see also Fig. \ref{PP_13}):
\begin{itemize}
\item[(i)] pair production by a high-energy photon propagating in a strong laser field (multiphoton Breit-Wheeler pair production);
\item[(ii)] pair production by a Coulomb field in the presence of a strong laser field;
\item[(iii)] pair production by two counterpropagating strong laser beams forming a standing light wave.
\end{itemize}

These processes share the common feature to possess different interaction regimes which are mainly characterized by the value of the parameter $\xi_0$. When $\xi_0\ll 1$ the presence of the laser field can be taken into account perturbatively and this yields a pair production rate $R$ of the form $R_{e^+\text{-}e^-}\sim m\xi_0^{2\ell_m}$, where $\ell_m$ is the minimum integer number which kinematically allows the process. For the process (i) it is $\ell_m(k_0k)> 2m^2$, with $k_0^{\mu}$ and $k^{\mu}$ being the four-momentum of the laser photon and the incoming photon, respectively. For the process (ii) it is $\ell_m\omega^{\star}_0>2m$, where $\omega^{\star}_0=\omega_0 u_{c,-}$ is the laser angular frequency in the rest-frame of the charge which produces the Coulomb field and which has four-velocity $u_c^{\mu}$. Finally, for the process (iii) it is $\ell_m\omega_0>2m$. Due to the specific dependence of the pair production rate on $\xi_0$, this regime of pair production is called multiphoton regime. In contrast, when $\xi_0\gg 1$ the presence of the laser field has to be taken into account exactly by performing the calculations in the Furry picture. As the condition $\xi_0\gg 1$ is realized for vanishing laser frequencies at a fixed laser amplitude, this regime is called quasistatic regime and the pair-production rate here is governed by a different parameter which depends on the process at hand. For process (i), for example, the form of the rate depends on the physical parameter $\varkappa_0$ introduced in Sec. \ref{FED_Q} and it scales as $\sim m\varkappa^{3/2}_0\exp(-8/3\varkappa_0)$ if $\varkappa_0\ll 1$ and as $\sim m\varkappa_0^{2/3}$ if $\varkappa_0\gg 1$ \cite{Reiss_1962,Nikishov_1964_a}. For process (ii), we distinguish the case in which the incoming particle is an electron, from that in which it is a heavier particle like a nucleus with charge number $Z$. In the first case the pair production rate depends on the parameter $\chi_0$, already introduced in Sec. \ref{FED_Q}, and the recoil due to the pair creation on the electron has to be taken into account (see also Sec. \ref{PP_L}). In the second case the motion of the nucleus is usually assumed not to be altered by the pair creation process and the nucleus itself is described as a background Coulomb field (see also Sec. \ref{BH}). The pair-production rate depends on the parameter $\chi_{0,n}=u_{n,-}(E_0/F_{\text{cr}})$, with $u^{\mu}_n$ being the four-velocity of the nucleus, and on the nuclear parameter $Z\alpha$. Specifically, the pair-production rate scales as $m(Z\alpha)^2\exp(-2\sqrt{3}/\chi_{0,n})$ if $\chi_{0,n}\ll 1$ and as $m(Z\alpha)^2\chi_{0,n}\ln \chi_{0,n}$ if $\chi_{0,n}\gg 1$ \cite{Yakovlev_1966,Milstein_2006}. Finally, for process (iii), the rate $R_{e^+\text{-}e^-}$ has been derived mainly by approximating the standing wave as an oscillating electric field (see also Sec. \ref{PP_SW}). It is found that $R_{e^+\text{-}e^-}$ depends on the ratio $\Upsilon_0=E_0/F_{\text{cr}}$, with $E_0$ being the amplitude of the standing wave in the (fixed) laboratory frame, and that it scales as $m\Upsilon_0^2\exp(-\pi/\Upsilon_0)$ if $\Upsilon_0\ll 1$ and as $m\Upsilon_0^2$ if $\Upsilon_0\gg 1$ \cite{Brezin_1970,Popov_1971,Popov_1972}. As expected, these scalings coincide with the corresponding ones in a constant electric field $E_0$ \cite{Schwinger_1951}.

The physical meaning of the three parameters $\varkappa_0$, $\chi_{0,n}$ and $\Upsilon_0$ can be qualitatively understood in the following way. For process (i) the dressing of the electron and positron mass (see Sec. \ref{FED_Q}) modifies the threshold of $e^+\text{-}e^-$ pair production at $\xi_0\gg 1$ according to $\ell_m(k_0k)\gtrsim 2m^2\xi_0^2$. Now, analogously to multiphoton Thomson and Compton scattering (see Sec. \ref{TCS_General}), the typical number of laser photons absorbed in pair production via photon-laser collision is of the order of $\xi_0^3$ \cite{Nikishov_1964_a} and the threshold condition becomes $\varkappa_0\gtrsim 1$. Concerning the process (ii), the appearance of the parameter $\chi_{0,n}$ in the quasistatic limit $\xi_0\gg 1$ can be understood by noting that the quantity $u_{n,-}E_0$ is the amplitude of the laser field in the rest-frame of the nucleus and that a constant and uniform electric field with strength of the order of $F_{\text{cr}}=m^2/|e|$ supplies a $e^+\text{-}e^-$ pair with its rest energy $2m$ along the typical length scale of QED $\lambda_C=1/m$ (see also Sec. \ref{FED_Q}). This last observation also demonstrates the presence of the parameter $\Upsilon_0$ for process (iii).  The typical exponential scaling of the pair production rate for $\xi_0\gg 1$, and at $\varkappa_0\ll 1$, $\chi_{0,n}\ll 1$ and $\Upsilon_0\ll 1$ for processes (i), (ii) and (iii), respectively, is reminiscent of a quantum tunneling process. Thus, one refers to this regime also as tunneling pair production (note, however, that the notion of ``tunneling'' in laser-induced processes should be regarded with special care beyond the dipole approximation \cite{Reiss_2008,Klaiber_2012}).

In all processes discussed above, the laser field is always participating directly in the pair creation step and fundamental properties of the quantum vacuum under extreme high-field conditions are probed. However, as will be seen shortly, lying at the border of experimental feasibility, the expected pair yields are generally rather small. It is worth mentioning here that lasers can also be applied for abundant generation of $e^+\text{-}e^-$ pairs \cite{Chen_2009, Chen_2010}. When a solid target is irradiated by an intense laser pulse, a plasma is formed and electrons are accelerated to high energies. They may emit radiation by bremsstrahlung which efficiently converts into $e^+\text{-}e^-$ pairs through the Bethe-Heitler process. The laser field plays an indirect role in the pair production here by serving solely as a particle accelerator. The prolific amount of antimatter generated this way may lead to interesting applications in various fields of science \cite{NaturePhot}. Abundant production of $e^+\text{-}e^-$ pairs and of high-energy photons in the collision of a multipetawatt laser beam and a solid target has been recently investigated in \onlinecite{Nakamura_2011} and \onlinecite{Ridgers_2012}. In particular, in \onlinecite{Nakamura_2011} it has been shown that almost all the laser pulse energy is converted after the collision into a well collimated high-power gamma-ray flash. Whereas, the numerical simulations in \onlinecite{Ridgers_2012} indicate that about $35\;\%$ of the energy of a $10\;\text{PW}$ laser pulse after the laser-target interaction is converted into a gamma-ray burst and that simultaneously a pure $e^+\text{-}e^-$ plasma is produced with a maximum positron density of $10^{26}\;\text{m$^{-3}$}$.

\subsection{Pair production in photon-laser and electron-laser collisions}
\label{PP_L}

Among the pair-production processes mentioned above, only laser-induced pair production for $\xi_0 < 1$ has been observed experimentally. Its feasibility has been shown, in fact, in the pioneering E-144 experiment at SLAC \cite{SLAC_PP,SLAC_PRD} (see \onlinecite{Reiss_1971} for a corresponding theoretical proposal). The experiment relied on collisions of the 46.6 GeV  electron beam from SLAC's linear accelerator with a counterpropagating intense laser pulse of photon energy of $\omega_0=2.4\;\text{eV}$ and intensity $1.3\times 10^{18}$\;W/cm$^2$ ($\xi_0\approx 0.4$). In the rest-frame of the electrons, the laser intensity and frequency are largely Doppler up-shifted to the required level and the pair generation probability is effectively enhanced. In principle both reactions (i) and (ii) contribute to pair production in this kind of collisions. Based on separate simulations of both production channels, reaction (i) was found to dominate. The high-energy photon originates from multiphoton Compton backscattering of a laser photon off the electron beam.

Despite the significance of the SLAC experiment, a unified description of pair creation in electron-laser collisions has been presented only recently, treating the competing mechanisms (i) and (ii) within the same formalism \cite{Hu_2010}. Good agreement with the experimental results has been obtained. Moreover, it was shown that the SLAC study observed the onset of nonperturbative pair creation dynamics, which adds even further significance to this benchmarking experiment (see also \onlinecite{Reiss_2009}). A formal treatment of the process has also been given in \onlinecite{Ilderton_2011}, where special emphasis is put on effects stemming from the finite duration of the laser pulse.

\begin{figure}
   \centering
   \includegraphics[width=0.8\linewidth]{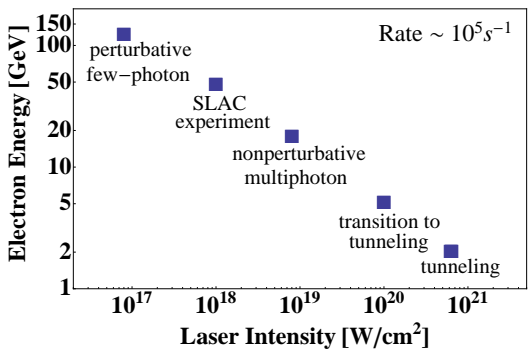}
   \caption{(Color online) Transition from the perturbative to the fully nonperturbative regimes of $e^+\text{-}e^-$ pair creation in electron-laser collisions. The laser photon energy is $2.4\;\text{eV}$. Adapted from \onlinecite{Hu_2010}.}
   \label{fig:2}
\end{figure}

Figure \ref{fig:2} shows a survey of various combinations of incoming electron energies and optical laser intensities which give rise to an observable pair yield. It covers the range from the perturbative few-photon regime ($\xi_0\approx 0.1$ at $I_0\approx 10^{17}\;\text{W/cm$^2$}$) to the highly nonperturbative domain ($\xi_0\approx 10$ at $I_0\approx 10^{21}\;\text{W/cm$^2$}$), where the contributions from thousands of photon absorption channels need to be included. We note that few-GeV electron beams can be produced today using compact laser-plasma accelerators \cite{Leemans_2006} (see also Sec. \ref{LAP}). Future pair creation studies may therefore rely on all-optical setups, where a laser-generated electron beam collides with a counterpropagating laser pulse. Another all-optical setup for pair creation by a seed electron exposed to two counterpropagating laser pulses was put forward in \onlinecite{Bell_2008}, which will be discussed in Sec. \ref{Cascade}.

Pair creation studies could also be conducted as a nonstandard application of the 17.5 GeV electron beam at the upcoming European XFEL beamline at DESY \cite{XFEL}, which will normally serve to generate coherent x-ray pulses. However, in combination with a table-top 10-TW optical laser system, it would also be very suitable to probe the various regimes of pair production. In particular, the production channel (ii) could be investigated by a suitable choice of beam parameters \cite{Hu_2010}.

Other aspects of pair creation by a high-energy photon and a strong laser field have been investigated in recent years. In \onlinecite{Heinzl_2010b}, process (i) was considered in the case where the laser pulse has finite duration. It was found that the finite pulse duration is imprinted on the spectra of created particles. Pair production by a high-energy photon and an ultrashort laser pulse was also considered in \onlinecite{Tuchin_2010}. Quantum interference effects can arise in photon-induced pair creation in a two-mode laser field of commensurate frequencies \cite{Narozhny_2000}.

In addition, the fundamental process (i) may allow for applications as a novel tool in ultrashort pulse spectrometry. A corresponding detection scheme for the characterization of short gamma-ray pulses of GeV photons down to the zeptosecond scale, called Streaking at High Energies with Electrons and Positrons (SHEEP),  has been proposed in \onlinecite{Ipp_2011}. The basic concept of SHEEP is based on $e^+\text{-}e^-$ pair production in vacuum by a photon of the test pulse, assisted by an auxiliary counter-propagating intense laser pulse. In contrast to conventional streak imaging, two particles with opposite charges, electron and positron, are created in the same relative phase within the third streaking pulse that co-propagates with the test pulse. By measuring simultaneously the energy and momentum of the electrons and the positrons originating from different positions within the test pulse, its length and, in principle, even its shape can be reconstructed. The time resolution of SHEEP for different classes of tests, streaking and strong pulses can range from femtosecond to zeptosecond duration.

\subsection{Pair production in nucleus-laser collisions}
\label{BH}
While in electron-laser collisions the contribution of reaction (ii) to pair production is in general small, it becomes accessible to experimental observation when the projectile electrons are replaced by heavier particles such as protons or other nuclei. The two-step production process via multiphoton Compton scattering will then be strongly suppressed by the large projectile mass. The recent commissioning of the LHC at CERN has stimulated substantial activities on pair production in combined laser and nuclear Coulomb fields, which may be viewed as a generalization of the well-known Bethe-Heitler process to strong fields (multiphoton Bethe-Heitler pair production). The large Lorentz factors $\gamma_n$ of the ultrarelativistic nuclear beams lead to efficient enhancement of the laser parameters in the projectile rest-frame.

Indeed, when a proton beam with Lorentz factor $\gamma_p\approx 3000$, as presently available at LHC, collides head-on with a superintense laser beam of intensity $I_0\approx 10^{22}$\;W/cm$^2$, the Lorentz-boosted laser field strength approaches the critical value $F_{\text{cr}}$. This circumstance motivated the first calculations of nonperturbative pair production in collisions of a relativistic nucleus with a superintense near-optical laser beam \cite{MVG2003NIMB,MVG2003PRA}. Smaller projectile Lorentz factors may be sufficient, when ultrastrong XFEL pulses are employed \cite{Avetissian_2003}. The calculations were based on an S-matrix treatment and assumed laser fields of circular polarization. Later on, also the case of linear field polarization was studied \cite{MVG2004PRA,Sieczka_2006, Kaminski_2006, Krajewska_2006}. This case is rendered more involved due to the appearance of generalized Bessel functions, which are of very high order when $\xi_0\gg 1$. The underlying S-matrix element is generally of the form
\begin{equation}
\label{Sff}
S_{p_+,\sigma_+,p_-,\sigma_-} = 
      -ie \int d^4x \Psi^{\dag}_{p_-,\sigma_-}(x)V_n(r) \Psi_{-p_+,-\sigma_+}(x).
\end{equation}
It describes the transition of an electron from the negative-energy Volkov state $\Psi_{-p_+,-\sigma_+}(x)$ to a positive-energy Volkov state $\Psi^{\dag}_{p_-,\sigma_-}(x)$, which is mediated by the Coulomb potential $V_n(r)=Z|e|/r$ of the projectile nucleus. An alternative approach to the problem based on the polarization operator in a plane electromagnetic wave has been developed in \onlinecite{Milstein_2006}. It allows to obtain total production rates analytically. Both approaches rely on the strong-field approximation and include the laser field exactly to all orders, whereas the nuclear field is treated at leading order in $Z\alpha$.

Since the high-intensity Bethe-Heitler process has not been observed in experiment yet, in recent years physicists have proposed scenarios which may allow to realize the various interaction regimes of the process by present-day technology. Few-photon Bethe-Heitler pair production in the perturbative domain could be realized in collisions of the LHC proton beam with an XUV pulse of angular frequency $\omega_X\approx 100$\;eV and of moderate intensity $I_X\sim 10^{14}$\;W/cm$^2$ \cite{Muller2009PLB}. Corresponding radiation sources of table-top dimension are available nowadays in many laboratories. They are based on HHG from atomic gas jets or solid surfaces (see Sec. \ref{HHG}). The rate $R_{e^+\text{-}e^-}$ of pair creation by two-photon absorption close to the energetic threshold (i.e., $\omega^{\star}_X= u_{n,-}\omega_X\gtrsim m$, for the angular frequency $\omega_X^{\star}$ of the XUV pulse in the rest-frame of the nucleus) is given by \cite{Milstein_2006}
\begin{equation}
\label{BH2lin}
R_{e^+\text{-}e^-} = \frac{1}{4^{3-j}}(Z\alpha)^2\xi_0^4\omega_X^{\star}\bigg(\frac{\omega_X^{\star}}{m}-1\bigg)^{j+2},
\end{equation}
with $j=0$ for linear polarization and $j=2$ for circular polarization.

In the quasistatic regime of the process sizable pair yields require superintense laser fields from a petawatt source in conjunction with an LHC proton beam \cite{MVG2003PRA,Sieczka_2006}. Such experiments will become feasible when petawatt laser pulses are made available by high-power devices of table-top size, rather than by immobile large-scale facilities as they exist at present. A method to enable tunneling pair production with more compact multiterawatt laser systems has been proposed in \onlinecite{DiPiazza_Lotstedt_2009,DiPiazza_Lotstedt_2010}. It relies on the application of an additional weak XUV field, which is superimposed on a powerful optical laser wave. In this two-color setup, the energy threshold for pair creation can be overcome by the absorption of one photon from the high-frequency field and several additional photons from the low-frequency field. As a result, by choosing the XUV frequency $\omega_X^{\star}$ such that the parameter $\delta=(2m - \omega_X^{\star})/m$ fulfills the conditions $0<\delta\ll 1$, the tunneling barrier can be substantially lowered and even controlled. The pair production rate in the quasistatic regime for $0<\delta\ll 1$ depends essentially only on the parameters $\delta$ and $\chi_{0,n}$, and on the classical nonlinearity parameter $\xi_X$ of the XUV field. For a circularly polarized strong laser field it becomes \cite{DiPiazza_Lotstedt_2009} 
\begin{equation}
R_{e^+\text{-}e^-}=\frac{1}{64\sqrt{\pi}}m(Z\alpha)^2\xi_X^2\chi_{0,n}^2\sqrt{\zeta_0}\exp\left(-\frac{2}{3}\frac{1}{\zeta_0} \right)
\label{terawatt}
\end{equation}
for $\zeta_0=\chi_{0,n}/2\delta^{3/2}\ll 1$, to be compared with the usual scaling $\sim(-2\sqrt{3}/\chi_{0,n})$ in the absence of the XUV field.

A related process is laser-assisted Bethe-Heitler pair creation, where the high-frequency photon energy satisfies $\omega_X^{\star}>2m$. A pronounced channeling of the $e^+\text{-}e^-$ pair due to the forces exerted by the laser field after their creation was found \cite{Lotstedt_2008,Lotstedt_2009}. Multiphoton Bethe-Heitler pair creation in a two-color laser wave was investigated in \onlinecite{Roshchupkin_2001}. 

Analytical formulas for positron energy spectra and angular distributions in the tunneling regime of the process were obtained in \onlinecite{KuchievPRA}. For pair production at $\xi_0\sim 1$ no analytical expressions are known because of the intermediate nature of this parameter regime. However, by performing a fitting procedure to numerically obtained results, a total pair production rate scaling as $m(Z\alpha)^2\exp(-3.49/\chi_{0,n})$ was obtained in \onlinecite{AboveThreshold}, which closely resembles the tunneling exponential behaviour $m(Z\alpha)^2\exp(-2\sqrt{3}/\chi_{0,n})$.

While in Eq.~(\ref{Sff}) the influence of the projectile is described by an external Coulomb field, the projectile can also be treated as a quantum particle which allows to study nuclear recoil effects \cite{SarahPRD,Krajewska_2010,Krajewska_2011}. Besides, in laser-nucleus collisions, bound-free pair creation can occur where the electron is created in a deeply bound atomic state of the nucleus. The process was studied first for circular laser polarization \cite{MVG2003PRL,Matveev_2005} and later on also for linear polarization \cite{CarlusPRA}, including contributions from the various atomic subshells.

\subsection{Pair production in a standing laser wave}
\label{PP_SW}
Purely light-induced pair production can occur when two noncopropagating laser waves are superimposed. The simplest field configuration consists of two counterpropagating laser pulses of equal frequency and intensity. The resulting field is a standing wave which is inhomogeneous both in space and time and a theoretical treatment of the process is very challenging. In order to render the problem tractable and since the production process mainly occurs where the electric field component of the background field is stronger than the magnetic one \cite{Dittrich_b_2000}, in the standard approach the resulting standing light wave is approximated by a purely electric field oscillating in time. This approximation is expected to be justified for a strong ($I_0> 10^{20}\;\text{W/cm$^2$}$), optical laser field where the typical spatial scale of the field variation $\lambda_0\sim 1\;\text{$\mu$m}$ is much larger than the pair formation length $m/|e|E_0=\lambda_C F_{\text{cr}}/E_0\approx 2.6\times 10^{-2}\;\text{$\mu$m}/\sqrt{I_0[10^{20}\;\text{W/cm$^2$}]}$ \cite{Ritus_1985}. Note the analogy between the formation length $m/|e|E_0$ for pair production and the tunneling length $l_{\text{tun}}\sim I_p/|e|E_0$ in atomic ionization (see Sec. \ref{HHG}), with $I_p$ being the ionization potential energy. 

Pair production in an oscillating electric field is a generalization of the Schwinger mechanism \cite{Schwinger_1951} to time-dependent fields and it has been considered by many theoreticians, starting from the seminal works \onlinecite{Brezin_1970} and \onlinecite{Popov_1971,Popov_1972} (for a comprehensive list of references until 2005, see \onlinecite{Salamin_2006}).

While in the laser-electron and laser-nucleus collisions of the previous subsections the Doppler boost of the laser parameters due to a highly relativistic Lorentz factor could be exploited, in laser-laser collisions this is not possible so that high field strengths $E_0$ or high frequencies $\omega_0$ are required in the laboratory frame. Theoreticians are therefore aiming to find ways for enhancing the pair production probability in order to render the process observable in the foreseeable future.

\begin{figure}
   \centering
   \includegraphics[width=0.8\linewidth]{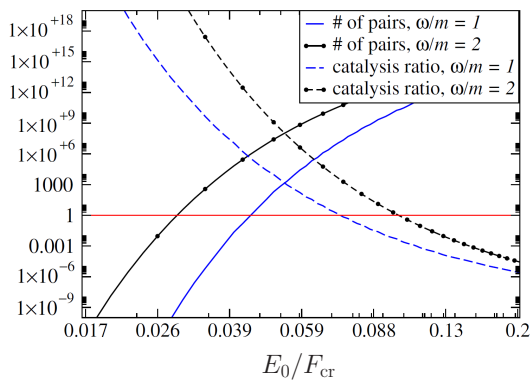}
   \caption{(Color) Number of pairs produced by the x-ray assisted Schwinger mechanism for two different values of the x-ray angular frequency $\omega_X$, indicated as $\omega$ in the figure, (blue and black solid lines) and the ratio of these catalyzed pairs to those produced by the standard Schwinger mechanism (blue and black dashed lines), both as functions of the optical field strength in units of $F_{\text{cr}}$. Adapted from \onlinecite{Dunne_2009}.}
   \label{Dunne}
\end{figure}

A first possibility to facilitate the observability of pair creation in a standing optical laser wave is to superimpose an x-ray photon (or any other high-frequency component) onto the high-field region \cite{Schutzhold_2008,Dunne_2009,Monin_2010}. In this way, the Schwinger mechanism is catalyzed so that the usual exponential suppression $\sim\exp(-\pi/\Upsilon_0)$ is significantly lowered. For example, in the limit when the x-ray energy approaches the threshold value $2m$, the pair production rate $R_{e^+\text{-}e^-}$ becomes
\begin{equation}
R_{e^+\text{-}e^-} \sim m\exp\left(-\frac{\pi -2}{\Upsilon_0}\right)
\label{catalyzed}
\end{equation}
assuming that the x-ray propagation direction is perpendicular to the electric field vector of the strong optical field. An overview of the pair production enhancement effect due to the x-ray assistance is shown in Fig.~\ref{Dunne}.

Another proposal to enhance the pair yield is the application of multiple colliding laser pulses instead of only two \cite{Bulanov_2010_a}. It has been demonstrated that the threshold laser energy necessary to produce a single pair, decreases when the number of colliding pulses is increased. The results are summarized in Table~\ref{Bulanov}. Pair production exceeds the threshold when eight laser pulses, with a total energy of 10 kJ, are simultaneously focused on one spot. Doubling (tripling) the number of pulses leads to an enhancement by two (six) orders of magnitude. The threshold energy drops from 40 kJ for two pulses to 5.1 kJ for 24 pulses, clearly indicating that the multiple-pulse geometry is strongly favorable. Besides, it was noticed that the pre-exponential volume factor in the pair creation probability can be very large and partially compensate for the exponential suppression \cite{Narozhny_2006}.

\begin{table}
\caption{Number $N_{e^+\text{-}e^-}$ of $e^+\text{-}e^-$ pairs produced by different numbers $n$ of laser pulses, with a total energy $W$ of 10 kJ. The threshold value total energy $W_{\rm th}$ needed to produce one $e^+\text{-}e^-$ pair is shown in the third column. The precise collision geometry and the pulse parameters can be found in \onlinecite{Bulanov_2010_a}. Adapted from \onlinecite{Bulanov_2010_a}.}
\begin{ruledtabular}
\begin{tabular}{ccc}
$n$ & $N_{e^+\text{-}e^-}$ at $W=10$\;kJ & $W_{\rm th}$ [kJ]\\ 
\hline
 2  & $< 10^{-18}$     & 40\\
 4  & $< 10^{-8}$      & 20\\
 8  & 4.0              & 10\\
 16 & $1.8\times 10^3$ & 8 \\
 24 & $4.2\times 10^6$ & 5.1\\
\end{tabular}
\end{ruledtabular}
\label{Bulanov}
\end{table}

Fine details of pair production in a time-dependent oscillating electric field are being studied nowadays because they might serve as characteristic signatures to discriminate the process of interest from potentially stronger background processes. For example, it was found that the momentum spectrum of the created particles is highly sensitive to a subcycle structure of the field \cite{Hebenstreit_2009} and that in the presence of an alternating-sign time-dependent electric field, coherent interference effects are observed in the Schwinger mechanism \cite{Akkermans_2012}. The observation in \onlinecite{Hebenstreit_2009} found an elegant mathematical explanation via the Stokes phenomenon \cite{Dumlu_PRL2010}. Further effects stemming from the precise shape of the external field were analyzed in \onlinecite{Dumlu_PRD2010,Dumlu_2011}. Also, the oscillating dynamics of the $e^+\text{-}e^-$ plasma created by a uniform electric field, including backreaction effects, was investigated \cite{Ruffini_2010, Ruffini_2011, Apostol_2011} (see also \onlinecite{Kim_2011,Bialynicki-Birula_2011}).

In addition to pair creation in superstrong laser pulses of low frequency, the process is also extensively discussed in connection with the upcoming XFEL facilities (see, e.g., \onlinecite{Ringwald_2001,Alkofer_2001}). Here the question arises as to what extent the spatial field dependence may influence the pair creation process, both in terms of total probabilities and particle momentum distributions. According to Noether's theorem, pair production in a time-dependent oscillating electric field occurs with conservation of the total momentum, as well as of the total spin. The problem therefore reduces effectively to a two-level system since the field couples negative and positive-energy electron states of same momentum and spin only. The production process exhibits resonance when the energy gap is an integer multiple of the laser frequency, leading to a characteristic Rabi flopping between the negative and positive-energy Dirac continua \cite{Popov_1971}. Due to the electron dressing by the oscillating field, the resonant laser frequencies are determined by the equation $\ell\omega_0 =2\langle \varepsilon\rangle$, where $\langle \varepsilon\rangle$ is the time-averaged electron energy in the time-dependent oscillating electric field. Accordingly, when the particle momentum is varied, several resonances occur corresponding to different photon numbers $\ell$. This gives rise to a characteristic ring structure in the momentum distribution \cite{Mocken_2010}.
\begin{figure}
   \centering
   \includegraphics[width=0.8\linewidth]{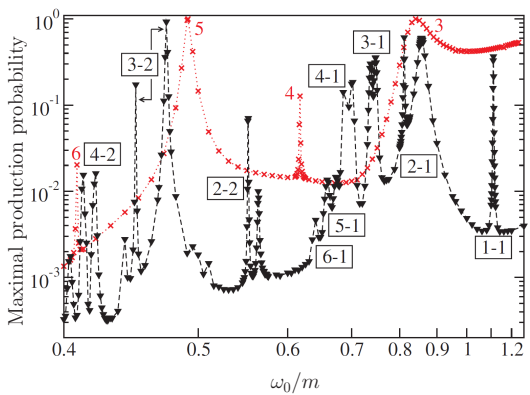}
   \caption{(Color online) Probability spectrum of $e^+\text{-}e^-$ pair production in two counterpropagating laser pulses, with the laser magnetic field included (black triangles) and neglected (red crosses). In the first case, the labeling $(\ell_r$-$\ell_l)$ signifies the number of absorbed photons from the right-left propagating wave; in the second case, the peak labels denote the total photon number ($\ell$). A vanishing initial momentum (i.e., positron momentum) and $\xi_0=1$ have been assumed. Adapted from \onlinecite{Ruf_2009}.}
   \label{resonances}
\end{figure}

Modifications of these well-established properties of the pair creation process, when the 
spatial field dependence and, thus, the laser magnetic-field component are accounted for, have been revealed in \onlinecite{Ruf_2009}. Utilizing an advanced computer code for solving the corresponding Dirac equation numerically, it was shown that the positions of the resonances are shifted, several new resonances occur, and the resonance lines are split due to the influence of the spatial field dependence (see Fig.~\ref{resonances}). The basic reason for these effects is that, in contrast to a uniform oscillating electric field, the photons in the counterpropagating laser pulses carry momentum along the beam axis. Therefore not only the total number $\ell$ of absorbed laser photons matters, but also how many of them have originated from the laser pulse travelling to the right and left, respectively. For example, for the multiphoton order $\ell=5$ two different resonance frequencies exist now, corresponding to $\ell_r=3$, $\ell_l=2$ on the one hand, and to $\ell_r=4$, $\ell_l=1$, on the other. Due to the photon momentum, the former two-level scheme is also broken into a $V$-type three-level scheme. This causes a splitting of the resonance lines, in analogy with the Autler-Townes effect known from atomic physics. 

For another numerical approach to space-time dependent problems in quantum field theory, we refer to the review \onlinecite{Cheng_2010}. Moreover, Schwinger pair production in a space-time dependent electric field pulse has been treated very recently within the Wigner formalism \cite{Hebenstreit_2011}. Here, a self-bunching effect of the created particles in phase space, due to the spatiotemporal structure of the pulse, was found.

Finally, we mention that, unlike a plane-wave field, a spatially focused laser beam is capable to produce $e^+\text{-}e^-$ pairs from vacuum and this process has been investigated in \onlinecite{Narozhny_2004} for different field polarizations. Spontaneous pair production may, in principle, also occur in a nuclear field for charge numbers $Z$ exceeding a critical value $Z_c$, which depends on the nuclear model. For example, $Z_c=173$ for a uniformly charged sphere with radius $1.2\times 10^{-12}\;\text{cm}$ \cite{Landau_b_4_1982}. See the reviews \onlinecite{Zeldovich_1972,Baur_2007} for more detailed information also on $e^+\text{-}e^-$ pair production in heavy ion collisions.

\subsection{Spin effects and other fundamental aspects of laser-induced pair creation}
A particularly interesting aspect of tunneling pair creation is the electron and positron spin-polarization. In general, pronounced spin signatures in a field-induced process may be expected when the background field strength approaches the critical value $F_{\text{cr}}$ \cite{Uggerhoj_2001,Walser_2002}. This indeed coincides with the condition for a sizable yield of tunneling pair production. Studies of spin effects in pair production by a high-energy photon and a strong laser field were performed in \onlinecite{Tsai_1993} and \onlinecite{Serbo_2005}, based on considerations on helicity amplitudes and on the spin-polarization vector, respectively. Characteristic differences between fermionic and bosonic particles have been revealed with respect to pair creation in an oscillating electric field \cite{Popov_1972} and in recent studies of the Klein paradox \cite{Grobe_Spin2004,Grobe_Spin2010,Cheng_2009b}. In the latter case it was shown that the existence of a fermionic (bosonic) particle in the initial state leads to suppression (enhancement) of the pair production probability due to the different quantum statistics. The enhancement in the bosonic case may be even exponential due to an avalanche process \cite{Wagner_Spin2010}. Concerning tunneling pair creation in combined laser and nuclear Coulomb fields, it has been shown that the internal spin-polarization vector is proportional in magnitude to $\chi_{0,n}$ and, to leading order, directed along the transverse momentum component of the electron \cite{DiPiazza_Spin2010}. A helicity analysis of pair production in laser-proton collisions revealed that: 1) right-handed leptons are emitted in the laboratory frame under slightly smaller angles with respect to the proton beam than left-handed ones; 2) the rate of pair creation of spin-$1/2$ particles exceeds by almost one order of magnitude the corresponding quantity for spin-0 particles \cite{Muller_Spin2011}.

Other fundamental aspects of laser-induced pair creation have been recently investigated as well. They comprise various kinds of $e^+\text{-}e^-$ correlations \cite{Krekora_2005,Fedorov_2006,Krajewska_2008}, multiple pair creation \cite{Cheng_2009}, questions of locality \cite{Cheng_2008} and vacuum decay times \cite{Labun_2009}, and consistency restrictions on the maximum laser field strength to guarantee the validity of the external-field approximation \cite{Gavrilov_2008}.

\section{QED cascades}
\label{Cascade}
As it was discussed in the previous section, the E-144 experiment at SLAC is the only one, so far, where laser-driven multiphoton $e^+\text{-}e^-$ pair production has been observed. Considering that about 100 positrons have been detected in $22000$ shots, each comprising the collision of about $10^7$ electrons with the laser beam, the process results to be rather inefficient. One could attribute this to the relatively low intensity $I_0$ of the laser system of $1.3\times 10^{18}\;\text{W/cm$^2$}$ ($\xi_0\approx 0.4$ as the laser photon energy was $\omega_0=2.4\;\text{eV}$). However, the extremely high energy $\varepsilon_0$ of the electron beam (about $46.6\;\text{GeV}$) ensured that that the nonlinear quantum parameter $\chi_0$ was about unity ($\chi_0\approx 0.3$). A recent investigation \cite{Sokolov_2010} has pointed out in general that in the mentioned setup, i.e., an electron beam colliding with a strong laser pulse, RR effects prevent the development of a cascade or avalanche process with an efficient, prolific production of $e^+\text{-}e^-$ pairs even at much larger laser intensities such that $\xi_0\gg 1$. By an avalanche or cascade process we mean here a process in which the incoming electrons emit high-energy photons in the laser field, which can interact with the field itself generating $e^+\text{-}e^-$ pairs, which, in turn, emit photons again and so on (of course a cascade process may also be initiated by a photon beam rather than by an electron beam). The above result has been obtained by numerically integrating the kinetic equations, which describe the evolution of the electron, the positron and the photon distributions in a plane-wave background field from a given initial electron distribution, and by accounting for the two basic processes that couple these distributions, i.e., multiphoton Compton scattering and multiphoton Breit-Wheeler pair production. The physical reason why an avalanche process cannot develop in a single plane-wave field can be understood in the following way. In the ultrarelativistic case $\xi_0\gg 1$, the above-mentioned basic processes in a plane-wave field are essentially controlled by the parameters $\chi_0$ and $\varkappa_0$, respectively (see also Secs. \ref{FED_Q}, \ref{TCS_General} and \ref{PP}). Now, at the $j^{\text{th}}$ step in which an electron/positron emits a photon or a photon transforms into an $e^+\text{-}e^-$ pair, the initial quantity $p^{(j)}_{0,-}$ or $k^{(j)}_{0,-}$ is conserved and it is distributed over the two final particles (an electron/positron and a photon in multiphoton Compton scattering and an $e^+\text{-}e^-$ pair in multiphoton Breit-Wheeler pair production). Thus, both resulting particles at each step will have a value of their own parameter $\chi^{\prime (j)}_0=(p^{\prime (j)}_{0,-}/m)(E_0/F_{\text{cr}})$ or $\varkappa^{\prime (j)}_0=(k^{\prime (j)}_{0,-}/m)(E_0/F_{\text{cr}})$ smaller than that of the incoming particle. Moreover, due to the special symmetry of the plane-wave field, the quantities $p^{(j)}_{0,-}$ or $k^{(j)}_{0,-}$ are also rigorously conserved between two steps (see also Sec. \ref{FED_C}). Then, the avalanche ends when the parameters $\chi^{(k)}_{0,i}$ and $\varkappa^{(k)}_{0,i}$ at a certain step $k$ are smaller than unity for $i\in [1,\ldots,N_k]$, with $N_k$ being the number of particles at that step.

\begin{figure}
\begin{center}
\includegraphics[width=0.8\linewidth]{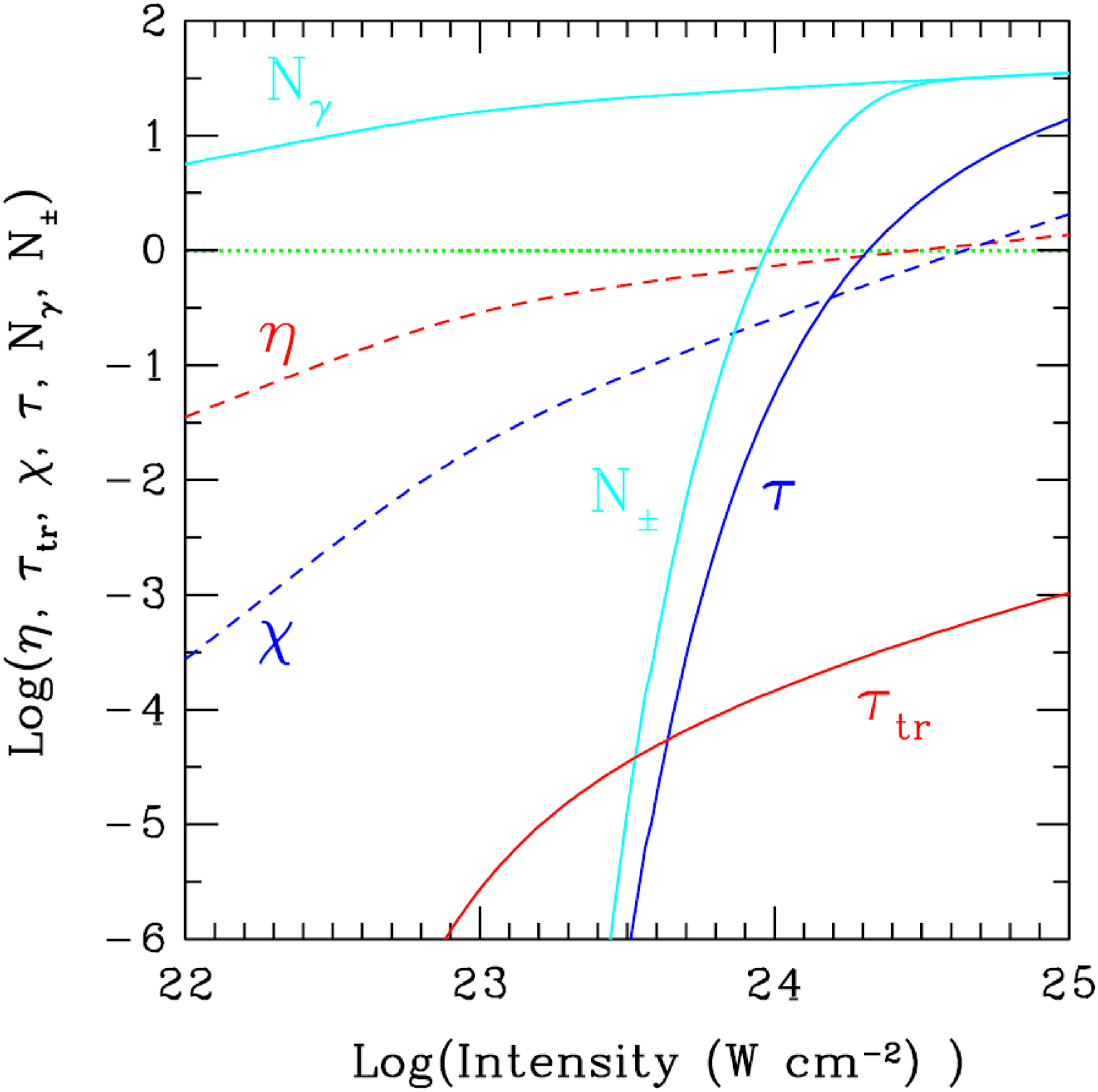}
\end{center}
\caption{(Color online) The number of $e^+\text{-}e^-$ pairs ($N_{\pm}$) and the number of photons ($N_{\gamma}$) created by an initial single electron in a rotating electric field as a function of the field intensity. The other plotted quantities are described in \onlinecite{Bell_2008}. From \onlinecite{Bell_2008}.}
\label{Bell_Cascade}
\end{figure}

The question arises as to whether other field configurations exist, where an avalanche process can be efficiently triggered (see the review \onlinecite{Aharonian_2003} for the development of QED cascades in matter, photon gas and magnetic field). A positive answer to this question has first been given in \onlinecite{Bell_2008}: even the presence of a single electron initially at rest in a standing wave generated by two identical counterpropagating circularly polarized laser fields can prime an avalanche process already at field intensities of the order of $10^{24}\;\text{W/cm$^2$}$. We note that in the presence of a single plane wave the same process would require an intensity of the order of $I_{\text{cr}}$, because for an electron initially at rest $\chi_0=E_0/F_{\text{cr}}$. From this point of view, the authors of \onlinecite{Bell_2008} explain qualitatively the advantage of employing two counterpropagating laser beams by means of an analogy taken from accelerator physics: a collision between two particles in their center-of-momenta is much more efficient than if one of the particles is initially at rest, because much more of the initial energy can be transferred, for example, to create new particles. The authors approximated the standing wave by a rotating electric field (see Sec. \ref{PP_SW}). In such a field and for an ultrarelativistic electron the controlling parameter is $\tilde{\chi}_0=(p_{\perp}/m)(E_0/F_{\text{cr}})$, where $p_{\perp}$ is the component of the electron momentum
perpendicular to the electric field. By estimating $p_{\perp}\sim m\xi_0$ (see also Eq. \eqref{Free_Sol_1_2}), one obtains $\tilde{\chi}_0\approx I_0[10^{24}\;\text{W/cm$^2$}]/\omega_0[\text{eV}]$ (here $\omega_0$, $E_0$ and $I_0$ are the standing-wave's angular frequency, electric field amplitude and intensity). The investigation in \onlinecite{Bell_2008} is based on the analysis of the trajectory of the electron in the rotating electric field including RR effects via the LL equation. Since the momentum of the electron oscillates around a value of the order of $m\xi_0$, the electron emits high-energy photons efficiently that can in turn trigger the cascade (see Fig. \ref{Bell_Cascade}). The possibility of describing
the evolution of the electron via its classical trajectory,
can be justified as follows. When an electron interacts with
a background electromagnetic field like that of a laser, 
quantum effects are essentially of two kinds \cite{Baier_b_1998}: 
the first one is associated with the quantum nature of
the electron motion and the second one with the recoil undergone by the electron
when it emits a photon. For an ultrarelativistic electron
it can be shown that, while the first kind of quantum effects is 
negligible, the second kind is large and has to be taken into
account \cite{Baier_b_1998}. Thus, the basic assumption is that, since the background laser
field is strong, the electron is promptly accelerated to ultrarelativistic
energies, the motion between two photon emissions is essentially
classical and, if necessary, only the emissions have to be treated quantum mechanically
by including the photon recoil. On the other hand, photons are assumed
to propagate in the field along straight lines.

The model employed in \onlinecite{Bell_2008} was improved in \onlinecite{Kirk_2009} by considering colliding pulsed fields with finite time duration and a realistic representation for the synchrotron spectrum emitted by a relativistic electron. The results in \onlinecite{Bell_2008} were essentially confirmed and numerical simulations with linearly polarized beams have shown a general insensitivity of the cascade development to the polarization of the beams. Another interesting finding in \onlinecite{Kirk_2009} is that the electrons in the standing wave tend to migrate to regions where the electric field vanishes and then they do not contribute to the pair-production process anymore. In both papers \onlinecite{Bell_2008,Kirk_2009} the emission of radiation by the electron was treated classically, i.e., the electron was supposed to loose energy and momentum continuously although in \onlinecite{Kirk_2009} the damping force-term in the LL equation was evaluated by employing the total emitted power calculated quantum mechanically. The stochastic nature of the emission of a photon has been taken into account in \onlinecite{Duclous_2011}. Analogously, the energy of the emitted photon is chosen randomly following the synchrotron spectral distribution and the momentum of the photon is always chosen to be parallel to that of the emitting electron. By contrast, in the pair production process by a photon, since the photon is not deflected by the laser field, it is assumed that after it has propagated one wavelength in the field, it decays into an $e^+\text{-}e^-$ pair. The main result in \onlinecite{Duclous_2011} is that at relatively low intensities of the order of $10^{23}\;\text{W/cm$^2$}$ the pair production rate is increased if the quantum nature of the photon emission is taken into account. The reason is that, due to the stochastic nature of the emission process, the electron can propagate for an unusually large distance before emitting. In this way it may gain an unusually large amount of energy and consequently emit a high-energy photon, that can be more easily converted into an $e^+\text{-}e^-$ pair. Moreover, the discontinuous nature of the (curvature of the) electron trajectory is shown to slow down the tendency of the electrons to migrate to regions where the electric field vanishes.

The intensity threshold of the avalanche process in a rotating electric field has also been investigated in \onlinecite{Fedotov_2010}. Denoting by $t_{\text{acc}}$ the time an electron needs to reach an energy corresponding to $\chi_0=1$ starting from rest in the given field, by $t_e$ ($t_{\gamma}$) the electron (photon) lifetime under photon emission ($e^+\text{-}e^-$ pair production) and by $t_{\text{esc}}$ the time after which the electron escapes from the laser field, the authors give the following conditions for the occurrence of the avalanche process: $t_{\text{acc}}\lesssim t_e,t_{\gamma}\ll t_{\text{esc}}$. Estimates based on the classical electron trajectory without including RR effects, lead to the simple condition $E_0\gtrsim \alpha F_{\text{cr}}$ for the avalanche to be primed in an optical field. The above estimate corresponds to an intensity of about $2.5\times 10^{25}\;\text{W/cm$^2$}$, i.e., one order of magnitude larger than what was found in \onlinecite{Bell_2008}. However, the main result of \onlinecite{Fedotov_2010} concerns the limitation, brought about on the maximal laser intensity that can be produced in the discussed field configuration by the starting of the avalanche process. In fact, the energy to accelerate the electrons and the positrons participating in the cascade has to come from the background electromagnetic field. By assuming an exponential increase of the number of electrons and positrons, it is found that already at laser intensities of the order of $10^{26}\;\text{W/cm$^2$}$ the created electrons and positrons have an energy which exceeds the initial total energy of the laser beams. This hints at the fact that at such intensities the colliding laser beams are completely depleted due to the avalanche process. The results obtained from qualitative estimates in \onlinecite{Fedotov_2010} have been scrutinized in \onlinecite{Elkina_2011} by means of more realistic numerical methods based on kinetic or cascade equations. In general, if $f_{\mp}(\bm{r},\bm{p},t)$ ($f_{\gamma}(\bm{r},\bm{k},t)$) is the electron/positron (photon) distribution function (upper and lower sign for electron and positron, respectively) in the phase-space $(\bm{r},\bm{p})$ ($(\bm{r},\bm{k})$) and $\varepsilon=\sqrt{m^2+\bm{p}^2}$  ($\omega=|\bm{k}|$), their evolution in the presence of a given classical electromagnetic field $(\bm{E}(\bm{r},t),\bm{B}(\bm{r},t))$ is described by the kinetic equations
\begin{align}
\label{f_ep}
\begin{split}
&\left[\frac{\partial}{\partial t}+\frac{\bm{p}}{\varepsilon}\cdot\frac{\partial }{\partial \bm{r}}\pm \bm{F}_L(\bm{r},\bm{p},t)\cdot\frac{\partial }{\partial \bm{p}}\right]f_{\mp}(\bm{r},\bm{p},t)\\
&=\int d\bm{k}\,w_{\text{rad}}(\bm{r},\bm{p}+\bm{k}\to\bm{k},t)f_{\mp}(\bm{r},\bm{p}+\bm{k},t)\\
&\quad-f_{\mp}(\bm{r},\bm{p},t)\int d\bm{k}\,w_{\text{rad}}(\bm{r},\bm{p}\to\bm{k},t)\\
&\quad+\int d\bm{k}\,w_{\text{cre}}(\bm{r},\bm{k}\to\bm{p},t)f_{\gamma}(\bm{r},\bm{k},t),
\end{split}\\
\label{f_gamma}
\begin{split}
&\left(\frac{\partial}{\partial t}+\frac{\bm{k}}{\omega}\cdot\frac{\partial }{\partial \bm{r}}\right)f_{\gamma}(\bm{r},\bm{p},t)\\
&=\int d\bm{p}\,w_{\text{rad}}(\bm{r},\bm{p}\to\bm{k},t)[f_+(\bm{r},\bm{p},t)+f_-(\bm{r},\bm{p},t)]\\
&\quad-f_{\gamma}(\bm{r},\bm{k},t)\int d\bm{p}\,w_{\text{cre}}(\bm{r},\bm{k}\to\bm{p},t),
\end{split}
\end{align}
where $\bm{F}_L(\bm{r},\bm{p},t)=e[\bm{E}(\bm{r},t)+(\bm{p}/\varepsilon)\times\bm{B}(\bm{r},t)]$ and where $w_{\text{rad}}(\bm{r},\bm{p}\to\bm{k},t)$ ($w_{\text{cre}}(\bm{r},\bm{k}\to\bm{p},t)$) is the local probability per unit time and unit momentum that an electron/positron (photon) with momentum $\bm{p}$ ($\bm{k}$) emits (creates) at the space-time point $(t,\bm{r})$ a photon with momentum $\bm{k}$ (an $e^+\text{-}e^-$ pair with the electron/positron having a momentum $\bm{p}$). It is worth pointing out here a connection between the development of a QED cascade and the quantum description of RR. In fact, as has been discussed in Sec. \ref{RR}, from a quantum point of view, RR corresponds to the multiple recoils experienced by the electron in the incoherent emission of many photons. Thus, in the kinetic approach RR is described by those terms in Eqs. \eqref{f_ep} and \eqref{f_gamma}, which do not involve $e^+\text{-}e^-$ pair production. In fact, in \onlinecite{Elkina_2011} it has been shown in the ultrarelativistic case that the equation of motion for the average momentum of the electron distribution, as derived from Eq. \eqref{f_ep}, coincides with the LL equation in the classical regime $\chi_0\ll 1$ (note that pair production is exponentially suppressed at $\chi_0\ll 1$).

In \onlinecite{Elkina_2011} the background field is approximated as a uniform, rotating electric field $\bm{E}(t)$ and $f_{\pm}(\bm{r},\bm{p},t)\to f_{\pm}(\bm{p},t)$ ($f_{\gamma}(\bm{r},\bm{k},t)\to f_{\gamma}(\bm{k},t)$). Analogously to \onlinecite{Duclous_2011}, it is assumed that the momentum of the photon emitted in multiphoton Compton scattering is parallel to that of the emitting electron (positron); in the same way, the momenta of the electron and positron created in multiphoton Breit-Wheeler pair production are assumed to be parallel to that of the creating photon. The evolution of the electron, positron and photon distributions has been investigated by numerically integrating the resulting cascade equations via a Monte Carlo method. Whereas, the instants of radiation and pair production have been randomly generated. In particular, the exponential increase of the number of $e^+\text{-}e^-$ pairs and the qualitative estimate, for example, of the typical energy of the electron at the moment of the photon emission carried out in \onlinecite{Fedotov_2010} have been confirmed (apart from discrepancies within one order of magnitude).

In both papers \onlinecite{Fedotov_2010,Elkina_2011} only the case of a rotating electric field was considered. In \onlinecite{Bulanov_2010} it is pointed out that the limitation on the maximal laser intensity reachable before the cascade is triggered, strongly depends on the polarization of the laser beams which create the standing wave. The paradigmatic cases of a rotating electric field and of an oscillating electric field are compared. The estimates presented in \onlinecite{Bulanov_2010} for the case of a rotating electric field essentially confirm that the avalanche starts at laser intensities of the order of $10^{25}\;\text{W/cm$^2$}$. The main physical reason why the cascade process in a circularly polarized standing wave starts at such an intensity is that in a rotating electric field the electron emits photons with typical energies of the order of $0.29\,\omega_0\gamma^3$, i.e., proportional to the cube of the Lorentz factor of the emitting electron $\gamma$ \cite{Bulanov_2010}. Whereas, in an oscillating electric field the typical emitted energy scales as $\gamma^2$, such that in order to radiate a hard photon with a given energy, a much more energetic electron is needed. Hence, the authors of \onlinecite{Bulanov_2010} conclude that in an oscillating electric field RR and quantum effects do not play a fundamental role at laser intensities smaller than $I_{\text{cr}}$ and that avalanche processes do not constitute a limitation. It is crucial however, for the conclusion in \onlinecite{Bulanov_2010} that the collision of the laser beams occurs in vacuum, i.e., the seed electrons and positrons which would trigger the cascade are supposed to be created in the collision itself.

\begin{figure}
\begin{center}
\includegraphics[width=0.8\linewidth]{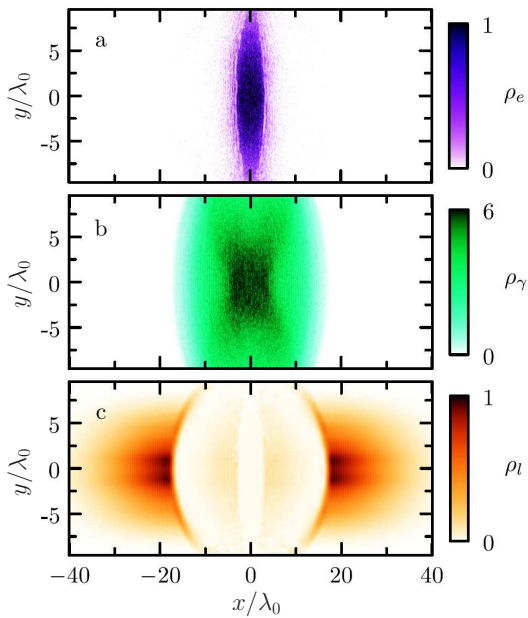}
\end{center}
\caption{(Color online) Snapshot of the normalized electron density $\rho_e$ (part a)), of the normalized photon density  $\rho_{\gamma}$ (part b)), and of the normalized laser intensity $\rho_l$ (part c)) $25.5$ laser periods after the two laser counterpropagating beams collided. The normalized density of positrons is approximately the same as that of the electrons. Also, the intensity of each colliding beam is $I_0=3\times 10^{24}\;\text{W/cm$^2$}$ and the common wavelength is $\lambda_0=0.8\;\text{$\mu$m}$. Adapted from \onlinecite{Nerush_2011}.}
\label{Nerush_Cascade}
\end{figure}

The question of the occurrence of the avalanche for two colliding linearly polarized pulses has also been addressed in \onlinecite{Nerush_2011}, where a detailed description of the system under investigation has been provided. In fact, previous models had assumed the background electromagnetic field as given, neglecting in this way the field generated by the electrons and positrons. The approach followed in \onlinecite{Nerush_2011} exploits the existence of two energy scales for the photons: one is that of the external laser field and of the plasma fields which is much smaller than $m$, and the other is that of the photons produced by the high-energy electrons which is, by contrast, much larger than $m$. The evolution of the low-energy photons is described by means of Maxwell's equations which are solved with a PIC code, i.e., the photons are treated as a classical electromagnetic field. Whereas, the production of hard photons as well as the creation of $e^+\text{-}e^{-}$ pairs is described as a stochastic process employing a Monte Carlo method. Unless low-energy ones, hard photons are treated as particles and their evolution is described via a distribution function. It is surprising that by considering a single seed electron initially at rest at a node of the magnetic field of a linearly polarized standing wave, an avalanche process is observed in the numerical simulation already for a laser intensity $I_0=3\times 10^{24}\;\text{W/cm$^2$}$ and a laser wavelength $\lambda_0=0.8\;\text{$\mu$m}$ (see Fig. \ref{Nerush_Cascade}).
The figure clearly shows the formation of an overdense $e^+\text{-}e^-$ plasma in the central region $|x|\lesssim \lambda_0$. In the same numerical simulations it is found that after about 20 laser periods almost half of the initial energy of the laser field has been transferred to the plasma. Disagreement with the predictions in \onlinecite{Bulanov_2010} is stated to be due to the formation of the avalanche in regions between the nodes of the electric and magnetic field, where the simplified analysis of the electron motion carried out in \onlinecite{Bulanov_2010} is not valid. Further analytical insight into the formation of the cascade has been reported in \onlinecite{Nerush_2011_b} by analyzing approximate solutions of the cascade equations in the presence of a rotating electric field.

\section{Muon-antimuon and pion-antipion pair production}
\label{MPP}
The production of $e^+\text{-}e^-$ pairs in strong laser fields has been discussed in Sec. \ref{PP}. In view of the ongoing technical progress the question arises as to whether also heavier particles such as muon-antimuon ($\mu^+\text{-}\mu^-$) or pion-antipion ($\pi^+\text{-}\pi^-$) pairs can be produced with the emerging near-future laser sources. The production of $\mu^+\text{-}\mu^-$ pairs from vacuum in the tunneling regime appears rather hopeless, though, since the required field needs to be close to $F_{\text{cr},\mu}=\varrho_{\mu}^2F_{\text{cr}}=5.6\times 10^{20}\;{\rm V/cm}$, with the ratio $\varrho_{\mu}=m_\mu/m\approx 207$ between the muon mass $m_{\mu}$ and the electron mass $m$. Even by boosting the effective laser fields with the Lorentz factors ($\sim 10^5$) of the most energetic electron beams available \cite{SLAC_PRD}, the value of $F_{\text{cr},\mu}$ seems out of reach. The tunneling production of $\pi^+\text{-}\pi^-$ pairs is even more difficult as $\varrho_{\pi}=m_{\pi}/m\approx 273$. However, $\mu^+\text{-}\mu^-$ and $\pi^+\text{-}\pi^-$ production can occur in microscopic collision processes in laser-generated or laser-driven plasmas, as well as by few-photon absorption from a high-frequency laser wave.

\subsection{Muon-antimuon and pion-antipion pair production in laser-driven collisions in plasmas}
\label{ML}
Energetic particle collisions in a plasma can in principle drive $\mu^+\text{-}\mu^-$ and $\pi^+\text{-}\pi^-$ production. The plasma may consist either of electrons and ions or of electrons and positrons. Both kinds of plasmas can be produced by intense laser beams interacting with a solid target. With respect to $e^+\text{-}e^-$ plasmas, this has been predicted by \onlinecite{Liang_1998}. As has been mentioned in Sec. \ref{PP}, abundant amounts of $e^+\text{-}e^-$ pairs have been recently produced in this manner at LLNL with pair densities of the order of $10^{16}$\;cm$^{-3}$ \cite{Chen_2009,Chen_2010} and much higher densities of the order of $10^{22}$\;cm$^{-3}$ have been also predicted \cite{Meyer-ter-Vehn_2001}. Theoreticians have therefore started to investigate the properties and time evolution of relativistic $e^+\text{-}e^-$ plasmas \cite{Thoma_RMP,Thoma_EPJD,Kuznetsova_2008,Ruffini_2007,Mustafa_2009,Hu_2011,Kuznetsova_2012}. In particular, it has been shown \cite{Kuznetsova_2008,Thoma_EPJD,Thoma_RMP} that in an $e^+\text{-}e^-$ plasma of 10 MeV temperature, $\mu^+\text{-}\mu^-$ pairs, $\pi^+\text{-}\pi^-$ pairs as well as neutral $\pi^0$ can be created in $e^+\text{-}e^-$ collisions. The required energy stems from the high-energy tails of the thermal distributions.

Also cold $e^+\text{-}e^-$ plasmas of high density can be generated nowadays due to dedicated positron accumulation and trapping techniques \cite{Cassidy_2005b}. When such a nonrelativistic low-energy plasma interacts with a superintense laser field, $\mu^+\text{-}\mu^-$ pair production can occur as well \cite{Muller_2006,Muller_2008b}. In this case, the plasma particles acquire the necessary energy by strong coupling to the external field which drives the electrons and positrons into violent collisions. The minimum laser peak intensity to ignite the reaction $e^+e^-\to\mu^+\mu^-$ amounts to about $7\times 10^{22}$\;W/cm$^2$ at a typical optical laser photon energy of $\omega_0=1\;\text{eV}$, corresponding to $\xi_{0,\rm min}=\varrho_{\mu}\approx 207$. The rate $R_{e^+e^-\to\mu^+\mu^-}$ of the process in the presence of a linearly polarized field reads
\begin{equation}
\label{Rfree}
R_{e^+e^-\to\mu^+\mu^-} \approx\frac{1}{2^3\pi^2} \frac{\alpha^2}{m^2\xi_0^4}
\sqrt{1-\frac{\xi_{0,\rm min}^2}{\xi_0^2}}\frac{N_+N_-}{V},
\end{equation}
with the number $N_{\pm}$ of electrons/positrons and the interaction volume $V$, which is determined by the laser focal spot size. Equation (\ref{Rfree}) may be made intuitively meaningful by introducing the invariant cross section $\sigma_{e^+e^-\to\mu^+\mu^-}$ of $\mu^+\text{-}\mu^-$ production in an $e^+\text{-}e^-$ collision in vacuum \cite{Peskin_1995}:
\begin{equation}
\label{sigma}
\sigma_{e^+e^-\to\mu^+\mu^-}=\frac{4\pi}{3}\frac{\alpha^2}{\varepsilon^{*\,2}}\sqrt{1-\frac{4m_{\mu}^2}{\varepsilon^{*\,2}}}\left(1+\frac{2m_{\mu}^2}{\varepsilon^{*\,2}}\right),
\end{equation}
where the upper index $^*$ indicates quantities in the center-of-mass system of the colliding electron and positron, as, e.g., the common electron and positron energy $\varepsilon^*$. Now, by exploiting the fact that the quantity $R_{e^+\text{-}e^-}/V$ is a Lorentz invariant and that in the present physical scenario $\varepsilon^*$ can be estimated as $m\xi_0$, Eq. \eqref{Rfree} implies the usual relation $R^*_{e^+e^-\to\mu^+\mu^-}/V^*\sim \sigma_{e^+e^-\to\mu^+\mu^-} n_+^* n_-^*$ between the number of events per unit volume and per unit time $R^*_{e^+e^-\to\mu^+\mu^-}/V^*$ and the cross section $\sigma_{e^+e^-\to\mu^+\mu^-}$; $n_\pm^*=N_\pm/V^*$ denote here the particle densities. The process $e^+e^-\to\mu^+\mu^-$ in the presence of an intense laser wave has also been considered in \onlinecite{Roshchupkin_2009}.

In laser-produced electron-ion plasmas resulting from intense laser-solid interactions,
$\mu^+\text{-}\mu^-$ and $\pi^+\text{-}\pi^-$ pairs can be generated by the cascade mechanism via energetic bremsstrahlung, like in the case of $e^+\text{-}e^-$ pair production mentioned in Sec. \ref{PP}. Assuming a laser-generated few-GeV electron beam, several hundreds to thousands of $\mu^+\text{-}\mu^-$ pairs arise from bremsstrahlung conversion in a high-$Z$ target material \cite{Kampfer_2009}. The production of $\pi^+\text{-}\pi^-$ pairs by laser-accelerated protons was considered in \onlinecite{Bychenkov_2001}, where a threshold laser intensity of $10^{21}$\;W/cm$^2$ for the process to occur was determined.

\subsection{Muon-antimuon and pion-antipion pair production in high-energy XFEL-nucleus collisions}
In this subsection another mechanism of $\mu^+\text{-}\mu^-$ and $\pi^+\text{-}\pi^-$ pair creation by laser fields will be pursued, which is based on the collision of an x-ray laser beam with an ultrarelativistic nuclear beam. This setup is similar to the one of Sec. \ref{BH}.

In the case of $\mu^+\text{-}\mu^-$ pair creation, by considering an x-ray photon energy of $\omega_0=12$\;keV and a nuclear relativistic Lorentz factor of $\gamma_n=7000$, the photon energy in the rest-frame of the nucleus amounts to $\omega^{\star}_0\approx 2\gamma_n \omega_0 = 168$\;MeV. The energy gap of $2m_\mu$ for $\mu^+\text{-}\mu^-$ pair production can thus be overcome by two-photon absorption \cite{Deneke_2008a,LPHYS2008}. Note that because of pronounced recoil effects, the Lorentz factor which would be required for two-photon $\mu^+\text{-}\mu^-$ production by a projectile electron is much larger: $\gamma\gtrsim 10^6$, corresponding to a currently unavailable electron energy in the TeV range.

At first sight, $e^+\text{-}e^-$ and $\mu^+\text{-}\mu^-$ pair production in combined laser and Coulomb fields seem to be very similar processes since the electron and muon only differ by their mass (and lifetime). In this picture, the corresponding production probabilities would coincide when the laser field strength and frequency are scaled in accordance with the mass ratio $\varrho_{\mu}$, i.e., $P_{\mu^+\text{-}\mu^-}(E_{0,\mu},\omega_{0,\mu})=P_{e^+\text{-}e^-}(E_0,\omega_0)$ for $E_{0,\mu}=\varrho_{\mu}^2E_0$ and $\omega_{0,\mu}=\varrho_{\mu}\omega_0$. This simple scaling argument does not apply, however, as the large muon mass is connected with a correspondingly small Compton wavelength $\lambda_{C,\mu}=\lambda_C/\varrho_{\mu}\approx 1.86\;$fm ($1\;\text{fm}=10^{-13}$ cm), which is smaller than the radius of most nuclei. As a result, while the nucleus can be approximately taken as pointlike in $e^+\text{-}e^-$ pair production ($\lambda_C\approx 386\;\text{fm}$), its finite extension must be taken into account in $\mu^+\text{-}\mu^-$ pair production. Pronounced nuclear size effects have also been found for $\mu^+\text{-}\mu^-$ production in relativistic heavy-ion collisions \cite{Baur_2007}.

Muon pair creation in XFEL-nucleus collisions can be calculated via the amplitude in Eq. \eqref{Sff}, with the nuclear potential $V_n(r)$ arising from an extended nucleus. It leads to the appearance of a nuclear form factor $F(\mathfrak{q}^2)$ which depends on the recoil momentum $\mathfrak{q}$. For example, $F(\mathfrak{q}^2) = \exp(-\mathfrak{q}^2a^2/6)$ for a Gaussian nuclear charge distribution of root-mean-square radius $a$. Since the typical recoil momentum is $\mathfrak{q}\sim m_\mu$,
the form factor leads to substantial suppression of the process. The fully differential production rate $dR_{\mu^+\text{-}\mu^-}=dR^{(\text{el})}_{\mu^+\text{-}\mu^-}+dR^{(\text{inel})}_{\mu^+\text{-}\mu^-}$ may be split into an elastic and an inelastic part, depending on whether the nucleus remains in its ground state or gets excited during the process. They read $dR^{(\text{el})}_{\mu^+\text{-}\mu^-}=dR^{(0)}_{\mu^+\text{-}\mu^-} Z^2 F^2(\mathfrak{q}^2)$ and $dR^{(\text{inel})}_{\mu^+\text{-}\mu^-}\approx dR^{(0)}_{\mu^+\text{-}\mu^-} Z [1-F^2(\mathfrak{q}^2)]$, respectively, with $dR^{(0)}_{\mu^+\text{-}\mu^-}$ being the production rate for a pointlike proton.

\begin{figure}
\begin{center}
\includegraphics[width=0.8\linewidth]{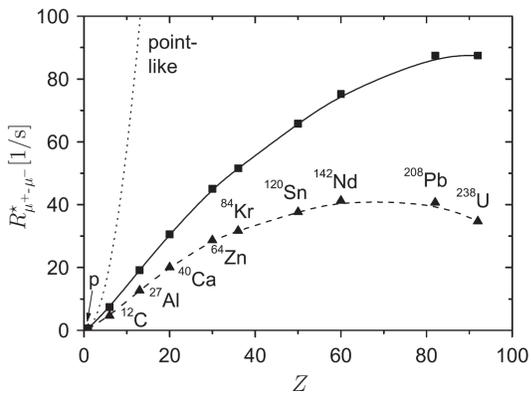}
\caption{\label{MuonTotal} Total rates for $\mu^+\text{-}\mu^-$ pair creation by two-photon absorption from an intense XFEL beam ($\omega_0=12$\;keV, $I_0=2.5\times 10^{22}$\;W/cm$^2$) colliding with various ultrarelativistic nuclei ($\gamma_n=7000$). The triangles show elastic rates, whereas the squares indicate total (``elastic + inelastic'') rates. The numerical data are connected by fit curves. The dotted line holds for a pointlike nucleus. The production rates are calculated in the rest-frame of the nucleus. Adapted from \onlinecite{Deneke_2008a}.}
\end{center} 
\end{figure}

Figure \ref{MuonTotal} shows total $\mu^+\text{-}\mu^-$ production rates $R^{\star}_{\mu^+\text{-}\mu^-}$ in the rest-frame of the nucleus for several nuclei colliding with an intense XFEL beam. For an extended nucleus, the elastic rate increases with its charge but decreases with its size. This interplay leads to the emergence of maximum elastic rates for medium-heavy ions. Figure \ref{MuonTotal} also implies that the total rate $R_{\mu^+\text{-}\mu^-}=R^{(\text{el})}_{\mu^+\text{-}\mu^-}+R^{(\text{inel})}_{\mu^+\text{-}\mu^-}$ in the laboratory frame saturates at high $Z$ values since $R^{(\text{inel})}_{\mu^+\text{-}\mu^-}$ increases with nuclear charge. For highly-charged nuclei the main contribution stems from the inelastic channel where the protons inside the nucleus act incoherently ($R^{(\text{inel})}_{\mu^+\text{-}\mu^-}\propto Z$). This implies that despite the high charges, high-order Coulomb corrections in $Z\alpha$ are of minor importance. In the collision, also $e^+\text{-}e^-$ pairs are produced by single-photon absorption in the nuclear field. However, this rather strong background process does not deplete the x-ray beam.

In XFEL-proton collisions, $\pi^+\text{-}\pi^-$ pairs can be generated as well. A corresponding calculation has been reported in \onlinecite{Anis_2011}, which includes both the electromagnetic and hadronic pion-proton interactions. The latter was described approximately by a phenomenological Yukawa potential. It was shown that, despite the larger pion mass, $\pi^+\text{-}\pi^-$ pair production by two-photon absorption from the XFEL field largely dominates over the corresponding process of $\mu^+\text{-}\mu^-$ pair production in the Doppler-boosted frequency range of $\omega^{\star}_0\approx 150\text{-} 210\;\text{MeV}$. This dominance is due to the much larger strength of the strong (hadronic) force compared with the electromagnetic force. As a consequence, in this energy range $\mu^+\text{-}\mu^-$ pairs are predominantly produced indirectly via two-photon $\pi^+\text{-}\pi^-$ production and subsequent pion decay, $\pi^+\to\mu^+ + \nu_\mu$ and $\pi^-\to\mu^- + {\bar \nu}_\mu$.

In relativistic laser-nucleus collisions, $\mu^+\text{-}\mu^-$ or $\pi^+\text{-}\pi^-$ pairs can also be produced indirectly within a two-step process \cite{Kuchiev_2007}. First, upon the collision, an $e^+\text{-}e^-$ pair is created via tunneling pair production. Afterwards, the pair, being still subject to the electromagnetic forces exerted by the laser field, is driven by the field into an energetic $e^+\text{-}e^-$ collision. If the collision energy is large enough, the reaction $e^+e^-\to\mu^+\mu^-$ may be triggered. This two-step mechanism thus represents a combination of the processes considered in Sec. \ref{BH} and \ref{ML}. Besides, it may be considered as a generalization of the well-established analogy between strong-field ionization and pair production (see Sec. \ref{PP}) to include also the recollision step.

\section{Nuclear physics}
\label{NP}
Influencing atomic nuclei with optical laser radiation is, in general, a difficult task because of the large nuclear level spacing $\Delta\mathcal{E}$ of the order of $1\;\text{keV}\text{-} 1\;\text{MeV}$, which exceeds typical laser photon energies $\omega_0\sim 1$ eV by orders of magnitude \cite{Matinyan_1998}. Also the laser's electric work performed over the tiny nuclear extension $r_n\sim 1\text{-} 5$ fm is usually too small to cause any sizable effect. In fact, the requirement $|e|E_0r_n\sim \Delta\mathcal{E}$ can only be satisfied for at laser-field amplitudes at least close to $F_{\text{cr}}$. Direct laser-nucleus interactions have therefore mostly been dismissed in the past.

On the other hand, laser-induced secondary reactions in nuclei have been explored especially in the late 1990s. Via laser-heated clusters and laser-produced plasmas, various nuclear reactions have been ignited, such as fission, fusion and neutron production. In all these cases, the interaction of the laser field with the target first produces secondary particles such as photo-electrons or bremsstrahlung photons which, in a subsequent step, trigger the nuclear reaction. For a recent review on this subject we refer to \onlinecite{Ledingham_2003,Ledingham_2010}.

In recent years, however, the interest in direct laser-nucleus coupling has been revived by the ongoing technological progress towards laser sources of increasingly high intensities as well as frequencies. Indeed, when suitable nuclear isotopes are considered, intense high-frequency fields or superstrong near-optical fields may be capable of affecting the nuclear structure and dynamics.

\subsection{Direct laser-nucleus interaction}

\subsubsection{Resonant laser-nucleus coupling}

There are several low-lying nuclear transitions in the keV range, and even a few in the eV range. Examples of the latter are $^{229}$Th ($\Delta\mathcal{E}\approx 7.6$ eV) and $^{235}$U ($\Delta\mathcal{E}\approx 76$ eV) \cite{Beck_2007}. These isotopes can be excited by the 5th harmonic of a Ti:Sa laser ($\omega_0=1.55\;\text{eV}$) and by pulses envisaged at the ELI attosecond source \cite{ELI}, respectively. Even higher frequencies can be attained by laser pulse reflection from relativistic flying mirrors of electrons extracted from an underdense plasma \cite{Bulanov_1994} or possibly also from an overdense plasma \cite{Habs_2008}. Otherwise keV-energy photons are generated by XFELs, which, as we have seen in Sec. \ref{x_ray}, are presently emerging as large-scale facilities, e.g., at SLAC \cite{LCLS} and DESY \cite{XFEL}, and which could be employed with focusing \cite{Mimura_2010} and reflection devices \cite{Shvydko_2011}. In addition, also XFEL facilities of table-top size \cite{Gruner_2007,Kneip_2010} and even fully coherent XFEL sources are envisaged such as the future XFEL Oscillator (XFELO) \cite{Kim_2008} or the Seeded XFEL (SXFEL) \cite{Feldhaus_1997,Linac_2011}. Brilliant gamma-ray beams with spectra peaked between 20 KeV and 150 KeV have been recently produced from resonant betatron motion of electrons in a plasma wake \cite{Cipiccia_2011}. A new material research center, the Matter-Radiation Interactions in Extremes (MaRIE) \cite{MaRIE_2011} is planned, allowing for both fully coherent XFEL light with photon energy up to 100 keV and accelerated ion beams. The photonuclear pillar of ELI to be set up near Bucharest (Romania) is planned to provide a compact XFEL along with an ion accelerator aiming for energies of $4\text{-}5$ GeV \cite{ELI}. In addition, coherent gamma-rays reaching few MeV energies via electron laser interaction are envisaged at this facility \cite{ELI,Habs_2009}.

With these sources of coherent high-frequency pulses, driving electric dipole (E1) transitions in nuclei is becoming feasible \cite{Buervenich_2006}. Table \ref{nuclei} displays a list of nuclei with suitable E1 transitions. Along with an appropriate moderate nucleus acceleration, resonance may be induced due to the Doppler shift via the factor $(1+v_n)\gamma_n$ in the counterpropagating setup, with $v_n$ and $\gamma_n$ being the nucleus velocity and its Lorentz factor, respectively \cite{Buervenich_2006}. For example, with $^{223}\mathrm{Ra}$ and an XFEL frequency of $12.4$ keV a factor of $(1+v_n)\gamma_n=4$ would be sufficient. In general such moderate pre-accelerations of the nuclei would be of great assistance since they increase the number of possible nuclear transitions for the limited number of available light frequencies. Note that the electric field strength of the laser pulse transforms analogously in the rest-frame of the nucleus, such that the applied laser field in the laboratory frame may correspondingly be weaker for a counterpropagating setup.
\begin{table}
\caption{The transition energy $\Delta\mathcal{E}$, the dipole moment $\mu$, the life-time $\tau_g$ ($\tau_e$) of the ground (excited) state of few relevant nuclear systems and 
E1 transitions~\cite{Aas_1999,NuclearDataBase}. Adapted from \onlinecite{Buervenich_2006}.}
\begin{ruledtabular}
\begin{tabular}{cccccc}
Nucleus & Transition & $\Delta\mathcal{E}$[keV] & $\mu$[$e$ fm]& $\tau_g$ & $\tau_e$[ps]
\\ \hline
$^{153}$Sm  & $3/2^- \! \to \! 3/2^+$ &  35.8  & 
$>0.75$\footnotemark[1]  & 47 h & $< 100$ \\
$^{181}$Ta  & $9/2^- \! \to \! 7/2^+$ &  6.2  & 
0.04\footnotemark[1]  & stable & $6\times 10^6$ \\
$^{225}$Ac  & $3/2^+ \! \to \! 3/2^-$ &  40.1  & 
0.24\footnotemark[1]  & 10.0 d & 720 \\
$^{223}$Ra  & $3/2^- \! \to \! 3/2^+$ &  50.1  & 
0.12  & 11.435 d & 730 \\
$^{227}$Th & $3/2^- \! \to \! 1/2^+$ & 37.9  & \footnotemark[2] 
& 18.68 d & \footnotemark[2] \\
$^{231}$Th & $5/2^- \! \to \! 5/2^+$ & 186  & 0.017 
& 25.52 h & 1030 \\
\end{tabular}
\end{ruledtabular}
\footnotetext[1]{Estimated via the Einstein A coefficient from $\tau_e$ and $\Delta \mathcal{E}$}
\footnotetext[2]{Not listed in the National Nuclear Data Center (NNDC) \cite{NuclearDataBase}}
\label{nuclei}
\end{table}

\begin{figure}
\includegraphics[width=0.8\linewidth]{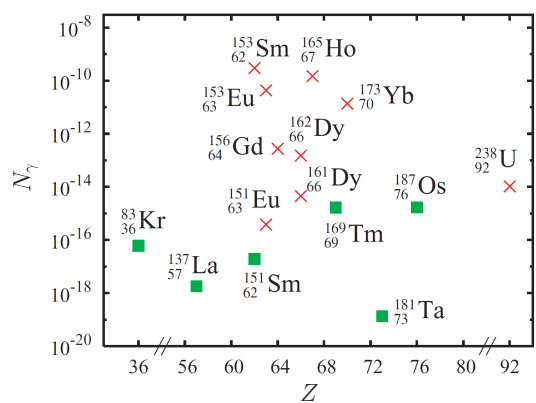}
\caption{(Color online) Number $N_{\gamma}$ of signal photons per nucleus per laser pulse for several isotopes with first excited states below 12.4~keV (green squares) and above  12.4~keV (red crosses). The results are plotted versus the atomic number $Z$. The considered European XFEL has a pulse duration of 100~fs and an average brilliance of 1.6$\times 10^{25}$ photons/(s\,mrad$^2$\,mm$^2$\,0.1\% bandwidth) \cite{XFEL}. Adapted from \onlinecite{Palffy_2008}.}
\label{Palffy1}
\end{figure}

For optimal coherence properties of the envisaged high-frequency facilities, subsequent pulse applications were shown to yield notable excitation of nuclei \cite{Buervenich_2006}. In addition, many low-energy electric quadrupole (E2) and magnetic dipole (M1) transitions are available. Here it is interesting to note that certain E2 or M1 transitions can indeed be competitive in strength with E1 transitions \cite{Palffy_2008,Palffy2_2008}. While the majority of transitions is available for high frequencies in the MeV domain (requiring substantial nucleus accelerations), Fig. \ref{Palffy1} displays also numerous suitable nuclear transitions below 12.4 keV along with the excitation efficiencies from realistic laser pulses.
While indeed experimental challenges are high, resonant direct interactions of laser radiation with nuclei is expected to pave the way for nuclear quantum optics. Especially control in exciting and deexciting certain long-living nuclear states would have dramatic implications for nuclear isomer research \cite{Walker_1999,Aprahamian_2005,Palffy_2007}. As an obvious application this would be of relevance for nuclear batteries \cite{Walker_1999,Aprahamian_2005}, i.e., for controlled pumping and release of energy stored in long-lived nuclear states. In atomic physics, the STimulated Raman Adiabatic Passage (STIRAP) technique has proven to be highly efficient in controlling populations robustly with high precision \cite{Bergmann_1998}. On the basis of currently envisaged accelerators and coherent high-frequency laser facilities, it has been recently shown that such an efficient coherent population transfer will also be feasible in nuclei \cite{Liao_2011}. Most recently, a nuclear control scheme with optimized pulse shapes and sequences has been developed in \onlinecite{Wong_2011}.

Serious challenges are certainly imposed by the nuclear linewidths that may either be too narrow to allow for sufficient interaction with the applied laser pulses or inhibit excitations and coherences due to large spontaneous decay. Decades of research in atomic physics allowing now for shaping atomic spectra via quantum interference \cite{Kiffner_2010,Evers_2002,Postavaru_2011} raise hopes that such obstacles may be overcome in the near future as well.

Direct photoexcitation of giant dipole resonances with few-MeV photons via laser-electron interaction was shown to be feasible \cite{Weidenmueller_2011} based on envisaged experimental facilities such as ELI. Finally, care has to be taken to compare the laser-induced nuclear channels with competing nuclear processes via, for example, bound electron transitions or electron captures in the atomic shells \cite{Palffy_2010,Palffy_2007}.

\subsubsection{Nonresonant laser-nucleus interactions}
\label{NR}
Already decades ago a lively debate was started on whether nuclear $\beta$-decay may be significantly affected by the presence of a strong laser pulse or not \cite{Nikishov_1964_b,Reiss1_1983,Becker2_1984,Akhmedov_1983} and a conclusive experimental answer to this issue is still to come. Most recently the notion of affecting nuclear $\alpha$-decay with strong laser pulses has been discussed showing that moderate changes of such nuclear reactions with the strongest envisaged laser pulses are indeed feasible \cite{Castaneda_2011,Castaneda_2012}.

When the laser intensity is high enough ($I_0 > 10^{26}$ W/cm$^2$) low-frequency laser fields are able to influence the nuclear structure without necessarily inducing nuclear reactions. In such ultrastrong fields, low-lying nuclear levels get modified by the dynamic (AC-) Stark shift \cite{Buervenich2_2006}. These AC-Stark shifts are of the same order as in typical atomic quantum optical systems relative to the respective transition frequencies. At even higher, supercritical intensities ($I_0 > 10^{29}$ W/cm$^2$) the laser field induces modifications to the proton root-mean-square radius and to the proton density distribution \cite{Buervenich2_2006}.

\subsection{Nuclear signatures in laser-driven atomic and molecular dynamics}

Muonic atoms represent traditional tools for nuclear spectroscopy by employing atomic physics techniques. Due to the large muon mass compared to that of the electron, $m_\mu\approx 207\, m$, and because of its correspondingly small Bohr radius $a_{B,\mu}=\lambda_{C,\mu}/\alpha\approx 255\;\text{fm}$, the muonic wave-function has a large overlap with the nucleus. Precise measurements of x-ray transitions between stationary muonic states are therefore sensitive to nuclear-structure features such as finite size, deformation, surface thickness, or polarization. 

\begin{figure}
\includegraphics[width=0.64\linewidth,angle=270]{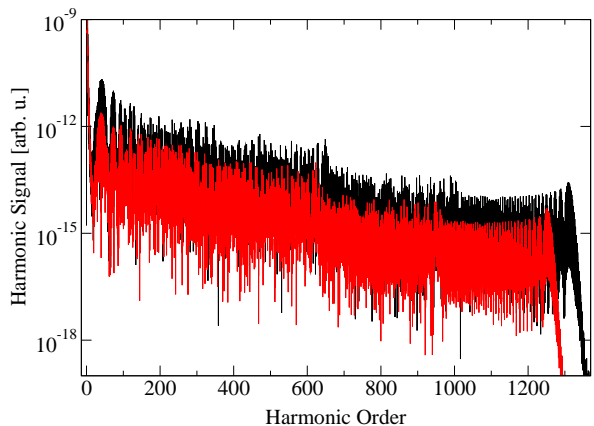}
\caption{(Color) HHG spectra emitted from muonic hydrogen (black) and muonic deuterium (red) in a laser field of intensity $I_0 \approx 10^{23}$ W/cm$^{2}$ and photon energy $\omega_0\approx 60$ eV. ``Arb. u.'' stands for ``arbitrary units''. From \onlinecite{Shahbaz_2007}.}
\label{muonicHHG}
\end{figure}

When a muonic atom is subjected to a strong laser field, the muon becomes a \emph{dynamic} nuclear probe which is periodically driven across the nucleus by the field. This can be inferred, for example, from the high-harmonic radiation emitted by such systems \cite{Shahbaz_2007,Shahbaz_2010}. Figure~\ref{muonicHHG} compares the HHG spectra from muonic hydrogen versus muonic deuterium subject to a very strong XUV laser field. Such fields are envisaged at the ELI attosecond source (see Fig. \ref{Megalaser}). Due to the different masses $M_n$ of the respective nuclei ($M_n=m_p$ for a hydrogen nucleus and $M_n\approx m_p+m_n$ for a deuterium nucleus by neglecting the binding energy, with $m_{p/n}$ being the proton/neutron mass), muonic hydrogen gives rise to a significantly larger harmonic cut-off energy. The reason can be understood by inspection of the ponderomotive energy
\begin{equation}
\label{UP}
U_{p}=\frac{e^{2}E_0^{2}}{4\omega_0^{2}M_{r}}=\frac{e^{2}E_0^{2}}{4\omega_0^{2}}\left(\frac{1}{m_\mu}+\frac{1}{M_{n}}\right),
\end{equation}
which depends in the present case on the reduced mass $M_r=m_\mu M_{n}/(m_\mu+ M_{n})$ of the muon-nucleus system. The reduced mass of muonic hydrogen ($\approx 93\;\text{MeV}$) is smaller than that  of muonic deuterium ($\approx 98\;\text{MeV}$) and this implies a larger ponderomotive energy and an enlarged \textit{plateau} extension. The influence of the nuclear mass can also be explained by the separated motions of the atomic binding partners. The muon and the nucleus are driven by the laser field into opposite directions along the laser's polarization axis. Upon recombination their kinetic energies sum up as indicated on the right-hand side of Eq. (\ref{UP}). Within this picture, the larger cut-off energy for muonic hydrogen results from the fact that, due to its smaller mass, the proton is more strongly accelerated by the laser field than the deuteron.

Due to the large muon mass, very high harmonic cut-off energies can be achieved via muonic atoms with charge number $Z$ in the nonrelativistic regime of interaction. Since the harmonic-conversion efficiency as well as the density of muonic atom samples are rather low, it is important to maximize the radiative signal strength. A sizable HHG signal requires efficient ionization on the one hand, and efficient recombination on the other. The former is guaranteed if the laser's electric field amplitude $E_0$ lies just below the border of over-barrier ionization,
\begin{equation}
\label{OBI}
E_0\lesssim \frac{M_{r}^{2}}{Q_{\rm eff}}\frac{(Z\alpha)^{3}}{16}.
\end{equation}
Here, $Q_{\rm eff} = |e|(Z/M_n+1/m_\mu)M_r$ represents an effective charge \cite{Shahbaz_2010,Reiss_1979}. Efficient recollision is guaranteed if the magnetic drift along the laser propagation direction can be neglected. Equation \eqref{drift} indicates that this is the case here, provided
\begin{equation}
\label{RP}
\left(\frac{Q_{\rm eff}E_0}{M_{r}\omega_0}\right)^3\lesssim \frac{16\omega_0}{\sqrt{2M_{r}I_p}}.
\end{equation}
The two above inequalities define a maximum laser intensity and a minimum laser angular frequency which are still in accordance with the conditions imposed. At these laser parameters, the maximum harmonic cut-off energies are attained and an efficient ionization-recollision process is guaranteed. For muonic hydrogen the corresponding lowest frequency lies in the Vacuum Ultraviolet (VUV) range ($\omega_0\approx 27$\;eV) and the maximum field intensity amounts to $1.6\times 10^{23}$\;W/cm$^{2}$. At these values, the harmonic spectrum extends to a maximum energy of approximately $0.55$ MeV. For light muonic atoms with nuclear charge number $Z>1$, the achievable cut-off energies are even higher, reaching several MeVs. This holds prospects for the production of coherent ultrashort gamma-ray pulses (see also \onlinecite{Xiang_2010}).

In principle, also nuclear size effects arise in the HHG signal from muonic atoms. This has been shown qualitatively in 1D numerical simulations, where a 50\% enhancement of the harmonic plateau height has been obtained for muonic hydrogen compared with muonic deuterium \cite{Shahbaz_2007,Shahbaz_2010}. This has been attributed to the enhanced final muon acceleration towards the hydrogenic core. For more precise predictions, 3D calculations are desirable.

Existing results also indicate that muonic atoms in high-intensity, high-frequency laser fields can be utilized to dynamically gain structure information on nuclear ground states via their high-harmonic response. Besides, the laser-driven muonic charge cloud, causing a time-dependent Coulomb field, may lead to nuclear excitation \cite{Shahbaz_2009}. The excitation probabilities are quite small, however, because of the large difference between the laser photon energy and the nuclear transition energy. Nuclear excitation can also be triggered by intense laser-induced recollisions of field-ionized high-energy electrons \cite{Mocken_2004,Milosevic_2004,Kornev_2007}.

Finally, muonic molecules are of particular interest for nuclear fusion studies. Modifications of muon-catalyzed fusion in strongly laser-driven muonic D$_2^+$ molecules have been investigated  \cite{Chelkowski_2004,Paramonov_2007}. It was found that applied field intensities of the order of $10^{23}$\;W/cm$^{2}$ can control the molecular recollision dynamics by triggering the nuclear reaction on a femtosecond time scale. Similar theoretical studies have recently been carried out on aligned (electronic) HT molecules, i.e., involving a Tritium atom \cite{Zhi_2009}.

\section{Laser colliders}
\label{LAP}
The fast advancement in laser technology is opening up the possibility of employing intense laser beams to efficiently accelerate charged particles and to make them collide for eventually even initiating high-energy reactions.

\subsection{Laser acceleration}
\label{LA}
Strong laser fields provide new mechanisms for particle acceleration alternative to conventional accelerator technology \cite{Tajima_1979}. With presently-available laser systems an enormous electron acceleration gradient $\sim 1$ GeV/cm can be achieved, which exceeds by three orders of magnitude that of conventional accelerators and which raises prospects for compact accelerators \cite{Mangles_2004,Geddes_2004,Faure_2004,Clayton_2010,Hafz_2008}. Different schemes of laser-electron acceleration have been proposed. These include the laser wakefield accelerator, the plasma beat wave accelerator, the self-modulated laser wakefield accelerator and plasma waves driven by multiple laser pulses (see the recent reviews \onlinecite{Esarey_2009,Malka_2011}). High-gradient plasma wakefields can also be generated with an ultrashort bunch of protons \cite{Caldwell_2009}, allowing electron acceleration to TeV energies in a single stage. The achievable current and emittance of presently-available laser-accelerated electron beams is sufficient to build synchrotron radiation sources or even to aim at compact XFEL lasers \cite{Schlenvoigt_2008}. 

Laser acceleration of ions provides quasimonoenergetic beams with energy of several MeVs per nucleon \cite{Hegelich_2006,Schwoerer_2006,Toncian_2006,Fuchs_2007,Haberberger_2012}. It mostly employs the interaction of high-intensity lasers with solid targets. One of the main goals of laser-ion acceleration is to create low-cost devices for medical applications, such as for hadron cancer therapy \cite{Combs_2009}. Several regimes have been identified for laser-ion acceleration (see also the forthcoming review \onlinecite{Macchi_2012}). For laser intensities in the range $10^{18}\text{-} 10^{21}$ W/cm$^2$ and for solid targets with a thickness ranging from a few to tens of micrometers, the so-called target-normal-sheath acceleration is the main mechanism \cite{Fuchs_2006}. A further laser-ion interaction process is the skin-layer ponderomotive acceleration  \cite{Badziak_2007}. By contrast, the radiation-pressure acceleration regime operates when the target thickness is decreased (see \onlinecite{Esirkepov_2004} and \onlinecite{Macchi_2009} for the so-called ``laser piston'' and ``light sail'' regimes, respectively). In \onlinecite{Galow_2011} a chirped ultrastrong laser pulse is applied to proton acceleration in a plasma. Chirping of the laser pulse ensures optimal phase synchronization of the protons with the laser field and leads to efficient proton energy gain from the field. In this way, a dense proton beam (with about $10^7$ protons per bunch) of high energy (250 MeV) and good quality (energy spread $\sim 1\%$) can be generated.

An alternative promising way for particle acceleration is direct laser acceleration in a tightly focused laser beam \cite{Salamin_2002a} or in crossed laser beams \cite{Salamin_2003}. Especially efficient accelerations \cite{Salamin_2006b,Gupta_2007,Bochkarev_2011} can be achieved in a radially polarized axicon laser beam \cite{Dorn_2003}. For example, the generation of mono-energetic GeV electrons from ionization in a radially polarized laser beam is theoretically demonstrated in \onlinecite{Salamin_2007b,Salamin_2010}. A setup for direct laser acceleration of protons and bare carbon nuclei is considered in \onlinecite{Salamin_2008}. It has been shown that laser pulses of $0.1\text{-} 10$ PW  can accelerate the nuclei directly to energies in the range required for hadron therapy. Simulations in \onlinecite{Galow_2010} and further optimization studies in \onlinecite{Harman_2011} indicate that protons stemming from laser-plasma processes can be efficiently post-accelerated employing single and crossed pulsed laser beams, focused to spot radii of the order of the laser wavelength. The protons in the resulting beam have kinetic energies exceeding 200 MeV and small energy spreads of about 1\%. The direct-acceleration method has proved to be efficient also for other applications. In \onlinecite{Salamin_2011} it is shown that 10 keV helium and carbon ions, injected into 1 TW-power crossed laser beams of radial polarization, can be accelerated in vacuum to energies of hundreds of keV necessary for ion lithography.

\subsection{Laser-plasma linear collider}

Laser-electron accelerators have already entered the GeV energy domain where the realm of particle physics starts \cite{Leemans_2006}. In fact, the strong interaction comes into play at distances $d$ of the order of $d \sim 1$ fm, which for relativistic processes corresponds to energies $\varepsilon \sim 1/d \sim 1$ GeV. Thus, a laser-based collider is in principle suitable for performing particle-physics experiments. However, in order to initiate high-energy reactions with sizable yield, not only GeV energies are required but also collision luminosities ${\mathcal L}$ at least as high as $10^{26}\text{-} 10^{27}$ cm$^{-2}$s$^{-1}$. Meanwhile, for the ultimate goal of being competitive with the next International Linear Collider \cite{ILC}, energies on the order of 1 TeV and luminosities of the order of $10^{34}$ cm$^{-2}$s$^{-1}$ are required \cite{Ellis_2001}.

The potential of the Laser-Plasma Accelerator (LPA) scheme to develop a laser-plasma linear collider is discussed in \onlinecite{Schroeder_2010}. Two LPA regimes are analyzed which are distinguished by the relationship between the laser beam waist size $w_0$ and the plasma frequency $\omega_p=\sqrt{4\pi n_pe^2/m}$, with $n_p$ being the plasma density: 1) the quasilinear regime at large radius of the laser beam $\omega_p^2w_0^2>2\xi_0^2/\sqrt{1+\xi_0^2/2}$ ($\xi_0\sim 1$); 2) the bubble regime, at $\omega_p w_0\lesssim 2\sqrt{\xi_0}$ ($\xi_0> 1$). The latter is less suitable for the collider application because the bubble cavity leads to defocusing for positrons; besides, the focusing forces and accelerating forces are not independently controllable here as both depend on the plasma density. Whereas, in the quasilinear regime this is possible due to the existence of a second control parameter given by the laser beam waist size. In the following, we will discuss qualitatively the scaling properties of the quasilinear regime. For more accurate expressions the reader is referred to \onlinecite{Schroeder_2010}.

In the standard LPA scheme, the electron plasma wave is driven by an intense laser pulse with duration $\tau_0$ of the order of the plasma wavelength $\lambda_p=2\pi/\omega_p$, which accelerates the electrons injected in the plasma wave by wave breaking \cite{Malka_2008,Esarey_2009,Leemans_2009}. The accelerating field $E_p$ of the plasma wave can be estimated from
\begin{equation}
E_p\sim \frac{m\omega_p}{|e|}\propto n_p^{1/2}.
\label{wakefield_E}
\end{equation}
In fact, in the plasma wave, the charge separation occurs on a length scale of the order of $\lambda_p$, producing a surface charge density $\sigma_p\sim |e|n_p\lambda_p$ and a field $E_p\sim 4\pi \sigma_p$, which corresponds to Eq. \eqref{wakefield_E}. The number of electrons $N_e$ that can be accelerated in a plasma wave is given approximately by the number of charged particles required to compensate for the laser-excited wakefield, having a longitudinal component $E_{\parallel}$. From the relation $E_{\parallel}\sim 4\pi N_e|e|/\pi w_0^2$, it follows that
\begin{equation}
N_e\sim \frac{\pi n_p}{\omega_p^3}\propto n_p^{-1/2},
\label{wakefield_N}
\end{equation}
because $w_0\omega_p\sim 1$ in the quasilinear regime. The interaction length of a single LPA stage is
limited by laser-diffraction effects, dephasing of the electrons with respect to the accelerating field, and laser-energy depletion. Laser-diffraction effects can be reduced by employing a plasma channel, and plasma-density tapering can be utilized to prevent dephasing. Therefore, the LPA interaction length will be determined by the energy depletion length $L_d$. We can estimate the latter by equating the energy spent for accelerating the $N_e$ electrons along $L_d$ ($\sim N_e|e|E_pL_d$) to the energy of the laser pulse ($\sim E_0^2\pi w_0^2\tau_0/8\pi$). Recalling that $w_0\omega_p\sim 1$, this yields
\begin{equation}
L_d\sim \frac{\omega_0^2}{\omega_p^2}\lambda_p\propto n_p^{-3/2}.
\label{wakefield_Ld}
\end{equation}
A staging of LPA is required to achieve high current densities along with high energies. The electron energy gain $\Delta\varepsilon_s$ in a single-stage LPA is
\begin{equation}
\Delta\varepsilon_s\sim |e|E_pL_d\sim m\frac{\omega_0^2}{\omega_p^2}\propto n_p^{-1}.
\label{wakefield_Ws}
\end{equation}
Therefore, the number of stages $N_s$ to achieve a total acceleration energy $\varepsilon_0$ is $N_s=\varepsilon_0/ \Delta\varepsilon_s\sim  (\varepsilon_0/m)(\omega_p/\omega_0)^2\propto n_p$. It corresponds to a total collider length $L_c$ of
\begin{equation}
L_c\sim N_sL_d\sim \frac{\varepsilon_0}{m}\lambda_p\propto n_p^{-1/2}.
\label{wakefield_Ns}
\end{equation}
When two identical beams each with $N$ particles and with horizontal (vertical) transverse beam size $\sigma_x$ ($\sigma_y$) collide with a frequency $f$, the luminosity $\mathcal{L}$ is defined as $\mathcal{L}=N^2f/4\pi\sigma_x\sigma_y$. In LPA $N=N_e$ and $f$ is the laser repetition rate, then
\begin{equation}
{\mathcal L}\sim\frac{1}{64\pi}\frac{1}{\omega_p^2\sigma_x\sigma_y}\frac{f}{r_0^2}.
\label{wakefield_Lum}
\end{equation}
The laser energy $W_s$ required in a single stage in the LPA collider is approximately given by $W_s\sim m(\lambda_p/r_0)(\omega_0/\omega_p)^2$ at $\xi_0\sim 1$ and the total required power $P_T$ amounts to $P_T \sim  N_sW_sf\sim  \varepsilon_0f(\lambda_p/r_0)$.

The above estimates show that, although the number $N_e$ of electrons in the bunch as well as the single-stage energy gain $\Delta\varepsilon_s$ increase at low plasma densities, the accelerating gradient $\Delta\varepsilon_s/L_d$ nevertheless decreases because the laser depletion length $L_d$ and the overall collider length $L_c$ increase as well. Limiting the total length of each LPA in a collider to about 100 m will require a plasma density $n_p\sim 10^{17}$ cm$^{-3}$ to provide a center-of mass energy of $\sim 1$ TeV for electrons, with $\sim 10$ GeV energy gain per stage \cite{Schroeder_2010}. For a number of electrons per bunch of $N_e\sim 10^9$, a laser repetition rate of $15\;\text{kHz}$ and a transverse beam size of about $10\;\text{nm}$ would be required to reach the goal-value of $10^{34}$ cm$^{-2}$s$^{-1}$ for the accelerator luminosity. At the usual condition $w_0\sim 1/\omega_p\sim 10\;\text{$\mu$m}$, instead, the luminosity amounts to ${\mathcal L}\sim 10^{27}$ cm$^{-2}$s$^{-1}$. In the above conditions each acceleration stage would be powered by a laser pulse with an energy of $30$ J corresponding to an average power of about $0.5\;\text{MW}$ at the required repetition rate of $15\;\text{kHz}$. Such high-average powers are beyond the performance of present-day lasers. The future hopes for high-average-power lasers are connected with diode-pumped lasers and new amplifier materials. Further challenges of LPA colliders are their complexity, as they involve plasma channels and density tapering, and, most importantly, the problem of how to accelerate positrons. 

\subsection{Laser micro-collider}
\label{MC}
We turn now to another scheme for a laser collider, which is based on principles quite different than those of the LPA scheme discussed above. In the LPA scheme, the electron is accelerated due to its synchronous motion with the propagating field. In this way the symmetry in the energy exchange process between the electron and the oscillating field is broken, as required by the Lawson-Woodward theorem \cite{Woodward_1947,Lawson_1979}. Another way to exploit the energy gain of the electron in the oscillating laser field is to initiate high-energy processes \textit{in situ}, i.e., inside the laser beam \cite{McDonald_1999}. In this case the temporary energy gain of the electron during interaction with a half cycle of the laser wave is used to trigger some processes during which the electron state may change (in particular, the electron may annihilate with a positron) and the desired asymmetry in the energy exchange can be achieved. In fact, this approach is widely employed in the nonrelativistic regime via laser-driven recollisions of an ionized electron with its parent ion (see Sec. \ref{HHG}). 

The question arises also as to whether the temporary energy gain of the electron in the laser beam can also be employed in the relativistic regime at ultrahigh energies. As pointed out in  Sec. \ref{RD}, an extension of the established recollision scheme with normal atoms into the relativistic regime is hindered by the relativistic drift. However, the drift will not cause any problem when positronium (Ps) atoms are used because its constituent particles, electron and positron, have the same absolute value of the charge-to-mass ratio (see Sec. \ref{FED_C}).
\begin{figure}
\includegraphics[width=0.8\linewidth]{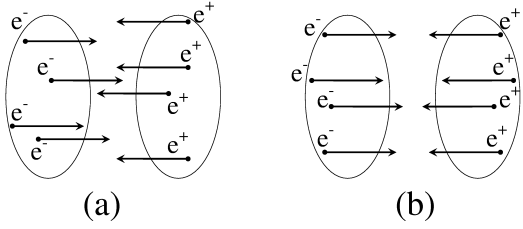}
\caption{(a) In conventional $e^+\text{-}e^-$ colliders bunches of accelerated electrons and positrons are focused to collide head-on-head {\it incoherently}, i.e., the bunches collide head-on-head but electrons and positrons in the bunch do not. (b) In the recollision-based collider, the electron and positron originating from the same Ps atom may collide head-on-head {\it coherently} \cite{Henrich_2004}. From \onlinecite{Hatsagortsyan_2006}.}
\label{microcollis}
\end{figure}

A corresponding realization of high-energy $e^+\text{-}e^-$ recollisions in the GeV domain aiming at particle reactions has been proposed in \onlinecite{Hatsagortsyan_2006}. It relies on (initially nonrelativistic) Ps atoms exposed to super-intense laser pulses. After almost instantaneous ionization of Ps in the strong laser field, the free electron and positron oscillate in opposite directions along the laser electric field and experience the same ponderomotive drift motion along the laser propagation direction. In this way, the particles acquire energy from the field and are driven into periodic $e^+\text{-}e^-$ collisions \cite{Henrich_2004}. Provided that the applied laser intensity is large enough, elementary particle reactions like heavy lepton-pair production can be induced in these recollisions. The common center-of-mass energy $\varepsilon^*$ of the electron and the positron at the recollision time arises mainly from the transversal momentum of the particles and it scales as $\varepsilon^*\sim m\xi_0$. A basic particle reaction which could be triggered in a laser-driven collider is $e^+\text{-}e^-$ annihilation with production of a $\mu^+\text{-}\mu^-$ pair, i.e., $e^+e^-\to\mu^+\mu^-$. The energy threshold for this process in the center-of-mass system is $2m_{\mu}\approx 210\,$MeV. It can be reached with a laser field such that $\xi_0 \sim 200$, corresponding to laser intensities of the order of $10^{22}$ W/cm$^2$, currently within reach.

In addition, the proposed recollision-based laser collider can yield high luminosities compared to conventional laser accelerators. In the latter, bunches of electrons and positrons are accelerated and brought into head-on-head collision. However, the particles in the bunch are distributed randomly such that each microscopic $e^+\text{-}e^-$ collision is not head-on-head but has a mean impact parameter $b_i\sim a_b$ determined by the beam radius $a_b$, characterizing the collision as {\it incoherent} (see Fig. \ref{microcollis}). Instead, in the recollision-based collider the electron and the positron stem from the same Ps atom with initial coordinates being confined within the range of one Bohr radius $a_B\approx 5.3\times 10^{-9}\;\text{cm}$. Since they are driven coherently by the laser field, they can recollide with a mean impact parameter $b_c\sim a_{wp}$ of the order of the electron wave packet size $a_{wp}$ (see Fig. \ref{microcollis}). Consequently, the luminosity contains a {\it coherent} component \cite{Hatsagortsyan_2006}:
\begin{equation}
\mathcal{L}=\left[\frac{N_p(N_p-1)}{b_i^2}+\frac{N_p}{b_c^2}\right]f,
\label{L}
\end{equation}
where $N_p$ is the number of particles in the bunch, and $f$ is the bunch repetition frequency. The coherent component $(N_p/b_c^2)f$ can lead to a substantial luminosity enhancement in the case when the particle number is low and the  particle's wave packet spreading is small, $N_pa_{wp}^2<a_b^2$. Note that the reaction ${\rm Ps}\to\mu^+\mu^-$ arising in a strongly laser-driven $e^+\text{-}e^-$ plasma may be considered as the coherent counterpart of the incoherent process $e^+e^-\to\mu^+\mu^-$ discussed in Sec. \ref{ML}.

Rigorous quantum-electrodynamical calculations for $\mu^+\text{-}\mu^-$ pair production in a laser field have been performed in \onlinecite{Muller_2006,Muller_2008a,Muller_2008b}. In agreement with Eq.~(\ref{L}), they enabled the development of a simple-man's model in which the rate of the laser-driven process can be expressed via a convolution of the rescattering electron wave packet with the field-free cross section $\sigma_{e^+e^-\to\mu^+\mu^-}$ (see Eq. \ref{sigma}). The latter attains the maximal value $\sigma^{(\text{max})}_{e^+e^-\to\mu^+\mu^-}\sim \alpha^2\lambda_{C,\mu}^2\sim 10^{-30}\;\text{cm$^2$}$ at $\varepsilon^*\approx 260\;\text{MeV}$ \cite{Peskin_1995}. However, when the field driving the Ps atoms is a single laser wave, the $e^+\text{-}e^-$ recollision times are long and the $\mu^+\text{-}\mu^-$ production process is substantially suppressed by extensive wave packet spreading. This obstacle can be overcome when two counterpropagating laser beams are employed.

\begin{figure}
\includegraphics[width=0.8\linewidth]{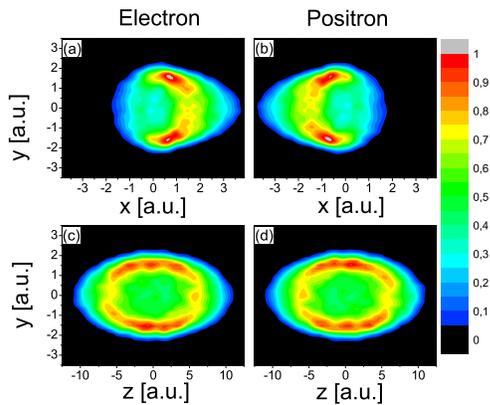}
\caption{(Color) The coordinate-space distributions of the electron and the positron wave-packets at the recollision time in focused counterpropagating pulses along the $z$ direction with $w_0 =10\;\mu$m, $\lambda_0=0.8\;\mu $m and with $I_0 = 4.7\times10^{22}$W/cm$^2$ (parts (a) and (c)) and $I_0 = 1.4\times10^{23}$W/cm$^2$ (parts (b) and (d)). The Ps atom is initially located at the origin. Spatial coordinates are given in ``atomic units'', with  $1\;\text{a.u.}=0.05\;\text{nm}$. Adapted from \onlinecite{Liu_2009}.}
\label{MCcrossed}
\end{figure}

The role of the spreading of the electron wave packet in counterpropagating focused laser beams of circular and linear polarization has been investigated in detail in \onlinecite{Liu_2009}. The advantage of the circular-polarization setup is the focusing of the recolliding electron wave packet. However, this advantage is reduced by a spatial offset in the $e^+\text{-}e^-$ collision when the initial coordinate of the Ps atom deviates from the symmetric position between the laser pulses. The latter imposes a severe restriction on the Ps gas size along the laser propagation direction. Thus, the linear-polarization setup is preferable when the offset at the recollision is very small and the wave packet size at the recollision is within acceptable limits. Results from a Monte Carlo simulation of the $e^+\text{-}e^-$ wave-packet dynamics in counterpropagating linearly polarized laser pulses are shown in Fig. \ref{MCcrossed}.

The luminosity ${\mathcal L}$ and the number of reaction events ${\mathcal N}$ for the recollision-based collider with counter-propagating laser pulses can be estimated as:
\begin{align}
{\mathcal L} \sim&  N_{\text{Ps}}\frac{1}{a_{wp}^3 }\tau_r f,\\
{\mathcal N} \sim& \frac{\sigma^{(\text{max})}_{e^+e^-\to\mu^+\mu^-}}{a_{wp}^3}\tau_r N_{\text{Ps}}N_L,
\label{Ps_Lum}
\end{align}
respectively, where $N_{\text{Ps}}$ is the number of Ps atoms, $\tau_r$ the recollision time of the order of the lasers period, $N_L$ the number of laser pulses, and $f$ the laser repetition rate. Taking  $N_{\text{Ps}}\approx 10^8$ \cite{Cassidy_2005}, $f=1$ Hz and the spatial extension of the $e^+\text{-}e^-$ pair from Fig. \ref{MCcrossed}, one estimates a luminosity of ${\mathcal L}\sim 10^{27}$ cm$^{-2}$s$^{-1}$ and about one $\mu^+\text{-}\mu^-$ pair production event every $10^3$ laser shots at a laser intensity of $4.7\times 10^{22}$ W/cm$^2$. 

In conclusion, the scheme of the recollision-based laser collider allows to realize high-energy and high-luminosity collisions in a microscopic setup. However, it is not easily scalable to the parameters of the ILC, namely, to TeV energies and luminosities of the order of $10^{34}$ cm$^{-2}$s$^{-1}$.

\section{Particle physics within and beyond the Standard Model}
\label{Part_Phys}
The sustained progress in laser technology towards higher and higher field intensities raises the question as to what extent ultrastrong laser fields may develop into a useful tool for particle physics beyond QED. Below we review theoretical predictions regarding the influence of super-intense laser waves on electroweak processes and their potential for probing new physics beyond the Standard Model.

\subsection{Electroweak sector of the Standard Model}

The energy scale of weak interactions is set by the masses of the $W^\pm$ and $Z^0$ exchange bosons, $m_W\approx m_Z\sim 100$\,GeV. Therefore, the influence of external laser fields, even if strong on the scale of QED, is generally rather small. An overview of weak interaction processes in the presence of intense electromagnetic fields has been given in \onlinecite{Kurilin_1999}. 

Various weak decay processes in the presence of intense laser fields have been considered. They can be divided into two classes: 1) laser-assisted processes which also exist in the absence of the field but may be modified due to its presence; 2) field-induced processes which can only proceed when a background field is present, providing an additional energy reservoir. With respect to processes from the first category, $\pi\to\mu +\nu$ and $\mu^- \to e^- + \bar{\nu}_e + \nu_\mu$ have already been examined \cite{Ritus_1985}. Laser-assisted muon decay has also been revisited recently \cite{Narozhny_2008, Dicus_2009, Farzinnia_2009}. $W^\pm$ and $Z^0$ boson decay into a fermion-antifermion pair was calculated in \onlinecite{Kurilin_2004, Kurilin_2009}. In all cases, the effect of the laser field was found to be small. As a general result, the presence of the laser field modifies the field-free decay rate $R_{M,0}$ of a particle with mass $M$ to $R_M = R_{0,M}(1+\Delta)$, with the correction $\Delta$ being of the order of $\chi_{0,M}^2$ in the range of parameters $\xi_{0,M}=\xi_0 m/M > 1$ and $\chi_{0,M}=\chi_0(m/M)^3\ll 1$. The elastic scattering of a muon neutrino and an electron in the presence of a strong laser field has been considered in \onlinecite{Bai_2012} and multiphoton effects in the cross section are predicted.

External fields can also induce decay processes, which are energetically forbidden otherwise. In \onlinecite{Kurilin_1999}, the field-induced lepton decay $l^-\to W^- + \nu_l$ was considered. Since the mass of the initial-state particle is smaller than the mass of the decay products, the process is clearly impossible in vacuum. The presence of the field does allow for such an exotic decay, but the probability $P_{l^-\to W^- + \nu_l}$ remains exponentially suppressed, i.e., $P_{l^-\to W^- + \nu_l}\sim \exp(-1/\chi_{0,W})$, where $\chi_{0,W}=\chi_0(m/m_W)^3$.

Finally, the production of an $e^+\text{-}e^-$ pair by high-energy neutrino impact on a strong laser pulse has been calculated in \onlinecite{Tinsley_2005}. The setup is similar to the $e^+\text{-}e^-$ pair production processes in QED discussed in Secs. \ref{PP_L} and \ref{BH}. However, as it was shown in \onlinecite{Tinsley_2005}, the laser-induced process $\nu \to \nu + e^+ + e^-$ is extremely unlikely. At a field intensity of about $I_0=3\times 10^{18}\;\text{W/cm$^2$}$, the production length is on the order of a light year, even for a neutrino energy of 1 PeV.

\subsection{Particle physics beyond the Standard Model}

Recently, a lot of attention has been devoted to the possibility of employing intense laser sources to test aspects of physical theories which go even beyond the Standard Model. In \onlinecite{Heinzl_2010c}, for example, it is envisaged that effects of the noncommutativity of space-time modify the kinematics of multiphoton Compton scattering by inducing a nonzero photon mass. We recall that in noncommutative quantum field theories operators $X^{\mu}$ are associated to spacetime coordinates $x^{\mu}$, which do not commute, but rather satisfy the commutation relations $[X^{\mu},X^{\nu}]=i\Theta^{\mu\nu}$, with $\Theta^{\mu\nu}$ being an antisymmetric constant tensor \cite{Douglas_2001}.

On a different side, one of the still open problems of the Standard Model is the so-called strong CP problem \cite{Kim_2010}. The nontrivial structure of the vacuum, as predicted by Quantum Chromodynamics (QCD), allows for the violation within QCD of the combined symmetry of charge conjugation (C) and parity (P). This implies a value for the neutron's electric dipole moment which, however, is already many orders of magnitude larger than experimental upper limits. One way of solving this problem was suggested in \onlinecite{Peccei_1977} which required the existence of a massive pseudoscalar boson, called axion. The axion has never been observed experimentally although some of its properties can be predicted on theoretical grounds: it should be electrically neutral and its mass should not exceed $1\;\text{eV}$ in order of magnitude. Although being electrically neutral, the axion is predicted to couple to the electromagnetic field $\mathcal{F}^{\mu\nu}(x)$ through a Lagrangian-density term
\begin{equation}
\label{L_ag}
\mathscr{L}_{a\gamma}(x)=\frac{g}{4}a(x)\mathcal{F}^{\mu\nu}(x)\tilde{\mathcal{F}}_{\mu\nu}(x), 
\end{equation}
with $g$ being the photon-axion coupling constant and $a(x)$ the axion field. 

The photon-axion Lagrangian density in Eq. \eqref{L_ag} has mainly two implications: 1) the existence of axions induces a change in the polarization of a light beam passing through a background electromagnetic field; 2) a photon can transform into an axion (and vice versa) in the presence of a background electromagnetic field. The first prediction has been tested in experiments like the Brookhaven-Fermilab-Rochester-Trieste (BFRT) \cite{BFRT} and the Polarizzazione del Vuoto con LASer (PVLAS) \cite{PVLAS}, where a linearly polarized probe laser field crossed a region in which a strong magnetic field was present  of $3.25\;\text{T}$ and $5\;\text{T}$ at BFRT and at PVLAS, respectively. Testing the second prediction is the aim of the so-called ``light shining through a wall'' experiments like the Any Light Particle Search (ALPS) \cite{ALPS}, the CERN Axion Solar Telescope (CAST) \cite{CAST} and Gamma to milli-eV particle search (GammeV) \cite{GammeV} (see also the detailed theoretical analysis in \onlinecite{Adler_2008}). In the GammeV experiment, for example, the light of a Nd:YAG laser passes through a region in which a $5\;\text{T}$ magnetic field is present. A mirror is positioned behind that region in order to reflect the laser light. The axions which would eventually be created in the magnetic-field region pass through the mirror undisturbed and can be reconverted to photons by means of a second magnetic field, activated after the mirror itself. So far these experiments have given negative results. An interesting experimental proposal has been put forward in \onlinecite{Rabadan_2006}, where the high-energy photon beam delivered by an XFEL facility has been suggested as a probe beam to test regions of parameters (like the axion mass $m_a$ or the photon-axion coupling constant $g$) which are inaccessible via optical laser light. 

The perspective for reaching ultrahigh intensities at future laser facilities has stimulated new proposals for employing such fields to elicit the photon-axion interaction \cite{Gies_2009}. In fact, an advantage of using strong uniform magnetic fields is that they can be kept strong for a macroscopically long time (of the order of hours) and on a macroscopic spatial region (of the order of $1\;\text{m}$) (for QED processes occurring in a strong magnetic field, see the standard review \onlinecite{Erber_1966} and the very current overview paper \onlinecite{Dunne_2012} for recent progresses in the field). On the other hand, laser beams deliver fields much stronger than those employed in the mentioned experiments (the magnetic field strength of a laser beam with the available intensity of $10^{22}\;\text{W/cm$^2$}$ amounts to about $6.5\times 10^5\;\text{T}$) but in a microscopic space-time region. However, it has been first realized in \onlinecite{Mendonca_2007} that envisaged ultrahigh intensities at future laser facilities may compensate for the tiny space-time extension of the laser spot region. In \onlinecite{Mendonca_2007} the coupled equations of the electromagnetic field $\mathcal{F}^{\mu\nu}(x)$ and the axion field $a(x)$
\begin{equation}
\label{a_F}
\left\{
\begin{aligned}
&\partial_{\mu}\partial^{\mu}a+m_a^2a=\frac{1}{4}g\mathcal{F}^{\mu\nu}\tilde{\mathcal{F}}_{\mu\nu}\\
&\partial_{\mu}\mathcal{F}^{\mu\nu}=g(\partial_{\mu}a)\tilde{\mathcal{F}}^{\mu\nu}
\end{aligned}
\right.
\end{equation}
are solved approximately. It is shown that if a probe laser field propagates through a strong plane-wave field, the axion field ``grows'' at the expense mainly of the probe field itself, whose intensity should be observed to decrease. Laser powers of the order of $1\;\text{PW}$ have already been shown to provide stronger hints for the presence of axions than magnetic-field-based experiments like the PVLAS. More realistic Gaussian laser beams are considered in \onlinecite{Dobrich_2010} where the starting point is also represented by Eq. \eqref{a_F}. The suggested experimental setup assumes a probe electromagnetic beam with angular frequency $\omega_{p,\text{in}}$ passing through a strong counterpropagating Gaussian beam with angular frequency $\omega_{0,\parallel}$ and another strong Gaussian beam propagating perpendicularly and with angular frequency $\omega_{0,\perp}$. By choosing $\omega_{0,\perp}=2\omega_{0,\parallel}$, it is found that after a photon-axion-photon double conversion, photons are generated with angular frequencies $\omega_{p,\text{out}}=\omega_{p,\text{in}}\pm \omega_{0,\parallel}$. The amplitudes of these processes are shown to be peaked at specific values of the axion mass $m_{a,\pm}=2\sqrt{\omega_{p,\text{in}}\omega_{0,\parallel}+\omega_{0,\parallel}^2(1\pm 1)/2}$. Since the optical photon energies are of the order of $1\;\text{eV}$, this setup allows for the investigation of values of the axion mass in this regime. This is very important because such a region of the axion mass is inaccessible to experiments based on strong magnetic fields, which can probe regions at most in the meV range.

In addition to electrically neutral new particles such as axions, yet unobserved particles with nonzero charge may also exist. The fact that they have so far escaped detection implies that they are either very heavy (rendering them a target for large-scale accelerator experiments), or that they are light but very weakly charged. In the latter case, these so-called minicharged particles, i.e., particles with absolute value of the electric charge much smaller than $|e|$, are suitable candidates for laser-based searches \cite{Gies_2009}. Let $m_\epsilon$ and $Q_\epsilon=\epsilon e$, with $0<\epsilon \ll 1$, denote the minicharged particle mass and charge, respectively. Then, the corresponding critical field scale $F_{\text{cr},\epsilon}=m_\epsilon^2/|Q_\epsilon|$ can be much lower than $F_{\text{cr}}$. As a consequence, vacuum nonlinearities associated with minicharged particles may be very pronounced in an external laser field with intensity much less than $I_{\text{cr}}\sim 10^{29}\;\text{W/cm$^2$}$. Moreover, even at optical photon energies $\sim 1\;\text{eV}$ the effective Lagrangian approach might become inappropriate to describe the relevant physics if $m_{\epsilon}\lesssim 1\;\text{eV}$ (see Sec. \ref{Diff_Low}). In \onlinecite{Gies_2006}, vacuum dichroism and birefringence effects due to the existence of minicharged particles were analyzed when a probe laser beam with $\omega_p>2m_\epsilon$ traverses a magnetic field. It was shown that polarization measurements in this setup would provide much stronger constraints on minicharged particles in the mass range below 0.1~eV than in previous laboratory searches.

\section{Conclusion and outlook}
\label{Conlc}
The fast development of laser technology has been paving the way to employ laser sources for investigating relativistic, quantum electrodynamical, nuclear and high-energy processes. Starting with the lowest required intensities, relativistic atomic processes are already within the reach of available laser systems, while the proposed methods to compensate the deteriorating effects of the relativistic drift still have to be tested experimentally. Moreover, a fully consistent theoretical interpretation of recent experimental results on correlation effects in relativistic multielectron tunneling is still missing.

Concerning the interaction of free electrons with intense laser beams, we have seen that experiments have been performed to explore the classical regime. Only the E-144 experiment at SLAC has so far been realized on multiphoton Compton scattering, although presently available lasers and electron beams would allow for probing this regime in full detail. We have also pointed out that at laser intensities of the order of $10^{22}\text{-} 10^{23}\;\text{W/cm$^2$}$, RR effects come into play at electron energies of the order of a few GeV. It is envisaged that the quantum radiation-dominated regime, where quantum and RR effects substantially alter the electron dynamics, could be one of the first extreme regimes of light-matter interaction to be probed with upcoming petawatt laser facilities. On the theoretical side, most of the calculations have been performed by approximating the laser field as a plane wave, as the Dirac equation in the presence of a focused background field cannot be solved analytically. Certainly, new methods have to be developed to calculate photon spectra including quantum effects and spatio-temporal focusing of the laser field in order to be able to quantitatively interpret upcoming experimental results.

Nonlinear quantum electrodynamical effects have been shown to become observable at future multipetawatt laser facilities, as well as at ELI and HiPER. Here the main challenges concern the measurability of tiny effects on the polarization of probe beams and on the detection of a typically very low number of signal photons out of large backgrounds. Similar challenges are envisaged to detect the presence of light and weakly-interacting hypothetical particles like axions and minicharged particles. The physical properties (mass, coupling constants, etc.) of such hypothetical particles are, of course, unknown. In this respect intense laser fields may be employed here to set bounds on physical quantities like the axion mass and, in particular, to scan regions of physical parameters, which are inaccessible to conventional methods based, for example, on astrophysical observations.

Different schemes have been proposed to observe $e^+\text{-}e^-$ pair production at intensities below the Schwinger limit, which seems now to be feasible in the near future, at least from a theoretical point of view. Corresponding studies would complement the results of the pioneering E-144 experiment and deepen our understanding of the QED vacuum in the presence of extreme electromagnetic fields. This is also connected with the recent investigations on the development of QED cascades in laser-laser collisions. In addition to being  intrinsically interesting, the development of QED cascades is expected to set a limit on the maximal attainable laser intensity. However, the study of quantum cascades in intense laser fields has started relatively recently and is still under vivid development. More advanced analytical and numerical methods are required in order to describe realistically and quantitatively such a complex system as an electron-positron-photon plasma in the presence of a strong driving laser field.

Nuclear quantum optics is also a new exciting and promising field. Since the energy difference between nuclear levels is typically in the multi-keV and MeV range, high-frequency laser pulses, especially in combination with accelerators, are preferable in controlling nuclear dynamics. As pointed out, especially table-top highly coherent x-ray light beams, envisaged for the future, open up perspectives for exciting applications including nuclear state preparation and nuclear batteries.

Finally, we want to point out that most of the considered processes have not yet been observed or tested experimentally. This is in our opinion one of the most challenging aspects of upcoming laser physics, not only from an experimental point of view, but also from the point of view of theoretical methods. Experimentally, the main reason is that in order to test, for example, nonlinear quantum electrodynamics or to investigate nuclear quantum optics, high-energy particle beams (including photon beams) are required to be available in the same laboratory as the strong laser. On the one hand, the combined expertise from different experimental physical communities is required to perform such complex but fundamental experiments. On the other hand, the fast technological development of laser-plasma accelerators is very promising and exciting, as this seems the most feasible way towards the realization of stable, table-top high-energy particle accelerators. Combining such high-energy probe beams with an ultrarelativistic laser beam in a single all-optical setup will certainly result in a unique tool for advancing our understanding of intense laser-matter interactions.

\begin{acknowledgments}
We are grateful to many students, colleagues and collaborators for inspiring discussions, ideas, and suggestions, in particular in joint publications during the here most relevant last six years, to H. Bauke, T. J. B{\"u}rvenich, A. Dadi, C. Deneke, J. Evers, B. Galow, R. Grobe, Z. Harman, H. Hu, A. Ipp, U. D. Jentschura, B. King, M. Klaiber, M. Kohler, G. Yu. Kryuchkyan, W.-T. Liao, C. Liu, E. L\"{o}tstedt, A. Macchi, F. Mackenroth, S. Meuren, A. I. Milstein, G. Mocken, S. J. M\"{u}ller, T.-O. M\"{u}ller, A. P\'{a}lffy, F. Pegoraro, M. Ruf, Y. I. Salamin, M. Tamburini, M. Verschl, and A. B. Voitkiv.
\end{acknowledgments}

\section*{List of frequently-used symbols}
\begin{tabular}{ll}
\hspace{-15pt} $E_0$  &  laser field amplitude\\
\hspace{-15pt} $F_{\text{cr}}=m^2/|e|$ & critical electromagnetic field \\
&($\approx 1.3\times 10^{16}\;\text{V/cm}$)\\
&($\approx 4.4\times 10^{14}\;\text{G}$)\\
\hspace{-15pt} $I_0=E_0^2/4\pi$  &  laser peak intensity\\
\hspace{-15pt} $I_{\text{cr}}=F_{\text{cr}}^2/4\pi$ & critical laser intensity \\
& ($\approx 4.6\times 10^{29}\;\text{W/cm$^2$}$)\\
\hspace{-15pt} $Z$  & nuclear charge number\\
\hspace{-15pt} $e$ \hspace{105pt} & electron charge \\
&($\approx -1/\sqrt{137}\approx -0.085$)\\
\hspace{-15pt} $k^{\mu}=(\omega,\bm{k})=\omega n^{\mu}$ & initial or incoming photon \\
&four-momentum\\
\hspace{-15pt} $k^{\prime\mu}=(\omega',\bm{k}')=\omega' n^{\prime\mu}$ & final or outgoing photon \\
&four-momentum\\
\hspace{-15pt} $k_0^{\mu}=\omega_0n_0^{\mu}=\omega_0(1,\bm{n}_0)$ & laser photon four-momentum\\
\hspace{-15pt} $m$ & electron mass\\
&($\approx 0.511\;\text{MeV}$)\\
\hspace{-15pt} $p_0^{\mu}=(\varepsilon_0,\bm{p}_0)=m\gamma_0(1,\bm{\beta}_0)$ & initial or incoming electron \\
&four-momentum\\
\hspace{-15pt} $\alpha=e^2$ & fine-structure constant \\
& ($\approx 1/137\approx 7.3\times 10^{-3}$)\\
\hspace{-15pt} $\chi_0=(p_{0,-}/m)(E_0/F_{\text{cr}})$  &  nonlinear electron quantum \\
&parameter \\
\hspace{-15pt} $\varkappa_0=(k_-/m)(E_0/F_{\text{cr}})$  &  nonlinear photon quantum \\
&parameter \\
\hspace{-15pt} $\lambda_0=T_0=2\pi/\omega_0$  & laser wavelength and period \\
\hspace{-15pt} $\lambda_C=1/m$  & Compton wavelength \\
& ($\approx 3.9\times 10^{-11}\;\text{cm}$)\\
\hspace{-15pt} $\xi_0=|e|E_0/m\omega_0$  &  classical relativistic \\
& parameter \\
\hspace{-15pt} $\omega_0$ \hspace{105pt}  & laser angular frequency\\
\end{tabular}

\bibliographystyle{apsrmp}
\bibliography{rmp-bibliography}

\begin{thebibliography}{519}
\expandafter\ifx\csname natexlab\endcsname\relax\def\natexlab#1{#1}\fi
\expandafter\ifx\csname bibnamefont\endcsname\relax
  \def\bibnamefont#1{#1}\fi
\expandafter\ifx\csname bibfnamefont\endcsname\relax
  \def\bibfnamefont#1{#1}\fi
\expandafter\ifx\csname citenamefont\endcsname\relax
  \def\citenamefont#1{#1}\fi
\expandafter\ifx\csname url\endcsname\relax
  \def\url#1{\texttt{#1}}\fi
\expandafter\ifx\csname urlprefix\endcsname\relax\def\urlprefix{URL }\fi
\providecommand{\bibinfo}[2]{#2}
\providecommand{\eprint}[2][]{\url{#2}}

\bibitem[{\citenamefont{Aas} \emph{et~al.}(1999)\citenamefont{Aas, Mach,
  Kvasil, Borge, Fogelberg, Grant, Gulda, Hageb\o, Hoff, Kurcewicz, Lindroth,
  L{\o}vh{\o}iden} \emph{et~al.}}]{Aas_1999}
\bibinfo{author}{\bibnamefont{Aas}, \bibfnamefont{A.~J.}},
  \bibinfo{author}{\bibfnamefont{H.}~\bibnamefont{Mach}},
  \bibinfo{author}{\bibfnamefont{J.}~\bibnamefont{Kvasil}},
  \bibinfo{author}{\bibfnamefont{M.~J.~G.} \bibnamefont{Borge}},
  \bibinfo{author}{\bibfnamefont{B.}~\bibnamefont{Fogelberg}},
  \bibinfo{author}{\bibfnamefont{I.~S.} \bibnamefont{Grant}},
  \bibinfo{author}{\bibfnamefont{K.}~\bibnamefont{Gulda}},
  \bibinfo{author}{\bibfnamefont{E.}~\bibnamefont{Hageb\o}},
  \bibinfo{author}{\bibfnamefont{P.}~\bibnamefont{Hoff}},
  \bibinfo{author}{\bibfnamefont{W.}~\bibnamefont{Kurcewicz}},
  \bibinfo{author}{\bibfnamefont{A.}~\bibnamefont{Lindroth}},
  \bibinfo{author}{\bibfnamefont{G.}~\bibnamefont{L{\o}vh{\o}iden}},
  \emph{et~al.}, \bibinfo{year}{1999}, \bibinfo{journal}{Nucl. Phys. A}
  \textbf{\bibinfo{volume}{654}}, \bibinfo{pages}{499}.

\bibitem[{\citenamefont{Abraham}(1905)}]{Abraham_b_1905}
\bibinfo{author}{\bibnamefont{Abraham}, \bibfnamefont{M.}},
  \bibinfo{year}{1905}, \emph{\bibinfo{title}{Theorie der Elektrizit{\"a}t}}
  (\bibinfo{publisher}{Teubner, Leipzig}).

\bibitem[{\citenamefont{Adler} \emph{et~al.}(2008)\citenamefont{Adler, Gamboa,
  M\'{e}ndez, and L\'{o}pez-Sarri\'{o}n}}]{Adler_2008}
\bibinfo{author}{\bibnamefont{Adler}, \bibfnamefont{S.~L.}},
  \bibinfo{author}{\bibfnamefont{J.}~\bibnamefont{Gamboa}},
  \bibinfo{author}{\bibfnamefont{F.}~\bibnamefont{M\'{e}ndez}}, and
  \bibinfo{author}{\bibfnamefont{J.}~\bibnamefont{L\'{o}pez-Sarri\'{o}n}},
  \bibinfo{year}{2008}, \bibinfo{journal}{Ann. Phys. (N.Y.)}
  \textbf{\bibinfo{volume}{323}}, \bibinfo{pages}{2851}.

\bibitem[{\citenamefont{Agostini and DiMauro}(2004)}]{Agostini_2004}
\bibinfo{author}{\bibnamefont{Agostini}, \bibfnamefont{P.}}, and
  \bibinfo{author}{\bibfnamefont{L.~F.} \bibnamefont{DiMauro}},
  \bibinfo{year}{2004}, \bibinfo{journal}{Rep. Prog. Phys.}
  \textbf{\bibinfo{volume}{67}}, \bibinfo{pages}{813}.

\bibitem[{\citenamefont{Aharonian and Plyasheshnikov}(2003)}]{Aharonian_2003}
\bibinfo{author}{\bibnamefont{Aharonian}, \bibfnamefont{F.~A.}}, and
  \bibinfo{author}{\bibfnamefont{A.~V.} \bibnamefont{Plyasheshnikov}},
  \bibinfo{year}{2003}, \bibinfo{journal}{Astropart. Phys.}
  \textbf{\bibinfo{volume}{19}}, \bibinfo{pages}{525}.

\bibitem[{\citenamefont{Akhiezer}(1937)}]{Akhiezer_1937}
\bibinfo{author}{\bibnamefont{Akhiezer}, \bibfnamefont{A.~I.}},
  \bibinfo{year}{1937}, \bibinfo{journal}{Phys. Z. Sowjetunion}
  \textbf{\bibinfo{volume}{11}}, \bibinfo{pages}{263}.

\bibitem[{\citenamefont{Akhmedov}(1983)}]{Akhmedov_1983}
\bibinfo{author}{\bibnamefont{Akhmedov}, \bibfnamefont{E.~K.}},
  \bibinfo{year}{1983}, \bibinfo{journal}{Sov. Phys. JETP}
  \textbf{\bibinfo{volume}{58}}, \bibinfo{pages}{883}.

\bibitem[{\citenamefont{Akkermans and Dunne}(2012)}]{Akkermans_2012}
\bibinfo{author}{\bibnamefont{Akkermans}, \bibfnamefont{E.}}, and
  \bibinfo{author}{\bibfnamefont{G.~V.} \bibnamefont{Dunne}},
  \bibinfo{year}{2012}, \bibinfo{journal}{Phys. Rev. Lett.}
  \textbf{\bibinfo{volume}{108}}, \bibinfo{pages}{030401}.

\bibitem[{\citenamefont{Aksenov} \emph{et~al.}(2007)\citenamefont{Aksenov,
  Ruffini, and Vereshchagin}}]{Ruffini_2007}
\bibinfo{author}{\bibnamefont{Aksenov}, \bibfnamefont{A.~G.}},
  \bibinfo{author}{\bibfnamefont{R.}~\bibnamefont{Ruffini}}, and
  \bibinfo{author}{\bibfnamefont{G.~V.} \bibnamefont{Vereshchagin}},
  \bibinfo{year}{2007}, \bibinfo{journal}{Phys. Rev. Lett.}
  \textbf{\bibinfo{volume}{99}}, \bibinfo{pages}{125003}.

\bibitem[{\citenamefont{Albert} \emph{et~al.}(2010)\citenamefont{Albert,
  Anderson, Gibson, Hagmann, Johnson, Messerly, Semenov, Shverdin, Rusnak,
  Tremaine, Hartemann, Siders} \emph{et~al.}}]{Albert_2010}
\bibinfo{author}{\bibnamefont{Albert}, \bibfnamefont{F.}},
  \bibinfo{author}{\bibfnamefont{S.~G.} \bibnamefont{Anderson}},
  \bibinfo{author}{\bibfnamefont{D.~J.} \bibnamefont{Gibson}},
  \bibinfo{author}{\bibfnamefont{C.~A.} \bibnamefont{Hagmann}},
  \bibinfo{author}{\bibfnamefont{M.~S.} \bibnamefont{Johnson}},
  \bibinfo{author}{\bibfnamefont{M.}~\bibnamefont{Messerly}},
  \bibinfo{author}{\bibfnamefont{V.}~\bibnamefont{Semenov}},
  \bibinfo{author}{\bibfnamefont{M.~Y.} \bibnamefont{Shverdin}},
  \bibinfo{author}{\bibfnamefont{B.}~\bibnamefont{Rusnak}},
  \bibinfo{author}{\bibfnamefont{A.~M.} \bibnamefont{Tremaine}},
  \bibinfo{author}{\bibfnamefont{F.~V.} \bibnamefont{Hartemann}},
  \bibinfo{author}{\bibfnamefont{C.~W.} \bibnamefont{Siders}}, \emph{et~al.},
  \bibinfo{year}{2010}, \bibinfo{journal}{Phys. Rev. ST Accel. Beams}
  \textbf{\bibinfo{volume}{13}}, \bibinfo{pages}{070704}.

\bibitem[{\citenamefont{Alkofer} \emph{et~al.}(2001)\citenamefont{Alkofer,
  Hecht, Roberts, Schmidt, and Vinnik}}]{Alkofer_2001}
\bibinfo{author}{\bibnamefont{Alkofer}, \bibfnamefont{R.}},
  \bibinfo{author}{\bibfnamefont{M.~B.} \bibnamefont{Hecht}},
  \bibinfo{author}{\bibfnamefont{C.~D.} \bibnamefont{Roberts}},
  \bibinfo{author}{\bibfnamefont{S.~M.} \bibnamefont{Schmidt}}, and
  \bibinfo{author}{\bibfnamefont{D.~V.} \bibnamefont{Vinnik}},
  \bibinfo{year}{2001}, \bibinfo{journal}{Phys. Rev. Lett.}
  \textbf{\bibinfo{volume}{87}}, \bibinfo{pages}{193902}.

\bibitem[{\citenamefont{{ALPS}}(2010)}]{ALPS}
\bibinfo{author}{\bibnamefont{{ALPS}}} (\bibinfo{collaboration}{Any Light
  Particle Search}), \bibinfo{year}{2010},
  \urlprefix\url{http://alps.desy.de/}.

\bibitem[{\citenamefont{Ammosov} \emph{et~al.}(1986)\citenamefont{Ammosov,
  Delone, and Krainov}}]{Ammosov_1986}
\bibinfo{author}{\bibnamefont{Ammosov}, \bibfnamefont{M.~V.}},
  \bibinfo{author}{\bibfnamefont{N.~B.} \bibnamefont{Delone}}, and
  \bibinfo{author}{\bibfnamefont{V.~P.} \bibnamefont{Krainov}},
  \bibinfo{year}{1986}, \bibinfo{journal}{Sov. Phys. JETP}
  \textbf{\bibinfo{volume}{64}}, \bibinfo{pages}{1191}.

\bibitem[{\citenamefont{Apostol}(2011)}]{Apostol_2011}
\bibinfo{author}{\bibnamefont{Apostol}, \bibfnamefont{M.}},
  \bibinfo{year}{2011}, \bibinfo{journal}{J. Mod. Opt.}
  \textbf{\bibinfo{volume}{58}}, \bibinfo{pages}{611}.

\bibitem[{\citenamefont{Aprahamian and Sun}(2005)}]{Aprahamian_2005}
\bibinfo{author}{\bibnamefont{Aprahamian}, \bibfnamefont{A.}}, and
  \bibinfo{author}{\bibfnamefont{Y.}~\bibnamefont{Sun}}, \bibinfo{year}{2005},
  \bibinfo{journal}{Nature Phys.} \textbf{\bibinfo{volume}{1}},
  \bibinfo{pages}{81}.

\bibitem[{\citenamefont{{Astra Gemini}}(2011)}]{AG}
\bibinfo{author}{\bibnamefont{{Astra Gemini}}}, \bibinfo{year}{2011},
  \urlprefix\url{http://www.clf.rl.ac.uk/Science/12258.aspx}.

\bibitem[{\citenamefont{Avetissian}
  \emph{et~al.}(2003)\citenamefont{Avetissian, Avetissian, Mkrtchian, and
  Sedrakian}}]{Avetissian_2003}
\bibinfo{author}{\bibnamefont{Avetissian}, \bibfnamefont{H.~K.}},
  \bibinfo{author}{\bibfnamefont{A.~K.} \bibnamefont{Avetissian}},
  \bibinfo{author}{\bibfnamefont{G.~F.} \bibnamefont{Mkrtchian}}, and
  \bibinfo{author}{\bibfnamefont{K.~V.} \bibnamefont{Sedrakian}},
  \bibinfo{year}{2003}, \bibinfo{journal}{Nucl. Instr. Meth. Phys. Res. A}
  \textbf{\bibinfo{volume}{507}}, \bibinfo{pages}{582}.

\bibitem[{\citenamefont{Babzien} \emph{et~al.}(2006)\citenamefont{Babzien,
  Ben-Zvi, Kusche, Pavlishin, Pogorelsky, Siddons, Yakimenko, Cline, Zhou,
  Hirose, Kamiya, Kumita} \emph{et~al.}}]{Babzien_2006}
\bibinfo{author}{\bibnamefont{Babzien}, \bibfnamefont{M.}},
  \bibinfo{author}{\bibfnamefont{I.}~\bibnamefont{Ben-Zvi}},
  \bibinfo{author}{\bibfnamefont{K.}~\bibnamefont{Kusche}},
  \bibinfo{author}{\bibfnamefont{I.~V.} \bibnamefont{Pavlishin}},
  \bibinfo{author}{\bibfnamefont{I.~V.} \bibnamefont{Pogorelsky}},
  \bibinfo{author}{\bibfnamefont{D.~P.} \bibnamefont{Siddons}},
  \bibinfo{author}{\bibfnamefont{V.}~\bibnamefont{Yakimenko}},
  \bibinfo{author}{\bibfnamefont{D.}~\bibnamefont{Cline}},
  \bibinfo{author}{\bibfnamefont{F.}~\bibnamefont{Zhou}},
  \bibinfo{author}{\bibfnamefont{T.}~\bibnamefont{Hirose}},
  \bibinfo{author}{\bibfnamefont{Y.}~\bibnamefont{Kamiya}},
  \bibinfo{author}{\bibfnamefont{T.}~\bibnamefont{Kumita}}, \emph{et~al.},
  \bibinfo{year}{2006}, \bibinfo{journal}{Phys. Rev. Lett.}
  \textbf{\bibinfo{volume}{96}}, \bibinfo{pages}{054802}.

\bibitem[{\citenamefont{Badziak}(2007)}]{Badziak_2007}
\bibinfo{author}{\bibnamefont{Badziak}, \bibfnamefont{J.}},
  \bibinfo{year}{2007}, \bibinfo{journal}{Opto-Electron. Rev.}
  \textbf{\bibinfo{volume}{15}}, \bibinfo{pages}{1}.

\bibitem[{\citenamefont{Bai} \emph{et~al.}(2012)\citenamefont{Bai, Zheng, and
  Wang}}]{Bai_2012}
\bibinfo{author}{\bibnamefont{Bai}, \bibfnamefont{L.}},
  \bibinfo{author}{\bibfnamefont{M.-Y.} \bibnamefont{Zheng}}, and
  \bibinfo{author}{\bibfnamefont{B.-H.} \bibnamefont{Wang}},
  \bibinfo{year}{2012}, \bibinfo{journal}{Phys. Rev. A}
  \textbf{\bibinfo{volume}{85}}, \bibinfo{pages}{013402}.

\bibitem[{\citenamefont{Baier} \emph{et~al.}(1974)\citenamefont{Baier, Fadin,
  Katkov, and Kuraev}}]{Baier_1974}
\bibinfo{author}{\bibnamefont{Baier}, \bibfnamefont{V.~N.}},
  \bibinfo{author}{\bibfnamefont{V.~S.} \bibnamefont{Fadin}},
  \bibinfo{author}{\bibfnamefont{V.~M.} \bibnamefont{Katkov}}, and
  \bibinfo{author}{\bibfnamefont{E.~A.} \bibnamefont{Kuraev}},
  \bibinfo{year}{1974}, \bibinfo{journal}{Phys. Lett. B}
  \textbf{\bibinfo{volume}{49}}, \bibinfo{pages}{385}.

\bibitem[{\citenamefont{Baier and Katkov}(2005)}]{Baier_2005}
\bibinfo{author}{\bibnamefont{Baier}, \bibfnamefont{V.~N.}}, and
  \bibinfo{author}{\bibfnamefont{V.~M.} \bibnamefont{Katkov}},
  \bibinfo{year}{2005}, \bibinfo{journal}{Phys. Rep.}
  \textbf{\bibinfo{volume}{409}}, \bibinfo{pages}{261}.

\bibitem[{\citenamefont{Baier}
  \emph{et~al.}(1976{\natexlab{a}})\citenamefont{Baier, Katkov, Milstein, and
  Strakhovenko}}]{Baier_1976_a}
\bibinfo{author}{\bibnamefont{Baier}, \bibfnamefont{V.~N.}},
  \bibinfo{author}{\bibfnamefont{V.~M.} \bibnamefont{Katkov}},
  \bibinfo{author}{\bibfnamefont{A.~I.} \bibnamefont{Milstein}}, and
  \bibinfo{author}{\bibfnamefont{V.~M.} \bibnamefont{Strakhovenko}},
  \bibinfo{year}{1976}{\natexlab{a}}, \bibinfo{journal}{Sov. Phys. JETP}
  \textbf{\bibinfo{volume}{42}}, \bibinfo{pages}{400}.

\bibitem[{\citenamefont{Baier}
  \emph{et~al.}(1975{\natexlab{a}})\citenamefont{Baier, Katkov, and
  Strakhovenko}}]{Baier_1975_a}
\bibinfo{author}{\bibnamefont{Baier}, \bibfnamefont{V.~N.}},
  \bibinfo{author}{\bibfnamefont{V.~M.} \bibnamefont{Katkov}}, and
  \bibinfo{author}{\bibfnamefont{V.~M.} \bibnamefont{Strakhovenko}},
  \bibinfo{year}{1975}{\natexlab{a}}, \bibinfo{journal}{Sov. Phys. JETP}
  \textbf{\bibinfo{volume}{40}}, \bibinfo{pages}{225}.

\bibitem[{\citenamefont{Baier}
  \emph{et~al.}(1975{\natexlab{b}})\citenamefont{Baier, Katkov, and
  Strakhovenko}}]{Baier_1975_b}
\bibinfo{author}{\bibnamefont{Baier}, \bibfnamefont{V.~N.}},
  \bibinfo{author}{\bibfnamefont{V.~M.} \bibnamefont{Katkov}}, and
  \bibinfo{author}{\bibfnamefont{V.~M.} \bibnamefont{Strakhovenko}},
  \bibinfo{year}{1975}{\natexlab{b}}, \bibinfo{journal}{Sov. Phys. JETP}
  \textbf{\bibinfo{volume}{41}}, \bibinfo{pages}{198}.

\bibitem[{\citenamefont{Baier} \emph{et~al.}(1998)\citenamefont{Baier, Katkov,
  and Strakhovenko}}]{Baier_b_1998}
\bibinfo{author}{\bibnamefont{Baier}, \bibfnamefont{V.~N.}},
  \bibinfo{author}{\bibfnamefont{V.~M.} \bibnamefont{Katkov}}, and
  \bibinfo{author}{\bibfnamefont{V.~M.} \bibnamefont{Strakhovenko}},
  \bibinfo{year}{1998}, \emph{\bibinfo{title}{Electromagnetic Processes at High
  Energies in Oriented Single Crystals}} (\bibinfo{publisher}{World Scientific,
  Singapore}).

\bibitem[{\citenamefont{Baier}
  \emph{et~al.}(1976{\natexlab{b}})\citenamefont{Baier, Milstein, and
  Strakhovenko}}]{Baier_1976_b}
\bibinfo{author}{\bibnamefont{Baier}, \bibfnamefont{V.~N.}},
  \bibinfo{author}{\bibfnamefont{A.~I.} \bibnamefont{Milstein}}, and
  \bibinfo{author}{\bibfnamefont{V.~M.} \bibnamefont{Strakhovenko}},
  \bibinfo{year}{1976}{\natexlab{b}}, \bibinfo{journal}{Sov. Phys. JETP}
  \textbf{\bibinfo{volume}{42}}, \bibinfo{pages}{961}.

\bibitem[{\citenamefont{Bamber} \emph{et~al.}(1999)\citenamefont{Bamber, Boege,
  Koffas, Kotseroglou, Melissinos, Meyerhofer, Reis, Ragg, Bula, McDonald,
  Prebys, Burke} \emph{et~al.}}]{SLAC_PRD}
\bibinfo{author}{\bibnamefont{Bamber}, \bibfnamefont{C.}},
  \bibinfo{author}{\bibfnamefont{S.~J.} \bibnamefont{Boege}},
  \bibinfo{author}{\bibfnamefont{T.}~\bibnamefont{Koffas}},
  \bibinfo{author}{\bibfnamefont{T.}~\bibnamefont{Kotseroglou}},
  \bibinfo{author}{\bibfnamefont{A.~C.} \bibnamefont{Melissinos}},
  \bibinfo{author}{\bibfnamefont{D.~D.} \bibnamefont{Meyerhofer}},
  \bibinfo{author}{\bibfnamefont{D.~A.} \bibnamefont{Reis}},
  \bibinfo{author}{\bibfnamefont{W.}~\bibnamefont{Ragg}},
  \bibinfo{author}{\bibfnamefont{C.}~\bibnamefont{Bula}},
  \bibinfo{author}{\bibfnamefont{K.~T.} \bibnamefont{McDonald}},
  \bibinfo{author}{\bibfnamefont{E.~J.} \bibnamefont{Prebys}},
  \bibinfo{author}{\bibfnamefont{D.~L.} \bibnamefont{Burke}}, \emph{et~al.},
  \bibinfo{year}{1999}, \bibinfo{journal}{Phys. Rev. D}
  \textbf{\bibinfo{volume}{60}}, \bibinfo{pages}{092004}.

\bibitem[{\citenamefont{Bauke} \emph{et~al.}(2011)\citenamefont{Bauke,
  Hetzheim, Mocken, Ruf, and Keitel}}]{Bauke_2011}
\bibinfo{author}{\bibnamefont{Bauke}, \bibfnamefont{H.}},
  \bibinfo{author}{\bibfnamefont{H.~G.} \bibnamefont{Hetzheim}},
  \bibinfo{author}{\bibfnamefont{G.~R.} \bibnamefont{Mocken}},
  \bibinfo{author}{\bibfnamefont{M.}~\bibnamefont{Ruf}}, and
  \bibinfo{author}{\bibfnamefont{C.~H.} \bibnamefont{Keitel}},
  \bibinfo{year}{2011}, \bibinfo{journal}{Phys. Rev. A}
  \textbf{\bibinfo{volume}{83}}, \bibinfo{pages}{063414}.

\bibitem[{\citenamefont{Bauke and Keitel}(2011)}]{Bauke_2011b}
\bibinfo{author}{\bibnamefont{Bauke}, \bibfnamefont{H.}}, and
  \bibinfo{author}{\bibfnamefont{C.~H.} \bibnamefont{Keitel}},
  \bibinfo{year}{2011}, \bibinfo{journal}{Comp. Phys. Commun.}
  \textbf{\bibinfo{volume}{182}}, \bibinfo{pages}{2454}.

\bibitem[{\citenamefont{Baur} \emph{et~al.}(2007)\citenamefont{Baur, Hencken,
  and Trautmann}}]{Baur_2007}
\bibinfo{author}{\bibnamefont{Baur}, \bibfnamefont{G.}},
  \bibinfo{author}{\bibfnamefont{K.}~\bibnamefont{Hencken}}, and
  \bibinfo{author}{\bibfnamefont{D.}~\bibnamefont{Trautmann}},
  \bibinfo{year}{2007}, \bibinfo{journal}{Phys. Rep.}
  \textbf{\bibinfo{volume}{453}}, \bibinfo{pages}{1}.

\bibitem[{\citenamefont{Beck} \emph{et~al.}(2007)\citenamefont{Beck, Becker,
  Beiersdorfer, Brown, Moody, Wilhelmy, Porter, Kilbourne, and
  Kelley}}]{Beck_2007}
\bibinfo{author}{\bibnamefont{Beck}, \bibfnamefont{B.~R.}},
  \bibinfo{author}{\bibfnamefont{J.~A.} \bibnamefont{Becker}},
  \bibinfo{author}{\bibfnamefont{P.}~\bibnamefont{Beiersdorfer}},
  \bibinfo{author}{\bibfnamefont{G.~V.} \bibnamefont{Brown}},
  \bibinfo{author}{\bibfnamefont{K.~J.} \bibnamefont{Moody}},
  \bibinfo{author}{\bibfnamefont{J.~B.} \bibnamefont{Wilhelmy}},
  \bibinfo{author}{\bibfnamefont{F.~S.} \bibnamefont{Porter}},
  \bibinfo{author}{\bibfnamefont{C.~A.} \bibnamefont{Kilbourne}}, and
  \bibinfo{author}{\bibfnamefont{R.~L.} \bibnamefont{Kelley}},
  \bibinfo{year}{2007}, \bibinfo{journal}{Phys. Rev. Lett.}
  \textbf{\bibinfo{volume}{98}}, \bibinfo{pages}{142501}.

\bibitem[{\citenamefont{Becker} \emph{et~al.}(2002)\citenamefont{Becker,
  Grasbon, Kopold, Milosevic, Paulus, and Walther}}]{Becker_2002}
\bibinfo{author}{\bibnamefont{Becker}, \bibfnamefont{W.}},
  \bibinfo{author}{\bibfnamefont{F.}~\bibnamefont{Grasbon}},
  \bibinfo{author}{\bibfnamefont{R.}~\bibnamefont{Kopold}},
  \bibinfo{author}{\bibfnamefont{D.~B.} \bibnamefont{Milosevic}},
  \bibinfo{author}{\bibfnamefont{G.~G.} \bibnamefont{Paulus}}, and
  \bibinfo{author}{\bibfnamefont{H.}~\bibnamefont{Walther}},
  \bibinfo{year}{2002}, \bibinfo{journal}{Adv. Atom. Mol. Opt. Phys.}
  \textbf{\bibinfo{volume}{48}}, \bibinfo{pages}{35}.

\bibitem[{\citenamefont{Becker} \emph{et~al.}(1984)\citenamefont{Becker,
  Schlicher, and Scully}}]{Becker2_1984}
\bibinfo{author}{\bibnamefont{Becker}, \bibfnamefont{W.}},
  \bibinfo{author}{\bibfnamefont{R.~R.} \bibnamefont{Schlicher}}, and
  \bibinfo{author}{\bibfnamefont{M.~O.} \bibnamefont{Scully}},
  \bibinfo{year}{1984}, \bibinfo{journal}{Nucl. Phys. A}
  \textbf{\bibinfo{volume}{426}}, \bibinfo{pages}{125}.

\bibitem[{\citenamefont{Bell and Kirk}(2008)}]{Bell_2008}
\bibinfo{author}{\bibnamefont{Bell}, \bibfnamefont{A.~R.}}, and
  \bibinfo{author}{\bibfnamefont{J.~G.} \bibnamefont{Kirk}},
  \bibinfo{year}{2008}, \bibinfo{journal}{Phys. Rev. Lett.}
  \textbf{\bibinfo{volume}{101}}, \bibinfo{pages}{200403}.

\bibitem[{\citenamefont{{BELLA}}(2012)}]{BELLA}
\bibinfo{author}{\bibnamefont{{BELLA}}} (\bibinfo{collaboration}{Berkeley Lab
  Laser Accelerator}), \bibinfo{year}{2012},
  \urlprefix\url{http://www.lbl.gov/publicinfo/newscenter/features/2008/apr/af-bella.html}.

\bibitem[{\citenamefont{Benedetti} \emph{et~al.}(2011)\citenamefont{Benedetti,
  Han, Ruffini, and Vereshchagin}}]{Ruffini_2011}
\bibinfo{author}{\bibnamefont{Benedetti}, \bibfnamefont{A.}},
  \bibinfo{author}{\bibfnamefont{W.-B.} \bibnamefont{Han}},
  \bibinfo{author}{\bibfnamefont{R.}~\bibnamefont{Ruffini}}, and
  \bibinfo{author}{\bibfnamefont{G.~V.} \bibnamefont{Vereshchagin}},
  \bibinfo{year}{2011}, \bibinfo{journal}{Phys. Lett. B}
  \textbf{\bibinfo{volume}{698}}, \bibinfo{pages}{75}.

\bibitem[{\citenamefont{Berestetskii}
  \emph{et~al.}(1982)\citenamefont{Berestetskii, Lifshitz, and
  Pitaevskii}}]{Landau_b_4_1982}
\bibinfo{author}{\bibnamefont{Berestetskii}, \bibfnamefont{V.~B.}},
  \bibinfo{author}{\bibfnamefont{E.~M.} \bibnamefont{Lifshitz}}, and
  \bibinfo{author}{\bibfnamefont{L.~P.} \bibnamefont{Pitaevskii}},
  \bibinfo{year}{1982}, \emph{\bibinfo{title}{Quantum electrodynamics}}
  (\bibinfo{publisher}{Elsevier Butterworth-Heinemann, Oxford}).

\bibitem[{\citenamefont{Bergmann} \emph{et~al.}(1998)\citenamefont{Bergmann,
  Theuer, and Shore}}]{Bergmann_1998}
\bibinfo{author}{\bibnamefont{Bergmann}, \bibfnamefont{K.}},
  \bibinfo{author}{\bibfnamefont{H.}~\bibnamefont{Theuer}}, and
  \bibinfo{author}{\bibfnamefont{B.~W.} \bibnamefont{Shore}},
  \bibinfo{year}{1998}, \bibinfo{journal}{Rev. Mod. Phys.}
  \textbf{\bibinfo{volume}{70}}, \bibinfo{pages}{1003}.

\bibitem[{\citenamefont{Bhattacharyya}
  \emph{et~al.}(2011)\citenamefont{Bhattacharyya, Mazumder, Chakrabarti, and
  Faisal}}]{Faisal_2011}
\bibinfo{author}{\bibnamefont{Bhattacharyya}, \bibfnamefont{S.}},
  \bibinfo{author}{\bibfnamefont{M.}~\bibnamefont{Mazumder}},
  \bibinfo{author}{\bibfnamefont{J.}~\bibnamefont{Chakrabarti}}, and
  \bibinfo{author}{\bibfnamefont{F.~H.~M.} \bibnamefont{Faisal}},
  \bibinfo{year}{2011}, \bibinfo{journal}{Phys. Rev. A}
  \textbf{\bibinfo{volume}{83}}, \bibinfo{pages}{043407}.

\bibitem[{\citenamefont{Bialynicki-Birula and
  Rudnicki}(2011)}]{Bialynicki-Birula_2011}
\bibinfo{author}{\bibnamefont{Bialynicki-Birula}, \bibfnamefont{I.}}, and
  \bibinfo{author}{\bibfnamefont{{\L}.}~\bibnamefont{Rudnicki}},
  \bibinfo{year}{2011}, \eprint{arXiv:1108.2615v1}.

\bibitem[{\citenamefont{Boca and Florescu}(2009)}]{Boca_2009}
\bibinfo{author}{\bibnamefont{Boca}, \bibfnamefont{M.}}, and
  \bibinfo{author}{\bibfnamefont{V.}~\bibnamefont{Florescu}},
  \bibinfo{year}{2009}, \bibinfo{journal}{Phys. Rev. A}
  \textbf{\bibinfo{volume}{80}}, \bibinfo{pages}{053403}.

\bibitem[{\citenamefont{Boca and Florescu}(2010)}]{Boca_2010}
\bibinfo{author}{\bibnamefont{Boca}, \bibfnamefont{M.}}, and
  \bibinfo{author}{\bibfnamefont{V.}~\bibnamefont{Florescu}},
  \bibinfo{year}{2010}, \bibinfo{journal}{Rom. J. Phys.}
  \textbf{\bibinfo{volume}{55}}, \bibinfo{pages}{511}.

\bibitem[{\citenamefont{Boca and Florescu}(2011)}]{Boca_2011}
\bibinfo{author}{\bibnamefont{Boca}, \bibfnamefont{M.}}, and
  \bibinfo{author}{\bibfnamefont{V.}~\bibnamefont{Florescu}},
  \bibinfo{year}{2011}, \bibinfo{journal}{Eur. Phys. J. D}
  \textbf{\bibinfo{volume}{61}}, \bibinfo{pages}{449}.

\bibitem[{\citenamefont{Boca and Oprea}(2011)}]{Boca_2011b}
\bibinfo{author}{\bibnamefont{Boca}, \bibfnamefont{M.}}, and
  \bibinfo{author}{\bibfnamefont{A.}~\bibnamefont{Oprea}},
  \bibinfo{year}{2011}, \bibinfo{journal}{Phys. Scr.}
  \textbf{\bibinfo{volume}{83}}, \bibinfo{pages}{055404}.

\bibitem[{\citenamefont{Bochkarev} \emph{et~al.}(2011)\citenamefont{Bochkarev,
  Popov, and Bychenkov}}]{Bochkarev_2011}
\bibinfo{author}{\bibnamefont{Bochkarev}, \bibfnamefont{S.}},
  \bibinfo{author}{\bibfnamefont{K.}~\bibnamefont{Popov}}, and
  \bibinfo{author}{\bibfnamefont{V.}~\bibnamefont{Bychenkov}},
  \bibinfo{year}{2011}, \bibinfo{journal}{Plasma Phys. Rep.}
  \textbf{\bibinfo{volume}{37}}, \bibinfo{pages}{603}.

\bibitem[{\citenamefont{Bonifacio} \emph{et~al.}(1984)\citenamefont{Bonifacio,
  Pellegrini, and Narducci}}]{Bonifacio_1984}
\bibinfo{author}{\bibnamefont{Bonifacio}, \bibfnamefont{R.}},
  \bibinfo{author}{\bibfnamefont{C.}~\bibnamefont{Pellegrini}}, and
  \bibinfo{author}{\bibfnamefont{L.~M.} \bibnamefont{Narducci}},
  \bibinfo{year}{1984}, \bibinfo{journal}{Opt. Commun.}
  \textbf{\bibinfo{volume}{50}}, \bibinfo{pages}{373}.

\bibitem[{\citenamefont{Breit and Wheeler}(1934)}]{Breit_1934}
\bibinfo{author}{\bibnamefont{Breit}, \bibfnamefont{G.}}, and
  \bibinfo{author}{\bibfnamefont{J.~A.} \bibnamefont{Wheeler}},
  \bibinfo{year}{1934}, \bibinfo{journal}{Phys. Rev.}
  \textbf{\bibinfo{volume}{46}}, \bibinfo{pages}{1087}.

\bibitem[{\citenamefont{Brezin and Itzykson}(1970)}]{Brezin_1970}
\bibinfo{author}{\bibnamefont{Brezin}, \bibfnamefont{E.}}, and
  \bibinfo{author}{\bibfnamefont{C.}~\bibnamefont{Itzykson}},
  \bibinfo{year}{1970}, \bibinfo{journal}{Phys. Rev. D}
  \textbf{\bibinfo{volume}{2}}, \bibinfo{pages}{1191}.

\bibitem[{\citenamefont{Brodin} \emph{et~al.}(2007)\citenamefont{Brodin,
  Marklund, Eliasson, and Shukla}}]{Brodin_2007}
\bibinfo{author}{\bibnamefont{Brodin}, \bibfnamefont{G.}},
  \bibinfo{author}{\bibfnamefont{M.}~\bibnamefont{Marklund}},
  \bibinfo{author}{\bibfnamefont{B.}~\bibnamefont{Eliasson}}, and
  \bibinfo{author}{\bibfnamefont{P.~K.} \bibnamefont{Shukla}},
  \bibinfo{year}{2007}, \bibinfo{journal}{Phys. Rev. Lett.}
  \textbf{\bibinfo{volume}{98}}, \bibinfo{pages}{125001}.

\bibitem[{\citenamefont{Brodin} \emph{et~al.}(2001)\citenamefont{Brodin,
  Marklund, and Stenflo}}]{Brodin_2001}
\bibinfo{author}{\bibnamefont{Brodin}, \bibfnamefont{G.}},
  \bibinfo{author}{\bibfnamefont{M.}~\bibnamefont{Marklund}}, and
  \bibinfo{author}{\bibfnamefont{L.}~\bibnamefont{Stenflo}},
  \bibinfo{year}{2001}, \bibinfo{journal}{Phys. Rev. Lett.}
  \textbf{\bibinfo{volume}{87}}, \bibinfo{pages}{171801}.

\bibitem[{\citenamefont{Brown and Kibble}(1964)}]{Brown_1964}
\bibinfo{author}{\bibnamefont{Brown}, \bibfnamefont{L.~S.}}, and
  \bibinfo{author}{\bibfnamefont{T.~W.~B.} \bibnamefont{Kibble}},
  \bibinfo{year}{1964}, \bibinfo{journal}{Phys. Rev.}
  \textbf{\bibinfo{volume}{133}}, \bibinfo{pages}{A705}.

\bibitem[{\citenamefont{Brown} \emph{et~al.}(2004)\citenamefont{Brown,
  Anderson, Barty, Betts, Booth, Crane, Cross, Fittinghoff, Gibson, Hartemann,
  Hartouni, Kuba} \emph{et~al.}}]{Rosenzweig_2004}
\bibinfo{author}{\bibnamefont{Brown}, \bibfnamefont{W.~J.}},
  \bibinfo{author}{\bibfnamefont{S.~G.} \bibnamefont{Anderson}},
  \bibinfo{author}{\bibfnamefont{C.~P.~J.} \bibnamefont{Barty}},
  \bibinfo{author}{\bibfnamefont{S.~M.} \bibnamefont{Betts}},
  \bibinfo{author}{\bibfnamefont{R.}~\bibnamefont{Booth}},
  \bibinfo{author}{\bibfnamefont{J.~K.} \bibnamefont{Crane}},
  \bibinfo{author}{\bibfnamefont{R.~R.} \bibnamefont{Cross}},
  \bibinfo{author}{\bibfnamefont{D.~N.} \bibnamefont{Fittinghoff}},
  \bibinfo{author}{\bibfnamefont{D.~J.} \bibnamefont{Gibson}},
  \bibinfo{author}{\bibfnamefont{F.~V.} \bibnamefont{Hartemann}},
  \bibinfo{author}{\bibfnamefont{E.~P.} \bibnamefont{Hartouni}},
  \bibinfo{author}{\bibfnamefont{J.}~\bibnamefont{Kuba}}, \emph{et~al.},
  \bibinfo{year}{2004}, \bibinfo{journal}{Phys. Rev. ST Accel. Beams}
  \textbf{\bibinfo{volume}{7}}, \bibinfo{pages}{060702}.

\bibitem[{\citenamefont{Brown and Hartemann}(2004)}]{Brown_2004}
\bibinfo{author}{\bibnamefont{Brown}, \bibfnamefont{W.~J.}}, and
  \bibinfo{author}{\bibfnamefont{F.~V.} \bibnamefont{Hartemann}},
  \bibinfo{year}{2004}, \bibinfo{journal}{Phys. Rev. ST Accel. Beams}
  \textbf{\bibinfo{volume}{7}}, \bibinfo{pages}{060703}.

\bibitem[{\citenamefont{Bu and Ji}(2010)}]{Bu_2010}
\bibinfo{author}{\bibnamefont{Bu}, \bibfnamefont{Z.}}, and
  \bibinfo{author}{\bibfnamefont{P.}~\bibnamefont{Ji}}, \bibinfo{year}{2010},
  \bibinfo{journal}{Phys. Plasmas} \textbf{\bibinfo{volume}{17}},
  \bibinfo{pages}{073103}.

\bibitem[{\citenamefont{Bula} \emph{et~al.}(1996)\citenamefont{Bula, McDonald,
  Prebys, Bamber, Boege, Kotseroglou, Melissinos, Meyerhofer, Ragg, Burke,
  Field, Horton-Smith} \emph{et~al.}}]{Bula_1996}
\bibinfo{author}{\bibnamefont{Bula}, \bibfnamefont{C.}},
  \bibinfo{author}{\bibfnamefont{K.~T.} \bibnamefont{McDonald}},
  \bibinfo{author}{\bibfnamefont{E.~J.} \bibnamefont{Prebys}},
  \bibinfo{author}{\bibfnamefont{C.}~\bibnamefont{Bamber}},
  \bibinfo{author}{\bibfnamefont{S.}~\bibnamefont{Boege}},
  \bibinfo{author}{\bibfnamefont{T.}~\bibnamefont{Kotseroglou}},
  \bibinfo{author}{\bibfnamefont{A.~C.} \bibnamefont{Melissinos}},
  \bibinfo{author}{\bibfnamefont{D.~D.} \bibnamefont{Meyerhofer}},
  \bibinfo{author}{\bibfnamefont{W.}~\bibnamefont{Ragg}},
  \bibinfo{author}{\bibfnamefont{D.~L.} \bibnamefont{Burke}},
  \bibinfo{author}{\bibfnamefont{R.~C.} \bibnamefont{Field}},
  \bibinfo{author}{\bibfnamefont{G.}~\bibnamefont{Horton-Smith}},
  \emph{et~al.}, \bibinfo{year}{1996}, \bibinfo{journal}{Phys. Rev. Lett.}
  \textbf{\bibinfo{volume}{76}}, \bibinfo{pages}{3116}.

\bibitem[{\citenamefont{Bulanov}
  \emph{et~al.}(2010{\natexlab{a}})\citenamefont{Bulanov, Mur, Narozhny, Nees,
  and Popov}}]{Bulanov_2010_a}
\bibinfo{author}{\bibnamefont{Bulanov}, \bibfnamefont{S.~S.}},
  \bibinfo{author}{\bibfnamefont{V.~D.} \bibnamefont{Mur}},
  \bibinfo{author}{\bibfnamefont{N.~B.} \bibnamefont{Narozhny}},
  \bibinfo{author}{\bibfnamefont{J.}~\bibnamefont{Nees}}, and
  \bibinfo{author}{\bibfnamefont{V.~S.} \bibnamefont{Popov}},
  \bibinfo{year}{2010}{\natexlab{a}}, \bibinfo{journal}{Phys. Rev. Lett.}
  \textbf{\bibinfo{volume}{104}}, \bibinfo{pages}{220404}.

\bibitem[{\citenamefont{Bulanov}
  \emph{et~al.}(2010{\natexlab{b}})\citenamefont{Bulanov, {T. Zh. Esirkepov},
  Thomas, Koga, and Bulanov}}]{Bulanov_2010}
\bibinfo{author}{\bibnamefont{Bulanov}, \bibfnamefont{S.~S.}},
  \bibinfo{author}{\bibnamefont{{T. Zh. Esirkepov}}},
  \bibinfo{author}{\bibfnamefont{A.~G.~R.} \bibnamefont{Thomas}},
  \bibinfo{author}{\bibfnamefont{J.~K.} \bibnamefont{Koga}}, and
  \bibinfo{author}{\bibfnamefont{S.~V.} \bibnamefont{Bulanov}},
  \bibinfo{year}{2010}{\natexlab{b}}, \bibinfo{journal}{Phys. Rev. Lett.}
  \textbf{\bibinfo{volume}{105}}, \bibinfo{pages}{220407}.

\bibitem[{\citenamefont{Bulanov} \emph{et~al.}(2011)\citenamefont{Bulanov,
  Esirkepov, Kando, Koga, and Bulanov}}]{Bulanov_2011}
\bibinfo{author}{\bibnamefont{Bulanov}, \bibfnamefont{S.~V.}},
  \bibinfo{author}{\bibfnamefont{T.~Z.} \bibnamefont{Esirkepov}},
  \bibinfo{author}{\bibfnamefont{M.}~\bibnamefont{Kando}},
  \bibinfo{author}{\bibfnamefont{J.~K.} \bibnamefont{Koga}}, and
  \bibinfo{author}{\bibfnamefont{S.~S.} \bibnamefont{Bulanov}},
  \bibinfo{year}{2011}, \bibinfo{journal}{Phys. Rev. E}
  \textbf{\bibinfo{volume}{84}}, \bibinfo{pages}{056605}.

\bibitem[{\citenamefont{Bulanov} \emph{et~al.}(1994)\citenamefont{Bulanov,
  Naumova, and Pegoraro}}]{Bulanov_1994}
\bibinfo{author}{\bibnamefont{Bulanov}, \bibfnamefont{S.~V.}},
  \bibinfo{author}{\bibfnamefont{N.~M.} \bibnamefont{Naumova}}, and
  \bibinfo{author}{\bibfnamefont{F.}~\bibnamefont{Pegoraro}},
  \bibinfo{year}{1994}, \bibinfo{journal}{Phys. Plasmas}
  \textbf{\bibinfo{volume}{1}}, \bibinfo{pages}{745}.

\bibitem[{\citenamefont{Burke} \emph{et~al.}(1997)\citenamefont{Burke, Field,
  Horton-Smith, Spencer, Walz, Berridge, Bugg, Shmakov, Weidemann, Bula,
  McDonald, Prebys} \emph{et~al.}}]{SLAC_PP}
\bibinfo{author}{\bibnamefont{Burke}, \bibfnamefont{D.~L.}},
  \bibinfo{author}{\bibfnamefont{R.~C.} \bibnamefont{Field}},
  \bibinfo{author}{\bibfnamefont{G.}~\bibnamefont{Horton-Smith}},
  \bibinfo{author}{\bibfnamefont{J.~E.} \bibnamefont{Spencer}},
  \bibinfo{author}{\bibfnamefont{D.}~\bibnamefont{Walz}},
  \bibinfo{author}{\bibfnamefont{S.~C.} \bibnamefont{Berridge}},
  \bibinfo{author}{\bibfnamefont{W.~M.} \bibnamefont{Bugg}},
  \bibinfo{author}{\bibfnamefont{K.}~\bibnamefont{Shmakov}},
  \bibinfo{author}{\bibfnamefont{A.~W.} \bibnamefont{Weidemann}},
  \bibinfo{author}{\bibfnamefont{C.}~\bibnamefont{Bula}},
  \bibinfo{author}{\bibfnamefont{K.~T.} \bibnamefont{McDonald}},
  \bibinfo{author}{\bibfnamefont{E.~J.} \bibnamefont{Prebys}}, \emph{et~al.},
  \bibinfo{year}{1997}, \bibinfo{journal}{Phys. Rev. Lett.}
  \textbf{\bibinfo{volume}{79}}, \bibinfo{pages}{1626}.

\bibitem[{\citenamefont{Burton} \emph{et~al.}(2007)\citenamefont{Burton,
  Gratus, and Tucker}}]{Burton_2007}
\bibinfo{author}{\bibnamefont{Burton}, \bibfnamefont{D.~A.}},
  \bibinfo{author}{\bibfnamefont{J.}~\bibnamefont{Gratus}}, and
  \bibinfo{author}{\bibfnamefont{R.~W.} \bibnamefont{Tucker}},
  \bibinfo{year}{2007}, \bibinfo{journal}{Ann. Phys. (N.Y.)}
  \textbf{\bibinfo{volume}{322}}, \bibinfo{pages}{599}.

\bibitem[{\citenamefont{B{\"u}rvenich}
  \emph{et~al.}(2006{\natexlab{a}})\citenamefont{B{\"u}rvenich, Evers, and
  Keitel}}]{Buervenich_2006}
\bibinfo{author}{\bibnamefont{B{\"u}rvenich}, \bibfnamefont{T.~J.}},
  \bibinfo{author}{\bibfnamefont{J.}~\bibnamefont{Evers}}, and
  \bibinfo{author}{\bibfnamefont{C.~H.} \bibnamefont{Keitel}},
  \bibinfo{year}{2006}{\natexlab{a}}, \bibinfo{journal}{Phys. Rev. Lett.}
  \textbf{\bibinfo{volume}{96}}, \bibinfo{pages}{142501}.

\bibitem[{\citenamefont{B{\"u}rvenich}
  \emph{et~al.}(2006{\natexlab{b}})\citenamefont{B{\"u}rvenich, Evers, and
  Keitel}}]{Buervenich2_2006}
\bibinfo{author}{\bibnamefont{B{\"u}rvenich}, \bibfnamefont{T.~J.}},
  \bibinfo{author}{\bibfnamefont{J.}~\bibnamefont{Evers}}, and
  \bibinfo{author}{\bibfnamefont{C.~H.} \bibnamefont{Keitel}},
  \bibinfo{year}{2006}{\natexlab{b}}, \bibinfo{journal}{Phys.~Rev.~C}
  \textbf{\bibinfo{volume}{74}}, \bibinfo{pages}{044601}.

\bibitem[{\citenamefont{Bychenkov} \emph{et~al.}(2001)\citenamefont{Bychenkov,
  Sentoku, Bulanov, Mima, Mourou, and Tolokonnikov}}]{Bychenkov_2001}
\bibinfo{author}{\bibnamefont{Bychenkov}, \bibfnamefont{V.}},
  \bibinfo{author}{\bibfnamefont{Y.}~\bibnamefont{Sentoku}},
  \bibinfo{author}{\bibfnamefont{S.}~\bibnamefont{Bulanov}},
  \bibinfo{author}{\bibfnamefont{K.}~\bibnamefont{Mima}},
  \bibinfo{author}{\bibfnamefont{G.}~\bibnamefont{Mourou}}, and
  \bibinfo{author}{\bibfnamefont{S.}~\bibnamefont{Tolokonnikov}},
  \bibinfo{year}{2001}, \bibinfo{journal}{JETP Letters}
  \textbf{\bibinfo{volume}{74}}, \bibinfo{pages}{586}.

\bibitem[{\citenamefont{Caldwell} \emph{et~al.}(2009)\citenamefont{Caldwell,
  Lotov, Pukhov, and Simon}}]{Caldwell_2009}
\bibinfo{author}{\bibnamefont{Caldwell}, \bibfnamefont{A.}},
  \bibinfo{author}{\bibfnamefont{K.}~\bibnamefont{Lotov}},
  \bibinfo{author}{\bibfnamefont{A.}~\bibnamefont{Pukhov}}, and
  \bibinfo{author}{\bibfnamefont{F.}~\bibnamefont{Simon}},
  \bibinfo{year}{2009}, \bibinfo{journal}{Nature Phys.}
  \textbf{\bibinfo{volume}{5}}, \bibinfo{pages}{563}.

\bibitem[{\citenamefont{Cameron} \emph{et~al.}(1993)\citenamefont{Cameron,
  Cantatore, Melissinos, Ruoso, Semertzidis, Halama, Lazarus, Prodell, Nezrick,
  Rizzo, and Zavattini}}]{BFRT}
\bibinfo{author}{\bibnamefont{Cameron}, \bibfnamefont{R.}},
  \bibinfo{author}{\bibfnamefont{G.}~\bibnamefont{Cantatore}},
  \bibinfo{author}{\bibfnamefont{A.~C.} \bibnamefont{Melissinos}},
  \bibinfo{author}{\bibfnamefont{G.}~\bibnamefont{Ruoso}},
  \bibinfo{author}{\bibfnamefont{Y.}~\bibnamefont{Semertzidis}},
  \bibinfo{author}{\bibfnamefont{H.~J.} \bibnamefont{Halama}},
  \bibinfo{author}{\bibfnamefont{D.~M.} \bibnamefont{Lazarus}},
  \bibinfo{author}{\bibfnamefont{A.~G.} \bibnamefont{Prodell}},
  \bibinfo{author}{\bibfnamefont{F.}~\bibnamefont{Nezrick}},
  \bibinfo{author}{\bibfnamefont{C.}~\bibnamefont{Rizzo}}, and
  \bibinfo{author}{\bibfnamefont{E.}~\bibnamefont{Zavattini}},
  \bibinfo{year}{1993}, \bibinfo{journal}{Phys. Rev. D}
  \textbf{\bibinfo{volume}{47}}, \bibinfo{pages}{3707}.

\bibitem[{\citenamefont{Cassidy} \emph{et~al.}(2005)\citenamefont{Cassidy,
  Deng, Greaves, Maruo, Nishiyama, Snyder, Tanaka, and Mills}}]{Cassidy_2005b}
\bibinfo{author}{\bibnamefont{Cassidy}, \bibfnamefont{D.~B.}},
  \bibinfo{author}{\bibfnamefont{S.~H.~M.} \bibnamefont{Deng}},
  \bibinfo{author}{\bibfnamefont{R.~G.} \bibnamefont{Greaves}},
  \bibinfo{author}{\bibfnamefont{T.}~\bibnamefont{Maruo}},
  \bibinfo{author}{\bibfnamefont{N.}~\bibnamefont{Nishiyama}},
  \bibinfo{author}{\bibfnamefont{J.~B.} \bibnamefont{Snyder}},
  \bibinfo{author}{\bibfnamefont{H.~K.~M.} \bibnamefont{Tanaka}}, and
  \bibinfo{author}{\bibfnamefont{A.~P.} \bibnamefont{Mills}},
  \bibinfo{year}{2005}, \bibinfo{journal}{Phys. Rev. Lett.}
  \textbf{\bibinfo{volume}{95}}, \bibinfo{pages}{195006}.

\bibitem[{\citenamefont{Cassidy and Mills}(2005)}]{Cassidy_2005}
\bibinfo{author}{\bibnamefont{Cassidy}, \bibfnamefont{D.~B.}}, and
  \bibinfo{author}{\bibfnamefont{A.~P.} \bibnamefont{Mills}},
  \bibinfo{year}{2005}, \bibinfo{journal}{Nature (London)}
  \textbf{\bibinfo{volume}{449}}, \bibinfo{pages}{195}.

\bibitem[{\citenamefont{{CAST}}(2008)}]{CAST}
\bibinfo{author}{\bibnamefont{{CAST}}} (\bibinfo{collaboration}{CERN Axion
  Solar Telescope}), \bibinfo{year}{2008},
  \urlprefix\url{http://cast.web.cern.ch/CAST/}.

\bibitem[{\citenamefont{{Casta\~neda~Cort{\'e}s}}
  \emph{et~al.}(2012)\citenamefont{{Casta\~neda~Cort{\'e}s}, P{\'a}lffy, and
  Keitel}}]{Castaneda_2012}
\bibinfo{author}{\bibnamefont{{Casta\~neda~Cort{\'e}s}}, \bibfnamefont{H.~M.}},
  \bibinfo{author}{\bibfnamefont{A.}~\bibnamefont{P{\'a}lffy}}, and
  \bibinfo{author}{\bibfnamefont{C.~H.} \bibnamefont{Keitel}},
  \bibinfo{year}{2012}, \bibinfo{pages}{to be published}.

\bibitem[{\citenamefont{{Casta\~neda~Cort{\'e}s}}
  \emph{et~al.}(2011)\citenamefont{{Casta\~neda~Cort{\'e}s}, Popruzhenko,
  Bauer, and P{\'a}lffy}}]{Castaneda_2011}
\bibinfo{author}{\bibnamefont{{Casta\~neda~Cort{\'e}s}}, \bibfnamefont{H.~M.}},
  \bibinfo{author}{\bibfnamefont{S.~V.} \bibnamefont{Popruzhenko}},
  \bibinfo{author}{\bibfnamefont{D.}~\bibnamefont{Bauer}}, and
  \bibinfo{author}{\bibfnamefont{A.}~\bibnamefont{P{\'a}lffy}},
  \bibinfo{year}{2011}, \bibinfo{journal}{New J. Phys.}
  \textbf{\bibinfo{volume}{13}}, \bibinfo{pages}{063007}.

\bibitem[{\citenamefont{Chambaret} \emph{et~al.}(2009)\citenamefont{Chambaret,
  Georges, Ch\'{e}riaux, Rey, Blanc, Audebert, Douillet, Paillard, Cavillac,
  Fournet, Mathieu, and Mourou}}]{Chambaret_2009}
\bibinfo{author}{\bibnamefont{Chambaret}, \bibfnamefont{J.~P.}},
  \bibinfo{author}{\bibfnamefont{P.}~\bibnamefont{Georges}},
  \bibinfo{author}{\bibfnamefont{G.}~\bibnamefont{Ch\'{e}riaux}},
  \bibinfo{author}{\bibfnamefont{G.}~\bibnamefont{Rey}},
  \bibinfo{author}{\bibfnamefont{C.~L.} \bibnamefont{Blanc}},
  \bibinfo{author}{\bibfnamefont{P.}~\bibnamefont{Audebert}},
  \bibinfo{author}{\bibfnamefont{D.}~\bibnamefont{Douillet}},
  \bibinfo{author}{\bibfnamefont{J.~L.} \bibnamefont{Paillard}},
  \bibinfo{author}{\bibfnamefont{P.}~\bibnamefont{Cavillac}},
  \bibinfo{author}{\bibfnamefont{D.}~\bibnamefont{Fournet}},
  \bibinfo{author}{\bibfnamefont{F.}~\bibnamefont{Mathieu}}, and
  \bibinfo{author}{\bibfnamefont{G.}~\bibnamefont{Mourou}},
  \bibinfo{year}{2009}, in \emph{\bibinfo{booktitle}{Frontiers in Optics}}, p.
  \bibinfo{pages}{FMI2}.

\bibitem[{\citenamefont{Chelkowski}
  \emph{et~al.}(2004)\citenamefont{Chelkowski, Bandrauk, and
  Corkum}}]{Chelkowski_2004}
\bibinfo{author}{\bibnamefont{Chelkowski}, \bibfnamefont{S.}},
  \bibinfo{author}{\bibfnamefont{A.~D.} \bibnamefont{Bandrauk}}, and
  \bibinfo{author}{\bibfnamefont{P.~B.} \bibnamefont{Corkum}},
  \bibinfo{year}{2004}, \bibinfo{journal}{Phys. Rev. Lett.}
  \textbf{\bibinfo{volume}{93}}, \bibinfo{pages}{083602}.

\bibitem[{\citenamefont{Chen} \emph{et~al.}(2009)\citenamefont{Chen, Wilks,
  Bonlie, Liang, Myatt, Price, Meyerhofer, and Beiersdorfer}}]{Chen_2009}
\bibinfo{author}{\bibnamefont{Chen}, \bibfnamefont{H.}},
  \bibinfo{author}{\bibfnamefont{S.~C.} \bibnamefont{Wilks}},
  \bibinfo{author}{\bibfnamefont{J.~D.} \bibnamefont{Bonlie}},
  \bibinfo{author}{\bibfnamefont{E.~P.} \bibnamefont{Liang}},
  \bibinfo{author}{\bibfnamefont{J.}~\bibnamefont{Myatt}},
  \bibinfo{author}{\bibfnamefont{D.~F.} \bibnamefont{Price}},
  \bibinfo{author}{\bibfnamefont{D.~D.} \bibnamefont{Meyerhofer}}, and
  \bibinfo{author}{\bibfnamefont{P.}~\bibnamefont{Beiersdorfer}},
  \bibinfo{year}{2009}, \bibinfo{journal}{Phys. Rev. Lett.}
  \textbf{\bibinfo{volume}{102}}, \bibinfo{pages}{105001}.

\bibitem[{\citenamefont{Chen} \emph{et~al.}(2011)\citenamefont{Chen, Pukhov,
  Yu, and Sheng}}]{Chen_2011}
\bibinfo{author}{\bibnamefont{Chen}, \bibfnamefont{M.}},
  \bibinfo{author}{\bibfnamefont{A.}~\bibnamefont{Pukhov}},
  \bibinfo{author}{\bibfnamefont{T.-P.} \bibnamefont{Yu}}, and
  \bibinfo{author}{\bibfnamefont{Z.-M.} \bibnamefont{Sheng}},
  \bibinfo{year}{2011}, \bibinfo{journal}{Plasma Phys. Control. Fusion}
  \textbf{\bibinfo{volume}{53}}, \bibinfo{pages}{014004}.

\bibitem[{\citenamefont{Chen} \emph{et~al.}(2010)\citenamefont{Chen, Arpin,
  Popmintchev, Gerrity, Zhang, Seaberg, Popmintchev, Murnane, and
  Kapteyn}}]{Chen_2010}
\bibinfo{author}{\bibnamefont{Chen}, \bibfnamefont{M.-C.}},
  \bibinfo{author}{\bibfnamefont{P.}~\bibnamefont{Arpin}},
  \bibinfo{author}{\bibfnamefont{T.}~\bibnamefont{Popmintchev}},
  \bibinfo{author}{\bibfnamefont{M.}~\bibnamefont{Gerrity}},
  \bibinfo{author}{\bibfnamefont{B.}~\bibnamefont{Zhang}},
  \bibinfo{author}{\bibfnamefont{M.}~\bibnamefont{Seaberg}},
  \bibinfo{author}{\bibfnamefont{D.}~\bibnamefont{Popmintchev}},
  \bibinfo{author}{\bibfnamefont{M.~M.} \bibnamefont{Murnane}}, and
  \bibinfo{author}{\bibfnamefont{H.~C.} \bibnamefont{Kapteyn}},
  \bibinfo{year}{2010}, \bibinfo{journal}{Phys. Rev. Lett.}
  \textbf{\bibinfo{volume}{105}}, \bibinfo{pages}{173901}.

\bibitem[{\citenamefont{Chen} \emph{et~al.}(1998)\citenamefont{Chen,
  Maksimchuk, and Umstadter}}]{Chen_1998}
\bibinfo{author}{\bibnamefont{Chen}, \bibfnamefont{S.-Y.}},
  \bibinfo{author}{\bibfnamefont{A.}~\bibnamefont{Maksimchuk}}, and
  \bibinfo{author}{\bibfnamefont{D.}~\bibnamefont{Umstadter}},
  \bibinfo{year}{1998}, \bibinfo{journal}{Nature (London)}
  \textbf{\bibinfo{volume}{396}}, \bibinfo{pages}{653}.

\bibitem[{\citenamefont{Cheng} \emph{et~al.}(2008)\citenamefont{Cheng, Bowen,
  Gerry, Su, and Grobe}}]{Cheng_2008}
\bibinfo{author}{\bibnamefont{Cheng}, \bibfnamefont{T.}},
  \bibinfo{author}{\bibfnamefont{S.~P.} \bibnamefont{Bowen}},
  \bibinfo{author}{\bibfnamefont{C.~C.} \bibnamefont{Gerry}},
  \bibinfo{author}{\bibfnamefont{Q.}~\bibnamefont{Su}}, and
  \bibinfo{author}{\bibfnamefont{R.}~\bibnamefont{Grobe}},
  \bibinfo{year}{2008}, \bibinfo{journal}{Phys. Rev. A}
  \textbf{\bibinfo{volume}{77}}, \bibinfo{pages}{032106}.

\bibitem[{\citenamefont{Cheng}
  \emph{et~al.}(2009{\natexlab{a}})\citenamefont{Cheng, Su, and
  Grobe}}]{Cheng_2009}
\bibinfo{author}{\bibnamefont{Cheng}, \bibfnamefont{T.}},
  \bibinfo{author}{\bibfnamefont{Q.}~\bibnamefont{Su}}, and
  \bibinfo{author}{\bibfnamefont{R.}~\bibnamefont{Grobe}},
  \bibinfo{year}{2009}{\natexlab{a}}, \bibinfo{journal}{Phys. Rev. A}
  \textbf{\bibinfo{volume}{80}}, \bibinfo{pages}{013410}.

\bibitem[{\citenamefont{Cheng} \emph{et~al.}(2010)\citenamefont{Cheng, Su, and
  Grobe}}]{Cheng_2010}
\bibinfo{author}{\bibnamefont{Cheng}, \bibfnamefont{T.}},
  \bibinfo{author}{\bibfnamefont{Q.}~\bibnamefont{Su}}, and
  \bibinfo{author}{\bibfnamefont{R.}~\bibnamefont{Grobe}},
  \bibinfo{year}{2010}, \bibinfo{journal}{Contemp. Phys.}
  \textbf{\bibinfo{volume}{51}}, \bibinfo{pages}{315}.

\bibitem[{\citenamefont{Cheng}
  \emph{et~al.}(2009{\natexlab{b}})\citenamefont{Cheng, Ware, Su, and
  Grobe}}]{Cheng_2009b}
\bibinfo{author}{\bibnamefont{Cheng}, \bibfnamefont{T.}},
  \bibinfo{author}{\bibfnamefont{M.~R.} \bibnamefont{Ware}},
  \bibinfo{author}{\bibfnamefont{Q.}~\bibnamefont{Su}}, and
  \bibinfo{author}{\bibfnamefont{R.}~\bibnamefont{Grobe}},
  \bibinfo{year}{2009}{\natexlab{b}}, \bibinfo{journal}{Phys. Rev. A}
  \textbf{\bibinfo{volume}{80}}, \bibinfo{pages}{062105}.

\bibitem[{\citenamefont{Chiril\u{a}}
  \emph{et~al.}(2004)\citenamefont{Chiril\u{a}, Joachain, Kylstra, and
  Potvliege}}]{Chirila_2004}
\bibinfo{author}{\bibnamefont{Chiril\u{a}}, \bibfnamefont{C.~C.}},
  \bibinfo{author}{\bibfnamefont{C.~J.} \bibnamefont{Joachain}},
  \bibinfo{author}{\bibfnamefont{N.~J.} \bibnamefont{Kylstra}}, and
  \bibinfo{author}{\bibfnamefont{R.~M.} \bibnamefont{Potvliege}},
  \bibinfo{year}{2004}, \bibinfo{journal}{Phys. Rev. Lett.}
  \textbf{\bibinfo{volume}{93}}, \bibinfo{pages}{243603}.

\bibitem[{\citenamefont{Chiril\u{a}}
  \emph{et~al.}(2002)\citenamefont{Chiril\u{a}, Kylstra, Potvliege, and
  Joachain}}]{Joachain_2002}
\bibinfo{author}{\bibnamefont{Chiril\u{a}}, \bibfnamefont{C.~C.}},
  \bibinfo{author}{\bibfnamefont{N.~J.} \bibnamefont{Kylstra}},
  \bibinfo{author}{\bibfnamefont{R.~M.} \bibnamefont{Potvliege}}, and
  \bibinfo{author}{\bibfnamefont{C.~J.} \bibnamefont{Joachain}},
  \bibinfo{year}{2002}, \bibinfo{journal}{Phys. Rev. A}
  \textbf{\bibinfo{volume}{66}}, \bibinfo{pages}{063411}.

\bibitem[{\citenamefont{Chouffani} \emph{et~al.}(2002)\citenamefont{Chouffani,
  Wells, Harmon, Jones, and Lancaster}}]{Chouffani_2002}
\bibinfo{author}{\bibnamefont{Chouffani}, \bibfnamefont{K.}},
  \bibinfo{author}{\bibfnamefont{D.}~\bibnamefont{Wells}},
  \bibinfo{author}{\bibfnamefont{F.}~\bibnamefont{Harmon}},
  \bibinfo{author}{\bibfnamefont{J.}~\bibnamefont{Jones}}, and
  \bibinfo{author}{\bibfnamefont{G.}~\bibnamefont{Lancaster}},
  \bibinfo{year}{2002}, \bibinfo{journal}{Nucl. Instr. Meth. Phys. Res. A}
  \textbf{\bibinfo{volume}{495}}, \bibinfo{pages}{95}.

\bibitem[{\citenamefont{Chowdhury} \emph{et~al.}(2001)\citenamefont{Chowdhury,
  Barty, and Walker}}]{Chowdhury_2001}
\bibinfo{author}{\bibnamefont{Chowdhury}, \bibfnamefont{E.~A.}},
  \bibinfo{author}{\bibfnamefont{C.~P.~J.} \bibnamefont{Barty}}, and
  \bibinfo{author}{\bibfnamefont{B.~C.} \bibnamefont{Walker}},
  \bibinfo{year}{2001}, \bibinfo{journal}{Phys. Rev. A}
  \textbf{\bibinfo{volume}{63}}, \bibinfo{pages}{042712}.

\bibitem[{\citenamefont{Cipiccia} \emph{et~al.}(2011)\citenamefont{Cipiccia,
  Islam, Ersfeld, Shanks, Brunetti, Vieux, Yang, Issac, Wiggins, Welsh, Anania,
  Maneuski} \emph{et~al.}}]{Cipiccia_2011}
\bibinfo{author}{\bibnamefont{Cipiccia}, \bibfnamefont{S.}},
  \bibinfo{author}{\bibfnamefont{M.~R.} \bibnamefont{Islam}},
  \bibinfo{author}{\bibfnamefont{B.}~\bibnamefont{Ersfeld}},
  \bibinfo{author}{\bibfnamefont{R.~P.} \bibnamefont{Shanks}},
  \bibinfo{author}{\bibfnamefont{E.}~\bibnamefont{Brunetti}},
  \bibinfo{author}{\bibfnamefont{G.}~\bibnamefont{Vieux}},
  \bibinfo{author}{\bibfnamefont{X.}~\bibnamefont{Yang}},
  \bibinfo{author}{\bibfnamefont{R.~C.} \bibnamefont{Issac}},
  \bibinfo{author}{\bibfnamefont{S.~M.} \bibnamefont{Wiggins}},
  \bibinfo{author}{\bibfnamefont{G.~H.} \bibnamefont{Welsh}},
  \bibinfo{author}{\bibfnamefont{M.-P.} \bibnamefont{Anania}},
  \bibinfo{author}{\bibfnamefont{D.}~\bibnamefont{Maneuski}}, \emph{et~al.},
  \bibinfo{year}{2011}, \bibinfo{journal}{Nature Phys.}
  \textbf{\bibinfo{volume}{7}}, \bibinfo{pages}{867}.

\bibitem[{\citenamefont{Clayton} \emph{et~al.}(2010)\citenamefont{Clayton,
  Ralph, Albert, Fonseca, Glenzer, Joshi, Lu, Marsh, Martins, Mori, Pak, Tsung}
  \emph{et~al.}}]{Clayton_2010}
\bibinfo{author}{\bibnamefont{Clayton}, \bibfnamefont{C.~E.}},
  \bibinfo{author}{\bibfnamefont{J.~E.} \bibnamefont{Ralph}},
  \bibinfo{author}{\bibfnamefont{F.}~\bibnamefont{Albert}},
  \bibinfo{author}{\bibfnamefont{R.~A.} \bibnamefont{Fonseca}},
  \bibinfo{author}{\bibfnamefont{S.~H.} \bibnamefont{Glenzer}},
  \bibinfo{author}{\bibfnamefont{C.}~\bibnamefont{Joshi}},
  \bibinfo{author}{\bibfnamefont{W.}~\bibnamefont{Lu}},
  \bibinfo{author}{\bibfnamefont{K.~A.} \bibnamefont{Marsh}},
  \bibinfo{author}{\bibfnamefont{S.~F.} \bibnamefont{Martins}},
  \bibinfo{author}{\bibfnamefont{W.~B.} \bibnamefont{Mori}},
  \bibinfo{author}{\bibfnamefont{A.}~\bibnamefont{Pak}},
  \bibinfo{author}{\bibfnamefont{F.~S.} \bibnamefont{Tsung}}, \emph{et~al.},
  \bibinfo{year}{2010}, \bibinfo{journal}{Phys. Rev. Lett.}
  \textbf{\bibinfo{volume}{105}}, \bibinfo{pages}{105003}.

\bibitem[{\citenamefont{Combs} \emph{et~al.}(2009)\citenamefont{Combs,
  Nikoghosyan, Jaekel, Karger, Haberer, M\"{u}nter, Huber, Debus, and
  Schulz-Ertner}}]{Combs_2009}
\bibinfo{author}{\bibnamefont{Combs}, \bibfnamefont{S.~E.}},
  \bibinfo{author}{\bibfnamefont{A.}~\bibnamefont{Nikoghosyan}},
  \bibinfo{author}{\bibfnamefont{O.}~\bibnamefont{Jaekel}},
  \bibinfo{author}{\bibfnamefont{C.~P.} \bibnamefont{Karger}},
  \bibinfo{author}{\bibfnamefont{T.}~\bibnamefont{Haberer}},
  \bibinfo{author}{\bibfnamefont{M.~W.} \bibnamefont{M\"{u}nter}},
  \bibinfo{author}{\bibfnamefont{P.~E.} \bibnamefont{Huber}},
  \bibinfo{author}{\bibfnamefont{J.}~\bibnamefont{Debus}}, and
  \bibinfo{author}{\bibfnamefont{D.}~\bibnamefont{Schulz-Ertner}},
  \bibinfo{year}{2009}, \bibinfo{journal}{Cancer}
  \textbf{\bibinfo{volume}{115}}, \bibinfo{pages}{1348}.

\bibitem[{\citenamefont{Corkum}(1993)}]{Corkum_1993}
\bibinfo{author}{\bibnamefont{Corkum}, \bibfnamefont{P.~B.}},
  \bibinfo{year}{1993}, \bibinfo{journal}{Phys. Rev. Lett.}
  \textbf{\bibinfo{volume}{71}}, \bibinfo{pages}{1994}.

\bibitem[{\citenamefont{Dadi and M\"uller}(2011)}]{Anis_2011}
\bibinfo{author}{\bibnamefont{Dadi}, \bibfnamefont{A.}}, and
  \bibinfo{author}{\bibfnamefont{C.}~\bibnamefont{M\"uller}},
  \bibinfo{year}{2011}, \bibinfo{journal}{Phys. Lett. B}
  \textbf{\bibinfo{volume}{697}}, \bibinfo{pages}{142}.

\bibitem[{\citenamefont{Dammasch} \emph{et~al.}(2001)\citenamefont{Dammasch,
  D\"orr, Eichmann, Lenz, and Sandner}}]{Dammasch_2001}
\bibinfo{author}{\bibnamefont{Dammasch}, \bibfnamefont{M.}},
  \bibinfo{author}{\bibfnamefont{M.}~\bibnamefont{D\"orr}},
  \bibinfo{author}{\bibfnamefont{U.}~\bibnamefont{Eichmann}},
  \bibinfo{author}{\bibfnamefont{E.}~\bibnamefont{Lenz}}, and
  \bibinfo{author}{\bibfnamefont{W.}~\bibnamefont{Sandner}},
  \bibinfo{year}{2001}, \bibinfo{journal}{Phys. Rev. A}
  \textbf{\bibinfo{volume}{64}}, \bibinfo{pages}{061402}.

\bibitem[{\citenamefont{Debus} \emph{et~al.}(2010)\citenamefont{Debus,
  Bussmann, Siebold, Jochmann, Schramm, Cowan, and Sauerbrey}}]{Debus_2010}
\bibinfo{author}{\bibnamefont{Debus}, \bibfnamefont{A.}},
  \bibinfo{author}{\bibfnamefont{M.}~\bibnamefont{Bussmann}},
  \bibinfo{author}{\bibfnamefont{M.}~\bibnamefont{Siebold}},
  \bibinfo{author}{\bibfnamefont{A.}~\bibnamefont{Jochmann}},
  \bibinfo{author}{\bibfnamefont{U.}~\bibnamefont{Schramm}},
  \bibinfo{author}{\bibfnamefont{T.}~\bibnamefont{Cowan}}, and
  \bibinfo{author}{\bibfnamefont{R.}~\bibnamefont{Sauerbrey}},
  \bibinfo{year}{2010}, \bibinfo{journal}{Appl. Phys. B}
  \textbf{\bibinfo{volume}{100}}, \bibinfo{pages}{61}.

\bibitem[{\citenamefont{Deneke and M\"uller}(2008)}]{CarlusPRA}
\bibinfo{author}{\bibnamefont{Deneke}, \bibfnamefont{C.}}, and
  \bibinfo{author}{\bibfnamefont{C.}~\bibnamefont{M\"uller}},
  \bibinfo{year}{2008}, \bibinfo{journal}{Phys. Rev. A}
  \textbf{\bibinfo{volume}{78}}, \bibinfo{pages}{033431}.

\bibitem[{\citenamefont{Di~Piazza}(2008)}]{Di_Piazza_2008_a}
\bibinfo{author}{\bibnamefont{Di~Piazza}, \bibfnamefont{A.}},
  \bibinfo{year}{2008}, \bibinfo{journal}{Lett. Math. Phys.}
  \textbf{\bibinfo{volume}{83}}, \bibinfo{pages}{305}.

\bibitem[{\citenamefont{Di~Piazza} \emph{et~al.}(2006)\citenamefont{Di~Piazza,
  Hatsagortsyan, and Keitel}}]{Di_Piazza_2006}
\bibinfo{author}{\bibnamefont{Di~Piazza}, \bibfnamefont{A.}},
  \bibinfo{author}{\bibfnamefont{K.~Z.} \bibnamefont{Hatsagortsyan}}, and
  \bibinfo{author}{\bibfnamefont{C.~H.} \bibnamefont{Keitel}},
  \bibinfo{year}{2006}, \bibinfo{journal}{Phys. Rev. Lett.}
  \textbf{\bibinfo{volume}{97}}, \bibinfo{pages}{083603}.

\bibitem[{\citenamefont{Di~Piazza}
  \emph{et~al.}(2007{\natexlab{a}})\citenamefont{Di~Piazza, Hatsagortsyan, and
  Keitel}}]{Di_Piazza_2007b}
\bibinfo{author}{\bibnamefont{Di~Piazza}, \bibfnamefont{A.}},
  \bibinfo{author}{\bibfnamefont{K.~Z.} \bibnamefont{Hatsagortsyan}}, and
  \bibinfo{author}{\bibfnamefont{C.~H.} \bibnamefont{Keitel}},
  \bibinfo{year}{2007}{\natexlab{a}}, \bibinfo{journal}{Laser Phys.}
  \textbf{\bibinfo{volume}{17}}, \bibinfo{pages}{345}.

\bibitem[{\citenamefont{Di~Piazza}
  \emph{et~al.}(2007{\natexlab{b}})\citenamefont{Di~Piazza, Hatsagortsyan, and
  Keitel}}]{Di_Piazza_2007_a}
\bibinfo{author}{\bibnamefont{Di~Piazza}, \bibfnamefont{A.}},
  \bibinfo{author}{\bibfnamefont{K.~Z.} \bibnamefont{Hatsagortsyan}}, and
  \bibinfo{author}{\bibfnamefont{C.~H.} \bibnamefont{Keitel}},
  \bibinfo{year}{2007}{\natexlab{b}}, \bibinfo{journal}{Phys. Plasmas}
  \textbf{\bibinfo{volume}{14}}, \bibinfo{pages}{032102}.

\bibitem[{\citenamefont{Di~Piazza}
  \emph{et~al.}(2008{\natexlab{a}})\citenamefont{Di~Piazza, Hatsagortsyan, and
  Keitel}}]{Di_Piazza_2008}
\bibinfo{author}{\bibnamefont{Di~Piazza}, \bibfnamefont{A.}},
  \bibinfo{author}{\bibfnamefont{K.~Z.} \bibnamefont{Hatsagortsyan}}, and
  \bibinfo{author}{\bibfnamefont{C.~H.} \bibnamefont{Keitel}},
  \bibinfo{year}{2008}{\natexlab{a}}, \bibinfo{journal}{Phys. Rev. Lett.}
  \textbf{\bibinfo{volume}{100}}, \bibinfo{pages}{010403}.

\bibitem[{\citenamefont{Di~Piazza}
  \emph{et~al.}(2008{\natexlab{b}})\citenamefont{Di~Piazza, Hatsagortsyan, and
  Keitel}}]{Di_Piazza_2008d}
\bibinfo{author}{\bibnamefont{Di~Piazza}, \bibfnamefont{A.}},
  \bibinfo{author}{\bibfnamefont{K.~Z.} \bibnamefont{Hatsagortsyan}}, and
  \bibinfo{author}{\bibfnamefont{C.~H.} \bibnamefont{Keitel}},
  \bibinfo{year}{2008}{\natexlab{b}}, \bibinfo{journal}{Phys. Rev. A}
  \textbf{\bibinfo{volume}{78}}, \bibinfo{pages}{062109}.

\bibitem[{\citenamefont{Di~Piazza}
  \emph{et~al.}(2009{\natexlab{a}})\citenamefont{Di~Piazza, Hatsagortsyan, and
  Keitel}}]{Di_Piazza_2009}
\bibinfo{author}{\bibnamefont{Di~Piazza}, \bibfnamefont{A.}},
  \bibinfo{author}{\bibfnamefont{K.~Z.} \bibnamefont{Hatsagortsyan}}, and
  \bibinfo{author}{\bibfnamefont{C.~H.} \bibnamefont{Keitel}},
  \bibinfo{year}{2009}{\natexlab{a}}, \bibinfo{journal}{Phys. Rev. Lett.}
  \textbf{\bibinfo{volume}{102}}, \bibinfo{pages}{254802}.

\bibitem[{\citenamefont{Di~Piazza}
  \emph{et~al.}(2010{\natexlab{a}})\citenamefont{Di~Piazza, Hatsagortsyan, and
  Keitel}}]{Di_Piazza_2010}
\bibinfo{author}{\bibnamefont{Di~Piazza}, \bibfnamefont{A.}},
  \bibinfo{author}{\bibfnamefont{K.~Z.} \bibnamefont{Hatsagortsyan}}, and
  \bibinfo{author}{\bibfnamefont{C.~H.} \bibnamefont{Keitel}},
  \bibinfo{year}{2010}{\natexlab{a}}, \bibinfo{journal}{Phys. Rev. Lett.}
  \textbf{\bibinfo{volume}{105}}, \bibinfo{pages}{220403}.

\bibitem[{\citenamefont{Di~Piazza}
  \emph{et~al.}(2009{\natexlab{b}})\citenamefont{Di~Piazza, L\"otstedt,
  Milstein, and Keitel}}]{DiPiazza_Lotstedt_2009}
\bibinfo{author}{\bibnamefont{Di~Piazza}, \bibfnamefont{A.}},
  \bibinfo{author}{\bibfnamefont{E.}~\bibnamefont{L\"otstedt}},
  \bibinfo{author}{\bibfnamefont{A.~I.} \bibnamefont{Milstein}}, and
  \bibinfo{author}{\bibfnamefont{C.~H.} \bibnamefont{Keitel}},
  \bibinfo{year}{2009}{\natexlab{b}}, \bibinfo{journal}{Phys. Rev. Lett.}
  \textbf{\bibinfo{volume}{103}}, \bibinfo{pages}{170403}.

\bibitem[{\citenamefont{Di~Piazza}
  \emph{et~al.}(2010{\natexlab{b}})\citenamefont{Di~Piazza, L\"otstedt,
  Milstein, and Keitel}}]{DiPiazza_Lotstedt_2010}
\bibinfo{author}{\bibnamefont{Di~Piazza}, \bibfnamefont{A.}},
  \bibinfo{author}{\bibfnamefont{E.}~\bibnamefont{L\"otstedt}},
  \bibinfo{author}{\bibfnamefont{A.~I.} \bibnamefont{Milstein}}, and
  \bibinfo{author}{\bibfnamefont{C.~H.} \bibnamefont{Keitel}},
  \bibinfo{year}{2010}{\natexlab{b}}, \bibinfo{journal}{Phys. Rev. A}
  \textbf{\bibinfo{volume}{81}}, \bibinfo{pages}{062122}.

\bibitem[{\citenamefont{Di~Piazza and Milstein}(2008)}]{Di_Piazza_2008c}
\bibinfo{author}{\bibnamefont{Di~Piazza}, \bibfnamefont{A.}}, and
  \bibinfo{author}{\bibfnamefont{A.~I.} \bibnamefont{Milstein}},
  \bibinfo{year}{2008}, \bibinfo{journal}{Phys. Rev. A}
  \textbf{\bibinfo{volume}{77}}, \bibinfo{pages}{042102}.

\bibitem[{\citenamefont{Di~Piazza}
  \emph{et~al.}(2007{\natexlab{c}})\citenamefont{Di~Piazza, Milstein, and
  Keitel}}]{Di_Piazza_2007}
\bibinfo{author}{\bibnamefont{Di~Piazza}, \bibfnamefont{A.}},
  \bibinfo{author}{\bibfnamefont{A.~I.} \bibnamefont{Milstein}}, and
  \bibinfo{author}{\bibfnamefont{C.~H.} \bibnamefont{Keitel}},
  \bibinfo{year}{2007}{\natexlab{c}}, \bibinfo{journal}{Phys. Rev. A}
  \textbf{\bibinfo{volume}{76}}, \bibinfo{pages}{032103}.

\bibitem[{\citenamefont{Di~Piazza}
  \emph{et~al.}(2010{\natexlab{c}})\citenamefont{Di~Piazza, Milstein, and
  M\"uller}}]{DiPiazza_Spin2010}
\bibinfo{author}{\bibnamefont{Di~Piazza}, \bibfnamefont{A.}},
  \bibinfo{author}{\bibfnamefont{A.~I.} \bibnamefont{Milstein}}, and
  \bibinfo{author}{\bibfnamefont{C.}~\bibnamefont{M\"uller}},
  \bibinfo{year}{2010}{\natexlab{c}}, \bibinfo{journal}{Phys. Rev. A}
  \textbf{\bibinfo{volume}{82}}, \bibinfo{pages}{062110}.

\bibitem[{\citenamefont{DiChiara} \emph{et~al.}(2008)\citenamefont{DiChiara,
  Ghebregziabher, Sauer, Waesche, Palaniyappan, Wen, and
  Walker}}]{DiChiara_2008}
\bibinfo{author}{\bibnamefont{DiChiara}, \bibfnamefont{A.~D.}},
  \bibinfo{author}{\bibfnamefont{I.}~\bibnamefont{Ghebregziabher}},
  \bibinfo{author}{\bibfnamefont{R.}~\bibnamefont{Sauer}},
  \bibinfo{author}{\bibfnamefont{J.}~\bibnamefont{Waesche}},
  \bibinfo{author}{\bibfnamefont{S.}~\bibnamefont{Palaniyappan}},
  \bibinfo{author}{\bibfnamefont{B.~L.} \bibnamefont{Wen}}, and
  \bibinfo{author}{\bibfnamefont{B.~C.} \bibnamefont{Walker}},
  \bibinfo{year}{2008}, \bibinfo{journal}{Phys. Rev. Lett.}
  \textbf{\bibinfo{volume}{101}}, \bibinfo{pages}{173002}.

\bibitem[{\citenamefont{DiChiara} \emph{et~al.}(2010)\citenamefont{DiChiara,
  Ghebregziabher, Waesche, Stanev, Ekanayake, Barclay, Wells, Watts, Videtto,
  Mancuso, and Walker}}]{DiChiara_2010}
\bibinfo{author}{\bibnamefont{DiChiara}, \bibfnamefont{A.~D.}},
  \bibinfo{author}{\bibfnamefont{I.}~\bibnamefont{Ghebregziabher}},
  \bibinfo{author}{\bibfnamefont{J.~M.} \bibnamefont{Waesche}},
  \bibinfo{author}{\bibfnamefont{T.}~\bibnamefont{Stanev}},
  \bibinfo{author}{\bibfnamefont{N.}~\bibnamefont{Ekanayake}},
  \bibinfo{author}{\bibfnamefont{L.~R.} \bibnamefont{Barclay}},
  \bibinfo{author}{\bibfnamefont{S.~J.} \bibnamefont{Wells}},
  \bibinfo{author}{\bibfnamefont{A.}~\bibnamefont{Watts}},
  \bibinfo{author}{\bibfnamefont{M.}~\bibnamefont{Videtto}},
  \bibinfo{author}{\bibfnamefont{C.~A.} \bibnamefont{Mancuso}}, and
  \bibinfo{author}{\bibfnamefont{B.~C.} \bibnamefont{Walker}},
  \bibinfo{year}{2010}, \bibinfo{journal}{Phys. Rev. A}
  \textbf{\bibinfo{volume}{81}}, \bibinfo{pages}{043417}.

\bibitem[{\citenamefont{Dicus} \emph{et~al.}(2009)\citenamefont{Dicus,
  Farzinnia, Repko, and Tinsley}}]{Dicus_2009}
\bibinfo{author}{\bibnamefont{Dicus}, \bibfnamefont{D.~A.}},
  \bibinfo{author}{\bibfnamefont{A.}~\bibnamefont{Farzinnia}},
  \bibinfo{author}{\bibfnamefont{W.~W.} \bibnamefont{Repko}}, and
  \bibinfo{author}{\bibfnamefont{T.~M.} \bibnamefont{Tinsley}},
  \bibinfo{year}{2009}, \bibinfo{journal}{Phys. Rev. D}
  \textbf{\bibinfo{volume}{79}}, \bibinfo{pages}{013004}.

\bibitem[{\citenamefont{Dimitrovski}
  \emph{et~al.}(2009)\citenamefont{Dimitrovski, F\o{}rre, and
  Madsen}}]{Dimitrovski_2009}
\bibinfo{author}{\bibnamefont{Dimitrovski}, \bibfnamefont{D.}},
  \bibinfo{author}{\bibfnamefont{M.}~\bibnamefont{F\o{}rre}}, and
  \bibinfo{author}{\bibfnamefont{L.~B.} \bibnamefont{Madsen}},
  \bibinfo{year}{2009}, \bibinfo{journal}{Phys. Rev. A}
  \textbf{\bibinfo{volume}{80}}, \bibinfo{pages}{053412}.

\bibitem[{\citenamefont{Dirac}(1928)}]{Dirac_1928}
\bibinfo{author}{\bibnamefont{Dirac}, \bibfnamefont{P.~A.~M.}},
  \bibinfo{year}{1928}, \bibinfo{journal}{Proc. R. Soc. London, Ser. A}
  \textbf{\bibinfo{volume}{117}}, \bibinfo{pages}{610}.

\bibitem[{\citenamefont{Dirac}(1938)}]{Dirac_1938}
\bibinfo{author}{\bibnamefont{Dirac}, \bibfnamefont{P.~A.~M.}},
  \bibinfo{year}{1938}, \bibinfo{journal}{Proc. R. Soc. London, Ser. A}
  \textbf{\bibinfo{volume}{167}}, \bibinfo{pages}{148}.

\bibitem[{\citenamefont{Dittrich and Gies}(2000)}]{Dittrich_b_2000}
\bibinfo{author}{\bibnamefont{Dittrich}, \bibfnamefont{W.}}, and
  \bibinfo{author}{\bibfnamefont{H.}~\bibnamefont{Gies}}, \bibinfo{year}{2000},
  \emph{\bibinfo{title}{Probing the Quantum Vacuum}}
  (\bibinfo{publisher}{Springer, Heidelberg}).

\bibitem[{\citenamefont{Dittrich and Reuter}(1985)}]{Dittrich_b_1985}
\bibinfo{author}{\bibnamefont{Dittrich}, \bibfnamefont{W.}}, and
  \bibinfo{author}{\bibfnamefont{M.}~\bibnamefont{Reuter}},
  \bibinfo{year}{1985}, \emph{\bibinfo{title}{Effective Lagrangians in Quantum
  Electrodynamics}} (\bibinfo{publisher}{Springer, Heidelberg}).

\bibitem[{\citenamefont{D{\"o}brich and Gies}(2010)}]{Dobrich_2010}
\bibinfo{author}{\bibnamefont{D{\"o}brich}, \bibfnamefont{B.}}, and
  \bibinfo{author}{\bibfnamefont{H.}~\bibnamefont{Gies}}, \bibinfo{year}{2010},
  \bibinfo{journal}{J. High Energy Phys.} \textbf{\bibinfo{volume}{10}},
  \bibinfo{pages}{022}.

\bibitem[{\citenamefont{Dorn} \emph{et~al.}(2003)\citenamefont{Dorn, Quabis,
  and Leuchs}}]{Dorn_2003}
\bibinfo{author}{\bibnamefont{Dorn}, \bibfnamefont{R.}},
  \bibinfo{author}{\bibfnamefont{S.}~\bibnamefont{Quabis}}, and
  \bibinfo{author}{\bibfnamefont{G.}~\bibnamefont{Leuchs}},
  \bibinfo{year}{2003}, \bibinfo{journal}{Phys. Rev. Lett.}
  \textbf{\bibinfo{volume}{91}}, \bibinfo{pages}{233901}.

\bibitem[{\citenamefont{Douglas and Nekrasov}(2001)}]{Douglas_2001}
\bibinfo{author}{\bibnamefont{Douglas}, \bibfnamefont{M.~R.}}, and
  \bibinfo{author}{\bibfnamefont{N.~A.} \bibnamefont{Nekrasov}},
  \bibinfo{year}{2001}, \bibinfo{journal}{Rev. Mod. Phys.}
  \textbf{\bibinfo{volume}{73}}, \bibinfo{pages}{977}.

\bibitem[{\citenamefont{Dromey} \emph{et~al.}(2006)\citenamefont{Dromey, Zepf,
  Gopal, Lancaster, Wei, Krushelnick, Tatarakis, Vakakis, Moustaizis, Kodama,
  Tampo, Stoeckl} \emph{et~al.}}]{Dromey_2006}
\bibinfo{author}{\bibnamefont{Dromey}, \bibfnamefont{B.}},
  \bibinfo{author}{\bibfnamefont{M.}~\bibnamefont{Zepf}},
  \bibinfo{author}{\bibfnamefont{A.}~\bibnamefont{Gopal}},
  \bibinfo{author}{\bibfnamefont{K.}~\bibnamefont{Lancaster}},
  \bibinfo{author}{\bibfnamefont{M.~S.} \bibnamefont{Wei}},
  \bibinfo{author}{\bibfnamefont{K.}~\bibnamefont{Krushelnick}},
  \bibinfo{author}{\bibfnamefont{M.}~\bibnamefont{Tatarakis}},
  \bibinfo{author}{\bibfnamefont{N.}~\bibnamefont{Vakakis}},
  \bibinfo{author}{\bibfnamefont{S.}~\bibnamefont{Moustaizis}},
  \bibinfo{author}{\bibfnamefont{R.}~\bibnamefont{Kodama}},
  \bibinfo{author}{\bibfnamefont{M.}~\bibnamefont{Tampo}},
  \bibinfo{author}{\bibfnamefont{C.}~\bibnamefont{Stoeckl}}, \emph{et~al.},
  \bibinfo{year}{2006}, \bibinfo{journal}{Nature Phys.}
  \textbf{\bibinfo{volume}{2}}, \bibinfo{pages}{456}.

\bibitem[{\citenamefont{Duclous} \emph{et~al.}(2011)\citenamefont{Duclous,
  Kirk, and Bell}}]{Duclous_2011}
\bibinfo{author}{\bibnamefont{Duclous}, \bibfnamefont{R.}},
  \bibinfo{author}{\bibfnamefont{J.~G.} \bibnamefont{Kirk}}, and
  \bibinfo{author}{\bibfnamefont{A.~R.} \bibnamefont{Bell}},
  \bibinfo{year}{2011}, \bibinfo{journal}{Plasma Phys. Control. Fusion}
  \textbf{\bibinfo{volume}{53}}, \bibinfo{pages}{015009}.

\bibitem[{\citenamefont{Dumlu}(2010)}]{Dumlu_PRD2010}
\bibinfo{author}{\bibnamefont{Dumlu}, \bibfnamefont{C.~K.}},
  \bibinfo{year}{2010}, \bibinfo{journal}{Phys. Rev. D}
  \textbf{\bibinfo{volume}{82}}, \bibinfo{pages}{045007}.

\bibitem[{\citenamefont{Dumlu and Dunne}(2010)}]{Dumlu_PRL2010}
\bibinfo{author}{\bibnamefont{Dumlu}, \bibfnamefont{C.~K.}}, and
  \bibinfo{author}{\bibfnamefont{G.~V.} \bibnamefont{Dunne}},
  \bibinfo{year}{2010}, \bibinfo{journal}{Phys. Rev. Lett.}
  \textbf{\bibinfo{volume}{104}}, \bibinfo{pages}{250402}.

\bibitem[{\citenamefont{Dumlu and Dunne}(2011)}]{Dumlu_2011}
\bibinfo{author}{\bibnamefont{Dumlu}, \bibfnamefont{C.~K.}}, and
  \bibinfo{author}{\bibfnamefont{G.~V.} \bibnamefont{Dunne}},
  \bibinfo{year}{2011}, \bibinfo{journal}{Phys. Rev. D}
  \textbf{\bibinfo{volume}{83}}, \bibinfo{pages}{065028}.

\bibitem[{\citenamefont{Dunne}(2012)}]{Dunne_2012}
\bibinfo{author}{\bibnamefont{Dunne}, \bibfnamefont{G.}}, \bibinfo{year}{2012},
  \eprint{arXiv:1202.1557v1}.

\bibitem[{\citenamefont{Dunne} \emph{et~al.}(2009)\citenamefont{Dunne, Gies,
  and Sch\"utzhold}}]{Dunne_2009}
\bibinfo{author}{\bibnamefont{Dunne}, \bibfnamefont{G.~V.}},
  \bibinfo{author}{\bibfnamefont{H.}~\bibnamefont{Gies}}, and
  \bibinfo{author}{\bibfnamefont{R.}~\bibnamefont{Sch\"utzhold}},
  \bibinfo{year}{2009}, \bibinfo{journal}{Phys. Rev. D}
  \textbf{\bibinfo{volume}{80}}, \bibinfo{pages}{111301(R)}.

\bibitem[{\citenamefont{Eberly}(1969)}]{Eberly_1969}
\bibinfo{author}{\bibnamefont{Eberly}, \bibfnamefont{J.~H.}},
  \bibinfo{year}{1969}, in \emph{\bibinfo{booktitle}{Progress in Optics, Vol.
  7}}, edited by \bibinfo{editor}{\bibfnamefont{E.}~\bibnamefont{Wolf}}
  (\bibinfo{publisher}{North Holland, Amsterdam}), p. \bibinfo{pages}{359}.

\bibitem[{\citenamefont{Ehlotzky} \emph{et~al.}(2009)\citenamefont{Ehlotzky,
  Krajewska, and Kaminski}}]{Ehlotzky_2009}
\bibinfo{author}{\bibnamefont{Ehlotzky}, \bibfnamefont{F.}},
  \bibinfo{author}{\bibfnamefont{K.}~\bibnamefont{Krajewska}}, and
  \bibinfo{author}{\bibfnamefont{J.~Z.} \bibnamefont{Kaminski}},
  \bibinfo{year}{2009}, \bibinfo{journal}{Rep. Progr. Phys.}
  \textbf{\bibinfo{volume}{72}}, \bibinfo{pages}{046401}.

\bibitem[{\citenamefont{{ELI}}(2011)}]{ELI}
\bibinfo{author}{\bibnamefont{{ELI}}} (\bibinfo{collaboration}{Extreme Light
  Infrastructure}), \bibinfo{year}{2011},
  \urlprefix\url{http://www.extreme-light-infrastructure.eu/}.

\bibitem[{\citenamefont{Elkina} \emph{et~al.}(2011)\citenamefont{Elkina,
  Fedotov, {I. Yu. Kostyukov}, Legkov, Narozhny, Nerush, and
  Ruhl}}]{Elkina_2011}
\bibinfo{author}{\bibnamefont{Elkina}, \bibfnamefont{N.~V.}},
  \bibinfo{author}{\bibfnamefont{A.~M.} \bibnamefont{Fedotov}},
  \bibinfo{author}{\bibnamefont{{I. Yu. Kostyukov}}},
  \bibinfo{author}{\bibfnamefont{M.~V.} \bibnamefont{Legkov}},
  \bibinfo{author}{\bibfnamefont{N.~B.} \bibnamefont{Narozhny}},
  \bibinfo{author}{\bibfnamefont{E.~N.} \bibnamefont{Nerush}}, and
  \bibinfo{author}{\bibfnamefont{H.}~\bibnamefont{Ruhl}}, \bibinfo{year}{2011},
  \bibinfo{journal}{Phys. Rev. ST Accel. Beams} \textbf{\bibinfo{volume}{14}},
  \bibinfo{pages}{054401}.

\bibitem[{\citenamefont{Ellis and Wilson}(2001)}]{Ellis_2001}
\bibinfo{author}{\bibnamefont{Ellis}, \bibfnamefont{J.}}, and
  \bibinfo{author}{\bibfnamefont{I.}~\bibnamefont{Wilson}},
  \bibinfo{year}{2001}, \bibinfo{journal}{Nature (London)}
  \textbf{\bibinfo{volume}{409}}, \bibinfo{pages}{431}.

\bibitem[{\citenamefont{Emma} \emph{et~al.}(2010)\citenamefont{Emma, Akre,
  Arthur, Bionta, Bostedt, Bozek, Brachmann, Bucksbaum, Coffee, Decker, Ding,
  Dowell} \emph{et~al.}}]{Emma_2010}
\bibinfo{author}{\bibnamefont{Emma}, \bibfnamefont{P.}},
  \bibinfo{author}{\bibfnamefont{R.}~\bibnamefont{Akre}},
  \bibinfo{author}{\bibfnamefont{J.}~\bibnamefont{Arthur}},
  \bibinfo{author}{\bibfnamefont{R.}~\bibnamefont{Bionta}},
  \bibinfo{author}{\bibfnamefont{C.}~\bibnamefont{Bostedt}},
  \bibinfo{author}{\bibfnamefont{J.}~\bibnamefont{Bozek}},
  \bibinfo{author}{\bibfnamefont{A.}~\bibnamefont{Brachmann}},
  \bibinfo{author}{\bibfnamefont{P.}~\bibnamefont{Bucksbaum}},
  \bibinfo{author}{\bibfnamefont{R.}~\bibnamefont{Coffee}},
  \bibinfo{author}{\bibfnamefont{F.-J.} \bibnamefont{Decker}},
  \bibinfo{author}{\bibfnamefont{Y.}~\bibnamefont{Ding}},
  \bibinfo{author}{\bibfnamefont{D.}~\bibnamefont{Dowell}}, \emph{et~al.},
  \bibinfo{year}{2010}, \bibinfo{journal}{Nature Photon.}
  \textbf{\bibinfo{volume}{4}}, \bibinfo{pages}{641}.

\bibitem[{\citenamefont{Englert and Rinehart}(1983)}]{Englert_1983}
\bibinfo{author}{\bibnamefont{Englert}, \bibfnamefont{T.~J.}}, and
  \bibinfo{author}{\bibfnamefont{E.~A.} \bibnamefont{Rinehart}},
  \bibinfo{year}{1983}, \bibinfo{journal}{Phys. Rev. A}
  \textbf{\bibinfo{volume}{28}}, \bibinfo{pages}{1539}.

\bibitem[{\citenamefont{Erber}(1966)}]{Erber_1966}
\bibinfo{author}{\bibnamefont{Erber}, \bibfnamefont{T.}}, \bibinfo{year}{1966},
  \bibinfo{journal}{Rev. Mod. Phys.} \textbf{\bibinfo{volume}{38}},
  \bibinfo{pages}{626}.

\bibitem[{\citenamefont{Eriksson} \emph{et~al.}(2004)\citenamefont{Eriksson,
  Brodin, Marklund, and Stenflo}}]{Eriksson_2004}
\bibinfo{author}{\bibnamefont{Eriksson}, \bibfnamefont{D.}},
  \bibinfo{author}{\bibfnamefont{G.}~\bibnamefont{Brodin}},
  \bibinfo{author}{\bibfnamefont{M.}~\bibnamefont{Marklund}}, and
  \bibinfo{author}{\bibfnamefont{L.}~\bibnamefont{Stenflo}},
  \bibinfo{year}{2004}, \bibinfo{journal}{Phys. Rev. A}
  \textbf{\bibinfo{volume}{70}}, \bibinfo{pages}{013808}.

\bibitem[{\citenamefont{Esarey} \emph{et~al.}(2009)\citenamefont{Esarey,
  Schroeder, and Leemans}}]{Esarey_2009}
\bibinfo{author}{\bibnamefont{Esarey}, \bibfnamefont{E.}},
  \bibinfo{author}{\bibfnamefont{C.~B.} \bibnamefont{Schroeder}}, and
  \bibinfo{author}{\bibfnamefont{W.~P.} \bibnamefont{Leemans}},
  \bibinfo{year}{2009}, \bibinfo{journal}{Rev. Mod. Phys.}
  \textbf{\bibinfo{volume}{81}}, \bibinfo{pages}{1229}.

\bibitem[{\citenamefont{Esirkepov} \emph{et~al.}(2004)\citenamefont{Esirkepov,
  Borghesi, Bulanov, Mourou, and Tajima}}]{Esirkepov_2004}
\bibinfo{author}{\bibnamefont{Esirkepov}, \bibfnamefont{T.}},
  \bibinfo{author}{\bibfnamefont{M.}~\bibnamefont{Borghesi}},
  \bibinfo{author}{\bibfnamefont{S.~V.} \bibnamefont{Bulanov}},
  \bibinfo{author}{\bibfnamefont{G.}~\bibnamefont{Mourou}}, and
  \bibinfo{author}{\bibfnamefont{T.}~\bibnamefont{Tajima}},
  \bibinfo{year}{2004}, \bibinfo{journal}{Phys. Rev. Lett.}
  \textbf{\bibinfo{volume}{92}}, \bibinfo{pages}{175003}.

\bibitem[{\citenamefont{Euler}(1936)}]{Euler_1936_a}
\bibinfo{author}{\bibnamefont{Euler}, \bibfnamefont{H.}}, \bibinfo{year}{1936},
  \bibinfo{journal}{Ann. Phys. (Leipzig)} \textbf{\bibinfo{volume}{26}},
  \bibinfo{pages}{398}.

\bibitem[{\citenamefont{{European XFEL}}(2011)}]{XFEL}
\bibinfo{author}{\bibnamefont{{European XFEL}}}, \bibinfo{year}{2011},
  \urlprefix\url{http://xfel.eu/}.

\bibitem[{\citenamefont{Evers and Keitel}(2002)}]{Evers_2002}
\bibinfo{author}{\bibnamefont{Evers}, \bibfnamefont{J.}}, and
  \bibinfo{author}{\bibfnamefont{C.~H.} \bibnamefont{Keitel}},
  \bibinfo{year}{2002}, \bibinfo{journal}{Phys. Rev. Lett.}
  \textbf{\bibinfo{volume}{89}}, \bibinfo{pages}{163601}.

\bibitem[{\citenamefont{Faisal}(1973)}]{Faisal_1973}
\bibinfo{author}{\bibnamefont{Faisal}, \bibfnamefont{F.~H.~M.}},
  \bibinfo{year}{1973}, \bibinfo{journal}{J. Phys. B}
  \textbf{\bibinfo{volume}{6}}, \bibinfo{pages}{L89}.

\bibitem[{\citenamefont{Farzinnia} \emph{et~al.}(2009)\citenamefont{Farzinnia,
  Dicus, Repko, and Tinsley}}]{Farzinnia_2009}
\bibinfo{author}{\bibnamefont{Farzinnia}, \bibfnamefont{A.}},
  \bibinfo{author}{\bibfnamefont{D.~A.} \bibnamefont{Dicus}},
  \bibinfo{author}{\bibfnamefont{W.~W.} \bibnamefont{Repko}}, and
  \bibinfo{author}{\bibfnamefont{T.~M.} \bibnamefont{Tinsley}},
  \bibinfo{year}{2009}, \bibinfo{journal}{Phys. Rev. D}
  \textbf{\bibinfo{volume}{80}}, \bibinfo{pages}{073004}.

\bibitem[{\citenamefont{Faure} \emph{et~al.}(2004)\citenamefont{Faure, Glinec,
  Pukhov, Kiselev, Gordienko, Lefebvre, Rousseau, Burgy, and
  Malka}}]{Faure_2004}
\bibinfo{author}{\bibnamefont{Faure}, \bibfnamefont{J.}},
  \bibinfo{author}{\bibfnamefont{Y.}~\bibnamefont{Glinec}},
  \bibinfo{author}{\bibfnamefont{A.}~\bibnamefont{Pukhov}},
  \bibinfo{author}{\bibfnamefont{S.}~\bibnamefont{Kiselev}},
  \bibinfo{author}{\bibfnamefont{S.}~\bibnamefont{Gordienko}},
  \bibinfo{author}{\bibfnamefont{E.}~\bibnamefont{Lefebvre}},
  \bibinfo{author}{\bibfnamefont{J.-P.} \bibnamefont{Rousseau}},
  \bibinfo{author}{\bibfnamefont{F.}~\bibnamefont{Burgy}}, and
  \bibinfo{author}{\bibfnamefont{V.}~\bibnamefont{Malka}},
  \bibinfo{year}{2004}, \bibinfo{journal}{Nature (London)}
  \textbf{\bibinfo{volume}{431}}, \bibinfo{pages}{541}.

\bibitem[{\citenamefont{Feder}(2010)}]{Feder_2010}
\bibinfo{author}{\bibnamefont{Feder}, \bibfnamefont{T.}}, \bibinfo{year}{2010},
  \bibinfo{journal}{Phys. Today} \textbf{\bibinfo{volume}{63}},
  \bibinfo{pages}{20}.

\bibitem[{\citenamefont{Fedorov} \emph{et~al.}(2006)\citenamefont{Fedorov,
  Efremova, and Volkov}}]{Fedorov_2006}
\bibinfo{author}{\bibnamefont{Fedorov}, \bibfnamefont{M.~V.}},
  \bibinfo{author}{\bibfnamefont{M.~A.} \bibnamefont{Efremova}}, and
  \bibinfo{author}{\bibfnamefont{P.~A.} \bibnamefont{Volkov}},
  \bibinfo{year}{2006}, \bibinfo{journal}{Opt. Commun.}
  \textbf{\bibinfo{volume}{264}}, \bibinfo{pages}{413}.

\bibitem[{\citenamefont{Fedotov} \emph{et~al.}(2010)\citenamefont{Fedotov,
  Narozhny, Mourou, and Korn}}]{Fedotov_2010}
\bibinfo{author}{\bibnamefont{Fedotov}, \bibfnamefont{A.~M.}},
  \bibinfo{author}{\bibfnamefont{N.~B.} \bibnamefont{Narozhny}},
  \bibinfo{author}{\bibfnamefont{G.}~\bibnamefont{Mourou}}, and
  \bibinfo{author}{\bibfnamefont{G.}~\bibnamefont{Korn}}, \bibinfo{year}{2010},
  \bibinfo{journal}{Phys. Rev. Lett.} \textbf{\bibinfo{volume}{105}},
  \bibinfo{pages}{080402}.

\bibitem[{\citenamefont{Feldhaus} \emph{et~al.}(1997)\citenamefont{Feldhaus,
  Saldin, Schneider, Schneidmiller, and Yurkov}}]{Feldhaus_1997}
\bibinfo{author}{\bibnamefont{Feldhaus}, \bibfnamefont{J.}},
  \bibinfo{author}{\bibfnamefont{E.~L.} \bibnamefont{Saldin}},
  \bibinfo{author}{\bibfnamefont{J.~R.} \bibnamefont{Schneider}},
  \bibinfo{author}{\bibfnamefont{E.~A.} \bibnamefont{Schneidmiller}}, and
  \bibinfo{author}{\bibfnamefont{M.~V.} \bibnamefont{Yurkov}},
  \bibinfo{year}{1997}, \bibinfo{journal}{Opt. Commun.}
  \textbf{\bibinfo{volume}{140}}, \bibinfo{pages}{341}.

\bibitem[{\citenamefont{Fennel} \emph{et~al.}(2010)\citenamefont{Fennel,
  Meiwes-Broer, Tiggesb\"aumker, Reinhard, Dinh, and Suraud}}]{Fennel_2010}
\bibinfo{author}{\bibnamefont{Fennel}, \bibfnamefont{{\relax Th}.}},
  \bibinfo{author}{\bibfnamefont{K.-H.} \bibnamefont{Meiwes-Broer}},
  \bibinfo{author}{\bibfnamefont{J.}~\bibnamefont{Tiggesb\"aumker}},
  \bibinfo{author}{\bibfnamefont{P.-G.} \bibnamefont{Reinhard}},
  \bibinfo{author}{\bibfnamefont{P.~M.} \bibnamefont{Dinh}}, and
  \bibinfo{author}{\bibfnamefont{E.}~\bibnamefont{Suraud}},
  \bibinfo{year}{2010}, \bibinfo{journal}{Rev. Mod. Phys.}
  \textbf{\bibinfo{volume}{82}}, \bibinfo{pages}{1793}.

\bibitem[{\citenamefont{Ferrando} \emph{et~al.}(2007)\citenamefont{Ferrando,
  Michinel, Seco, and Tommasini}}]{Ferrando_2007}
\bibinfo{author}{\bibnamefont{Ferrando}, \bibfnamefont{A.}},
  \bibinfo{author}{\bibfnamefont{H.}~\bibnamefont{Michinel}},
  \bibinfo{author}{\bibfnamefont{M.}~\bibnamefont{Seco}}, and
  \bibinfo{author}{\bibfnamefont{D.}~\bibnamefont{Tommasini}},
  \bibinfo{year}{2007}, \bibinfo{journal}{Phys. Rev. Lett.}
  \textbf{\bibinfo{volume}{99}}, \bibinfo{pages}{150404}.

\bibitem[{\citenamefont{Ferraro}(2010)}]{Ferraro_2010}
\bibinfo{author}{\bibnamefont{Ferraro}, \bibfnamefont{R.}},
  \bibinfo{year}{2010}, \bibinfo{journal}{J. Phys. A}
  \textbf{\bibinfo{volume}{43}}, \bibinfo{pages}{195202}.

\bibitem[{\citenamefont{Ferris and Gratus}(2011)}]{Ferris_2011}
\bibinfo{author}{\bibnamefont{Ferris}, \bibfnamefont{M.~R.}}, and
  \bibinfo{author}{\bibfnamefont{J.}~\bibnamefont{Gratus}},
  \bibinfo{year}{2011}, \bibinfo{journal}{J. Math. Phys.}
  \textbf{\bibinfo{volume}{52}}, \bibinfo{pages}{092902}.

\bibitem[{\citenamefont{Fillion-Gourdeau}
  \emph{et~al.}(2012)\citenamefont{Fillion-Gourdeau, Lorin, and
  Bandrauk}}]{Fillion-Gourdeau_2012}
\bibinfo{author}{\bibnamefont{Fillion-Gourdeau}, \bibfnamefont{F.}},
  \bibinfo{author}{\bibfnamefont{E.}~\bibnamefont{Lorin}}, and
  \bibinfo{author}{\bibfnamefont{A.~D.} \bibnamefont{Bandrauk}},
  \bibinfo{year}{2012}, \bibinfo{journal}{Comp. Phys. Commun.}
  \textbf{\bibinfo{volume}{183}}, \bibinfo{pages}{1403}.

\bibitem[{\citenamefont{Fischer} \emph{et~al.}(2006)\citenamefont{Fischer,
  Lein, and Keitel}}]{Fischer_2006}
\bibinfo{author}{\bibnamefont{Fischer}, \bibfnamefont{R.}},
  \bibinfo{author}{\bibfnamefont{M.}~\bibnamefont{Lein}}, and
  \bibinfo{author}{\bibfnamefont{C.~H.} \bibnamefont{Keitel}},
  \bibinfo{year}{2006}, \bibinfo{journal}{Phys. Rev. Lett.}
  \textbf{\bibinfo{volume}{97}}, \bibinfo{pages}{143901}.

\bibitem[{\citenamefont{Fischer} \emph{et~al.}(2007)\citenamefont{Fischer,
  Lein, and Keitel}}]{Fischer_2007}
\bibinfo{author}{\bibnamefont{Fischer}, \bibfnamefont{R.}},
  \bibinfo{author}{\bibfnamefont{M.}~\bibnamefont{Lein}}, and
  \bibinfo{author}{\bibfnamefont{C.~H.} \bibnamefont{Keitel}},
  \bibinfo{year}{2007}, \bibinfo{journal}{J. Phys. B}
  \textbf{\bibinfo{volume}{40}}, \bibinfo{pages}{F113}.

\bibitem[{\citenamefont{{FLASH}}(2011)}]{FLASH}
\bibinfo{author}{\bibnamefont{{FLASH}}} (\bibinfo{collaboration}{Free-Electron
  Laser in Hamburg}), \bibinfo{year}{2011},
  \urlprefix\url{http://hasylab.desy.de/facilities/flash/index_eng.html}.

\bibitem[{\citenamefont{F\o{}rre} \emph{et~al.}(2006)\citenamefont{F\o{}rre,
  Hansen, Kocbach, Selst\o{}, and Madsen}}]{Foerre_2006}
\bibinfo{author}{\bibnamefont{F\o{}rre}, \bibfnamefont{M.}},
  \bibinfo{author}{\bibfnamefont{J.~P.} \bibnamefont{Hansen}},
  \bibinfo{author}{\bibfnamefont{L.}~\bibnamefont{Kocbach}},
  \bibinfo{author}{\bibfnamefont{S.}~\bibnamefont{Selst\o{}}}, and
  \bibinfo{author}{\bibfnamefont{L.~B.} \bibnamefont{Madsen}},
  \bibinfo{year}{2006}, \bibinfo{journal}{Phys. Rev. Lett.}
  \textbf{\bibinfo{volume}{97}}, \bibinfo{pages}{043601}.

\bibitem[{\citenamefont{F\o{}rre} \emph{et~al.}(2007)\citenamefont{F\o{}rre,
  Selst\o{}, Hansen, Kjeldsen, and Madsen}}]{Forre_2007}
\bibinfo{author}{\bibnamefont{F\o{}rre}, \bibfnamefont{M.}},
  \bibinfo{author}{\bibfnamefont{S.}~\bibnamefont{Selst\o{}}},
  \bibinfo{author}{\bibfnamefont{J.~P.} \bibnamefont{Hansen}},
  \bibinfo{author}{\bibfnamefont{T.~K.} \bibnamefont{Kjeldsen}}, and
  \bibinfo{author}{\bibfnamefont{L.~B.} \bibnamefont{Madsen}},
  \bibinfo{year}{2007}, \bibinfo{journal}{Phys. Rev. A}
  \textbf{\bibinfo{volume}{76}}, \bibinfo{pages}{033415}.

\bibitem[{\citenamefont{Fradkin} \emph{et~al.}(1991)\citenamefont{Fradkin,
  Gitman, and Shvartsman}}]{Fradkin_b_1991}
\bibinfo{author}{\bibnamefont{Fradkin}, \bibfnamefont{E.~S.}},
  \bibinfo{author}{\bibfnamefont{D.~M.} \bibnamefont{Gitman}}, and
  \bibinfo{author}{\bibfnamefont{{\relax Sh}.~M.} \bibnamefont{Shvartsman}},
  \bibinfo{year}{1991}, \emph{\bibinfo{title}{Quantum Electrodynamics with
  Unstable Vacuum}} (\bibinfo{publisher}{Springer, Berlin}).

\bibitem[{\citenamefont{Fuchs} \emph{et~al.}(2006)\citenamefont{Fuchs, Antici,
  d'Humi\`{e}res, Lefebvre, Borghesi, Brambrink, Cecchetti, Kaluza, Malka,
  Manclossi, Meyroneinc, Mora} \emph{et~al.}}]{Fuchs_2006}
\bibinfo{author}{\bibnamefont{Fuchs}, \bibfnamefont{J.}},
  \bibinfo{author}{\bibfnamefont{P.}~\bibnamefont{Antici}},
  \bibinfo{author}{\bibfnamefont{E.}~\bibnamefont{d'Humi\`{e}res}},
  \bibinfo{author}{\bibfnamefont{E.}~\bibnamefont{Lefebvre}},
  \bibinfo{author}{\bibfnamefont{M.}~\bibnamefont{Borghesi}},
  \bibinfo{author}{\bibfnamefont{E.}~\bibnamefont{Brambrink}},
  \bibinfo{author}{\bibfnamefont{C.~A.} \bibnamefont{Cecchetti}},
  \bibinfo{author}{\bibfnamefont{M.}~\bibnamefont{Kaluza}},
  \bibinfo{author}{\bibfnamefont{V.}~\bibnamefont{Malka}},
  \bibinfo{author}{\bibfnamefont{M.}~\bibnamefont{Manclossi}},
  \bibinfo{author}{\bibfnamefont{S.}~\bibnamefont{Meyroneinc}},
  \bibinfo{author}{\bibfnamefont{P.}~\bibnamefont{Mora}}, \emph{et~al.},
  \bibinfo{year}{2006}, \bibinfo{journal}{Nature Phys.}
  \textbf{\bibinfo{volume}{2}}, \bibinfo{pages}{48}.

\bibitem[{\citenamefont{Fuchs} \emph{et~al.}(2007)\citenamefont{Fuchs,
  Cecchetti, Borghesi, Grismayer, d'Humi\`eres, Antici, Atzeni, Mora, Pipahl,
  Romagnani, Schiavi, Sentoku} \emph{et~al.}}]{Fuchs_2007}
\bibinfo{author}{\bibnamefont{Fuchs}, \bibfnamefont{J.}},
  \bibinfo{author}{\bibfnamefont{C.~A.} \bibnamefont{Cecchetti}},
  \bibinfo{author}{\bibfnamefont{M.}~\bibnamefont{Borghesi}},
  \bibinfo{author}{\bibfnamefont{T.}~\bibnamefont{Grismayer}},
  \bibinfo{author}{\bibfnamefont{E.}~\bibnamefont{d'Humi\`eres}},
  \bibinfo{author}{\bibfnamefont{P.}~\bibnamefont{Antici}},
  \bibinfo{author}{\bibfnamefont{S.}~\bibnamefont{Atzeni}},
  \bibinfo{author}{\bibfnamefont{P.}~\bibnamefont{Mora}},
  \bibinfo{author}{\bibfnamefont{A.}~\bibnamefont{Pipahl}},
  \bibinfo{author}{\bibfnamefont{L.}~\bibnamefont{Romagnani}},
  \bibinfo{author}{\bibfnamefont{A.}~\bibnamefont{Schiavi}},
  \bibinfo{author}{\bibfnamefont{Y.}~\bibnamefont{Sentoku}}, \emph{et~al.},
  \bibinfo{year}{2007}, \bibinfo{journal}{Phys. Rev. Lett.}
  \textbf{\bibinfo{volume}{99}}, \bibinfo{pages}{015002}.

\bibitem[{\citenamefont{Furry}(1951)}]{Furry_1951}
\bibinfo{author}{\bibnamefont{Furry}, \bibfnamefont{W.~H.}},
  \bibinfo{year}{1951}, \bibinfo{journal}{Phys. Rev.}
  \textbf{\bibinfo{volume}{81}}, \bibinfo{pages}{115}.

\bibitem[{\citenamefont{Galow} \emph{et~al.}(2010)\citenamefont{Galow, Harman,
  and Keitel}}]{Galow_2010}
\bibinfo{author}{\bibnamefont{Galow}, \bibfnamefont{B.~J.}},
  \bibinfo{author}{\bibfnamefont{Z.}~\bibnamefont{Harman}}, and
  \bibinfo{author}{\bibfnamefont{C.~H.} \bibnamefont{Keitel}},
  \bibinfo{year}{2010}, \bibinfo{journal}{Opt. Express}
  \textbf{\bibinfo{volume}{18}}, \bibinfo{pages}{25950}.

\bibitem[{\citenamefont{Galow} \emph{et~al.}(2011)\citenamefont{Galow, Salamin,
  Liseykina, Harman, and Keitel}}]{Galow_2011}
\bibinfo{author}{\bibnamefont{Galow}, \bibfnamefont{B.~J.}},
  \bibinfo{author}{\bibfnamefont{Y.~I.} \bibnamefont{Salamin}},
  \bibinfo{author}{\bibfnamefont{T.~V.} \bibnamefont{Liseykina}},
  \bibinfo{author}{\bibfnamefont{Z.}~\bibnamefont{Harman}}, and
  \bibinfo{author}{\bibfnamefont{C.~H.} \bibnamefont{Keitel}},
  \bibinfo{year}{2011}, \bibinfo{journal}{Phys. Rev. Lett.}
  \textbf{\bibinfo{volume}{107}}, \bibinfo{pages}{185002}.

\bibitem[{\citenamefont{{GammeV}}(2011)}]{GammeV}
\bibinfo{author}{\bibnamefont{{GammeV}}} (\bibinfo{collaboration}{Gamma to
  milli-eV particle search}), \bibinfo{year}{2011},
  \urlprefix\url{http://gammev.fnal.gov/}.

\bibitem[{\citenamefont{Gavrilov and Gitman}(2008)}]{Gavrilov_2008}
\bibinfo{author}{\bibnamefont{Gavrilov}, \bibfnamefont{S.~P.}}, and
  \bibinfo{author}{\bibfnamefont{D.~M.} \bibnamefont{Gitman}},
  \bibinfo{year}{2008}, \bibinfo{journal}{Phys. Rev. Lett.}
  \textbf{\bibinfo{volume}{101}}, \bibinfo{pages}{130403}.

\bibitem[{\citenamefont{Geddes} \emph{et~al.}(2004)\citenamefont{Geddes, Toth,
  van Tilborg, Esarey, Schroeder, Bruhwiler, Nieter, Cary, and
  Leemans}}]{Geddes_2004}
\bibinfo{author}{\bibnamefont{Geddes}, \bibfnamefont{C.~G.~R.}},
  \bibinfo{author}{\bibfnamefont{C.}~\bibnamefont{Toth}},
  \bibinfo{author}{\bibfnamefont{J.}~\bibnamefont{van Tilborg}},
  \bibinfo{author}{\bibfnamefont{E.}~\bibnamefont{Esarey}},
  \bibinfo{author}{\bibfnamefont{C.~B.} \bibnamefont{Schroeder}},
  \bibinfo{author}{\bibfnamefont{D.}~\bibnamefont{Bruhwiler}},
  \bibinfo{author}{\bibfnamefont{C.}~\bibnamefont{Nieter}},
  \bibinfo{author}{\bibfnamefont{J.}~\bibnamefont{Cary}}, and
  \bibinfo{author}{\bibfnamefont{W.~P.} \bibnamefont{Leemans}},
  \bibinfo{year}{2004}, \bibinfo{journal}{Nature (London)}
  \textbf{\bibinfo{volume}{431}}, \bibinfo{pages}{538}.

\bibitem[{\citenamefont{{GEKKO EXA}}(2011)}]{Gekko_EXA}
\bibinfo{author}{\bibnamefont{{GEKKO EXA}}}, \bibinfo{year}{2011},
  \bibinfo{note}{(Homepage mostly in Japanese)},
  \urlprefix\url{www.physics.harvard.edu/~wilson/energypmp/2010_Ongena2}.

\bibitem[{\citenamefont{Gerstner}(2007)}]{Gerstner_2007}
\bibinfo{author}{\bibnamefont{Gerstner}, \bibfnamefont{E.}},
  \bibinfo{year}{2007}, \bibinfo{journal}{Nature (London)}
  \textbf{\bibinfo{volume}{446}}, \bibinfo{pages}{17}.

\bibitem[{\citenamefont{Gibson} \emph{et~al.}(2010)\citenamefont{Gibson,
  Albert, Anderson, Betts, Messerly, Phan, Semenov, Shverdin, Tremaine,
  Hartemann, Siders, McNabb} \emph{et~al.}}]{Gibson_2010}
\bibinfo{author}{\bibnamefont{Gibson}, \bibfnamefont{D.~J.}},
  \bibinfo{author}{\bibfnamefont{F.}~\bibnamefont{Albert}},
  \bibinfo{author}{\bibfnamefont{S.~G.} \bibnamefont{Anderson}},
  \bibinfo{author}{\bibfnamefont{S.~M.} \bibnamefont{Betts}},
  \bibinfo{author}{\bibfnamefont{M.~J.} \bibnamefont{Messerly}},
  \bibinfo{author}{\bibfnamefont{H.~H.} \bibnamefont{Phan}},
  \bibinfo{author}{\bibfnamefont{V.~A.} \bibnamefont{Semenov}},
  \bibinfo{author}{\bibfnamefont{M.~Y.} \bibnamefont{Shverdin}},
  \bibinfo{author}{\bibfnamefont{A.~M.} \bibnamefont{Tremaine}},
  \bibinfo{author}{\bibfnamefont{F.~V.} \bibnamefont{Hartemann}},
  \bibinfo{author}{\bibfnamefont{C.~W.} \bibnamefont{Siders}},
  \bibinfo{author}{\bibfnamefont{D.~P.} \bibnamefont{McNabb}}, \emph{et~al.},
  \bibinfo{year}{2010}, \bibinfo{journal}{Phys. Rev. ST Accel. Beams}
  \textbf{\bibinfo{volume}{13}}, \bibinfo{pages}{070703}.

\bibitem[{\citenamefont{Gibson} \emph{et~al.}(2004)\citenamefont{Gibson,
  Anderson, Barty, Betts, Booth, Brown, Crane, Cross, Fittinghoff, Hartemann,
  Kuba, Sage} \emph{et~al.}}]{Gibson_2004}
\bibinfo{author}{\bibnamefont{Gibson}, \bibfnamefont{D.~J.}},
  \bibinfo{author}{\bibfnamefont{S.~G.} \bibnamefont{Anderson}},
  \bibinfo{author}{\bibfnamefont{C.~P.~J.} \bibnamefont{Barty}},
  \bibinfo{author}{\bibfnamefont{S.~M.} \bibnamefont{Betts}},
  \bibinfo{author}{\bibfnamefont{R.}~\bibnamefont{Booth}},
  \bibinfo{author}{\bibfnamefont{W.~J.} \bibnamefont{Brown}},
  \bibinfo{author}{\bibfnamefont{J.~K.} \bibnamefont{Crane}},
  \bibinfo{author}{\bibfnamefont{R.~R.} \bibnamefont{Cross}},
  \bibinfo{author}{\bibfnamefont{D.~N.} \bibnamefont{Fittinghoff}},
  \bibinfo{author}{\bibfnamefont{F.~V.} \bibnamefont{Hartemann}},
  \bibinfo{author}{\bibfnamefont{J.}~\bibnamefont{Kuba}},
  \bibinfo{author}{\bibfnamefont{G.~P.~L.} \bibnamefont{Sage}}, \emph{et~al.},
  \bibinfo{year}{2004}, \bibinfo{journal}{Phys. Plasmas}
  \textbf{\bibinfo{volume}{11}}, \bibinfo{pages}{2857}.

\bibitem[{\citenamefont{Gies}(2009)}]{Gies_2009}
\bibinfo{author}{\bibnamefont{Gies}, \bibfnamefont{H.}}, \bibinfo{year}{2009},
  \bibinfo{journal}{Eur. Phys. J. D} \textbf{\bibinfo{volume}{55}},
  \bibinfo{pages}{311}.

\bibitem[{\citenamefont{Gies} \emph{et~al.}(2006)\citenamefont{Gies, Jaeckel,
  and Ringwald}}]{Gies_2006}
\bibinfo{author}{\bibnamefont{Gies}, \bibfnamefont{H.}},
  \bibinfo{author}{\bibfnamefont{J.}~\bibnamefont{Jaeckel}}, and
  \bibinfo{author}{\bibfnamefont{A.}~\bibnamefont{Ringwald}},
  \bibinfo{year}{2006}, \bibinfo{journal}{Phys. Rev. Lett.}
  \textbf{\bibinfo{volume}{97}}, \bibinfo{pages}{140402}.

\bibitem[{\citenamefont{Goldman}(1964)}]{Goldman_1964}
\bibinfo{author}{\bibnamefont{Goldman}, \bibfnamefont{I.}},
  \bibinfo{year}{1964}, \bibinfo{journal}{Sov. Phys. JETP}
  \textbf{\bibinfo{volume}{46}}, \bibinfo{pages}{1412}.

\bibitem[{\citenamefont{Goulielmakis}
  \emph{et~al.}(2008)\citenamefont{Goulielmakis, Schultze, Hofstetter,
  Yakovlev, Gagnon, Uiberacker, Aquila, Gullikson, Attwood, Kienberger, Krausz,
  and Kleineberg}}]{Goulielmakis_2008}
\bibinfo{author}{\bibnamefont{Goulielmakis}, \bibfnamefont{E.}},
  \bibinfo{author}{\bibfnamefont{M.}~\bibnamefont{Schultze}},
  \bibinfo{author}{\bibfnamefont{M.}~\bibnamefont{Hofstetter}},
  \bibinfo{author}{\bibfnamefont{V.~S.} \bibnamefont{Yakovlev}},
  \bibinfo{author}{\bibfnamefont{J.}~\bibnamefont{Gagnon}},
  \bibinfo{author}{\bibfnamefont{M.}~\bibnamefont{Uiberacker}},
  \bibinfo{author}{\bibfnamefont{A.~L.} \bibnamefont{Aquila}},
  \bibinfo{author}{\bibfnamefont{E.~M.} \bibnamefont{Gullikson}},
  \bibinfo{author}{\bibfnamefont{D.~T.} \bibnamefont{Attwood}},
  \bibinfo{author}{\bibfnamefont{R.}~\bibnamefont{Kienberger}},
  \bibinfo{author}{\bibfnamefont{F.}~\bibnamefont{Krausz}}, and
  \bibinfo{author}{\bibfnamefont{U.}~\bibnamefont{Kleineberg}},
  \bibinfo{year}{2008}, \bibinfo{journal}{Science}
  \textbf{\bibinfo{volume}{320}}, \bibinfo{pages}{1614}.

\bibitem[{\citenamefont{Gralla} \emph{et~al.}(2009)\citenamefont{Gralla, Harte,
  and Wald}}]{Gralla_2009}
\bibinfo{author}{\bibnamefont{Gralla}, \bibfnamefont{S.~E.}},
  \bibinfo{author}{\bibfnamefont{A.~I.} \bibnamefont{Harte}}, and
  \bibinfo{author}{\bibfnamefont{R.~M.} \bibnamefont{Wald}},
  \bibinfo{year}{2009}, \bibinfo{journal}{Phys. Rev. D}
  \textbf{\bibinfo{volume}{80}}, \bibinfo{pages}{024031}.

\bibitem[{\citenamefont{Gr\"{u}ner}
  \emph{et~al.}(2007)\citenamefont{Gr\"{u}ner, Becker, Schramm, Eichner, Fuchs,
  Weingartner, Habs, Meyer-ter Vehn, Geissler, Ferrario, Serafini, van~der
  Geer} \emph{et~al.}}]{Gruner_2007}
\bibinfo{author}{\bibnamefont{Gr\"{u}ner}, \bibfnamefont{F.}},
  \bibinfo{author}{\bibfnamefont{S.}~\bibnamefont{Becker}},
  \bibinfo{author}{\bibfnamefont{U.}~\bibnamefont{Schramm}},
  \bibinfo{author}{\bibfnamefont{T.}~\bibnamefont{Eichner}},
  \bibinfo{author}{\bibfnamefont{M.}~\bibnamefont{Fuchs}},
  \bibinfo{author}{\bibfnamefont{R.}~\bibnamefont{Weingartner}},
  \bibinfo{author}{\bibfnamefont{D.}~\bibnamefont{Habs}},
  \bibinfo{author}{\bibfnamefont{J.}~\bibnamefont{Meyer-ter Vehn}},
  \bibinfo{author}{\bibfnamefont{M.}~\bibnamefont{Geissler}},
  \bibinfo{author}{\bibfnamefont{M.}~\bibnamefont{Ferrario}},
  \bibinfo{author}{\bibfnamefont{L.}~\bibnamefont{Serafini}},
  \bibinfo{author}{\bibfnamefont{B.}~\bibnamefont{van~der Geer}},
  \emph{et~al.}, \bibinfo{year}{2007}, \bibinfo{journal}{Appl. Phys. B}
  \textbf{\bibinfo{volume}{86}}, \bibinfo{pages}{431}.

\bibitem[{\citenamefont{Gubbini} \emph{et~al.}(2005)\citenamefont{Gubbini,
  Eichmann, Kalashnikov, and Sandner}}]{Gubbini_2005}
\bibinfo{author}{\bibnamefont{Gubbini}, \bibfnamefont{E.}},
  \bibinfo{author}{\bibfnamefont{U.}~\bibnamefont{Eichmann}},
  \bibinfo{author}{\bibfnamefont{M.}~\bibnamefont{Kalashnikov}}, and
  \bibinfo{author}{\bibfnamefont{W.}~\bibnamefont{Sandner}},
  \bibinfo{year}{2005}, \bibinfo{journal}{Phys. Rev. Lett.}
  \textbf{\bibinfo{volume}{94}}, \bibinfo{pages}{053602}.

\bibitem[{\citenamefont{Gupta} \emph{et~al.}(2007)\citenamefont{Gupta, Kant,
  Kim, and Suk}}]{Gupta_2007}
\bibinfo{author}{\bibnamefont{Gupta}, \bibfnamefont{D.~N.}},
  \bibinfo{author}{\bibfnamefont{N.}~\bibnamefont{Kant}},
  \bibinfo{author}{\bibfnamefont{D.~E.} \bibnamefont{Kim}}, and
  \bibinfo{author}{\bibfnamefont{H.}~\bibnamefont{Suk}}, \bibinfo{year}{2007},
  \bibinfo{journal}{Phys. Lett. A} \textbf{\bibinfo{volume}{368}},
  \bibinfo{pages}{402}.

\bibitem[{\citenamefont{Haberberger}
  \emph{et~al.}(2012)\citenamefont{Haberberger, Tochitsky, Fiuza, Gong,
  Fonseca, Silva, Mori, and Joshi}}]{Haberberger_2012}
\bibinfo{author}{\bibnamefont{Haberberger}, \bibfnamefont{D.}},
  \bibinfo{author}{\bibfnamefont{S.}~\bibnamefont{Tochitsky}},
  \bibinfo{author}{\bibfnamefont{F.}~\bibnamefont{Fiuza}},
  \bibinfo{author}{\bibfnamefont{C.}~\bibnamefont{Gong}},
  \bibinfo{author}{\bibfnamefont{R.~A.} \bibnamefont{Fonseca}},
  \bibinfo{author}{\bibfnamefont{L.~O.} \bibnamefont{Silva}},
  \bibinfo{author}{\bibfnamefont{W.~B.} \bibnamefont{Mori}}, and
  \bibinfo{author}{\bibfnamefont{C.}~\bibnamefont{Joshi}},
  \bibinfo{year}{2012}, \bibinfo{journal}{Nature Phys.}
  \textbf{\bibinfo{volume}{8}}, \bibinfo{pages}{95}.

\bibitem[{\citenamefont{Habs} \emph{et~al.}(2008)\citenamefont{Habs, Hegelich,
  Schreiber, Gross, Henig, Kiefer, and Jung}}]{Habs_2008}
\bibinfo{author}{\bibnamefont{Habs}, \bibfnamefont{D.}},
  \bibinfo{author}{\bibfnamefont{M.}~\bibnamefont{Hegelich}},
  \bibinfo{author}{\bibfnamefont{J.}~\bibnamefont{Schreiber}},
  \bibinfo{author}{\bibfnamefont{M.}~\bibnamefont{Gross}},
  \bibinfo{author}{\bibfnamefont{A.}~\bibnamefont{Henig}},
  \bibinfo{author}{\bibfnamefont{D.}~\bibnamefont{Kiefer}}, and
  \bibinfo{author}{\bibfnamefont{D.}~\bibnamefont{Jung}}, \bibinfo{year}{2008},
  \bibinfo{journal}{Appl. Phys. B} \textbf{\bibinfo{volume}{93}},
  \bibinfo{pages}{349}.

\bibitem[{\citenamefont{Habs} \emph{et~al.}(2009)\citenamefont{Habs, Tajima,
  Schreiber, Barty, Fujiwara, and Thirolf}}]{Habs_2009}
\bibinfo{author}{\bibnamefont{Habs}, \bibfnamefont{D.}},
  \bibinfo{author}{\bibfnamefont{T.}~\bibnamefont{Tajima}},
  \bibinfo{author}{\bibfnamefont{J.}~\bibnamefont{Schreiber}},
  \bibinfo{author}{\bibfnamefont{C.~P.~J.} \bibnamefont{Barty}},
  \bibinfo{author}{\bibfnamefont{M.}~\bibnamefont{Fujiwara}}, and
  \bibinfo{author}{\bibfnamefont{P.~G.} \bibnamefont{Thirolf}},
  \bibinfo{year}{2009}, \bibinfo{journal}{Eur. Phys. J. D}
  \textbf{\bibinfo{volume}{55}}, \bibinfo{pages}{279}.

\bibitem[{\citenamefont{Hadad} \emph{et~al.}(2010)\citenamefont{Hadad, Labun,
  Rafelski, Elkina, Klier, and Ruhl}}]{Hadad_2010}
\bibinfo{author}{\bibnamefont{Hadad}, \bibfnamefont{Y.}},
  \bibinfo{author}{\bibfnamefont{L.}~\bibnamefont{Labun}},
  \bibinfo{author}{\bibfnamefont{J.}~\bibnamefont{Rafelski}},
  \bibinfo{author}{\bibfnamefont{N.}~\bibnamefont{Elkina}},
  \bibinfo{author}{\bibfnamefont{C.}~\bibnamefont{Klier}}, and
  \bibinfo{author}{\bibfnamefont{H.}~\bibnamefont{Ruhl}}, \bibinfo{year}{2010},
  \bibinfo{journal}{Phys. Rev. D} \textbf{\bibinfo{volume}{82}},
  \bibinfo{pages}{096012}.

\bibitem[{\citenamefont{Hafz} \emph{et~al.}(2008)\citenamefont{Hafz, Jeong,
  Choi, Lee, Pae, Kulagin, Sung, Yu, Hong, Hosokai, Cary, Ko}
  \emph{et~al.}}]{Hafz_2008}
\bibinfo{author}{\bibnamefont{Hafz}, \bibfnamefont{N.~A.~M.}},
  \bibinfo{author}{\bibfnamefont{T.~M.} \bibnamefont{Jeong}},
  \bibinfo{author}{\bibfnamefont{I.~W.} \bibnamefont{Choi}},
  \bibinfo{author}{\bibfnamefont{S.~K.} \bibnamefont{Lee}},
  \bibinfo{author}{\bibfnamefont{K.~H.} \bibnamefont{Pae}},
  \bibinfo{author}{\bibfnamefont{V.~V.} \bibnamefont{Kulagin}},
  \bibinfo{author}{\bibfnamefont{J.~H.} \bibnamefont{Sung}},
  \bibinfo{author}{\bibfnamefont{T.~J.} \bibnamefont{Yu}},
  \bibinfo{author}{\bibfnamefont{K.-H.} \bibnamefont{Hong}},
  \bibinfo{author}{\bibfnamefont{T.}~\bibnamefont{Hosokai}},
  \bibinfo{author}{\bibfnamefont{J.~R.} \bibnamefont{Cary}},
  \bibinfo{author}{\bibfnamefont{D.-K.} \bibnamefont{Ko}}, \emph{et~al.},
  \bibinfo{year}{2008}, \bibinfo{journal}{Nature Photon.}
  \textbf{\bibinfo{volume}{2}}, \bibinfo{pages}{571}.

\bibitem[{\citenamefont{Hammond}(2010)}]{Hammond_2010_b}
\bibinfo{author}{\bibnamefont{Hammond}, \bibfnamefont{R.~T.}},
  \bibinfo{year}{2010}, \bibinfo{journal}{Electron. J. Theor. Phys.}
  \textbf{\bibinfo{volume}{7}}, \bibinfo{pages}{221}.

\bibitem[{\citenamefont{Han} \emph{et~al.}(2010)\citenamefont{Han, Ruffini, and
  Xue}}]{Ruffini_2010}
\bibinfo{author}{\bibnamefont{Han}, \bibfnamefont{W.-B.}},
  \bibinfo{author}{\bibfnamefont{R.}~\bibnamefont{Ruffini}}, and
  \bibinfo{author}{\bibfnamefont{S.-S.} \bibnamefont{Xue}},
  \bibinfo{year}{2010}, \bibinfo{journal}{Phys. Lett. B}
  \textbf{\bibinfo{volume}{691}}, \bibinfo{pages}{99}.

\bibitem[{\citenamefont{Har-Shemesh and Di~Piazza}(2012)}]{Har-Shemesh_2012}
\bibinfo{author}{\bibnamefont{Har-Shemesh}, \bibfnamefont{O.}}, and
  \bibinfo{author}{\bibfnamefont{A.}~\bibnamefont{Di~Piazza}},
  \bibinfo{year}{2012}, \bibinfo{journal}{Opt. Lett.}
  \textbf{\bibinfo{volume}{37}}, \bibinfo{pages}{1352}.

\bibitem[{\citenamefont{Harman} \emph{et~al.}(2011)\citenamefont{Harman,
  Salamin, Galow, and Keitel}}]{Harman_2011}
\bibinfo{author}{\bibnamefont{Harman}, \bibfnamefont{Z.}},
  \bibinfo{author}{\bibfnamefont{Y.~I.} \bibnamefont{Salamin}},
  \bibinfo{author}{\bibfnamefont{B.~J.} \bibnamefont{Galow}}, and
  \bibinfo{author}{\bibfnamefont{C.~H.} \bibnamefont{Keitel}},
  \bibinfo{year}{2011}, \bibinfo{journal}{Phys. Rev. A}
  \textbf{\bibinfo{volume}{84}}, \bibinfo{pages}{053814}.

\bibitem[{\citenamefont{Hartemann} \emph{et~al.}(2004)\citenamefont{Hartemann,
  Tremaine, Anderson, Barty, Betts, Booth, Brown, Crane, Cross, Gibson,
  Fittinghoff, Kuba} \emph{et~al.}}]{Hartemann_2004}
\bibinfo{author}{\bibnamefont{Hartemann}, \bibfnamefont{F.}},
  \bibinfo{author}{\bibfnamefont{A.}~\bibnamefont{Tremaine}},
  \bibinfo{author}{\bibfnamefont{S.}~\bibnamefont{Anderson}},
  \bibinfo{author}{\bibfnamefont{C.}~\bibnamefont{Barty}},
  \bibinfo{author}{\bibfnamefont{S.}~\bibnamefont{Betts}},
  \bibinfo{author}{\bibfnamefont{R.}~\bibnamefont{Booth}},
  \bibinfo{author}{\bibfnamefont{W.}~\bibnamefont{Brown}},
  \bibinfo{author}{\bibfnamefont{J.}~\bibnamefont{Crane}},
  \bibinfo{author}{\bibfnamefont{R.}~\bibnamefont{Cross}},
  \bibinfo{author}{\bibfnamefont{D.}~\bibnamefont{Gibson}},
  \bibinfo{author}{\bibfnamefont{D.}~\bibnamefont{Fittinghoff}},
  \bibinfo{author}{\bibfnamefont{J.}~\bibnamefont{Kuba}}, \emph{et~al.},
  \bibinfo{year}{2004}, \bibinfo{journal}{Laser Part. Beams}
  \textbf{\bibinfo{volume}{22}}, \bibinfo{pages}{221}.

\bibitem[{\citenamefont{Hartemann}(2001)}]{Hartemann_b_2001}
\bibinfo{author}{\bibnamefont{Hartemann}, \bibfnamefont{F.~V.}},
  \bibinfo{year}{2001}, \emph{\bibinfo{title}{High-Field Electrodynamics}}
  (\bibinfo{publisher}{CRC Press, Boca Raton}).

\bibitem[{\citenamefont{Hartemann} \emph{et~al.}(2005)\citenamefont{Hartemann,
  Brown, Gibson, Anderson, Tremaine, Springer, Wootton, Hartouni, and
  Barty}}]{Hartemann_2005}
\bibinfo{author}{\bibnamefont{Hartemann}, \bibfnamefont{F.~V.}},
  \bibinfo{author}{\bibfnamefont{W.~J.} \bibnamefont{Brown}},
  \bibinfo{author}{\bibfnamefont{D.~J.} \bibnamefont{Gibson}},
  \bibinfo{author}{\bibfnamefont{S.~G.} \bibnamefont{Anderson}},
  \bibinfo{author}{\bibfnamefont{A.~M.} \bibnamefont{Tremaine}},
  \bibinfo{author}{\bibfnamefont{P.~T.} \bibnamefont{Springer}},
  \bibinfo{author}{\bibfnamefont{A.~J.} \bibnamefont{Wootton}},
  \bibinfo{author}{\bibfnamefont{E.~P.} \bibnamefont{Hartouni}}, and
  \bibinfo{author}{\bibfnamefont{C.~P.~J.} \bibnamefont{Barty}},
  \bibinfo{year}{2005}, \bibinfo{journal}{Phys. Rev. ST Accel. Beams}
  \textbf{\bibinfo{volume}{8}}, \bibinfo{pages}{100702}.

\bibitem[{\citenamefont{Hartemann} \emph{et~al.}(2007)\citenamefont{Hartemann,
  Gibson, Brown, Rousse, Phuoc, Malka, Faure, and Pukhov}}]{Hartemann_2007}
\bibinfo{author}{\bibnamefont{Hartemann}, \bibfnamefont{F.~V.}},
  \bibinfo{author}{\bibfnamefont{D.~J.} \bibnamefont{Gibson}},
  \bibinfo{author}{\bibfnamefont{W.~J.} \bibnamefont{Brown}},
  \bibinfo{author}{\bibfnamefont{A.}~\bibnamefont{Rousse}},
  \bibinfo{author}{\bibfnamefont{K.~T.} \bibnamefont{Phuoc}},
  \bibinfo{author}{\bibfnamefont{V.}~\bibnamefont{Malka}},
  \bibinfo{author}{\bibfnamefont{J.}~\bibnamefont{Faure}}, and
  \bibinfo{author}{\bibfnamefont{A.}~\bibnamefont{Pukhov}},
  \bibinfo{year}{2007}, \bibinfo{journal}{Phys. Rev. ST Accel. Beams}
  \textbf{\bibinfo{volume}{10}}, \bibinfo{pages}{011301}.

\bibitem[{\citenamefont{Hartemann and Kerman}(1996)}]{Hartemann_1996}
\bibinfo{author}{\bibnamefont{Hartemann}, \bibfnamefont{F.~V.}}, and
  \bibinfo{author}{\bibfnamefont{A.~K.} \bibnamefont{Kerman}},
  \bibinfo{year}{1996}, \bibinfo{journal}{Phys. Rev. Lett.}
  \textbf{\bibinfo{volume}{76}}, \bibinfo{pages}{624}.

\bibitem[{\citenamefont{Hartemann} \emph{et~al.}(2008)\citenamefont{Hartemann,
  Siders, and Barty}}]{Hartemann_2008}
\bibinfo{author}{\bibnamefont{Hartemann}, \bibfnamefont{F.~V.}},
  \bibinfo{author}{\bibfnamefont{C.~W.} \bibnamefont{Siders}}, and
  \bibinfo{author}{\bibfnamefont{C.~P.~J.} \bibnamefont{Barty}},
  \bibinfo{year}{2008}, \bibinfo{journal}{Phys. Rev. Lett.}
  \textbf{\bibinfo{volume}{100}}, \bibinfo{pages}{125001}.

\bibitem[{\citenamefont{Harvey}
  \emph{et~al.}(2011{\natexlab{a}})\citenamefont{Harvey, Heinzl, Iji, and
  Langfeld}}]{Harvey_2011}
\bibinfo{author}{\bibnamefont{Harvey}, \bibfnamefont{C.}},
  \bibinfo{author}{\bibfnamefont{T.}~\bibnamefont{Heinzl}},
  \bibinfo{author}{\bibfnamefont{N.}~\bibnamefont{Iji}}, and
  \bibinfo{author}{\bibfnamefont{K.}~\bibnamefont{Langfeld}},
  \bibinfo{year}{2011}{\natexlab{a}}, \bibinfo{journal}{Phys. Rev. D}
  \textbf{\bibinfo{volume}{83}}, \bibinfo{pages}{076013}.

\bibitem[{\citenamefont{Harvey} \emph{et~al.}(2009)\citenamefont{Harvey,
  Heinzl, and Ilderton}}]{Harvey_2009}
\bibinfo{author}{\bibnamefont{Harvey}, \bibfnamefont{C.}},
  \bibinfo{author}{\bibfnamefont{T.}~\bibnamefont{Heinzl}}, and
  \bibinfo{author}{\bibfnamefont{A.}~\bibnamefont{Ilderton}},
  \bibinfo{year}{2009}, \bibinfo{journal}{Phys. Rev. A}
  \textbf{\bibinfo{volume}{79}}, \bibinfo{pages}{063407}.

\bibitem[{\citenamefont{Harvey}
  \emph{et~al.}(2011{\natexlab{b}})\citenamefont{Harvey, Heinzl, and
  Marklund}}]{Harvey_2011b}
\bibinfo{author}{\bibnamefont{Harvey}, \bibfnamefont{C.}},
  \bibinfo{author}{\bibfnamefont{T.}~\bibnamefont{Heinzl}}, and
  \bibinfo{author}{\bibfnamefont{M.}~\bibnamefont{Marklund}},
  \bibinfo{year}{2011}{\natexlab{b}}, \bibinfo{journal}{Phys. Rev. D}
  \textbf{\bibinfo{volume}{84}}, \bibinfo{pages}{116005}.

\bibitem[{\citenamefont{Harvey and Marklund}(2012)}]{Harvey_2011c}
\bibinfo{author}{\bibnamefont{Harvey}, \bibfnamefont{C.}}, and
  \bibinfo{author}{\bibfnamefont{M.}~\bibnamefont{Marklund}},
  \bibinfo{year}{2012}, \bibinfo{journal}{Phys. Rev. A}
  \textbf{\bibinfo{volume}{85}}, \bibinfo{pages}{013412}.

\bibitem[{\citenamefont{Hatsagortsyan}
  \emph{et~al.}(2008)\citenamefont{Hatsagortsyan, Klaiber, M\"{u}ller, Kohler,
  and Keitel}}]{Hatsagortsyan_2008}
\bibinfo{author}{\bibnamefont{Hatsagortsyan}, \bibfnamefont{K.~Z.}},
  \bibinfo{author}{\bibfnamefont{M.}~\bibnamefont{Klaiber}},
  \bibinfo{author}{\bibfnamefont{C.}~\bibnamefont{M\"{u}ller}},
  \bibinfo{author}{\bibfnamefont{M.~C.} \bibnamefont{Kohler}}, and
  \bibinfo{author}{\bibfnamefont{C.~H.} \bibnamefont{Keitel}},
  \bibinfo{year}{2008}, \bibinfo{journal}{J. Opt. Soc. Am. B}
  \textbf{\bibinfo{volume}{25}}, \bibinfo{pages}{B92}.

\bibitem[{\citenamefont{Hatsagortsyan}
  \emph{et~al.}(2006)\citenamefont{Hatsagortsyan, M\"uller, and
  Keitel}}]{Hatsagortsyan_2006}
\bibinfo{author}{\bibnamefont{Hatsagortsyan}, \bibfnamefont{K.~Z.}},
  \bibinfo{author}{\bibfnamefont{C.}~\bibnamefont{M\"uller}}, and
  \bibinfo{author}{\bibfnamefont{C.~H.} \bibnamefont{Keitel}},
  \bibinfo{year}{2006}, \bibinfo{journal}{Europhys. Lett.}
  \textbf{\bibinfo{volume}{76}}, \bibinfo{pages}{29}.

\bibitem[{\citenamefont{Hebenstreit}
  \emph{et~al.}(2009)\citenamefont{Hebenstreit, Alkofer, Dunne, and
  Gies}}]{Hebenstreit_2009}
\bibinfo{author}{\bibnamefont{Hebenstreit}, \bibfnamefont{F.}},
  \bibinfo{author}{\bibfnamefont{R.}~\bibnamefont{Alkofer}},
  \bibinfo{author}{\bibfnamefont{G.~V.} \bibnamefont{Dunne}}, and
  \bibinfo{author}{\bibfnamefont{H.}~\bibnamefont{Gies}}, \bibinfo{year}{2009},
  \bibinfo{journal}{Phys. Rev. Lett.} \textbf{\bibinfo{volume}{102}},
  \bibinfo{pages}{150404}.

\bibitem[{\citenamefont{Hebenstreit}
  \emph{et~al.}(2011)\citenamefont{Hebenstreit, Alkofer, and
  Gies}}]{Hebenstreit_2011}
\bibinfo{author}{\bibnamefont{Hebenstreit}, \bibfnamefont{F.}},
  \bibinfo{author}{\bibfnamefont{R.}~\bibnamefont{Alkofer}}, and
  \bibinfo{author}{\bibfnamefont{H.}~\bibnamefont{Gies}}, \bibinfo{year}{2011},
  \bibinfo{journal}{Phys. Rev. Lett.} \textbf{\bibinfo{volume}{107}},
  \bibinfo{pages}{180403}.

\bibitem[{\citenamefont{Hegelich} \emph{et~al.}(2006)\citenamefont{Hegelich,
  Albright, Cobble, Flippo, Letzring, Paffett, Ruhl, Schreiber, Schulze, and
  Fern\'{a}ndez}}]{Hegelich_2006}
\bibinfo{author}{\bibnamefont{Hegelich}, \bibfnamefont{B.~M.}},
  \bibinfo{author}{\bibfnamefont{B.~J.} \bibnamefont{Albright}},
  \bibinfo{author}{\bibfnamefont{J.}~\bibnamefont{Cobble}},
  \bibinfo{author}{\bibfnamefont{K.}~\bibnamefont{Flippo}},
  \bibinfo{author}{\bibfnamefont{S.}~\bibnamefont{Letzring}},
  \bibinfo{author}{\bibfnamefont{M.}~\bibnamefont{Paffett}},
  \bibinfo{author}{\bibfnamefont{H.}~\bibnamefont{Ruhl}},
  \bibinfo{author}{\bibfnamefont{J.}~\bibnamefont{Schreiber}},
  \bibinfo{author}{\bibfnamefont{R.~K.} \bibnamefont{Schulze}}, and
  \bibinfo{author}{\bibfnamefont{J.~C.} \bibnamefont{Fern\'{a}ndez}},
  \bibinfo{year}{2006}, \bibinfo{journal}{Nature (London)}
  \textbf{\bibinfo{volume}{439}}, \bibinfo{pages}{441}.

\bibitem[{\citenamefont{Hein} \emph{et~al.}(2010)\citenamefont{Hein, Hornung,
  B{\"o}defeld, Podleska, S\"{a}vert, Wachs, Kessler, Keppler, Wolf, Polz,
  J\"{a}ckel, Nicolai} \emph{et~al.}}]{Hein_2010}
\bibinfo{author}{\bibnamefont{Hein}, \bibfnamefont{J.}},
  \bibinfo{author}{\bibfnamefont{M.}~\bibnamefont{Hornung}},
  \bibinfo{author}{\bibfnamefont{R.}~\bibnamefont{B{\"o}defeld}},
  \bibinfo{author}{\bibfnamefont{S.}~\bibnamefont{Podleska}},
  \bibinfo{author}{\bibfnamefont{A.}~\bibnamefont{S\"{a}vert}},
  \bibinfo{author}{\bibfnamefont{R.}~\bibnamefont{Wachs}},
  \bibinfo{author}{\bibfnamefont{A.}~\bibnamefont{Kessler}},
  \bibinfo{author}{\bibfnamefont{S.}~\bibnamefont{Keppler}},
  \bibinfo{author}{\bibfnamefont{M.}~\bibnamefont{Wolf}},
  \bibinfo{author}{\bibfnamefont{J.}~\bibnamefont{Polz}},
  \bibinfo{author}{\bibfnamefont{O.}~\bibnamefont{J\"{a}ckel}},
  \bibinfo{author}{\bibfnamefont{M.}~\bibnamefont{Nicolai}}, \emph{et~al.},
  \bibinfo{year}{2010}, \bibinfo{journal}{AIP Conf. Proc.}
  \textbf{\bibinfo{volume}{1228}}, \bibinfo{pages}{159}.

\bibitem[{\citenamefont{Heinzl and Ilderton}(2009)}]{Heinzl_2009}
\bibinfo{author}{\bibnamefont{Heinzl}, \bibfnamefont{T.}}, and
  \bibinfo{author}{\bibfnamefont{A.}~\bibnamefont{Ilderton}},
  \bibinfo{year}{2009}, \bibinfo{journal}{Opt. Commun.}
  \textbf{\bibinfo{volume}{282}}, \bibinfo{pages}{1879}.

\bibitem[{\citenamefont{Heinzl}
  \emph{et~al.}(2010{\natexlab{a}})\citenamefont{Heinzl, Ilderton, and
  Marklund}}]{Heinzl_2010b}
\bibinfo{author}{\bibnamefont{Heinzl}, \bibfnamefont{T.}},
  \bibinfo{author}{\bibfnamefont{A.}~\bibnamefont{Ilderton}}, and
  \bibinfo{author}{\bibfnamefont{M.}~\bibnamefont{Marklund}},
  \bibinfo{year}{2010}{\natexlab{a}}, \bibinfo{journal}{Phys. Lett. B}
  \textbf{\bibinfo{volume}{692}}, \bibinfo{pages}{250}.

\bibitem[{\citenamefont{Heinzl}
  \emph{et~al.}(2010{\natexlab{b}})\citenamefont{Heinzl, Ilderton, and
  Marklund}}]{Heinzl_2010c}
\bibinfo{author}{\bibnamefont{Heinzl}, \bibfnamefont{T.}},
  \bibinfo{author}{\bibfnamefont{A.}~\bibnamefont{Ilderton}}, and
  \bibinfo{author}{\bibfnamefont{M.}~\bibnamefont{Marklund}},
  \bibinfo{year}{2010}{\natexlab{b}}, \bibinfo{journal}{Phys. Rev. D}
  \textbf{\bibinfo{volume}{81}}, \bibinfo{pages}{051902(R)}.

\bibitem[{\citenamefont{Heinzl} \emph{et~al.}(2006)\citenamefont{Heinzl,
  Liesfeld, Amthor, Schwoerer, Sauerbrey, and Wipf}}]{Heinzl_2006}
\bibinfo{author}{\bibnamefont{Heinzl}, \bibfnamefont{T.}},
  \bibinfo{author}{\bibfnamefont{B.}~\bibnamefont{Liesfeld}},
  \bibinfo{author}{\bibfnamefont{K.-U.} \bibnamefont{Amthor}},
  \bibinfo{author}{\bibfnamefont{H.}~\bibnamefont{Schwoerer}},
  \bibinfo{author}{\bibfnamefont{R.}~\bibnamefont{Sauerbrey}}, and
  \bibinfo{author}{\bibfnamefont{A.}~\bibnamefont{Wipf}}, \bibinfo{year}{2006},
  \bibinfo{journal}{Opt. Commun.} \textbf{\bibinfo{volume}{267}},
  \bibinfo{pages}{318}.

\bibitem[{\citenamefont{Heinzl}
  \emph{et~al.}(2010{\natexlab{c}})\citenamefont{Heinzl, Seipt, and
  K\"ampfer}}]{Heinzl_2010}
\bibinfo{author}{\bibnamefont{Heinzl}, \bibfnamefont{T.}},
  \bibinfo{author}{\bibfnamefont{D.}~\bibnamefont{Seipt}}, and
  \bibinfo{author}{\bibfnamefont{B.}~\bibnamefont{K\"ampfer}},
  \bibinfo{year}{2010}{\natexlab{c}}, \bibinfo{journal}{Phys. Rev. A}
  \textbf{\bibinfo{volume}{81}}, \bibinfo{pages}{022125}.

\bibitem[{\citenamefont{Heisenberg and Euler}(1936)}]{Heisenberg_1936}
\bibinfo{author}{\bibnamefont{Heisenberg}, \bibfnamefont{W.}}, and
  \bibinfo{author}{\bibfnamefont{H.}~\bibnamefont{Euler}},
  \bibinfo{year}{1936}, \bibinfo{journal}{Z. Phys.}
  \textbf{\bibinfo{volume}{98}}, \bibinfo{pages}{714}.

\bibitem[{\citenamefont{Heitler}(1984)}]{Heitler_b_1984}
\bibinfo{author}{\bibnamefont{Heitler}, \bibfnamefont{W.}},
  \bibinfo{year}{1984}, \emph{\bibinfo{title}{The Quantum Theory of Radiation}}
  (\bibinfo{publisher}{Dover Publications, New York}).

\bibitem[{\citenamefont{Henneberger}(1968)}]{Henneberger_1968}
\bibinfo{author}{\bibnamefont{Henneberger}, \bibfnamefont{W.~C.}},
  \bibinfo{year}{1968}, \bibinfo{journal}{Phys. Rev. Lett.}
  \textbf{\bibinfo{volume}{21}}, \bibinfo{pages}{838}.

\bibitem[{\citenamefont{Henrich} \emph{et~al.}(2004)\citenamefont{Henrich,
  Hatsagortsyan, and Keitel}}]{Henrich_2004}
\bibinfo{author}{\bibnamefont{Henrich}, \bibfnamefont{B.}},
  \bibinfo{author}{\bibfnamefont{K.~Z.} \bibnamefont{Hatsagortsyan}}, and
  \bibinfo{author}{\bibfnamefont{C.~H.} \bibnamefont{Keitel}},
  \bibinfo{year}{2004}, \bibinfo{journal}{Phys. Rev. Lett.}
  \textbf{\bibinfo{volume}{93}}, \bibinfo{pages}{013601}.

\bibitem[{\citenamefont{Hetzheim and Keitel}(2009)}]{Hetzheim_2009}
\bibinfo{author}{\bibnamefont{Hetzheim}, \bibfnamefont{H.~G.}}, and
  \bibinfo{author}{\bibfnamefont{C.~H.} \bibnamefont{Keitel}},
  \bibinfo{year}{2009}, \bibinfo{journal}{Phys. Rev. Lett.}
  \textbf{\bibinfo{volume}{102}}, \bibinfo{pages}{083003}.

\bibitem[{\citenamefont{{HiPER}}(2011)}]{HiPER}
\bibinfo{author}{\bibnamefont{{HiPER}}} (\bibinfo{collaboration}{High Power
  laser Energy Research}), \bibinfo{year}{2011},
  \urlprefix\url{http://www.hiper-laser.org/}.

\bibitem[{\citenamefont{Homma} \emph{et~al.}(2011)\citenamefont{Homma, Habs,
  and Tajima}}]{Homma_2011}
\bibinfo{author}{\bibnamefont{Homma}, \bibfnamefont{K.}},
  \bibinfo{author}{\bibfnamefont{D.}~\bibnamefont{Habs}}, and
  \bibinfo{author}{\bibfnamefont{T.}~\bibnamefont{Tajima}},
  \bibinfo{year}{2011}, \bibinfo{journal}{Appl. Phys. B}
  \textbf{\bibinfo{volume}{104}}, \bibinfo{pages}{769}.

\bibitem[{\citenamefont{Hu and M\"uller}(2011)}]{Hu_2011}
\bibinfo{author}{\bibnamefont{Hu}, \bibfnamefont{H.}}, and
  \bibinfo{author}{\bibfnamefont{C.}~\bibnamefont{M\"uller}},
  \bibinfo{year}{2011}, \bibinfo{journal}{Phys. Rev. Lett.}
  \textbf{\bibinfo{volume}{107}}, \bibinfo{pages}{090402}.

\bibitem[{\citenamefont{Hu} \emph{et~al.}(2010)\citenamefont{Hu, M\"uller, and
  Keitel}}]{Hu_2010}
\bibinfo{author}{\bibnamefont{Hu}, \bibfnamefont{H.}},
  \bibinfo{author}{\bibfnamefont{C.}~\bibnamefont{M\"uller}}, and
  \bibinfo{author}{\bibfnamefont{C.~H.} \bibnamefont{Keitel}},
  \bibinfo{year}{2010}, \bibinfo{journal}{Phys. Rev. Lett.}
  \textbf{\bibinfo{volume}{105}}, \bibinfo{pages}{080401}.

\bibitem[{\citenamefont{Hu and Keitel}(1999)}]{Hu_1999}
\bibinfo{author}{\bibnamefont{Hu}, \bibfnamefont{S.~X.}}, and
  \bibinfo{author}{\bibfnamefont{C.~H.} \bibnamefont{Keitel}},
  \bibinfo{year}{1999}, \bibinfo{journal}{Phys. Rev. Lett.}
  \textbf{\bibinfo{volume}{83}}, \bibinfo{pages}{4709}.

\bibitem[{\citenamefont{Hu and Keitel}(2001)}]{Hu_2001}
\bibinfo{author}{\bibnamefont{Hu}, \bibfnamefont{S.~X.}}, and
  \bibinfo{author}{\bibfnamefont{C.~H.} \bibnamefont{Keitel}},
  \bibinfo{year}{2001}, \bibinfo{journal}{Phys. Rev. A}
  \textbf{\bibinfo{volume}{63}}, \bibinfo{pages}{053402}.

\bibitem[{\citenamefont{Hugenschmidt}
  \emph{et~al.}(2012)\citenamefont{Hugenschmidt, Schreckenbach, Habs, and
  Thirolf}}]{Hugenschmidt_2012}
\bibinfo{author}{\bibnamefont{Hugenschmidt}, \bibfnamefont{C.}},
  \bibinfo{author}{\bibfnamefont{K.}~\bibnamefont{Schreckenbach}},
  \bibinfo{author}{\bibfnamefont{D.}~\bibnamefont{Habs}}, and
  \bibinfo{author}{\bibfnamefont{P.}~\bibnamefont{Thirolf}},
  \bibinfo{year}{2012}, \bibinfo{journal}{Appl. Phys. B}
  \textbf{\bibinfo{volume}{106}}, \bibinfo{pages}{241}.

\bibitem[{\citenamefont{{ILC}}(2011)}]{ILC}
\bibinfo{author}{\bibnamefont{{ILC}}} (\bibinfo{collaboration}{International
  Linear Collider}), \bibinfo{year}{2011},
  \urlprefix\url{http://www.linearcollider.org/}.

\bibitem[{\citenamefont{Ilderton}(2011)}]{Ilderton_2011}
\bibinfo{author}{\bibnamefont{Ilderton}, \bibfnamefont{A.}},
  \bibinfo{year}{2011}, \bibinfo{journal}{Phys. Rev. Lett.}
  \textbf{\bibinfo{volume}{106}}, \bibinfo{pages}{020404}.

\bibitem[{\citenamefont{Ipp} \emph{et~al.}(2011)\citenamefont{Ipp, Evers,
  Keitel, and Hatsagortsyan}}]{Ipp_2011}
\bibinfo{author}{\bibnamefont{Ipp}, \bibfnamefont{A.}},
  \bibinfo{author}{\bibfnamefont{J.}~\bibnamefont{Evers}},
  \bibinfo{author}{\bibfnamefont{C.~H.} \bibnamefont{Keitel}}, and
  \bibinfo{author}{\bibfnamefont{K.~Z.} \bibnamefont{Hatsagortsyan}},
  \bibinfo{year}{2011}, \bibinfo{journal}{Phys. Lett. B}
  \textbf{\bibinfo{volume}{702}}, \bibinfo{pages}{383}.

\bibitem[{\citenamefont{Ivanov} \emph{et~al.}(2005)\citenamefont{Ivanov,
  Kotkin, and Serbo}}]{Serbo_2005}
\bibinfo{author}{\bibnamefont{Ivanov}, \bibfnamefont{D.~Y.}},
  \bibinfo{author}{\bibfnamefont{G.~L.} \bibnamefont{Kotkin}}, and
  \bibinfo{author}{\bibfnamefont{V.~G.} \bibnamefont{Serbo}},
  \bibinfo{year}{2005}, \bibinfo{journal}{Eur. Phys. J. C}
  \textbf{\bibinfo{volume}{40}}, \bibinfo{pages}{27}.

\bibitem[{\citenamefont{Ivanov} \emph{et~al.}(2004)\citenamefont{Ivanov,
  Kotkin, and Serbo}}]{Ivanov_2004}
\bibinfo{author}{\bibnamefont{Ivanov}, \bibfnamefont{D.~{\relax Yu}.}},
  \bibinfo{author}{\bibfnamefont{G.~L.} \bibnamefont{Kotkin}}, and
  \bibinfo{author}{\bibfnamefont{V.~G.} \bibnamefont{Serbo}},
  \bibinfo{year}{2004}, \bibinfo{journal}{Eur. Phys. J. C}
  \textbf{\bibinfo{volume}{36}}, \bibinfo{pages}{127}.

\bibitem[{\citenamefont{Jackson}(1975)}]{Jackson_b_1975}
\bibinfo{author}{\bibnamefont{Jackson}, \bibfnamefont{J.~D.}},
  \bibinfo{year}{1975}, \emph{\bibinfo{title}{Classical Electrodynamics}}
  (\bibinfo{publisher}{Wiley, New York}).

\bibitem[{\citenamefont{Jauch and Rohrlich}(1976)}]{Jauch_b_1976}
\bibinfo{author}{\bibnamefont{Jauch}, \bibfnamefont{J.~M.}}, and
  \bibinfo{author}{\bibfnamefont{F.}~\bibnamefont{Rohrlich}},
  \bibinfo{year}{1976}, \emph{\bibinfo{title}{The Theory of Photons and
  Electrons}} (\bibinfo{publisher}{Springer, Berlin}).

\bibitem[{\citenamefont{Kami{\'n}ski}
  \emph{et~al.}(2006)\citenamefont{Kami{\'n}ski, Krajewska, and
  Ehlotzky}}]{Kaminski_2006}
\bibinfo{author}{\bibnamefont{Kami{\'n}ski}, \bibfnamefont{J.~Z.}},
  \bibinfo{author}{\bibfnamefont{K.}~\bibnamefont{Krajewska}}, and
  \bibinfo{author}{\bibfnamefont{F.}~\bibnamefont{Ehlotzky}},
  \bibinfo{year}{2006}, \bibinfo{journal}{Phys. Rev. A}
  \textbf{\bibinfo{volume}{74}}, \bibinfo{pages}{033402}.

\bibitem[{\citenamefont{Kaneyasu} \emph{et~al.}(2011)\citenamefont{Kaneyasu,
  Takabayashi, Iwasaki, and Koda}}]{Kaneyasu_2011}
\bibinfo{author}{\bibnamefont{Kaneyasu}, \bibfnamefont{T.}},
  \bibinfo{author}{\bibfnamefont{Y.}~\bibnamefont{Takabayashi}},
  \bibinfo{author}{\bibfnamefont{Y.}~\bibnamefont{Iwasaki}}, and
  \bibinfo{author}{\bibfnamefont{S.}~\bibnamefont{Koda}}, \bibinfo{year}{2011},
  \bibinfo{journal}{Nucl. Instr. Meth. Phys. Res. A}
  \textbf{\bibinfo{volume}{659}}, \bibinfo{pages}{30}.

\bibitem[{\citenamefont{Karagodsky}
  \emph{et~al.}(2010)\citenamefont{Karagodsky, Schieber, and
  Sch\"achter}}]{Karagodsky_2010}
\bibinfo{author}{\bibnamefont{Karagodsky}, \bibfnamefont{V.}},
  \bibinfo{author}{\bibfnamefont{D.}~\bibnamefont{Schieber}}, and
  \bibinfo{author}{\bibfnamefont{L.}~\bibnamefont{Sch\"achter}},
  \bibinfo{year}{2010}, \bibinfo{journal}{Phys. Rev. Lett.}
  \textbf{\bibinfo{volume}{104}}, \bibinfo{pages}{024801}.

\bibitem[{\citenamefont{Karplus and Neuman}(1950)}]{Karplus_1951}
\bibinfo{author}{\bibnamefont{Karplus}, \bibfnamefont{R.}}, and
  \bibinfo{author}{\bibfnamefont{M.}~\bibnamefont{Neuman}},
  \bibinfo{year}{1950}, \bibinfo{journal}{Phys. Rev.}
  \textbf{\bibinfo{volume}{80}}, \bibinfo{pages}{380}.

\bibitem[{\citenamefont{Kazinski and Shipulya}(2011)}]{Kazinski_2011}
\bibinfo{author}{\bibnamefont{Kazinski}, \bibfnamefont{P.~O.}}, and
  \bibinfo{author}{\bibfnamefont{M.~A.} \bibnamefont{Shipulya}},
  \bibinfo{year}{2011}, \bibinfo{journal}{Phys. Rev. E}
  \textbf{\bibinfo{volume}{83}}, \bibinfo{pages}{066606}.

\bibitem[{\citenamefont{Keitel}(2001)}]{Keitel_2001}
\bibinfo{author}{\bibnamefont{Keitel}, \bibfnamefont{C.~H.}},
  \bibinfo{year}{2001}, \bibinfo{journal}{Contemp. Phys.}
  \textbf{\bibinfo{volume}{42}}, \bibinfo{pages}{353}.

\bibitem[{\citenamefont{Keitel and Hu}(2002)}]{Hu_2002}
\bibinfo{author}{\bibnamefont{Keitel}, \bibfnamefont{C.~H.}}, and
  \bibinfo{author}{\bibfnamefont{S.~X.} \bibnamefont{Hu}},
  \bibinfo{year}{2002}, \bibinfo{journal}{Appl. Phys. Lett.}
  \textbf{\bibinfo{volume}{80}}, \bibinfo{pages}{541}.

\bibitem[{\citenamefont{Keitel} \emph{et~al.}(1993)\citenamefont{Keitel,
  Knight, and Burnett}}]{Keitel_1993}
\bibinfo{author}{\bibnamefont{Keitel}, \bibfnamefont{C.~H.}},
  \bibinfo{author}{\bibfnamefont{P.~L.} \bibnamefont{Knight}}, and
  \bibinfo{author}{\bibfnamefont{K.}~\bibnamefont{Burnett}},
  \bibinfo{year}{1993}, \bibinfo{journal}{Europhys. Lett.}
  \textbf{\bibinfo{volume}{24}}, \bibinfo{pages}{539}.

\bibitem[{\citenamefont{Keitel} \emph{et~al.}(1998)\citenamefont{Keitel,
  Szymanowski, Knight, and Maquet}}]{Keitel_1998}
\bibinfo{author}{\bibnamefont{Keitel}, \bibfnamefont{C.~H.}},
  \bibinfo{author}{\bibfnamefont{C.}~\bibnamefont{Szymanowski}},
  \bibinfo{author}{\bibfnamefont{P.~L.} \bibnamefont{Knight}}, and
  \bibinfo{author}{\bibfnamefont{A.}~\bibnamefont{Maquet}},
  \bibinfo{year}{1998}, \bibinfo{journal}{J. Phys. B}
  \textbf{\bibinfo{volume}{31}}, \bibinfo{pages}{L75}.

\bibitem[{\citenamefont{Keldysh}(1965)}]{Keldysh_1965}
\bibinfo{author}{\bibnamefont{Keldysh}, \bibfnamefont{L.~V.}},
  \bibinfo{year}{1965}, \bibinfo{journal}{Sov. Phys. JETP}
  \textbf{\bibinfo{volume}{20}}, \bibinfo{pages}{1307}.

\bibitem[{\citenamefont{Kiffner} \emph{et~al.}(2010)\citenamefont{Kiffner,
  Macovei, Evers, and Keitel}}]{Kiffner_2010}
\bibinfo{author}{\bibnamefont{Kiffner}, \bibfnamefont{M.}},
  \bibinfo{author}{\bibfnamefont{M.}~\bibnamefont{Macovei}},
  \bibinfo{author}{\bibfnamefont{J.}~\bibnamefont{Evers}}, and
  \bibinfo{author}{\bibfnamefont{C.~H.} \bibnamefont{Keitel}},
  \bibinfo{year}{2010}, \bibinfo{journal}{Prog. Opt.}
  \textbf{\bibinfo{volume}{55}}, \bibinfo{pages}{85}.

\bibitem[{\citenamefont{Kim} \emph{et~al.}(2009)\citenamefont{Kim, Lee, Chung,
  and Lee}}]{Kim_2009}
\bibinfo{author}{\bibnamefont{Kim}, \bibfnamefont{D.}},
  \bibinfo{author}{\bibfnamefont{H.}~\bibnamefont{Lee}},
  \bibinfo{author}{\bibfnamefont{S.}~\bibnamefont{Chung}}, and
  \bibinfo{author}{\bibfnamefont{K.}~\bibnamefont{Lee}}, \bibinfo{year}{2009},
  \bibinfo{journal}{New J. Phys.} \textbf{\bibinfo{volume}{11}},
  \bibinfo{pages}{063050}.

\bibitem[{\citenamefont{Kim and Carosi}(2010)}]{Kim_2010}
\bibinfo{author}{\bibnamefont{Kim}, \bibfnamefont{J.~E.}}, and
  \bibinfo{author}{\bibfnamefont{G.}~\bibnamefont{Carosi}},
  \bibinfo{year}{2010}, \bibinfo{journal}{Rev. Mod. Phys.}
  \textbf{\bibinfo{volume}{82}}, \bibinfo{pages}{557}.

\bibitem[{\citenamefont{Kim} \emph{et~al.}(2008)\citenamefont{Kim, Shvyd'ko,
  and Reiche}}]{Kim_2008}
\bibinfo{author}{\bibnamefont{Kim}, \bibfnamefont{K.-J.}},
  \bibinfo{author}{\bibfnamefont{{\relax Yu}.~V.} \bibnamefont{Shvyd'ko}}, and
  \bibinfo{author}{\bibfnamefont{S.}~\bibnamefont{Reiche}},
  \bibinfo{year}{2008}, \bibinfo{journal}{Phys. Rev. Lett.}
  \textbf{\bibinfo{volume}{100}}, \bibinfo{pages}{244802}.

\bibitem[{\citenamefont{Kim and Schubert}(2011)}]{Kim_2011}
\bibinfo{author}{\bibnamefont{Kim}, \bibfnamefont{S.~P.}}, and
  \bibinfo{author}{\bibfnamefont{C.}~\bibnamefont{Schubert}},
  \bibinfo{year}{2011}, \bibinfo{journal}{Phys. Rev. D}
  \textbf{\bibinfo{volume}{84}}, \bibinfo{pages}{125028}.

\bibitem[{\citenamefont{King}
  \emph{et~al.}(2010{\natexlab{a}})\citenamefont{King, Di~Piazza, and
  Keitel}}]{King_2010}
\bibinfo{author}{\bibnamefont{King}, \bibfnamefont{B.}},
  \bibinfo{author}{\bibfnamefont{A.}~\bibnamefont{Di~Piazza}}, and
  \bibinfo{author}{\bibfnamefont{C.~H.} \bibnamefont{Keitel}},
  \bibinfo{year}{2010}{\natexlab{a}}, \bibinfo{journal}{Nature Photon.}
  \textbf{\bibinfo{volume}{4}}, \bibinfo{pages}{92}.

\bibitem[{\citenamefont{King}
  \emph{et~al.}(2010{\natexlab{b}})\citenamefont{King, Di~Piazza, and
  Keitel}}]{King_2010_a}
\bibinfo{author}{\bibnamefont{King}, \bibfnamefont{B.}},
  \bibinfo{author}{\bibfnamefont{A.}~\bibnamefont{Di~Piazza}}, and
  \bibinfo{author}{\bibfnamefont{C.~H.} \bibnamefont{Keitel}},
  \bibinfo{year}{2010}{\natexlab{b}}, \bibinfo{journal}{Phys. Rev. A}
  \textbf{\bibinfo{volume}{82}}, \bibinfo{pages}{032114}.

\bibitem[{\citenamefont{King and Keitel}(2012)}]{King_2012}
\bibinfo{author}{\bibnamefont{King}, \bibfnamefont{B.}}, and
  \bibinfo{author}{\bibfnamefont{C.~H.} \bibnamefont{Keitel}},
  \bibinfo{year}{2012}, \eprint{arXiv:1202.3339v1}.

\bibitem[{\citenamefont{Kirk} \emph{et~al.}(2009)\citenamefont{Kirk, Bell, and
  Arka}}]{Kirk_2009}
\bibinfo{author}{\bibnamefont{Kirk}, \bibfnamefont{J.~G.}},
  \bibinfo{author}{\bibfnamefont{A.~R.} \bibnamefont{Bell}}, and
  \bibinfo{author}{\bibfnamefont{I.}~\bibnamefont{Arka}}, \bibinfo{year}{2009},
  \bibinfo{journal}{Plasma Phys. Control. Fusion}
  \textbf{\bibinfo{volume}{51}}, \bibinfo{pages}{085008}.

\bibitem[{\citenamefont{Kirsebom} \emph{et~al.}(2001)\citenamefont{Kirsebom,
  Mikkelsen, Uggerh\o{}j, Elsener, Ballestrero, Sona, and
  Vilakazi}}]{Uggerhoj_2001}
\bibinfo{author}{\bibnamefont{Kirsebom}, \bibfnamefont{K.}},
  \bibinfo{author}{\bibfnamefont{U.}~\bibnamefont{Mikkelsen}},
  \bibinfo{author}{\bibfnamefont{E.}~\bibnamefont{Uggerh\o{}j}},
  \bibinfo{author}{\bibfnamefont{K.}~\bibnamefont{Elsener}},
  \bibinfo{author}{\bibfnamefont{S.}~\bibnamefont{Ballestrero}},
  \bibinfo{author}{\bibfnamefont{P.}~\bibnamefont{Sona}}, and
  \bibinfo{author}{\bibfnamefont{Z.~Z.} \bibnamefont{Vilakazi}},
  \bibinfo{year}{2001}, \bibinfo{journal}{Phys. Rev. Lett.}
  \textbf{\bibinfo{volume}{87}}, \bibinfo{pages}{054801}.

\bibitem[{\citenamefont{Klaiber} \emph{et~al.}(2006)\citenamefont{Klaiber,
  Hatsagortsyan, and Keitel}}]{Klaiber_2006}
\bibinfo{author}{\bibnamefont{Klaiber}, \bibfnamefont{M.}},
  \bibinfo{author}{\bibfnamefont{K.~Z.} \bibnamefont{Hatsagortsyan}}, and
  \bibinfo{author}{\bibfnamefont{C.~H.} \bibnamefont{Keitel}},
  \bibinfo{year}{2006}, \bibinfo{journal}{Phys. Rev. A}
  \textbf{\bibinfo{volume}{74}}, \bibinfo{pages}{051803}.

\bibitem[{\citenamefont{Klaiber} \emph{et~al.}(2007)\citenamefont{Klaiber,
  Hatsagortsyan, and Keitel}}]{Klaiber_2007}
\bibinfo{author}{\bibnamefont{Klaiber}, \bibfnamefont{M.}},
  \bibinfo{author}{\bibfnamefont{K.~Z.} \bibnamefont{Hatsagortsyan}}, and
  \bibinfo{author}{\bibfnamefont{C.~H.} \bibnamefont{Keitel}},
  \bibinfo{year}{2007}, \bibinfo{journal}{Phys. Rev. A}
  \textbf{\bibinfo{volume}{75}}, \bibinfo{pages}{063413}.

\bibitem[{\citenamefont{Klaiber} \emph{et~al.}(2008)\citenamefont{Klaiber,
  Hatsagortsyan, M\"{u}ller, and Keitel}}]{Klaiber_2008}
\bibinfo{author}{\bibnamefont{Klaiber}, \bibfnamefont{M.}},
  \bibinfo{author}{\bibfnamefont{K.~Z.} \bibnamefont{Hatsagortsyan}},
  \bibinfo{author}{\bibfnamefont{C.}~\bibnamefont{M\"{u}ller}}, and
  \bibinfo{author}{\bibfnamefont{C.~H.} \bibnamefont{Keitel}},
  \bibinfo{year}{2008}, \bibinfo{journal}{Opt. Lett.}
  \textbf{\bibinfo{volume}{33}}, \bibinfo{pages}{411}.

\bibitem[{\citenamefont{Klaiber} \emph{et~al.}(2012)\citenamefont{Klaiber,
  Yakaboylu, Bauke, Hatsagortsyan, and Keitel}}]{Klaiber_2012}
\bibinfo{author}{\bibnamefont{Klaiber}, \bibfnamefont{M.}},
  \bibinfo{author}{\bibfnamefont{E.}~\bibnamefont{Yakaboylu}},
  \bibinfo{author}{\bibfnamefont{H.}~\bibnamefont{Bauke}},
  \bibinfo{author}{\bibfnamefont{K.~Z.} \bibnamefont{Hatsagortsyan}}, and
  \bibinfo{author}{\bibfnamefont{C.~H.} \bibnamefont{Keitel}},
  \bibinfo{year}{2012}, \bibinfo{pages}{to be published}.

\bibitem[{\citenamefont{Kneip} \emph{et~al.}(2010)\citenamefont{Kneip,
  McGuffey, Martins, Martins, Bellei, Chvykov, Dollar, Fonseca, Huntington,
  Kalintchenko, Maksimchuk, Mangles} \emph{et~al.}}]{Kneip_2010}
\bibinfo{author}{\bibnamefont{Kneip}, \bibfnamefont{S.}},
  \bibinfo{author}{\bibfnamefont{C.}~\bibnamefont{McGuffey}},
  \bibinfo{author}{\bibfnamefont{J.~L.} \bibnamefont{Martins}},
  \bibinfo{author}{\bibfnamefont{S.~F.} \bibnamefont{Martins}},
  \bibinfo{author}{\bibfnamefont{C.}~\bibnamefont{Bellei}},
  \bibinfo{author}{\bibfnamefont{V.}~\bibnamefont{Chvykov}},
  \bibinfo{author}{\bibfnamefont{F.}~\bibnamefont{Dollar}},
  \bibinfo{author}{\bibfnamefont{R.}~\bibnamefont{Fonseca}},
  \bibinfo{author}{\bibfnamefont{C.}~\bibnamefont{Huntington}},
  \bibinfo{author}{\bibfnamefont{G.}~\bibnamefont{Kalintchenko}},
  \bibinfo{author}{\bibfnamefont{A.}~\bibnamefont{Maksimchuk}},
  \bibinfo{author}{\bibfnamefont{S.~P.~D.} \bibnamefont{Mangles}},
  \emph{et~al.}, \bibinfo{year}{2010}, \bibinfo{journal}{Nature Phys.}
  \textbf{\bibinfo{volume}{6}}, \bibinfo{pages}{980}.

\bibitem[{\citenamefont{Koga} \emph{et~al.}(2005)\citenamefont{Koga, {T. Zh.
  Esirkepov}, and Bulanov}}]{Koga_2005}
\bibinfo{author}{\bibnamefont{Koga}, \bibfnamefont{J.}},
  \bibinfo{author}{\bibnamefont{{T. Zh. Esirkepov}}}, and
  \bibinfo{author}{\bibfnamefont{S.~V.} \bibnamefont{Bulanov}},
  \bibinfo{year}{2005}, \bibinfo{journal}{Phys. Plasmas}
  \textbf{\bibinfo{volume}{12}}, \bibinfo{pages}{093106}.

\bibitem[{\citenamefont{Kohler} \emph{et~al.}(2011)\citenamefont{Kohler,
  Klaiber, Hatsagortsyan, and Keitel}}]{Kohler_2011}
\bibinfo{author}{\bibnamefont{Kohler}, \bibfnamefont{M.~C.}},
  \bibinfo{author}{\bibfnamefont{M.}~\bibnamefont{Klaiber}},
  \bibinfo{author}{\bibfnamefont{K.~Z.} \bibnamefont{Hatsagortsyan}}, and
  \bibinfo{author}{\bibfnamefont{C.~H.} \bibnamefont{Keitel}},
  \bibinfo{year}{2011}, \bibinfo{journal}{Europhys. Lett.}
  \textbf{\bibinfo{volume}{94}}, \bibinfo{pages}{14002}.

\bibitem[{\citenamefont{Kohler} \emph{et~al.}(2012)\citenamefont{Kohler,
  Pfeifer, Hatsagortsyan, and Keitel}}]{Kohler_2012}
\bibinfo{author}{\bibnamefont{Kohler}, \bibfnamefont{M.~C.}},
  \bibinfo{author}{\bibfnamefont{T.}~\bibnamefont{Pfeifer}},
  \bibinfo{author}{\bibfnamefont{K.~Z.} \bibnamefont{Hatsagortsyan}}, and
  \bibinfo{author}{\bibfnamefont{C.~H.} \bibnamefont{Keitel}},
  \bibinfo{year}{2012}, \bibinfo{journal}{Adv. At. Mol. Opt. Phys.} ,
  \bibinfo{pages}{in press}.

\bibitem[{\citenamefont{Kornev} \emph{et~al.}(2009)\citenamefont{Kornev,
  Tulenko, and Zon}}]{Zon_2009}
\bibinfo{author}{\bibnamefont{Kornev}, \bibfnamefont{A.~S.}},
  \bibinfo{author}{\bibfnamefont{E.~B.} \bibnamefont{Tulenko}}, and
  \bibinfo{author}{\bibfnamefont{B.~A.} \bibnamefont{Zon}},
  \bibinfo{year}{2009}, \bibinfo{journal}{Phys. Rev. A}
  \textbf{\bibinfo{volume}{79}}, \bibinfo{pages}{063405}.

\bibitem[{\citenamefont{Kornev and Zon}(2007)}]{Kornev_2007}
\bibinfo{author}{\bibnamefont{Kornev}, \bibfnamefont{A.~S.}}, and
  \bibinfo{author}{\bibfnamefont{B.~A.} \bibnamefont{Zon}},
  \bibinfo{year}{2007}, \bibinfo{journal}{Laser Phys. Lett.}
  \textbf{\bibinfo{volume}{4}}, \bibinfo{pages}{588}.

\bibitem[{\citenamefont{Korzhimanov}
  \emph{et~al.}(2011)\citenamefont{Korzhimanov, Gonoskov, Khazanov, and
  Sergeev}}]{Korzhimanov_2011}
\bibinfo{author}{\bibnamefont{Korzhimanov}, \bibfnamefont{A.~V.}},
  \bibinfo{author}{\bibfnamefont{A.~A.} \bibnamefont{Gonoskov}},
  \bibinfo{author}{\bibfnamefont{E.~A.} \bibnamefont{Khazanov}}, and
  \bibinfo{author}{\bibfnamefont{A.~M.} \bibnamefont{Sergeev}},
  \bibinfo{year}{2011}, \bibinfo{journal}{Phys. Usp.}
  \textbf{\bibinfo{volume}{54}}, \bibinfo{pages}{9}.

\bibitem[{\citenamefont{Krajewska and Kami{\'n}ski}(2008)}]{Krajewska_2008}
\bibinfo{author}{\bibnamefont{Krajewska}, \bibfnamefont{K.}}, and
  \bibinfo{author}{\bibfnamefont{J.~Z.} \bibnamefont{Kami{\'n}ski}},
  \bibinfo{year}{2008}, \bibinfo{journal}{Laser Phys.}
  \textbf{\bibinfo{volume}{18}}, \bibinfo{pages}{185}.

\bibitem[{\citenamefont{Krajewska and Kami{\'n}ski}(2010)}]{Krajewska_2010}
\bibinfo{author}{\bibnamefont{Krajewska}, \bibfnamefont{K.}}, and
  \bibinfo{author}{\bibfnamefont{J.~Z.} \bibnamefont{Kami{\'n}ski}},
  \bibinfo{year}{2010}, \bibinfo{journal}{Phys. Rev. A}
  \textbf{\bibinfo{volume}{82}}, \bibinfo{pages}{013420}.

\bibitem[{\citenamefont{Krajewska and Kami{\'n}ski}(2011)}]{Krajewska_2011}
\bibinfo{author}{\bibnamefont{Krajewska}, \bibfnamefont{K.}}, and
  \bibinfo{author}{\bibfnamefont{J.~Z.} \bibnamefont{Kami{\'n}ski}},
  \bibinfo{year}{2011}, \bibinfo{journal}{Phys. Rev. A}
  \textbf{\bibinfo{volume}{84}}, \bibinfo{pages}{033416}.

\bibitem[{\citenamefont{Krajewska} \emph{et~al.}(2006)\citenamefont{Krajewska,
  Kami{\'n}ski, and Ehlotzky}}]{Krajewska_2006}
\bibinfo{author}{\bibnamefont{Krajewska}, \bibfnamefont{K.}},
  \bibinfo{author}{\bibfnamefont{J.~Z.} \bibnamefont{Kami{\'n}ski}}, and
  \bibinfo{author}{\bibfnamefont{F.}~\bibnamefont{Ehlotzky}},
  \bibinfo{year}{2006}, \bibinfo{journal}{Laser Phys.}
  \textbf{\bibinfo{volume}{16}}, \bibinfo{pages}{272}.

\bibitem[{\citenamefont{Kramers}(1956)}]{Kramers_1956}
\bibinfo{author}{\bibnamefont{Kramers}, \bibfnamefont{H.~A.}},
  \bibinfo{year}{1956}, \emph{\bibinfo{title}{Collected Scientific Papers}}
  (\bibinfo{publisher}{North-Holland, Amsterdam}).

\bibitem[{\citenamefont{Krausz and Ivanov}(2009)}]{Krausz_2009}
\bibinfo{author}{\bibnamefont{Krausz}, \bibfnamefont{F.}}, and
  \bibinfo{author}{\bibfnamefont{M.~{\relax Yu}.} \bibnamefont{Ivanov}},
  \bibinfo{year}{2009}, \bibinfo{journal}{Rev. Mod. Phys.}
  \textbf{\bibinfo{volume}{81}}, \bibinfo{pages}{163}.

\bibitem[{\citenamefont{Krekora} \emph{et~al.}(2004)\citenamefont{Krekora, Su,
  and Grobe}}]{Grobe_Spin2004}
\bibinfo{author}{\bibnamefont{Krekora}, \bibfnamefont{P.}},
  \bibinfo{author}{\bibfnamefont{Q.}~\bibnamefont{Su}}, and
  \bibinfo{author}{\bibfnamefont{R.}~\bibnamefont{Grobe}},
  \bibinfo{year}{2004}, \bibinfo{journal}{Phys. Rev. Lett.}
  \textbf{\bibinfo{volume}{92}}, \bibinfo{pages}{040406}.

\bibitem[{\citenamefont{Krekora} \emph{et~al.}(2005)\citenamefont{Krekora, Su,
  and Grobe}}]{Krekora_2005}
\bibinfo{author}{\bibnamefont{Krekora}, \bibfnamefont{P.}},
  \bibinfo{author}{\bibfnamefont{Q.}~\bibnamefont{Su}}, and
  \bibinfo{author}{\bibfnamefont{R.}~\bibnamefont{Grobe}},
  \bibinfo{year}{2005}, \bibinfo{journal}{J. Mod. Opt.}
  \textbf{\bibinfo{volume}{52}}, \bibinfo{pages}{489}.

\bibitem[{\citenamefont{Krivitskii and Tsytovich}(1991)}]{Krivitskii_1991}
\bibinfo{author}{\bibnamefont{Krivitskii}, \bibfnamefont{V.~S.}}, and
  \bibinfo{author}{\bibfnamefont{V.~N.} \bibnamefont{Tsytovich}},
  \bibinfo{year}{1991}, \bibinfo{journal}{Sov. Phys. Usp.}
  \textbf{\bibinfo{volume}{34}}, \bibinfo{pages}{250}.

\bibitem[{\citenamefont{Kryuchkyan and Hatsagortsyan}(2011)}]{Kryuchkyan_2011}
\bibinfo{author}{\bibnamefont{Kryuchkyan}, \bibfnamefont{G.~{\relax Yu}.}}, and
  \bibinfo{author}{\bibfnamefont{K.~Z.} \bibnamefont{Hatsagortsyan}},
  \bibinfo{year}{2011}, \bibinfo{journal}{Phys. Rev. Lett.}
  \textbf{\bibinfo{volume}{107}}, \bibinfo{pages}{053604}.

\bibitem[{\citenamefont{Kuchiev}(1987)}]{Kuchiev_1987}
\bibinfo{author}{\bibnamefont{Kuchiev}, \bibfnamefont{M.~{\relax Yu}.}},
  \bibinfo{year}{1987}, \bibinfo{journal}{JETP Lett.}
  \textbf{\bibinfo{volume}{45}}, \bibinfo{pages}{404}.

\bibitem[{\citenamefont{Kuchiev}(2007)}]{Kuchiev_2007}
\bibinfo{author}{\bibnamefont{Kuchiev}, \bibfnamefont{M.~{\relax Yu}.}},
  \bibinfo{year}{2007}, \bibinfo{journal}{Phys. Rev. Lett.}
  \textbf{\bibinfo{volume}{99}}, \bibinfo{pages}{130404}.

\bibitem[{\citenamefont{Kuchiev and Robinson}(2007)}]{KuchievPRA}
\bibinfo{author}{\bibnamefont{Kuchiev}, \bibfnamefont{M.~{\relax Yu}.}}, and
  \bibinfo{author}{\bibfnamefont{D.~J.} \bibnamefont{Robinson}},
  \bibinfo{year}{2007}, \bibinfo{journal}{Phys. Rev. A}
  \textbf{\bibinfo{volume}{76}}, \bibinfo{pages}{012107}.

\bibitem[{\citenamefont{Kurilin}(1999)}]{Kurilin_1999}
\bibinfo{author}{\bibnamefont{Kurilin}, \bibfnamefont{A.~V.}},
  \bibinfo{year}{1999}, \bibinfo{journal}{Nuovo Cim.}
  \textbf{\bibinfo{volume}{112}}, \bibinfo{pages}{977}.

\bibitem[{\citenamefont{Kurilin}(2004)}]{Kurilin_2004}
\bibinfo{author}{\bibnamefont{Kurilin}, \bibfnamefont{A.~V.}},
  \bibinfo{year}{2004}, \bibinfo{journal}{Phys. Atom. Nucl.}
  \textbf{\bibinfo{volume}{67}}, \bibinfo{pages}{2095}.

\bibitem[{\citenamefont{Kurilin}(2009)}]{Kurilin_2009}
\bibinfo{author}{\bibnamefont{Kurilin}, \bibfnamefont{A.~V.}},
  \bibinfo{year}{2009}, \bibinfo{journal}{Phys. Atom. Nucl.}
  \textbf{\bibinfo{volume}{72}}, \bibinfo{pages}{1034}.

\bibitem[{\citenamefont{Kuznetsova}
  \emph{et~al.}(2008)\citenamefont{Kuznetsova, Habs, and
  Rafelski}}]{Kuznetsova_2008}
\bibinfo{author}{\bibnamefont{Kuznetsova}, \bibfnamefont{I.}},
  \bibinfo{author}{\bibfnamefont{D.}~\bibnamefont{Habs}}, and
  \bibinfo{author}{\bibfnamefont{J.}~\bibnamefont{Rafelski}},
  \bibinfo{year}{2008}, \bibinfo{journal}{Phys. Rev. D}
  \textbf{\bibinfo{volume}{78}}, \bibinfo{pages}{014027}.

\bibitem[{\citenamefont{Kuznetsova and Rafelski}(2012)}]{Kuznetsova_2012}
\bibinfo{author}{\bibnamefont{Kuznetsova}, \bibfnamefont{I.}}, and
  \bibinfo{author}{\bibfnamefont{J.}~\bibnamefont{Rafelski}},
  \bibinfo{year}{2012}, \eprint{arXiv:1109.3546v2}.

\bibitem[{\citenamefont{Kylstra} \emph{et~al.}(2000)\citenamefont{Kylstra,
  Worthington, Patel, Knight, V\'azquez~de Aldana, and Roso}}]{Kylstra_2000}
\bibinfo{author}{\bibnamefont{Kylstra}, \bibfnamefont{N.~J.}},
  \bibinfo{author}{\bibfnamefont{R.~A.} \bibnamefont{Worthington}},
  \bibinfo{author}{\bibfnamefont{A.}~\bibnamefont{Patel}},
  \bibinfo{author}{\bibfnamefont{P.~L.} \bibnamefont{Knight}},
  \bibinfo{author}{\bibfnamefont{J.~R.} \bibnamefont{V\'azquez~de Aldana}}, and
  \bibinfo{author}{\bibfnamefont{L.}~\bibnamefont{Roso}}, \bibinfo{year}{2000},
  \bibinfo{journal}{Phys. Rev. Lett.} \textbf{\bibinfo{volume}{85}},
  \bibinfo{pages}{1835}.

\bibitem[{\citenamefont{Labun and Rafelski}(2009)}]{Labun_2009}
\bibinfo{author}{\bibnamefont{Labun}, \bibfnamefont{L.}}, and
  \bibinfo{author}{\bibfnamefont{J.}~\bibnamefont{Rafelski}},
  \bibinfo{year}{2009}, \bibinfo{journal}{Phys. Rev. D}
  \textbf{\bibinfo{volume}{79}}, \bibinfo{pages}{057901}.

\bibitem[{\citenamefont{Landau and Lifshitz}(1975)}]{Landau_b_2_1975}
\bibinfo{author}{\bibnamefont{Landau}, \bibfnamefont{L.~D.}}, and
  \bibinfo{author}{\bibfnamefont{E.~M.} \bibnamefont{Lifshitz}},
  \bibinfo{year}{1975}, \emph{\bibinfo{title}{The Classical Theory of Fields}}
  (\bibinfo{publisher}{Elsevier, Oxford}).

\bibitem[{\citenamefont{Lawson}(1979)}]{Lawson_1979}
\bibinfo{author}{\bibnamefont{Lawson}, \bibfnamefont{J.~D.}},
  \bibinfo{year}{1979}, \bibinfo{journal}{IEEE Trans. Nucl. Sci.}
  \textbf{\bibinfo{volume}{NS-26}}, \bibinfo{pages}{4217}.

\bibitem[{\citenamefont{{LCLS}}(2011)}]{LCLS}
\bibinfo{author}{\bibnamefont{{LCLS}}} (\bibinfo{collaboration}{Linac Coherent
  Laser Source}), \bibinfo{year}{2011},
  \urlprefix\url{https://slacportal.slac.stanford.edu/sites/lcls_public/Pages/Default.aspx}.

\bibitem[{\citenamefont{{LCLS II}}(2011)}]{Linac_2011}
\bibinfo{author}{\bibnamefont{{LCLS II}}} (\bibinfo{collaboration}{Linac
  Coherent Light Source II}), \bibinfo{year}{2011},
  \urlprefix\url{https://slacportal.slac.stanford.edu/sites/lcls_public/lcls_ii}.

\bibitem[{\citenamefont{Ledingham and Galster}(2010)}]{Ledingham_2010}
\bibinfo{author}{\bibnamefont{Ledingham}, \bibfnamefont{K.~W.~D.}}, and
  \bibinfo{author}{\bibfnamefont{W.}~\bibnamefont{Galster}},
  \bibinfo{year}{2010}, \bibinfo{journal}{New J. Phys.}
  \textbf{\bibinfo{volume}{12}}, \bibinfo{pages}{045005}.

\bibitem[{\citenamefont{Ledingham} \emph{et~al.}(2003)\citenamefont{Ledingham,
  McKenna, and Singhal}}]{Ledingham_2003}
\bibinfo{author}{\bibnamefont{Ledingham}, \bibfnamefont{K.~W.~D.}},
  \bibinfo{author}{\bibfnamefont{P.}~\bibnamefont{McKenna}}, and
  \bibinfo{author}{\bibfnamefont{R.~P.} \bibnamefont{Singhal}},
  \bibinfo{year}{2003}, \bibinfo{journal}{Science}
  \textbf{\bibinfo{volume}{300}}, \bibinfo{pages}{1107}.

\bibitem[{\citenamefont{Lee} \emph{et~al.}(2003)\citenamefont{Lee, Maslennikov,
  Milstein, Strakhovenko, and {Yu. A. Tikhonov}}}]{Lee_2003}
\bibinfo{author}{\bibnamefont{Lee}, \bibfnamefont{R.~N.}},
  \bibinfo{author}{\bibfnamefont{A.~L.} \bibnamefont{Maslennikov}},
  \bibinfo{author}{\bibfnamefont{A.~I.} \bibnamefont{Milstein}},
  \bibinfo{author}{\bibfnamefont{V.~M.} \bibnamefont{Strakhovenko}}, and
  \bibinfo{author}{\bibnamefont{{Yu. A. Tikhonov}}}, \bibinfo{year}{2003},
  \bibinfo{journal}{Phys. Rep.} \textbf{\bibinfo{volume}{373}},
  \bibinfo{pages}{213}.

\bibitem[{\citenamefont{Leemans and Esarey}(2009)}]{Leemans_2009}
\bibinfo{author}{\bibnamefont{Leemans}, \bibfnamefont{W.}}, and
  \bibinfo{author}{\bibfnamefont{E.}~\bibnamefont{Esarey}},
  \bibinfo{year}{2009}, \bibinfo{journal}{Phys. Today}
  \textbf{\bibinfo{volume}{62}}, \bibinfo{pages}{44}.

\bibitem[{\citenamefont{Leemans} \emph{et~al.}(2006)\citenamefont{Leemans,
  Nagler, Gonsalves, T\'{o}th, K.~Nakamura~and, Esarey, Schroeder, and
  Hooker}}]{Leemans_2006}
\bibinfo{author}{\bibnamefont{Leemans}, \bibfnamefont{W.~P.}},
  \bibinfo{author}{\bibfnamefont{B.}~\bibnamefont{Nagler}},
  \bibinfo{author}{\bibfnamefont{A.~J.} \bibnamefont{Gonsalves}},
  \bibinfo{author}{\bibfnamefont{C.}~\bibnamefont{T\'{o}th}},
  \bibinfo{author}{\bibfnamefont{C.~G. R.~G.} \bibnamefont{K.~Nakamura~and}},
  \bibinfo{author}{\bibfnamefont{E.}~\bibnamefont{Esarey}},
  \bibinfo{author}{\bibfnamefont{C.~B.} \bibnamefont{Schroeder}}, and
  \bibinfo{author}{\bibfnamefont{S.~M.} \bibnamefont{Hooker}},
  \bibinfo{year}{2006}, \bibinfo{journal}{Nature Phys.}
  \textbf{\bibinfo{volume}{2}}, \bibinfo{pages}{696}.

\bibitem[{\citenamefont{Leemans} \emph{et~al.}(1996)\citenamefont{Leemans,
  Schoenlein, Volfbeyn, Chin, Glover, Balling, Zolotorev, Kim, Chattopadhyay,
  and Shank}}]{Leemans_1996}
\bibinfo{author}{\bibnamefont{Leemans}, \bibfnamefont{W.~P.}},
  \bibinfo{author}{\bibfnamefont{R.~W.} \bibnamefont{Schoenlein}},
  \bibinfo{author}{\bibfnamefont{P.}~\bibnamefont{Volfbeyn}},
  \bibinfo{author}{\bibfnamefont{A.~H.} \bibnamefont{Chin}},
  \bibinfo{author}{\bibfnamefont{T.~E.} \bibnamefont{Glover}},
  \bibinfo{author}{\bibfnamefont{P.}~\bibnamefont{Balling}},
  \bibinfo{author}{\bibfnamefont{M.}~\bibnamefont{Zolotorev}},
  \bibinfo{author}{\bibfnamefont{K.~J.} \bibnamefont{Kim}},
  \bibinfo{author}{\bibfnamefont{S.}~\bibnamefont{Chattopadhyay}}, and
  \bibinfo{author}{\bibfnamefont{C.~V.} \bibnamefont{Shank}},
  \bibinfo{year}{1996}, \bibinfo{journal}{Phys. Rev. Lett.}
  \textbf{\bibinfo{volume}{77}}, \bibinfo{pages}{4182}.

\bibitem[{\citenamefont{Lehmann and Spatschek}(2011)}]{Lehmann_2011}
\bibinfo{author}{\bibnamefont{Lehmann}, \bibfnamefont{G.}}, and
  \bibinfo{author}{\bibfnamefont{K.~H.} \bibnamefont{Spatschek}},
  \bibinfo{year}{2011}, \bibinfo{journal}{Phys. Rev. E}
  \textbf{\bibinfo{volume}{84}}, \bibinfo{pages}{046409}.

\bibitem[{\citenamefont{Lewenstein}
  \emph{et~al.}(1994)\citenamefont{Lewenstein, Balcou, {M. Yu. Ivanov},
  L'Huillier, and Corkum}}]{Lewenstein_1994}
\bibinfo{author}{\bibnamefont{Lewenstein}, \bibfnamefont{M.}},
  \bibinfo{author}{\bibfnamefont{P.}~\bibnamefont{Balcou}},
  \bibinfo{author}{\bibnamefont{{M. Yu. Ivanov}}},
  \bibinfo{author}{\bibfnamefont{A.}~\bibnamefont{L'Huillier}}, and
  \bibinfo{author}{\bibfnamefont{P.~B.} \bibnamefont{Corkum}},
  \bibinfo{year}{1994}, \bibinfo{journal}{Phys. Rev. A}
  \textbf{\bibinfo{volume}{49}}, \bibinfo{pages}{2117}.

\bibitem[{\citenamefont{{LHC}}(2011)}]{LHC}
\bibinfo{author}{\bibnamefont{{LHC}}}, \bibinfo{year}{2011},
  \urlprefix\url{http://lhc.web.cern.ch/lhc/}.

\bibitem[{\citenamefont{Liang} \emph{et~al.}(1998)\citenamefont{Liang, Wilks,
  and Tabak}}]{Liang_1998}
\bibinfo{author}{\bibnamefont{Liang}, \bibfnamefont{E.~P.}},
  \bibinfo{author}{\bibfnamefont{S.~C.} \bibnamefont{Wilks}}, and
  \bibinfo{author}{\bibfnamefont{M.}~\bibnamefont{Tabak}},
  \bibinfo{year}{1998}, \bibinfo{journal}{Phys. Rev. Lett.}
  \textbf{\bibinfo{volume}{81}}, \bibinfo{pages}{4887}.

\bibitem[{\citenamefont{Liao} \emph{et~al.}(2011)\citenamefont{Liao,
  P{\'a}lffy, and Keitel}}]{Liao_2011}
\bibinfo{author}{\bibnamefont{Liao}, \bibfnamefont{W.}},
  \bibinfo{author}{\bibfnamefont{A.}~\bibnamefont{P{\'a}lffy}}, and
  \bibinfo{author}{\bibfnamefont{C.~H.} \bibnamefont{Keitel}},
  \bibinfo{year}{2011}, \bibinfo{journal}{Phys. Lett. B}
  \textbf{\bibinfo{volume}{705}}, \bibinfo{pages}{134}.

\bibitem[{\citenamefont{Lin} \emph{et~al.}(2006)\citenamefont{Lin, Li, and
  Becker}}]{Lin_06}
\bibinfo{author}{\bibnamefont{Lin}, \bibfnamefont{Q.}},
  \bibinfo{author}{\bibfnamefont{S.}~\bibnamefont{Li}}, and
  \bibinfo{author}{\bibfnamefont{W.}~\bibnamefont{Becker}},
  \bibinfo{year}{2006}, \bibinfo{journal}{Opt. Lett.}
  \textbf{\bibinfo{volume}{31}}, \bibinfo{pages}{2163}.

\bibitem[{\citenamefont{Liu} \emph{et~al.}(2009)\citenamefont{Liu, Kohler,
  Hatsagortsyan, M\"{u}ller, and Keitel}}]{Liu_2009}
\bibinfo{author}{\bibnamefont{Liu}, \bibfnamefont{C.}},
  \bibinfo{author}{\bibfnamefont{M.}~\bibnamefont{Kohler}},
  \bibinfo{author}{\bibfnamefont{K.~Z.} \bibnamefont{Hatsagortsyan}},
  \bibinfo{author}{\bibfnamefont{C.}~\bibnamefont{M\"{u}ller}}, and
  \bibinfo{author}{\bibfnamefont{C.~H.} \bibnamefont{Keitel}},
  \bibinfo{year}{2009}, \bibinfo{journal}{New J. Phys.}
  \textbf{\bibinfo{volume}{11}}, \bibinfo{pages}{105045}.

\bibitem[{\citenamefont{{LMJ}}(2011)}]{LMJ_2011}
\bibinfo{author}{\bibnamefont{{LMJ}}} (\bibinfo{collaboration}{Laser
  M\'{e}gaJoule}), \bibinfo{year}{2011}, \bibinfo{note}{(Homepage in French)},
  \urlprefix\url{http://www-lmj.cea.fr/index.htm}.

\bibitem[{\citenamefont{Lorentz}(1909)}]{Lorentz_b_1909}
\bibinfo{author}{\bibnamefont{Lorentz}, \bibfnamefont{H.~A.}},
  \bibinfo{year}{1909}, \emph{\bibinfo{title}{The Theory of Electrons}}
  (\bibinfo{publisher}{Teubner, Leipzig}).

\bibitem[{\citenamefont{L\"otstedt}
  \emph{et~al.}(2008)\citenamefont{L\"otstedt, Jentschura, and
  Keitel}}]{Lotstedt_2008}
\bibinfo{author}{\bibnamefont{L\"otstedt}, \bibfnamefont{E.}},
  \bibinfo{author}{\bibfnamefont{U.~D.} \bibnamefont{Jentschura}}, and
  \bibinfo{author}{\bibfnamefont{C.~H.} \bibnamefont{Keitel}},
  \bibinfo{year}{2008}, \bibinfo{journal}{Phys. Rev. Lett.}
  \textbf{\bibinfo{volume}{101}}, \bibinfo{pages}{203001}.

\bibitem[{\citenamefont{L\"otstedt}
  \emph{et~al.}(2009)\citenamefont{L\"otstedt, Jentschura, and
  Keitel}}]{Lotstedt_2009}
\bibinfo{author}{\bibnamefont{L\"otstedt}, \bibfnamefont{E.}},
  \bibinfo{author}{\bibfnamefont{U.~D.} \bibnamefont{Jentschura}}, and
  \bibinfo{author}{\bibfnamefont{C.~H.} \bibnamefont{Keitel}},
  \bibinfo{year}{2009}, \bibinfo{journal}{New J. Phys.}
  \textbf{\bibinfo{volume}{11}}, \bibinfo{pages}{013054}.

\bibitem[{\citenamefont{Lundin} \emph{et~al.}(2007)\citenamefont{Lundin,
  Stenflo, Brodin, Marklund, and Shukla}}]{Lundin_2007}
\bibinfo{author}{\bibnamefont{Lundin}, \bibfnamefont{J.}},
  \bibinfo{author}{\bibfnamefont{L.}~\bibnamefont{Stenflo}},
  \bibinfo{author}{\bibfnamefont{G.}~\bibnamefont{Brodin}},
  \bibinfo{author}{\bibfnamefont{M.}~\bibnamefont{Marklund}}, and
  \bibinfo{author}{\bibfnamefont{P.~K.} \bibnamefont{Shukla}},
  \bibinfo{year}{2007}, \textbf{\bibinfo{volume}{14}}, \bibinfo{pages}{064503}.

\bibitem[{\citenamefont{Lundstr\"om}
  \emph{et~al.}(2006)\citenamefont{Lundstr\"om, Brodin, Lundin, Marklund,
  Bingham, Collier, Mendon\c{c}a, and Norreys}}]{Lundstroem_2006}
\bibinfo{author}{\bibnamefont{Lundstr\"om}, \bibfnamefont{E.}},
  \bibinfo{author}{\bibfnamefont{G.}~\bibnamefont{Brodin}},
  \bibinfo{author}{\bibfnamefont{J.}~\bibnamefont{Lundin}},
  \bibinfo{author}{\bibfnamefont{M.}~\bibnamefont{Marklund}},
  \bibinfo{author}{\bibfnamefont{R.}~\bibnamefont{Bingham}},
  \bibinfo{author}{\bibfnamefont{J.}~\bibnamefont{Collier}},
  \bibinfo{author}{\bibfnamefont{J.~T.} \bibnamefont{Mendon\c{c}a}}, and
  \bibinfo{author}{\bibfnamefont{P.}~\bibnamefont{Norreys}},
  \bibinfo{year}{2006}, \bibinfo{journal}{Phys. Rev. Lett.}
  \textbf{\bibinfo{volume}{96}}, \bibinfo{pages}{083602}.

\bibitem[{\citenamefont{Macchi} \emph{et~al.}(2012)\citenamefont{Macchi,
  Borghesi, and Passoni}}]{Macchi_2012}
\bibinfo{author}{\bibnamefont{Macchi}, \bibfnamefont{A.}},
  \bibinfo{author}{\bibfnamefont{M.}~\bibnamefont{Borghesi}}, and
  \bibinfo{author}{\bibfnamefont{M.}~\bibnamefont{Passoni}},
  \bibinfo{year}{2012}, \bibinfo{journal}{Rev. Mod. Phys.} , \bibinfo{pages}{to
  be published}.

\bibitem[{\citenamefont{Macchi} \emph{et~al.}(2009)\citenamefont{Macchi,
  Veghini, and Pegoraro}}]{Macchi_2009}
\bibinfo{author}{\bibnamefont{Macchi}, \bibfnamefont{A.}},
  \bibinfo{author}{\bibfnamefont{S.}~\bibnamefont{Veghini}}, and
  \bibinfo{author}{\bibfnamefont{F.}~\bibnamefont{Pegoraro}},
  \bibinfo{year}{2009}, \bibinfo{journal}{Phys. Rev. Lett.}
  \textbf{\bibinfo{volume}{103}}, \bibinfo{pages}{085003}.

\bibitem[{\citenamefont{Mackenroth and Di~Piazza}(2011)}]{Mackenroth_2011}
\bibinfo{author}{\bibnamefont{Mackenroth}, \bibfnamefont{F.}}, and
  \bibinfo{author}{\bibfnamefont{A.}~\bibnamefont{Di~Piazza}},
  \bibinfo{year}{2011}, \bibinfo{journal}{Phys. Rev. A}
  \textbf{\bibinfo{volume}{83}}, \bibinfo{pages}{032106}.

\bibitem[{\citenamefont{Mackenroth}
  \emph{et~al.}(2010)\citenamefont{Mackenroth, Di~Piazza, and
  Keitel}}]{Mackenroth_2010}
\bibinfo{author}{\bibnamefont{Mackenroth}, \bibfnamefont{F.}},
  \bibinfo{author}{\bibfnamefont{A.}~\bibnamefont{Di~Piazza}}, and
  \bibinfo{author}{\bibfnamefont{C.~H.} \bibnamefont{Keitel}},
  \bibinfo{year}{2010}, \bibinfo{journal}{Phys. Rev. Lett.}
  \textbf{\bibinfo{volume}{105}}, \bibinfo{pages}{063903}.

\bibitem[{\citenamefont{Maiman}(1960)}]{Maiman_1960}
\bibinfo{author}{\bibnamefont{Maiman}, \bibfnamefont{T.~H.}},
  \bibinfo{year}{1960}, \bibinfo{journal}{nature}
  \textbf{\bibinfo{volume}{187}}, \bibinfo{pages}{493}.

\bibitem[{\citenamefont{Major} \emph{et~al.}(2010)\citenamefont{Major,
  Klingebiel, Skrobol, Ahmad, Wandt, Trushin, Krausz, and Karsch}}]{Major_2010}
\bibinfo{author}{\bibnamefont{Major}, \bibfnamefont{Z.}},
  \bibinfo{author}{\bibfnamefont{S.}~\bibnamefont{Klingebiel}},
  \bibinfo{author}{\bibfnamefont{C.}~\bibnamefont{Skrobol}},
  \bibinfo{author}{\bibfnamefont{I.}~\bibnamefont{Ahmad}},
  \bibinfo{author}{\bibfnamefont{C.}~\bibnamefont{Wandt}},
  \bibinfo{author}{\bibfnamefont{S.~A.} \bibnamefont{Trushin}},
  \bibinfo{author}{\bibfnamefont{F.}~\bibnamefont{Krausz}}, and
  \bibinfo{author}{\bibfnamefont{S.}~\bibnamefont{Karsch}},
  \bibinfo{year}{2010}, \bibinfo{journal}{AIP Conf. Proc.}
  \textbf{\bibinfo{volume}{1228}}, \bibinfo{pages}{117}.

\bibitem[{\citenamefont{Malka}(2011)}]{Malka_2011}
\bibinfo{author}{\bibnamefont{Malka}, \bibfnamefont{V.}}, \bibinfo{year}{2011},
  \eprint{arXiv:1112.5054v1}.

\bibitem[{\citenamefont{Malka} \emph{et~al.}(2008)\citenamefont{Malka, Faure,
  Gauduel, Lefebvre, Rousse, and Phuoc}}]{Malka_2008}
\bibinfo{author}{\bibnamefont{Malka}, \bibfnamefont{V.}},
  \bibinfo{author}{\bibfnamefont{J.}~\bibnamefont{Faure}},
  \bibinfo{author}{\bibfnamefont{Y.~A.} \bibnamefont{Gauduel}},
  \bibinfo{author}{\bibfnamefont{E.}~\bibnamefont{Lefebvre}},
  \bibinfo{author}{\bibfnamefont{A.}~\bibnamefont{Rousse}}, and
  \bibinfo{author}{\bibfnamefont{K.~T.} \bibnamefont{Phuoc}},
  \bibinfo{year}{2008}, \bibinfo{journal}{Nature Phys.}
  \textbf{\bibinfo{volume}{4}}, \bibinfo{pages}{447}.

\bibitem[{\citenamefont{Mangles} \emph{et~al.}(2004)\citenamefont{Mangles,
  Murphy, Najmudin, Thomas, Collier, Dangor, Divall, Foster, Gallacher, Hooker,
  Jaroszynski, Langley} \emph{et~al.}}]{Mangles_2004}
\bibinfo{author}{\bibnamefont{Mangles}, \bibfnamefont{S.~P.~D.}},
  \bibinfo{author}{\bibfnamefont{C.~D.} \bibnamefont{Murphy}},
  \bibinfo{author}{\bibfnamefont{Z.}~\bibnamefont{Najmudin}},
  \bibinfo{author}{\bibfnamefont{A.~G.~R.} \bibnamefont{Thomas}},
  \bibinfo{author}{\bibfnamefont{J.~L.} \bibnamefont{Collier}},
  \bibinfo{author}{\bibfnamefont{A.~E.} \bibnamefont{Dangor}},
  \bibinfo{author}{\bibfnamefont{E.~J.} \bibnamefont{Divall}},
  \bibinfo{author}{\bibfnamefont{P.~S.} \bibnamefont{Foster}},
  \bibinfo{author}{\bibfnamefont{J.~G.} \bibnamefont{Gallacher}},
  \bibinfo{author}{\bibfnamefont{C.~J.} \bibnamefont{Hooker}},
  \bibinfo{author}{\bibfnamefont{D.~A.} \bibnamefont{Jaroszynski}},
  \bibinfo{author}{\bibfnamefont{A.~J.} \bibnamefont{Langley}}, \emph{et~al.},
  \bibinfo{year}{2004}, \bibinfo{journal}{Nature (London)}
  \textbf{\bibinfo{volume}{431}}, \bibinfo{pages}{535}.

\bibitem[{\citenamefont{Mao} \emph{et~al.}(2010)\citenamefont{Mao, Kong, Ho,
  Che, Ban, Gu, and Kawata}}]{Mao_2010}
\bibinfo{author}{\bibnamefont{Mao}, \bibfnamefont{Q.~Q.}},
  \bibinfo{author}{\bibfnamefont{Q.}~\bibnamefont{Kong}},
  \bibinfo{author}{\bibfnamefont{Y.~K.} \bibnamefont{Ho}},
  \bibinfo{author}{\bibfnamefont{H.~O.} \bibnamefont{Che}},
  \bibinfo{author}{\bibfnamefont{H.~Y.} \bibnamefont{Ban}},
  \bibinfo{author}{\bibfnamefont{Y.~J.} \bibnamefont{Gu}}, and
  \bibinfo{author}{\bibfnamefont{S.}~\bibnamefont{Kawata}},
  \bibinfo{year}{2010}, \bibinfo{journal}{Laser Part. Beams}
  \textbf{\bibinfo{volume}{28}}, \bibinfo{pages}{83}.

\bibitem[{\citenamefont{{MaRIE}}(2011)}]{MaRIE_2011}
\bibinfo{author}{\bibnamefont{{MaRIE}}}
  (\bibinfo{collaboration}{Matter-Radiation Interactions in Extremes
  Experimental}), \bibinfo{year}{2011}, \urlprefix\url{http://marie.lanl.gov/}.

\bibitem[{\citenamefont{Marklund}(2010)}]{Marklund_2010}
\bibinfo{author}{\bibnamefont{Marklund}, \bibfnamefont{M.}},
  \bibinfo{year}{2010}, \bibinfo{journal}{Nature Photon.}
  \textbf{\bibinfo{volume}{4}}, \bibinfo{pages}{72}.

\bibitem[{\citenamefont{Marklund} \emph{et~al.}(2005)\citenamefont{Marklund,
  Brodin, Stenflo, and Shukla}}]{Marklund_2005}
\bibinfo{author}{\bibnamefont{Marklund}, \bibfnamefont{M.}},
  \bibinfo{author}{\bibfnamefont{G.}~\bibnamefont{Brodin}},
  \bibinfo{author}{\bibfnamefont{L.}~\bibnamefont{Stenflo}}, and
  \bibinfo{author}{\bibfnamefont{P.~K.} \bibnamefont{Shukla}},
  \bibinfo{year}{2005}, \bibinfo{journal}{New J. Phys.}
  \textbf{\bibinfo{volume}{7}}, \bibinfo{pages}{70}.

\bibitem[{\citenamefont{Marklund and Shukla}(2006)}]{Marklund_2006}
\bibinfo{author}{\bibnamefont{Marklund}, \bibfnamefont{M.}}, and
  \bibinfo{author}{\bibfnamefont{P.~K.} \bibnamefont{Shukla}},
  \bibinfo{year}{2006}, \bibinfo{journal}{Rev. Mod. Phys.}
  \textbf{\bibinfo{volume}{78}}, \bibinfo{pages}{591}.

\bibitem[{\citenamefont{Marx} \emph{et~al.}(2011)\citenamefont{Marx, Uschmann,
  H{\"o}fer, L{\"o}tzsch, Wehrhan, F{\"o}rster, Kaluza, St{\"o}hlker, Gies,
  Detlefs, Roth, H{\"a}rtwig} \emph{et~al.}}]{Marx_2011}
\bibinfo{author}{\bibnamefont{Marx}, \bibfnamefont{B.}},
  \bibinfo{author}{\bibfnamefont{I.}~\bibnamefont{Uschmann}},
  \bibinfo{author}{\bibfnamefont{S.}~\bibnamefont{H{\"o}fer}},
  \bibinfo{author}{\bibfnamefont{R.}~\bibnamefont{L{\"o}tzsch}},
  \bibinfo{author}{\bibfnamefont{O.}~\bibnamefont{Wehrhan}},
  \bibinfo{author}{\bibfnamefont{E.}~\bibnamefont{F{\"o}rster}},
  \bibinfo{author}{\bibfnamefont{M.}~\bibnamefont{Kaluza}},
  \bibinfo{author}{\bibfnamefont{T.}~\bibnamefont{St{\"o}hlker}},
  \bibinfo{author}{\bibfnamefont{H.}~\bibnamefont{Gies}},
  \bibinfo{author}{\bibfnamefont{C.}~\bibnamefont{Detlefs}},
  \bibinfo{author}{\bibfnamefont{T.}~\bibnamefont{Roth}},
  \bibinfo{author}{\bibfnamefont{J.}~\bibnamefont{H{\"a}rtwig}}, \emph{et~al.},
  \bibinfo{year}{2011}, \bibinfo{journal}{Opt. Commun.}
  \textbf{\bibinfo{volume}{284}}, \bibinfo{pages}{915}.

\bibitem[{\citenamefont{Matinyan}(1998)}]{Matinyan_1998}
\bibinfo{author}{\bibnamefont{Matinyan}, \bibfnamefont{S.}},
  \bibinfo{year}{1998}, \bibinfo{journal}{Phys. Rep.}
  \textbf{\bibinfo{volume}{298}}, \bibinfo{pages}{199}.

\bibitem[{\citenamefont{Matveev} \emph{et~al.}(2005)\citenamefont{Matveev,
  Gusarevich, and Pashev}}]{Matveev_2005}
\bibinfo{author}{\bibnamefont{Matveev}, \bibfnamefont{V.~I.}},
  \bibinfo{author}{\bibfnamefont{E.~S.} \bibnamefont{Gusarevich}}, and
  \bibinfo{author}{\bibfnamefont{I.~N.} \bibnamefont{Pashev}},
  \bibinfo{year}{2005}, \bibinfo{journal}{J. Exp. Theor. Phys.}
  \textbf{\bibinfo{volume}{100}}, \bibinfo{pages}{1043}.

\bibitem[{\citenamefont{McDonald and Shmakov}(1999)}]{McDonald_1999}
\bibinfo{author}{\bibnamefont{McDonald}, \bibfnamefont{K.~T.}}, and
  \bibinfo{author}{\bibfnamefont{K.}~\bibnamefont{Shmakov}},
  \bibinfo{year}{1999}, \bibinfo{journal}{Phys. Rev. ST Accel. Beams}
  \textbf{\bibinfo{volume}{2}}, \bibinfo{pages}{121301}.

\bibitem[{\citenamefont{Mendon\mbox{\c{c}}a}(2007)}]{Mendonca_2007}
\bibinfo{author}{\bibnamefont{Mendon\mbox{\c{c}}a}, \bibfnamefont{J.~T.}},
  \bibinfo{year}{2007}, \bibinfo{journal}{Europhys. Lett.}
  \textbf{\bibinfo{volume}{79}}, \bibinfo{pages}{21001}.

\bibitem[{\citenamefont{Mendon\mbox{\c{c}}a}
  \emph{et~al.}(2006)\citenamefont{Mendon\mbox{\c{c}}a, Marklund, Shukla, and
  Brodin}}]{Mendonca_2006}
\bibinfo{author}{\bibnamefont{Mendon\mbox{\c{c}}a}, \bibfnamefont{J.~T.}},
  \bibinfo{author}{\bibfnamefont{M.}~\bibnamefont{Marklund}},
  \bibinfo{author}{\bibfnamefont{P.~K.} \bibnamefont{Shukla}}, and
  \bibinfo{author}{\bibfnamefont{G.}~\bibnamefont{Brodin}},
  \bibinfo{year}{2006}, \bibinfo{journal}{Phys. Lett. A}
  \textbf{\bibinfo{volume}{359}}, \bibinfo{pages}{700}.

\bibitem[{\citenamefont{Meuren and Di~Piazza}(2011)}]{Meuren_2011}
\bibinfo{author}{\bibnamefont{Meuren}, \bibfnamefont{S.}}, and
  \bibinfo{author}{\bibfnamefont{A.}~\bibnamefont{Di~Piazza}},
  \bibinfo{year}{2011}, \bibinfo{journal}{Phys. Rev. Lett.}
  \textbf{\bibinfo{volume}{107}}, \bibinfo{pages}{260401}.

\bibitem[{\citenamefont{Milosevic} \emph{et~al.}(2004)\citenamefont{Milosevic,
  Corkum, and Brabec}}]{Milosevic_2004}
\bibinfo{author}{\bibnamefont{Milosevic}, \bibfnamefont{N.}},
  \bibinfo{author}{\bibfnamefont{P.~B.} \bibnamefont{Corkum}}, and
  \bibinfo{author}{\bibfnamefont{T.}~\bibnamefont{Brabec}},
  \bibinfo{year}{2004}, \bibinfo{journal}{Phys. Rev. Lett.}
  \textbf{\bibinfo{volume}{92}}, \bibinfo{pages}{013002}.

\bibitem[{\citenamefont{Milosevic} \emph{et~al.}(2002)\citenamefont{Milosevic,
  Krainov, and Brabec}}]{Milosevic_2002}
\bibinfo{author}{\bibnamefont{Milosevic}, \bibfnamefont{N.}},
  \bibinfo{author}{\bibfnamefont{V.~P.} \bibnamefont{Krainov}}, and
  \bibinfo{author}{\bibfnamefont{T.}~\bibnamefont{Brabec}},
  \bibinfo{year}{2002}, \bibinfo{journal}{Phys. Rev. Lett.}
  \textbf{\bibinfo{volume}{89}}, \bibinfo{pages}{193001}.

\bibitem[{\citenamefont{Milstein} \emph{et~al.}(2006)\citenamefont{Milstein,
  M\"uller, Hatsagortsyan, Jentschura, and Keitel}}]{Milstein_2006}
\bibinfo{author}{\bibnamefont{Milstein}, \bibfnamefont{A.~I.}},
  \bibinfo{author}{\bibfnamefont{C.}~\bibnamefont{M\"uller}},
  \bibinfo{author}{\bibfnamefont{K.~Z.} \bibnamefont{Hatsagortsyan}},
  \bibinfo{author}{\bibfnamefont{U.~D.} \bibnamefont{Jentschura}}, and
  \bibinfo{author}{\bibfnamefont{C.~H.} \bibnamefont{Keitel}},
  \bibinfo{year}{2006}, \bibinfo{journal}{Phys. Rev. A}
  \textbf{\bibinfo{volume}{73}}, \bibinfo{pages}{062106}.

\bibitem[{\citenamefont{Milstein and Schumacher}(1994)}]{Milstein_1994}
\bibinfo{author}{\bibnamefont{Milstein}, \bibfnamefont{A.~I.}}, and
  \bibinfo{author}{\bibfnamefont{M.}~\bibnamefont{Schumacher}},
  \bibinfo{year}{1994}, \bibinfo{journal}{Phys. Rep.}
  \textbf{\bibinfo{volume}{243}}, \bibinfo{pages}{183}.

\bibitem[{\citenamefont{Mimura} \emph{et~al.}(2010)\citenamefont{Mimura, Handa,
  Kimura, Yumoto, Yamakawa, Yokoyama, Matsuyama, Inagaki, Yamamura, Sano,
  Tamasaku, Nishino} \emph{et~al.}}]{Mimura_2010}
\bibinfo{author}{\bibnamefont{Mimura}, \bibfnamefont{H.}},
  \bibinfo{author}{\bibfnamefont{S.}~\bibnamefont{Handa}},
  \bibinfo{author}{\bibfnamefont{T.}~\bibnamefont{Kimura}},
  \bibinfo{author}{\bibfnamefont{H.}~\bibnamefont{Yumoto}},
  \bibinfo{author}{\bibfnamefont{D.}~\bibnamefont{Yamakawa}},
  \bibinfo{author}{\bibfnamefont{H.}~\bibnamefont{Yokoyama}},
  \bibinfo{author}{\bibfnamefont{S.}~\bibnamefont{Matsuyama}},
  \bibinfo{author}{\bibfnamefont{K.}~\bibnamefont{Inagaki}},
  \bibinfo{author}{\bibfnamefont{K.}~\bibnamefont{Yamamura}},
  \bibinfo{author}{\bibfnamefont{Y.}~\bibnamefont{Sano}},
  \bibinfo{author}{\bibfnamefont{K.}~\bibnamefont{Tamasaku}},
  \bibinfo{author}{\bibfnamefont{Y.}~\bibnamefont{Nishino}}, \emph{et~al.},
  \bibinfo{year}{2010}, \bibinfo{journal}{Nature Phys.}
  \textbf{\bibinfo{volume}{6}}, \bibinfo{pages}{122}.

\bibitem[{\citenamefont{Mocken and Keitel}(2008)}]{Mocken_2008}
\bibinfo{author}{\bibnamefont{Mocken}, \bibfnamefont{G.}}, and
  \bibinfo{author}{\bibfnamefont{C.~H.} \bibnamefont{Keitel}},
  \bibinfo{year}{2008}, \bibinfo{journal}{Comp. Phys. Commun.}
  \textbf{\bibinfo{volume}{178}}, \bibinfo{pages}{868}.

\bibitem[{\citenamefont{Mocken and Keitel}(2004)}]{Mocken_2004}
\bibinfo{author}{\bibnamefont{Mocken}, \bibfnamefont{G.~R.}}, and
  \bibinfo{author}{\bibfnamefont{C.~H.} \bibnamefont{Keitel}},
  \bibinfo{year}{2004}, \bibinfo{journal}{J. Phys. B}
  \textbf{\bibinfo{volume}{37}}, \bibinfo{pages}{L275}.

\bibitem[{\citenamefont{Mocken} \emph{et~al.}(2010)\citenamefont{Mocken, Ruf,
  M\"uller, and Keitel}}]{Mocken_2010}
\bibinfo{author}{\bibnamefont{Mocken}, \bibfnamefont{G.~R.}},
  \bibinfo{author}{\bibfnamefont{M.}~\bibnamefont{Ruf}},
  \bibinfo{author}{\bibfnamefont{C.}~\bibnamefont{M\"uller}}, and
  \bibinfo{author}{\bibfnamefont{C.~H.} \bibnamefont{Keitel}},
  \bibinfo{year}{2010}, \bibinfo{journal}{Phys. Rev. A}
  \textbf{\bibinfo{volume}{81}}, \bibinfo{pages}{022122}.

\bibitem[{\citenamefont{Monden and Kodama}(2011)}]{Monden_2011}
\bibinfo{author}{\bibnamefont{Monden}, \bibfnamefont{Y.}}, and
  \bibinfo{author}{\bibfnamefont{R.}~\bibnamefont{Kodama}},
  \bibinfo{year}{2011}, \bibinfo{journal}{Phys. Rev. Lett.}
  \textbf{\bibinfo{volume}{107}}, \bibinfo{pages}{073602}.

\bibitem[{\citenamefont{Monin and Voloshin}(2010)}]{Monin_2010}
\bibinfo{author}{\bibnamefont{Monin}, \bibfnamefont{A.}}, and
  \bibinfo{author}{\bibfnamefont{M.~B.} \bibnamefont{Voloshin}},
  \bibinfo{year}{2010}, \bibinfo{journal}{Phys. Rev. D}
  \textbf{\bibinfo{volume}{81}}, \bibinfo{pages}{085014}.

\bibitem[{\citenamefont{Moniz and Sharp}(1977)}]{Moniz_1977}
\bibinfo{author}{\bibnamefont{Moniz}, \bibfnamefont{E.~J.}}, and
  \bibinfo{author}{\bibfnamefont{D.~H.} \bibnamefont{Sharp}},
  \bibinfo{year}{1977}, \bibinfo{journal}{Phys. Rev. D}
  \textbf{\bibinfo{volume}{15}}, \bibinfo{pages}{2850}.

\bibitem[{\citenamefont{Moore} \emph{et~al.}(1999)\citenamefont{Moore, Ting,
  McNaught, Qiu, Burris, and Sprangle}}]{Moore_1999}
\bibinfo{author}{\bibnamefont{Moore}, \bibfnamefont{C.~I.}},
  \bibinfo{author}{\bibfnamefont{A.}~\bibnamefont{Ting}},
  \bibinfo{author}{\bibfnamefont{S.~J.} \bibnamefont{McNaught}},
  \bibinfo{author}{\bibfnamefont{J.}~\bibnamefont{Qiu}},
  \bibinfo{author}{\bibfnamefont{H.~R.} \bibnamefont{Burris}}, and
  \bibinfo{author}{\bibfnamefont{P.}~\bibnamefont{Sprangle}},
  \bibinfo{year}{1999}, \bibinfo{journal}{Phys. Rev. Lett.}
  \textbf{\bibinfo{volume}{82}}, \bibinfo{pages}{1688}.

\bibitem[{\citenamefont{Mourou}(2010)}]{Mourou_2010}
\bibinfo{author}{\bibnamefont{Mourou}, \bibfnamefont{G.}},
  \bibinfo{year}{2010}, \bibinfo{journal}{Interview with Optik \& Photonik}
  \textbf{\bibinfo{volume}{5}}, \bibinfo{pages}{7}.

\bibitem[{\citenamefont{Mourou and Tajima}(2011)}]{Mourou_2011}
\bibinfo{author}{\bibnamefont{Mourou}, \bibfnamefont{G.}}, and
  \bibinfo{author}{\bibfnamefont{T.}~\bibnamefont{Tajima}},
  \bibinfo{year}{2011}, \bibinfo{journal}{Science}
  \textbf{\bibinfo{volume}{331}}, \bibinfo{pages}{41}.

\bibitem[{\citenamefont{Mourou} \emph{et~al.}(2006)\citenamefont{Mourou,
  Tajima, and Bulanov}}]{Mourou_2006}
\bibinfo{author}{\bibnamefont{Mourou}, \bibfnamefont{G.~A.}},
  \bibinfo{author}{\bibfnamefont{T.}~\bibnamefont{Tajima}}, and
  \bibinfo{author}{\bibfnamefont{S.~V.} \bibnamefont{Bulanov}},
  \bibinfo{year}{2006}, \bibinfo{journal}{Rev. Mod. Phys.}
  \textbf{\bibinfo{volume}{78}}, \bibinfo{pages}{309}.

\bibitem[{\citenamefont{M\"uller}(2009)}]{Muller2009PLB}
\bibinfo{author}{\bibnamefont{M\"uller}, \bibfnamefont{C.}},
  \bibinfo{year}{2009}, \bibinfo{journal}{Phys. Lett. B}
  \textbf{\bibinfo{volume}{672}}, \bibinfo{pages}{56}.

\bibitem[{\citenamefont{M\"uller}
  \emph{et~al.}(2008{\natexlab{a}})\citenamefont{M\"uller, Deneke, and
  Keitel}}]{Deneke_2008a}
\bibinfo{author}{\bibnamefont{M\"uller}, \bibfnamefont{C.}},
  \bibinfo{author}{\bibfnamefont{C.}~\bibnamefont{Deneke}}, and
  \bibinfo{author}{\bibfnamefont{C.~H.} \bibnamefont{Keitel}},
  \bibinfo{year}{2008}{\natexlab{a}}, \bibinfo{journal}{Phys. Rev. Lett.}
  \textbf{\bibinfo{volume}{101}}, \bibinfo{pages}{060402}.

\bibitem[{\citenamefont{M\"uller}
  \emph{et~al.}(2009{\natexlab{a}})\citenamefont{M\"uller, Deneke, Ruf, Mocken,
  Hatsagortsyan, and Keitel}}]{LPHYS2008}
\bibinfo{author}{\bibnamefont{M\"uller}, \bibfnamefont{C.}},
  \bibinfo{author}{\bibfnamefont{C.}~\bibnamefont{Deneke}},
  \bibinfo{author}{\bibfnamefont{M.}~\bibnamefont{Ruf}},
  \bibinfo{author}{\bibfnamefont{G.~R.} \bibnamefont{Mocken}},
  \bibinfo{author}{\bibfnamefont{K.~Z.} \bibnamefont{Hatsagortsyan}}, and
  \bibinfo{author}{\bibfnamefont{C.~H.} \bibnamefont{Keitel}},
  \bibinfo{year}{2009}{\natexlab{a}}, \bibinfo{journal}{Laser Phys.}
  \textbf{\bibinfo{volume}{19}}, \bibinfo{pages}{791}.

\bibitem[{\citenamefont{M\"uller} \emph{et~al.}(2006)\citenamefont{M\"uller,
  Hatsagortsyan, and Keitel}}]{Muller_2006}
\bibinfo{author}{\bibnamefont{M\"uller}, \bibfnamefont{C.}},
  \bibinfo{author}{\bibfnamefont{K.~Z.} \bibnamefont{Hatsagortsyan}}, and
  \bibinfo{author}{\bibfnamefont{C.~H.} \bibnamefont{Keitel}},
  \bibinfo{year}{2006}, \bibinfo{journal}{Phys. Rev. D}
  \textbf{\bibinfo{volume}{74}}, \bibinfo{pages}{074017}.

\bibitem[{\citenamefont{M\"uller}
  \emph{et~al.}(2008{\natexlab{b}})\citenamefont{M\"uller, Hatsagortsyan, and
  Keitel}}]{Muller_2008b}
\bibinfo{author}{\bibnamefont{M\"uller}, \bibfnamefont{C.}},
  \bibinfo{author}{\bibfnamefont{K.~Z.} \bibnamefont{Hatsagortsyan}}, and
  \bibinfo{author}{\bibfnamefont{C.~H.} \bibnamefont{Keitel}},
  \bibinfo{year}{2008}{\natexlab{b}}, \bibinfo{journal}{Phys. Rev. A}
  \textbf{\bibinfo{volume}{78}}, \bibinfo{pages}{033408}.

\bibitem[{\citenamefont{M\"uller}
  \emph{et~al.}(2008{\natexlab{c}})\citenamefont{M\"uller, Hatsagortsyan, and
  Keitel}}]{Muller_2008a}
\bibinfo{author}{\bibnamefont{M\"uller}, \bibfnamefont{C.}},
  \bibinfo{author}{\bibfnamefont{K.~Z.} \bibnamefont{Hatsagortsyan}}, and
  \bibinfo{author}{\bibfnamefont{C.~H.} \bibnamefont{Keitel}},
  \bibinfo{year}{2008}{\natexlab{c}}, \bibinfo{journal}{Phys. Lett. B}
  \textbf{\bibinfo{volume}{659}}, \bibinfo{pages}{209}.

\bibitem[{\citenamefont{M\"uller}
  \emph{et~al.}(2009{\natexlab{b}})\citenamefont{M\"uller, Hatsagortsyan, Ruf,
  M\"{u}ller, Hetzheim, Kohler, and Keitel}}]{AboveThreshold}
\bibinfo{author}{\bibnamefont{M\"uller}, \bibfnamefont{C.}},
  \bibinfo{author}{\bibfnamefont{K.~Z.} \bibnamefont{Hatsagortsyan}},
  \bibinfo{author}{\bibfnamefont{M.}~\bibnamefont{Ruf}},
  \bibinfo{author}{\bibfnamefont{S.}~\bibnamefont{M\"{u}ller}},
  \bibinfo{author}{\bibfnamefont{H.~G.} \bibnamefont{Hetzheim}},
  \bibinfo{author}{\bibfnamefont{M.~C.} \bibnamefont{Kohler}}, and
  \bibinfo{author}{\bibfnamefont{C.~H.} \bibnamefont{Keitel}},
  \bibinfo{year}{2009}{\natexlab{b}}, \bibinfo{journal}{Laser Phys.}
  \textbf{\bibinfo{volume}{19}}, \bibinfo{pages}{1743}.

\bibitem[{\citenamefont{M\"uller and Keitel}(2009)}]{NaturePhot}
\bibinfo{author}{\bibnamefont{M\"uller}, \bibfnamefont{C.}}, and
  \bibinfo{author}{\bibfnamefont{C.~H.} \bibnamefont{Keitel}},
  \bibinfo{year}{2009}, \bibinfo{journal}{Nature Photon.}
  \textbf{\bibinfo{volume}{3}}, \bibinfo{pages}{245}.

\bibitem[{\citenamefont{M\"uller}
  \emph{et~al.}(2003{\natexlab{a}})\citenamefont{M\"uller, Voitkiv, and
  Gr\"un}}]{MVG2003NIMB}
\bibinfo{author}{\bibnamefont{M\"uller}, \bibfnamefont{C.}},
  \bibinfo{author}{\bibfnamefont{A.~B.} \bibnamefont{Voitkiv}}, and
  \bibinfo{author}{\bibfnamefont{N.}~\bibnamefont{Gr\"un}},
  \bibinfo{year}{2003}{\natexlab{a}}, \bibinfo{journal}{Nucl. Instr. Meth.
  Phys. Res. B} \textbf{\bibinfo{volume}{205}}, \bibinfo{pages}{306}.

\bibitem[{\citenamefont{M\"uller}
  \emph{et~al.}(2003{\natexlab{b}})\citenamefont{M\"uller, Voitkiv, and
  Gr\"un}}]{MVG2003PRA}
\bibinfo{author}{\bibnamefont{M\"uller}, \bibfnamefont{C.}},
  \bibinfo{author}{\bibfnamefont{A.~B.} \bibnamefont{Voitkiv}}, and
  \bibinfo{author}{\bibfnamefont{N.}~\bibnamefont{Gr\"un}},
  \bibinfo{year}{2003}{\natexlab{b}}, \bibinfo{journal}{Phys. Rev. A}
  \textbf{\bibinfo{volume}{67}}, \bibinfo{pages}{063407}.

\bibitem[{\citenamefont{M\"uller}
  \emph{et~al.}(2003{\natexlab{c}})\citenamefont{M\"uller, Voitkiv, and
  Gr\"un}}]{MVG2003PRL}
\bibinfo{author}{\bibnamefont{M\"uller}, \bibfnamefont{C.}},
  \bibinfo{author}{\bibfnamefont{A.~B.} \bibnamefont{Voitkiv}}, and
  \bibinfo{author}{\bibfnamefont{N.}~\bibnamefont{Gr\"un}},
  \bibinfo{year}{2003}{\natexlab{c}}, \bibinfo{journal}{Phys. Rev. Lett.}
  \textbf{\bibinfo{volume}{91}}, \bibinfo{pages}{223601}.

\bibitem[{\citenamefont{M\"uller} \emph{et~al.}(2004)\citenamefont{M\"uller,
  Voitkiv, and Gr\"un}}]{MVG2004PRA}
\bibinfo{author}{\bibnamefont{M\"uller}, \bibfnamefont{C.}},
  \bibinfo{author}{\bibfnamefont{A.~B.} \bibnamefont{Voitkiv}}, and
  \bibinfo{author}{\bibfnamefont{N.}~\bibnamefont{Gr\"un}},
  \bibinfo{year}{2004}, \bibinfo{journal}{Phys. Rev. A}
  \textbf{\bibinfo{volume}{70}}, \bibinfo{pages}{023412}.

\bibitem[{\citenamefont{M\"{u}ller}
  \emph{et~al.}(2009)\citenamefont{M\"{u}ller, Voitkiv, and
  Najjari}}]{Mueller_2009c}
\bibinfo{author}{\bibnamefont{M\"{u}ller}, \bibfnamefont{C.}},
  \bibinfo{author}{\bibfnamefont{A.~B.} \bibnamefont{Voitkiv}}, and
  \bibinfo{author}{\bibfnamefont{B.}~\bibnamefont{Najjari}},
  \bibinfo{year}{2009}, \bibinfo{journal}{J. Phys. B}
  \textbf{\bibinfo{volume}{42}}, \bibinfo{pages}{221001}.

\bibitem[{\citenamefont{M\"uller and M\"uller}(2009)}]{SarahPRD}
\bibinfo{author}{\bibnamefont{M\"uller}, \bibfnamefont{S.}}, and
  \bibinfo{author}{\bibfnamefont{C.}~\bibnamefont{M\"uller}},
  \bibinfo{year}{2009}, \bibinfo{journal}{Phys. Rev. D}
  \textbf{\bibinfo{volume}{80}}, \bibinfo{pages}{053014}.

\bibitem[{\citenamefont{M\"uller and M\"uller}(2011)}]{Muller_Spin2011}
\bibinfo{author}{\bibnamefont{M\"uller}, \bibfnamefont{T.-O.}}, and
  \bibinfo{author}{\bibfnamefont{C.}~\bibnamefont{M\"uller}},
  \bibinfo{year}{2011}, \bibinfo{journal}{Phys. Lett. B}
  \textbf{\bibinfo{volume}{696}}, \bibinfo{pages}{201}.

\bibitem[{\citenamefont{Mulser and Bauer}(2010)}]{Mulser_b_2010}
\bibinfo{author}{\bibnamefont{Mulser}, \bibfnamefont{P.}}, and
  \bibinfo{author}{\bibfnamefont{D.}~\bibnamefont{Bauer}},
  \bibinfo{year}{2010}, \emph{\bibinfo{title}{High Power Laser-Matter
  Interaction}} (\bibinfo{publisher}{Springer, Berlin}).

\bibitem[{\citenamefont{Mustafa and K\"ampfer}(2009)}]{Mustafa_2009}
\bibinfo{author}{\bibnamefont{Mustafa}, \bibfnamefont{M.~G.}}, and
  \bibinfo{author}{\bibfnamefont{B.}~\bibnamefont{K\"ampfer}},
  \bibinfo{year}{2009}, \bibinfo{journal}{Phys. Rev. A}
  \textbf{\bibinfo{volume}{79}}, \bibinfo{pages}{020103(R)}.

\bibitem[{Nakamura \emph{et~al.}(2010)\citenamefont{Nakamura}
  \emph{et~al.}}]{PDG_2010}
\bibinfo{author}{\bibnamefont{Nakamura}, \bibfnamefont{K.}}, \emph{et~al.}
  (\bibinfo{collaboration}{Particle Data Group}), \bibinfo{year}{2010},
  \bibinfo{journal}{J. Phys. G} \textbf{\bibinfo{volume}{37}},
  \bibinfo{pages}{075021}.

\bibitem[{\citenamefont{Nakamura} \emph{et~al.}(2011)\citenamefont{Nakamura,
  Koga, {T. Zh. Esirkepov}, Kando, and Georg~Korn}}]{Nakamura_2011}
\bibinfo{author}{\bibnamefont{Nakamura}, \bibfnamefont{T.}},
  \bibinfo{author}{\bibfnamefont{J.~K.} \bibnamefont{Koga}},
  \bibinfo{author}{\bibnamefont{{T. Zh. Esirkepov}}},
  \bibinfo{author}{\bibfnamefont{M.}~\bibnamefont{Kando}}, and
  \bibinfo{author}{\bibfnamefont{S.~V.~B.} \bibnamefont{Georg~Korn}},
  \bibinfo{year}{2011}, \eprint{arXiv:1111.5678v3}.

\bibitem[{\citenamefont{Narozhny} \emph{et~al.}(2004)\citenamefont{Narozhny,
  Bulanov, Mur, and Popov}}]{Narozhny_2004}
\bibinfo{author}{\bibnamefont{Narozhny}, \bibfnamefont{N.~B.}},
  \bibinfo{author}{\bibfnamefont{S.~S.} \bibnamefont{Bulanov}},
  \bibinfo{author}{\bibfnamefont{V.~D.} \bibnamefont{Mur}}, and
  \bibinfo{author}{\bibfnamefont{V.~S.} \bibnamefont{Popov}},
  \bibinfo{year}{2004}, \bibinfo{journal}{Phys. Lett. A}
  \textbf{\bibinfo{volume}{330}}, \bibinfo{pages}{1}.

\bibitem[{\citenamefont{Narozhny} \emph{et~al.}(2006)\citenamefont{Narozhny,
  Bulanov, Mur, and Popov}}]{Narozhny_2006}
\bibinfo{author}{\bibnamefont{Narozhny}, \bibfnamefont{N.~B.}},
  \bibinfo{author}{\bibfnamefont{S.~S.} \bibnamefont{Bulanov}},
  \bibinfo{author}{\bibfnamefont{V.~D.} \bibnamefont{Mur}}, and
  \bibinfo{author}{\bibfnamefont{V.~S.} \bibnamefont{Popov}},
  \bibinfo{year}{2006}, \bibinfo{journal}{J. Exp. Theor. Phys.}
  \textbf{\bibinfo{volume}{102}}, \bibinfo{pages}{9}.

\bibitem[{\citenamefont{Narozhny and Fedotov}(2008)}]{Narozhny_2008}
\bibinfo{author}{\bibnamefont{Narozhny}, \bibfnamefont{N.~B.}}, and
  \bibinfo{author}{\bibfnamefont{A.~M.} \bibnamefont{Fedotov}},
  \bibinfo{year}{2008}, \bibinfo{journal}{Phys. Rev. Lett.}
  \textbf{\bibinfo{volume}{100}}, \bibinfo{pages}{219101}.

\bibitem[{\citenamefont{Narozhny and Fofanov}(2000)}]{Narozhny_2000}
\bibinfo{author}{\bibnamefont{Narozhny}, \bibfnamefont{N.~B.}}, and
  \bibinfo{author}{\bibfnamefont{M.~S.} \bibnamefont{Fofanov}},
  \bibinfo{year}{2000}, \bibinfo{journal}{J. Exp. Theor. Phys.}
  \textbf{\bibinfo{volume}{90}}, \bibinfo{pages}{415}.

\bibitem[{\citenamefont{Naumova} \emph{et~al.}(2009)\citenamefont{Naumova,
  Schlegel, Tikhonchuk, Labaune, Sokolov, and Mourou}}]{Naumova_2009}
\bibinfo{author}{\bibnamefont{Naumova}, \bibfnamefont{N.}},
  \bibinfo{author}{\bibfnamefont{T.}~\bibnamefont{Schlegel}},
  \bibinfo{author}{\bibfnamefont{V.~T.} \bibnamefont{Tikhonchuk}},
  \bibinfo{author}{\bibfnamefont{C.}~\bibnamefont{Labaune}},
  \bibinfo{author}{\bibfnamefont{I.~V.} \bibnamefont{Sokolov}}, and
  \bibinfo{author}{\bibfnamefont{G.}~\bibnamefont{Mourou}},
  \bibinfo{year}{2009}, \bibinfo{journal}{Phys. Rev. Lett.}
  \textbf{\bibinfo{volume}{102}}, \bibinfo{pages}{025002}.

\bibitem[{\citenamefont{Nedoreshta}
  \emph{et~al.}(2009)\citenamefont{Nedoreshta, Roshchupkin, and
  Voroshilo}}]{Roshchupkin_2009}
\bibinfo{author}{\bibnamefont{Nedoreshta}, \bibfnamefont{V.~N.}},
  \bibinfo{author}{\bibfnamefont{S.~P.} \bibnamefont{Roshchupkin}}, and
  \bibinfo{author}{\bibfnamefont{A.~I.} \bibnamefont{Voroshilo}},
  \bibinfo{year}{2009}, \bibinfo{journal}{Laser Phys.}
  \textbf{\bibinfo{volume}{19}}, \bibinfo{pages}{531}.

\bibitem[{\citenamefont{Nerush}
  \emph{et~al.}(2011{\natexlab{a}})\citenamefont{Nerush, Bashmakov, and
  Kostyukov}}]{Nerush_2011_b}
\bibinfo{author}{\bibnamefont{Nerush}, \bibfnamefont{E.~N.}},
  \bibinfo{author}{\bibfnamefont{V.~F.} \bibnamefont{Bashmakov}}, and
  \bibinfo{author}{\bibnamefont{Kostyukov}},
  \bibinfo{year}{2011}{\natexlab{a}}, \bibinfo{journal}{Phys. Plasmas}
  \textbf{\bibinfo{volume}{18}}, \bibinfo{pages}{083107}.

\bibitem[{\citenamefont{Nerush}
  \emph{et~al.}(2011{\natexlab{b}})\citenamefont{Nerush, {I. Yu. Kostyukov},
  Fedotov, Narozhny, Elkina, and Ruhl}}]{Nerush_2011}
\bibinfo{author}{\bibnamefont{Nerush}, \bibfnamefont{E.~N.}},
  \bibinfo{author}{\bibnamefont{{I. Yu. Kostyukov}}},
  \bibinfo{author}{\bibfnamefont{A.~M.} \bibnamefont{Fedotov}},
  \bibinfo{author}{\bibfnamefont{N.~B.} \bibnamefont{Narozhny}},
  \bibinfo{author}{\bibfnamefont{N.~V.} \bibnamefont{Elkina}}, and
  \bibinfo{author}{\bibfnamefont{H.}~\bibnamefont{Ruhl}},
  \bibinfo{year}{2011}{\natexlab{b}}, \bibinfo{journal}{Phys. Rev. Lett.}
  \textbf{\bibinfo{volume}{106}}, \bibinfo{pages}{035001}.

\bibitem[{\citenamefont{{NIF}}(2011)}]{NIF}
\bibinfo{author}{\bibnamefont{{NIF}}} (\bibinfo{collaboration}{National
  Ignition Facility}), \bibinfo{year}{2011},
  \urlprefix\url{https://lasers.llnl.gov/about/nif/}.

\bibitem[{\citenamefont{Nikishov and
  Ritus}(1964{\natexlab{a}})}]{Nikishov_1964_a}
\bibinfo{author}{\bibnamefont{Nikishov}, \bibfnamefont{A.~I.}}, and
  \bibinfo{author}{\bibfnamefont{V.~I.} \bibnamefont{Ritus}},
  \bibinfo{year}{1964}{\natexlab{a}}, \bibinfo{journal}{Sov. Phys. JETP}
  \textbf{\bibinfo{volume}{19}}, \bibinfo{pages}{529}.

\bibitem[{\citenamefont{Nikishov and
  Ritus}(1964{\natexlab{b}})}]{Nikishov_1964_b}
\bibinfo{author}{\bibnamefont{Nikishov}, \bibfnamefont{A.~I.}}, and
  \bibinfo{author}{\bibfnamefont{V.~I.} \bibnamefont{Ritus}},
  \bibinfo{year}{1964}{\natexlab{b}}, \bibinfo{journal}{Sov. Phys. JETP}
  \textbf{\bibinfo{volume}{19}}, \bibinfo{pages}{1191}.

\bibitem[{\citenamefont{{NNDC}}(2011)}]{NuclearDataBase}
\bibinfo{author}{\bibnamefont{{NNDC}}} (\bibinfo{collaboration}{National
  Nuclear Data Center}), \bibinfo{year}{2011},
  \urlprefix\url{http://www.nndc.bnl.gov/}.

\bibitem[{\citenamefont{Noble} \emph{et~al.}(2011)\citenamefont{Noble, Gratus,
  Burton, Ersfeld, Islam, Kravets, Raj, and Jaroszynski}}]{Noble_2011}
\bibinfo{author}{\bibnamefont{Noble}, \bibfnamefont{A.}},
  \bibinfo{author}{\bibfnamefont{J.}~\bibnamefont{Gratus}},
  \bibinfo{author}{\bibfnamefont{D.}~\bibnamefont{Burton}},
  \bibinfo{author}{\bibfnamefont{B.}~\bibnamefont{Ersfeld}},
  \bibinfo{author}{\bibfnamefont{M.~R.} \bibnamefont{Islam}},
  \bibinfo{author}{\bibfnamefont{Y.}~\bibnamefont{Kravets}},
  \bibinfo{author}{\bibfnamefont{G.}~\bibnamefont{Raj}}, and
  \bibinfo{author}{\bibfnamefont{D.}~\bibnamefont{Jaroszynski}},
  \bibinfo{year}{2011}, \bibinfo{journal}{Proc. of SPIE}
  \textbf{\bibinfo{volume}{8079}}, \bibinfo{pages}{80790L}.

\bibitem[{\citenamefont{Olver} \emph{et~al.}(2010)\citenamefont{Olver, Lozier,
  Boisvert, and Clark}}]{NIST_b_2010}
\bibinfo{editor}{\bibnamefont{Olver}, \bibfnamefont{F.~W.~J.}},
  \bibinfo{editor}{\bibfnamefont{D.~W.} \bibnamefont{Lozier}},
  \bibinfo{editor}{\bibfnamefont{R.~F.} \bibnamefont{Boisvert}}, and
  \bibinfo{editor}{\bibfnamefont{C.~W.} \bibnamefont{Clark}} (eds.),
  \bibinfo{year}{2010}, \emph{\bibinfo{title}{NIST Handbook of Mathematical
  Functions}} (\bibinfo{publisher}{Cambridge University Press, Cambridge}).

\bibitem[{\citenamefont{{OMEGA EP}}(2011)}]{OMEGA_EP}
\bibinfo{author}{\bibnamefont{{OMEGA EP}}}, \bibinfo{year}{2011},
  \urlprefix\url{http://www.lle.rochester.edu/omega_facility/omega_ep/index.php}.

\bibitem[{\citenamefont{Omori} \emph{et~al.}(2003)\citenamefont{Omori, Aoki,
  Dobashi, Hirose, Kurihara, Okugi, Sakai, Tsunemi, Urakawa, Washio, and
  Yokoya}}]{Omori_2003}
\bibinfo{author}{\bibnamefont{Omori}, \bibfnamefont{T.}},
  \bibinfo{author}{\bibfnamefont{T.}~\bibnamefont{Aoki}},
  \bibinfo{author}{\bibfnamefont{K.}~\bibnamefont{Dobashi}},
  \bibinfo{author}{\bibfnamefont{T.}~\bibnamefont{Hirose}},
  \bibinfo{author}{\bibfnamefont{Y.}~\bibnamefont{Kurihara}},
  \bibinfo{author}{\bibfnamefont{T.}~\bibnamefont{Okugi}},
  \bibinfo{author}{\bibfnamefont{I.}~\bibnamefont{Sakai}},
  \bibinfo{author}{\bibfnamefont{A.}~\bibnamefont{Tsunemi}},
  \bibinfo{author}{\bibfnamefont{J.}~\bibnamefont{Urakawa}},
  \bibinfo{author}{\bibfnamefont{M.}~\bibnamefont{Washio}}, and
  \bibinfo{author}{\bibfnamefont{K.}~\bibnamefont{Yokoya}},
  \bibinfo{year}{2003}, \bibinfo{journal}{Nucl. Instr. Meth. Phys. Res. A}
  \textbf{\bibinfo{volume}{500}}, \bibinfo{pages}{232}.

\bibitem[{\citenamefont{Palaniyappan}
  \emph{et~al.}(2006)\citenamefont{Palaniyappan, Ghebregziabher, DiChiara,
  MacDonald, and Walker}}]{Walker_2006}
\bibinfo{author}{\bibnamefont{Palaniyappan}, \bibfnamefont{S.}},
  \bibinfo{author}{\bibfnamefont{I.}~\bibnamefont{Ghebregziabher}},
  \bibinfo{author}{\bibfnamefont{A.~D.} \bibnamefont{DiChiara}},
  \bibinfo{author}{\bibfnamefont{J.}~\bibnamefont{MacDonald}}, and
  \bibinfo{author}{\bibfnamefont{B.~C.} \bibnamefont{Walker}},
  \bibinfo{year}{2006}, \bibinfo{journal}{Phys. Rev. A}
  \textbf{\bibinfo{volume}{74}}, \bibinfo{pages}{033403}.

\bibitem[{\citenamefont{Palaniyappan}
  \emph{et~al.}(2008)\citenamefont{Palaniyappan, Mitchell, Sauer,
  Ghebregziabher, White, Decamp, and Walker}}]{Palaniyappan_2008}
\bibinfo{author}{\bibnamefont{Palaniyappan}, \bibfnamefont{S.}},
  \bibinfo{author}{\bibfnamefont{R.}~\bibnamefont{Mitchell}},
  \bibinfo{author}{\bibfnamefont{R.}~\bibnamefont{Sauer}},
  \bibinfo{author}{\bibfnamefont{I.}~\bibnamefont{Ghebregziabher}},
  \bibinfo{author}{\bibfnamefont{S.~L.} \bibnamefont{White}},
  \bibinfo{author}{\bibfnamefont{M.~F.} \bibnamefont{Decamp}}, and
  \bibinfo{author}{\bibfnamefont{B.~C.} \bibnamefont{Walker}},
  \bibinfo{year}{2008}, \bibinfo{journal}{Phys. Rev. Lett.}
  \textbf{\bibinfo{volume}{100}}, \bibinfo{pages}{183001}.

\bibitem[{\citenamefont{P{\'a}lffy}(2008)}]{Palffy2_2008}
\bibinfo{author}{\bibnamefont{P{\'a}lffy}, \bibfnamefont{A.}},
  \bibinfo{year}{2008}, \bibinfo{journal}{J. Mod. Opt.}
  \textbf{\bibinfo{volume}{55}}, \bibinfo{pages}{2603}.

\bibitem[{\citenamefont{P{\'a}lffy}(2010)}]{Palffy_2010}
\bibinfo{author}{\bibnamefont{P{\'a}lffy}, \bibfnamefont{A.}},
  \bibinfo{year}{2010}, \bibinfo{journal}{Contemp. Phys.}
  \textbf{\bibinfo{volume}{51}}, \bibinfo{pages}{471}.

\bibitem[{\citenamefont{P{\'a}lffy}
  \emph{et~al.}(2007)\citenamefont{P{\'a}lffy, Evers, and
  Keitel}}]{Palffy_2007}
\bibinfo{author}{\bibnamefont{P{\'a}lffy}, \bibfnamefont{A.}},
  \bibinfo{author}{\bibfnamefont{J.}~\bibnamefont{Evers}}, and
  \bibinfo{author}{\bibfnamefont{C.~H.} \bibnamefont{Keitel}},
  \bibinfo{year}{2007}, \bibinfo{journal}{Phys. Rev. Lett.}
  \textbf{\bibinfo{volume}{99}}, \bibinfo{pages}{172502}.

\bibitem[{\citenamefont{P{\'a}lffy}
  \emph{et~al.}(2008)\citenamefont{P{\'a}lffy, Evers, and
  Keitel}}]{Palffy_2008}
\bibinfo{author}{\bibnamefont{P{\'a}lffy}, \bibfnamefont{A.}},
  \bibinfo{author}{\bibfnamefont{J.}~\bibnamefont{Evers}}, and
  \bibinfo{author}{\bibfnamefont{C.~H.} \bibnamefont{Keitel}},
  \bibinfo{year}{2008}, \bibinfo{journal}{Phys. Rev. C}
  \textbf{\bibinfo{volume}{77}}, \bibinfo{pages}{044602}.

\bibitem[{\citenamefont{Paramonov}(2007)}]{Paramonov_2007}
\bibinfo{author}{\bibnamefont{Paramonov}, \bibfnamefont{G.~K.}},
  \bibinfo{year}{2007}, \bibinfo{journal}{Chem. Phys.}
  \textbf{\bibinfo{volume}{338}}, \bibinfo{pages}{329}.

\bibitem[{\citenamefont{Peatross} \emph{et~al.}(2008)\citenamefont{Peatross,
  M\"uller, Hatsagortsyan, and Keitel}}]{Peatross_2008}
\bibinfo{author}{\bibnamefont{Peatross}, \bibfnamefont{J.}},
  \bibinfo{author}{\bibfnamefont{C.}~\bibnamefont{M\"uller}},
  \bibinfo{author}{\bibfnamefont{K.~Z.} \bibnamefont{Hatsagortsyan}}, and
  \bibinfo{author}{\bibfnamefont{C.~H.} \bibnamefont{Keitel}},
  \bibinfo{year}{2008}, \bibinfo{journal}{Phys. Rev. Lett.}
  \textbf{\bibinfo{volume}{100}}, \bibinfo{pages}{153601}.

\bibitem[{\citenamefont{Peccei and Quinn}(1977)}]{Peccei_1977}
\bibinfo{author}{\bibnamefont{Peccei}, \bibfnamefont{R.~D.}}, and
  \bibinfo{author}{\bibfnamefont{H.~R.} \bibnamefont{Quinn}},
  \bibinfo{year}{1977}, \bibinfo{journal}{Phys. Rev. Lett.}
  \textbf{\bibinfo{volume}{38}}, \bibinfo{pages}{1440}.

\bibitem[{\citenamefont{Perelomov}(1967)}]{Perelomov_1967}
\bibinfo{author}{\bibnamefont{Perelomov}, \bibfnamefont{A.~M.}},
  \bibinfo{year}{1967}, \bibinfo{journal}{Sov. Phys. JETP}
  \textbf{\bibinfo{volume}{25}}, \bibinfo{pages}{336}.

\bibitem[{\citenamefont{Peskin and Schroeder}(1995)}]{Peskin_1995}
\bibinfo{author}{\bibnamefont{Peskin}, \bibfnamefont{M.~E.}}, and
  \bibinfo{author}{\bibfnamefont{D.~V.} \bibnamefont{Schroeder}},
  \bibinfo{year}{1995}, \emph{\bibinfo{title}{An Introduction to Quantum Field
  Theory}} (\bibinfo{publisher}{Addison-Wesley, Reading}).

\bibitem[{\citenamefont{{PETAL}}(2011)}]{PETAL}
\bibinfo{author}{\bibnamefont{{PETAL}}} (\bibinfo{collaboration}{PETawatt
  Aquitaine Laser}), \bibinfo{year}{2011},
  \urlprefix\url{http://petal.aquitaine.fr/-The-PETAL-laser-facility-.html}.

\bibitem[{\citenamefont{{PFS}}(2011)}]{PFS}
\bibinfo{author}{\bibnamefont{{PFS}}} (\bibinfo{collaboration}{Petawatt Field
  Sinthesizer}), \bibinfo{year}{2011},
  \urlprefix\url{http://www.attoworld.de/Home/attoworld/High-fieldPhysics/ThePetawattFieldSynthesizer/index.html}.

\bibitem[{\citenamefont{{PHELIX}}(2011)}]{PHELIX}
\bibinfo{author}{\bibnamefont{{PHELIX}}} (\bibinfo{collaboration}{Petawatt
  High-Energy Laser for heavy Ion eXperiments}), \bibinfo{year}{2011},
  \urlprefix\url{http://www.gsi.de/forschung/pp/phelix/index_e.html}.

\bibitem[{\citenamefont{Piskarskas}
  \emph{et~al.}(1986)\citenamefont{Piskarskas, Stabinis, and
  Yankauskas}}]{Piskarskas_1986}
\bibinfo{author}{\bibnamefont{Piskarskas}, \bibfnamefont{A.}},
  \bibinfo{author}{\bibfnamefont{A.}~\bibnamefont{Stabinis}}, and
  \bibinfo{author}{\bibfnamefont{A.}~\bibnamefont{Yankauskas}},
  \bibinfo{year}{1986}, \bibinfo{journal}{Sov. Phys. Usp.}
  \textbf{\bibinfo{volume}{29}}, \bibinfo{pages}{869}.

\bibitem[{\citenamefont{Pogorelsky}
  \emph{et~al.}(2000)\citenamefont{Pogorelsky, Ben-Zvi, Hirose, Kashiwagi,
  Yakimenko, Kusche, Siddons, Skaritka, Kumita, Tsunemi, Omori, Urakawa}
  \emph{et~al.}}]{Pogorelsky_2000}
\bibinfo{author}{\bibnamefont{Pogorelsky}, \bibfnamefont{I.~V.}},
  \bibinfo{author}{\bibfnamefont{I.}~\bibnamefont{Ben-Zvi}},
  \bibinfo{author}{\bibfnamefont{T.}~\bibnamefont{Hirose}},
  \bibinfo{author}{\bibfnamefont{S.}~\bibnamefont{Kashiwagi}},
  \bibinfo{author}{\bibfnamefont{V.}~\bibnamefont{Yakimenko}},
  \bibinfo{author}{\bibfnamefont{K.}~\bibnamefont{Kusche}},
  \bibinfo{author}{\bibfnamefont{P.}~\bibnamefont{Siddons}},
  \bibinfo{author}{\bibfnamefont{J.}~\bibnamefont{Skaritka}},
  \bibinfo{author}{\bibfnamefont{T.}~\bibnamefont{Kumita}},
  \bibinfo{author}{\bibfnamefont{A.}~\bibnamefont{Tsunemi}},
  \bibinfo{author}{\bibfnamefont{T.}~\bibnamefont{Omori}},
  \bibinfo{author}{\bibfnamefont{J.}~\bibnamefont{Urakawa}}, \emph{et~al.},
  \bibinfo{year}{2000}, \bibinfo{journal}{Phys. Rev. ST Accel. Beams}
  \textbf{\bibinfo{volume}{3}}, \bibinfo{pages}{090702}.

\bibitem[{\citenamefont{Popmintchev}
  \emph{et~al.}(2009)\citenamefont{Popmintchev, Chen, Bahabad, Gerrity,
  Sidorenko, Cohen, Christov, Murnane, and Kapteyn}}]{Popmintchev_2009}
\bibinfo{author}{\bibnamefont{Popmintchev}, \bibfnamefont{T.}},
  \bibinfo{author}{\bibfnamefont{M.-C.} \bibnamefont{Chen}},
  \bibinfo{author}{\bibfnamefont{A.}~\bibnamefont{Bahabad}},
  \bibinfo{author}{\bibfnamefont{M.}~\bibnamefont{Gerrity}},
  \bibinfo{author}{\bibfnamefont{P.}~\bibnamefont{Sidorenko}},
  \bibinfo{author}{\bibfnamefont{O.}~\bibnamefont{Cohen}},
  \bibinfo{author}{\bibfnamefont{I.~P.} \bibnamefont{Christov}},
  \bibinfo{author}{\bibfnamefont{M.~M.} \bibnamefont{Murnane}}, and
  \bibinfo{author}{\bibfnamefont{H.~C.} \bibnamefont{Kapteyn}},
  \bibinfo{year}{2009}, \bibinfo{journal}{Proc. Natl. Acad. Sci. U.S.A.}
  \textbf{\bibinfo{volume}{106}}, \bibinfo{pages}{10516}.

\bibitem[{\citenamefont{Popov} \emph{et~al.}(1997)\citenamefont{Popov, Mur, and
  Karnakov}}]{Popov_1997}
\bibinfo{author}{\bibnamefont{Popov}, \bibfnamefont{V.}},
  \bibinfo{author}{\bibfnamefont{V.}~\bibnamefont{Mur}}, and
  \bibinfo{author}{\bibfnamefont{B.}~\bibnamefont{Karnakov}},
  \bibinfo{year}{1997}, \bibinfo{journal}{JETP Letters}
  \textbf{\bibinfo{volume}{66}}, \bibinfo{pages}{229}.

\bibitem[{\citenamefont{Popov}(1971)}]{Popov_1971}
\bibinfo{author}{\bibnamefont{Popov}, \bibfnamefont{V.~S.}},
  \bibinfo{year}{1971}, \bibinfo{journal}{JETP Lett.}
  \textbf{\bibinfo{volume}{13}}, \bibinfo{pages}{185}.

\bibitem[{\citenamefont{Popov}(1972)}]{Popov_1972}
\bibinfo{author}{\bibnamefont{Popov}, \bibfnamefont{V.~S.}},
  \bibinfo{year}{1972}, \bibinfo{journal}{Sov. Phys. JETP}
  \textbf{\bibinfo{volume}{34}}, \bibinfo{pages}{709}.

\bibitem[{\citenamefont{Popov}(2004)}]{Popov_2004}
\bibinfo{author}{\bibnamefont{Popov}, \bibfnamefont{V.~S.}},
  \bibinfo{year}{2004}, \bibinfo{journal}{Phys. Usp.}
  \textbf{\bibinfo{volume}{47}}, \bibinfo{pages}{855}.

\bibitem[{\citenamefont{Popov} \emph{et~al.}(2006)\citenamefont{Popov,
  Karnakov, Mur, and Pozdnyakov}}]{Popov_2006}
\bibinfo{author}{\bibnamefont{Popov}, \bibfnamefont{V.~S.}},
  \bibinfo{author}{\bibfnamefont{B.~M.} \bibnamefont{Karnakov}},
  \bibinfo{author}{\bibfnamefont{V.~D.} \bibnamefont{Mur}}, and
  \bibinfo{author}{\bibfnamefont{S.~G.} \bibnamefont{Pozdnyakov}},
  \bibinfo{year}{2006}, \bibinfo{journal}{J. Exp. Theor. Phys.}
  \textbf{\bibinfo{volume}{102}}, \bibinfo{pages}{760}.

\bibitem[{\citenamefont{Postavaru} \emph{et~al.}(2011)\citenamefont{Postavaru,
  Harman, and Keitel}}]{Postavaru_2011}
\bibinfo{author}{\bibnamefont{Postavaru}, \bibfnamefont{O.}},
  \bibinfo{author}{\bibfnamefont{Z.}~\bibnamefont{Harman}}, and
  \bibinfo{author}{\bibfnamefont{C.~H.} \bibnamefont{Keitel}},
  \bibinfo{year}{2011}, \bibinfo{journal}{Phys. Rev. Lett.}
  \textbf{\bibinfo{volume}{106}}, \bibinfo{pages}{033001}.

\bibitem[{\citenamefont{Protopapas}
  \emph{et~al.}(1997)\citenamefont{Protopapas, Keitel, and
  Knight}}]{Protopapas_1997}
\bibinfo{author}{\bibnamefont{Protopapas}, \bibfnamefont{M.}},
  \bibinfo{author}{\bibfnamefont{C.~H.} \bibnamefont{Keitel}}, and
  \bibinfo{author}{\bibfnamefont{P.~L.} \bibnamefont{Knight}},
  \bibinfo{year}{1997}, \bibinfo{journal}{Rep. Progr. Phys.}
  \textbf{\bibinfo{volume}{60}}, \bibinfo{pages}{389}.

\bibitem[{\citenamefont{{PVLAS}}(2011)}]{PVLAS}
\bibinfo{author}{\bibnamefont{{PVLAS}}} (\bibinfo{collaboration}{Polarizzazione
  del Vuoto con LASer}), \bibinfo{year}{2011},
  \urlprefix\url{http://w3.ts.infn.it/experiments/pvlas/}.

\bibitem[{\citenamefont{Rabad\'an} \emph{et~al.}(2006)\citenamefont{Rabad\'an,
  Ringwald, and Sigurdson}}]{Rabadan_2006}
\bibinfo{author}{\bibnamefont{Rabad\'an}, \bibfnamefont{R.}},
  \bibinfo{author}{\bibfnamefont{A.}~\bibnamefont{Ringwald}}, and
  \bibinfo{author}{\bibfnamefont{K.}~\bibnamefont{Sigurdson}},
  \bibinfo{year}{2006}, \bibinfo{journal}{Phys. Rev. Lett.}
  \textbf{\bibinfo{volume}{96}}, \bibinfo{pages}{110407}.

\bibitem[{\citenamefont{Reiss}(1962)}]{Reiss_1962}
\bibinfo{author}{\bibnamefont{Reiss}, \bibfnamefont{H.~R.}},
  \bibinfo{year}{1962}, \bibinfo{journal}{J. Math. Phys.}
  \textbf{\bibinfo{volume}{3}}, \bibinfo{pages}{59}.

\bibitem[{\citenamefont{Reiss}(1971)}]{Reiss_1971}
\bibinfo{author}{\bibnamefont{Reiss}, \bibfnamefont{H.~R.}},
  \bibinfo{year}{1971}, \bibinfo{journal}{Phys. Rev. Lett.}
  \textbf{\bibinfo{volume}{26}}, \bibinfo{pages}{1072}.

\bibitem[{\citenamefont{Reiss}(1979)}]{Reiss_1979}
\bibinfo{author}{\bibnamefont{Reiss}, \bibfnamefont{H.~R.}},
  \bibinfo{year}{1979}, \bibinfo{journal}{Phys. Rev. A}
  \textbf{\bibinfo{volume}{19}}, \bibinfo{pages}{1140}.

\bibitem[{\citenamefont{Reiss}(1980)}]{Reiss_1980}
\bibinfo{author}{\bibnamefont{Reiss}, \bibfnamefont{H.~R.}},
  \bibinfo{year}{1980}, \bibinfo{journal}{Phys. Rev. A}
  \textbf{\bibinfo{volume}{22}}, \bibinfo{pages}{1786}.

\bibitem[{\citenamefont{Reiss}(1983)}]{Reiss1_1983}
\bibinfo{author}{\bibnamefont{Reiss}, \bibfnamefont{H.~R.}},
  \bibinfo{year}{1983}, \bibinfo{journal}{Phys. Rev. C}
  \textbf{\bibinfo{volume}{27}}, \bibinfo{pages}{1199}.

\bibitem[{\citenamefont{Reiss}(1990{\natexlab{a}})}]{Reiss_1990}
\bibinfo{author}{\bibnamefont{Reiss}, \bibfnamefont{H.~R.}},
  \bibinfo{year}{1990}{\natexlab{a}}, \bibinfo{journal}{Phys. Rev. A}
  \textbf{\bibinfo{volume}{42}}, \bibinfo{pages}{1476}.

\bibitem[{\citenamefont{Reiss}(1990{\natexlab{b}})}]{Reiss_1990_JOSAB}
\bibinfo{author}{\bibnamefont{Reiss}, \bibfnamefont{H.~R.}},
  \bibinfo{year}{1990}{\natexlab{b}}, \bibinfo{journal}{J. Opt. Soc. Am. B}
  \textbf{\bibinfo{volume}{7}}, \bibinfo{pages}{574}.

\bibitem[{\citenamefont{Reiss}(2008)}]{Reiss_2008}
\bibinfo{author}{\bibnamefont{Reiss}, \bibfnamefont{H.~R.}},
  \bibinfo{year}{2008}, \bibinfo{journal}{Phys. Rev. Lett.}
  \textbf{\bibinfo{volume}{101}}, \bibinfo{pages}{043002}.

\bibitem[{\citenamefont{Reiss}(2009)}]{Reiss_2009}
\bibinfo{author}{\bibnamefont{Reiss}, \bibfnamefont{H.~R.}},
  \bibinfo{year}{2009}, \bibinfo{journal}{Eur. Phys. J. D}
  \textbf{\bibinfo{volume}{55}}, \bibinfo{pages}{365}.

\bibitem[{\citenamefont{Ridgers} \emph{et~al.}(2012)\citenamefont{Ridgers,
  Brady, Duclous, Kirk, Bennett, Arber, Robinson, and Bell}}]{Ridgers_2012}
\bibinfo{author}{\bibnamefont{Ridgers}, \bibfnamefont{C.~P.}},
  \bibinfo{author}{\bibfnamefont{C.~S.} \bibnamefont{Brady}},
  \bibinfo{author}{\bibfnamefont{R.}~\bibnamefont{Duclous}},
  \bibinfo{author}{\bibfnamefont{J.~G.} \bibnamefont{Kirk}},
  \bibinfo{author}{\bibfnamefont{K.}~\bibnamefont{Bennett}},
  \bibinfo{author}{\bibfnamefont{T.~D.} \bibnamefont{Arber}},
  \bibinfo{author}{\bibfnamefont{A.~P.~L.} \bibnamefont{Robinson}}, and
  \bibinfo{author}{\bibfnamefont{A.~R.} \bibnamefont{Bell}},
  \bibinfo{year}{2012}, \bibinfo{journal}{Phys. Rev. Lett.}
  \textbf{\bibinfo{volume}{108}}, \bibinfo{pages}{165006}.

\bibitem[{\citenamefont{Ringwald}(2001)}]{Ringwald_2001}
\bibinfo{author}{\bibnamefont{Ringwald}, \bibfnamefont{A.}},
  \bibinfo{year}{2001}, \bibinfo{journal}{Phys. Lett. B}
  \textbf{\bibinfo{volume}{510}}, \bibinfo{pages}{107}.

\bibitem[{\citenamefont{Ritus}(1972)}]{Ritus_1972}
\bibinfo{author}{\bibnamefont{Ritus}, \bibfnamefont{V.~I.}},
  \bibinfo{year}{1972}, \bibinfo{journal}{Ann. Phys. (N.Y.)}
  \textbf{\bibinfo{volume}{69}}, \bibinfo{pages}{555}.

\bibitem[{\citenamefont{Ritus}(1985)}]{Ritus_1985}
\bibinfo{author}{\bibnamefont{Ritus}, \bibfnamefont{V.~I.}},
  \bibinfo{year}{1985}, \bibinfo{journal}{J. Sov. Laser Res.}
  \textbf{\bibinfo{volume}{6}}, \bibinfo{pages}{497}.

\bibitem[{\citenamefont{Rohrlich}(2007)}]{Rohrlich_b_2007}
\bibinfo{author}{\bibnamefont{Rohrlich}, \bibfnamefont{F.}},
  \bibinfo{year}{2007}, \emph{\bibinfo{title}{Classical Charged Particles}}
  (\bibinfo{publisher}{World Scientific, Singapore}).

\bibitem[{\citenamefont{Rohrlich}(2008)}]{Rohrlich_2008}
\bibinfo{author}{\bibnamefont{Rohrlich}, \bibfnamefont{F.}},
  \bibinfo{year}{2008}, \bibinfo{journal}{Phys. Rev. E}
  \textbf{\bibinfo{volume}{77}}, \bibinfo{pages}{046609}.

\bibitem[{\citenamefont{Roshchupkin}(2001)}]{Roshchupkin_2001}
\bibinfo{author}{\bibnamefont{Roshchupkin}, \bibfnamefont{S.~P.}},
  \bibinfo{year}{2001}, \bibinfo{journal}{Phys. At. Nucl.}
  \textbf{\bibinfo{volume}{64}}, \bibinfo{pages}{243}.

\bibitem[{\citenamefont{Ruf} \emph{et~al.}(2009)\citenamefont{Ruf, Mocken,
  M\"uller, Hatsagortsyan, and Keitel}}]{Ruf_2009}
\bibinfo{author}{\bibnamefont{Ruf}, \bibfnamefont{M.}},
  \bibinfo{author}{\bibfnamefont{G.~R.} \bibnamefont{Mocken}},
  \bibinfo{author}{\bibfnamefont{C.}~\bibnamefont{M\"uller}},
  \bibinfo{author}{\bibfnamefont{K.~Z.} \bibnamefont{Hatsagortsyan}}, and
  \bibinfo{author}{\bibfnamefont{C.~H.} \bibnamefont{Keitel}},
  \bibinfo{year}{2009}, \bibinfo{journal}{Phys. Rev. Lett.}
  \textbf{\bibinfo{volume}{102}}, \bibinfo{pages}{080402}.

\bibitem[{\citenamefont{Ruffini} \emph{et~al.}(2010)\citenamefont{Ruffini,
  Vereshchagin, and Xue}}]{Ruffini_2010_b}
\bibinfo{author}{\bibnamefont{Ruffini}, \bibfnamefont{R.}},
  \bibinfo{author}{\bibfnamefont{G.}~\bibnamefont{Vereshchagin}}, and
  \bibinfo{author}{\bibfnamefont{S.-S.} \bibnamefont{Xue}},
  \bibinfo{year}{2010}, \bibinfo{journal}{Phys. Rep.}
  \textbf{\bibinfo{volume}{487}}, \bibinfo{pages}{1}.

\bibitem[{\citenamefont{{SACLA}}(2011)}]{SACLA}
\bibinfo{author}{\bibnamefont{{SACLA}}} (\bibinfo{collaboration}{SPring-8
  Angstrom Compact free electron LAser}), \bibinfo{year}{2011},
  \urlprefix\url{http://xfel.riken.jp/eng/}.

\bibitem[{\citenamefont{Sakai} \emph{et~al.}(2003)\citenamefont{Sakai, Aoki,
  Dobashi, Fukuda, Higurashi, Hirose, Iimura, Kurihara, Okugi, Omori, Urakawa,
  Washio} \emph{et~al.}}]{Sakai_2003}
\bibinfo{author}{\bibnamefont{Sakai}, \bibfnamefont{I.}},
  \bibinfo{author}{\bibfnamefont{T.}~\bibnamefont{Aoki}},
  \bibinfo{author}{\bibfnamefont{K.}~\bibnamefont{Dobashi}},
  \bibinfo{author}{\bibfnamefont{M.}~\bibnamefont{Fukuda}},
  \bibinfo{author}{\bibfnamefont{A.}~\bibnamefont{Higurashi}},
  \bibinfo{author}{\bibfnamefont{T.}~\bibnamefont{Hirose}},
  \bibinfo{author}{\bibfnamefont{T.}~\bibnamefont{Iimura}},
  \bibinfo{author}{\bibfnamefont{Y.}~\bibnamefont{Kurihara}},
  \bibinfo{author}{\bibfnamefont{T.}~\bibnamefont{Okugi}},
  \bibinfo{author}{\bibfnamefont{T.}~\bibnamefont{Omori}},
  \bibinfo{author}{\bibfnamefont{J.}~\bibnamefont{Urakawa}},
  \bibinfo{author}{\bibfnamefont{M.}~\bibnamefont{Washio}}, \emph{et~al.},
  \bibinfo{year}{2003}, \bibinfo{journal}{Phys. Rev. ST Accel. Beams}
  \textbf{\bibinfo{volume}{6}}, \bibinfo{pages}{091001}.

\bibitem[{\citenamefont{Salamin}(2006)}]{Salamin_2006b}
\bibinfo{author}{\bibnamefont{Salamin}, \bibfnamefont{Y.~I.}},
  \bibinfo{year}{2006}, \bibinfo{journal}{Phys. Rev. A}
  \textbf{\bibinfo{volume}{73}}, \bibinfo{pages}{043402}.

\bibitem[{\citenamefont{Salamin}(2007)}]{Salamin_2007b}
\bibinfo{author}{\bibnamefont{Salamin}, \bibfnamefont{Y.~I.}},
  \bibinfo{year}{2007}, \bibinfo{journal}{Opt. Lett.}
  \textbf{\bibinfo{volume}{32}}, \bibinfo{pages}{90}.

\bibitem[{\citenamefont{Salamin}(2010)}]{Salamin_2010}
\bibinfo{author}{\bibnamefont{Salamin}, \bibfnamefont{Y.~I.}},
  \bibinfo{year}{2010}, \bibinfo{journal}{Phys. Rev. A}
  \textbf{\bibinfo{volume}{82}}, \bibinfo{pages}{013823}.

\bibitem[{\citenamefont{Salamin}(2011)}]{Salamin_2011}
\bibinfo{author}{\bibnamefont{Salamin}, \bibfnamefont{Y.~I.}},
  \bibinfo{year}{2011}, \bibinfo{journal}{Phys. Rev. ST Accel. Beams}
  \textbf{\bibinfo{volume}{14}}, \bibinfo{pages}{071302}.

\bibitem[{\citenamefont{Salamin and Faisal}(1998)}]{Salamin_1998}
\bibinfo{author}{\bibnamefont{Salamin}, \bibfnamefont{Y.~I.}}, and
  \bibinfo{author}{\bibfnamefont{F.~H.~M.} \bibnamefont{Faisal}},
  \bibinfo{year}{1998}, \bibinfo{journal}{Phys. Rev. A}
  \textbf{\bibinfo{volume}{54}}, \bibinfo{pages}{4383}.

\bibitem[{\citenamefont{Salamin} \emph{et~al.}(2008)\citenamefont{Salamin,
  Harman, and Keitel}}]{Salamin_2008}
\bibinfo{author}{\bibnamefont{Salamin}, \bibfnamefont{Y.~I.}},
  \bibinfo{author}{\bibfnamefont{Z.}~\bibnamefont{Harman}}, and
  \bibinfo{author}{\bibfnamefont{C.~H.} \bibnamefont{Keitel}},
  \bibinfo{year}{2008}, \bibinfo{journal}{Phys. Rev. Lett.}
  \textbf{\bibinfo{volume}{100}}, \bibinfo{pages}{155004}.

\bibitem[{\citenamefont{Salamin} \emph{et~al.}(2006)\citenamefont{Salamin, Hu,
  Hatsagortsyan, and Keitel}}]{Salamin_2006}
\bibinfo{author}{\bibnamefont{Salamin}, \bibfnamefont{Y.~I.}},
  \bibinfo{author}{\bibfnamefont{S.~X.} \bibnamefont{Hu}},
  \bibinfo{author}{\bibfnamefont{K.~Z.} \bibnamefont{Hatsagortsyan}}, and
  \bibinfo{author}{\bibfnamefont{C.~H.} \bibnamefont{Keitel}},
  \bibinfo{year}{2006}, \bibinfo{journal}{Phys. Rep.}
  \textbf{\bibinfo{volume}{427}}, \bibinfo{pages}{41}.

\bibitem[{\citenamefont{Salamin and Keitel}(2002)}]{Salamin_2002a}
\bibinfo{author}{\bibnamefont{Salamin}, \bibfnamefont{Y.~I.}}, and
  \bibinfo{author}{\bibfnamefont{C.~H.} \bibnamefont{Keitel}},
  \bibinfo{year}{2002}, \bibinfo{journal}{Phys. Rev. Lett.}
  \textbf{\bibinfo{volume}{88}}, \bibinfo{pages}{095005}.

\bibitem[{\citenamefont{Salamin} \emph{et~al.}(2002)\citenamefont{Salamin,
  Mocken, and Keitel}}]{Salamin_2002b}
\bibinfo{author}{\bibnamefont{Salamin}, \bibfnamefont{Y.~I.}},
  \bibinfo{author}{\bibfnamefont{G.~R.} \bibnamefont{Mocken}}, and
  \bibinfo{author}{\bibfnamefont{C.~H.} \bibnamefont{Keitel}},
  \bibinfo{year}{2002}, \bibinfo{journal}{Phys. Rev. ST Accel. Beams}
  \textbf{\bibinfo{volume}{5}}, \bibinfo{pages}{101301}.

\bibitem[{\citenamefont{Salamin} \emph{et~al.}(2003)\citenamefont{Salamin,
  Mocken, and Keitel}}]{Salamin_2003}
\bibinfo{author}{\bibnamefont{Salamin}, \bibfnamefont{Y.~I.}},
  \bibinfo{author}{\bibfnamefont{G.~R.} \bibnamefont{Mocken}}, and
  \bibinfo{author}{\bibfnamefont{C.~H.} \bibnamefont{Keitel}},
  \bibinfo{year}{2003}, \bibinfo{journal}{Phys. Rev. E}
  \textbf{\bibinfo{volume}{67}}, \bibinfo{pages}{016501}.

\bibitem[{\citenamefont{Sansone} \emph{et~al.}(2006)\citenamefont{Sansone,
  Benedetti, Calegari, Vozzi, Avaldi, Flammini, Poletto, Villoresi, Altucci,
  Velotta, Stagira, De~Silvestri} \emph{et~al.}}]{Sansone_2006}
\bibinfo{author}{\bibnamefont{Sansone}, \bibfnamefont{G.}},
  \bibinfo{author}{\bibfnamefont{E.}~\bibnamefont{Benedetti}},
  \bibinfo{author}{\bibfnamefont{F.}~\bibnamefont{Calegari}},
  \bibinfo{author}{\bibfnamefont{C.}~\bibnamefont{Vozzi}},
  \bibinfo{author}{\bibfnamefont{L.}~\bibnamefont{Avaldi}},
  \bibinfo{author}{\bibfnamefont{R.}~\bibnamefont{Flammini}},
  \bibinfo{author}{\bibfnamefont{L.}~\bibnamefont{Poletto}},
  \bibinfo{author}{\bibfnamefont{P.}~\bibnamefont{Villoresi}},
  \bibinfo{author}{\bibfnamefont{C.}~\bibnamefont{Altucci}},
  \bibinfo{author}{\bibfnamefont{R.}~\bibnamefont{Velotta}},
  \bibinfo{author}{\bibfnamefont{S.}~\bibnamefont{Stagira}},
  \bibinfo{author}{\bibfnamefont{S.}~\bibnamefont{De~Silvestri}},
  \emph{et~al.}, \bibinfo{year}{2006}, \bibinfo{journal}{Science}
  \textbf{\bibinfo{volume}{314}}, \bibinfo{pages}{443}.

\bibitem[{\citenamefont{Sarachik and Schappert}(1970)}]{Sarachik_1970}
\bibinfo{author}{\bibnamefont{Sarachik}, \bibfnamefont{E.~S.}}, and
  \bibinfo{author}{\bibfnamefont{G.~T.} \bibnamefont{Schappert}},
  \bibinfo{year}{1970}, \bibinfo{journal}{Phys. Rev. D}
  \textbf{\bibinfo{volume}{1}}, \bibinfo{pages}{2738}.

\bibitem[{\citenamefont{Sauter}(1931)}]{Sauter_1931}
\bibinfo{author}{\bibnamefont{Sauter}, \bibfnamefont{F.}},
  \bibinfo{year}{1931}, \bibinfo{journal}{Z. Phys.}
  \textbf{\bibinfo{volume}{69}}, \bibinfo{pages}{742}.

\bibitem[{\citenamefont{{SCAPA}}(2012)}]{SCAPA_2012}
\bibinfo{author}{\bibnamefont{{SCAPA}}} (\bibinfo{collaboration}{Scottish
  Centre for the Application of Plasma-based Accelerators}),
  \bibinfo{year}{2012}, \urlprefix\url{http://www.scapa.ac.uk/}.

\bibitem[{\citenamefont{Schafer} \emph{et~al.}(1993)\citenamefont{Schafer,
  Yang, DiMauro, and Kulander}}]{Schafer_1993}
\bibinfo{author}{\bibnamefont{Schafer}, \bibfnamefont{K.~J.}},
  \bibinfo{author}{\bibfnamefont{B.}~\bibnamefont{Yang}},
  \bibinfo{author}{\bibfnamefont{L.~F.} \bibnamefont{DiMauro}}, and
  \bibinfo{author}{\bibfnamefont{K.~C.} \bibnamefont{Kulander}},
  \bibinfo{year}{1993}, \bibinfo{journal}{Phys. Rev. Lett.}
  \textbf{\bibinfo{volume}{70}}, \bibinfo{pages}{1599}.

\bibitem[{\citenamefont{Schlegel} \emph{et~al.}(2009)\citenamefont{Schlegel,
  Naumova, Tikhonchuk, Labaune, Sokolov, and Mourou}}]{Schlegel_2009}
\bibinfo{author}{\bibnamefont{Schlegel}, \bibfnamefont{T.}},
  \bibinfo{author}{\bibfnamefont{N.}~\bibnamefont{Naumova}},
  \bibinfo{author}{\bibfnamefont{V.~T.} \bibnamefont{Tikhonchuk}},
  \bibinfo{author}{\bibfnamefont{C.}~\bibnamefont{Labaune}},
  \bibinfo{author}{\bibfnamefont{I.~V.} \bibnamefont{Sokolov}}, and
  \bibinfo{author}{\bibfnamefont{G.}~\bibnamefont{Mourou}},
  \bibinfo{year}{2009}, \bibinfo{journal}{Phys. Plasmas}
  \textbf{\bibinfo{volume}{16}}, \bibinfo{pages}{083103}.

\bibitem[{\citenamefont{Schlenvoigt}
  \emph{et~al.}(2008)\citenamefont{Schlenvoigt, Haupt, Debus, Budde, J\"ackel,
  Pfotenhauer, Schwoerer, Rohwer, Gallacher, Brunetti, Shanks, Wiggins}
  \emph{et~al.}}]{Schlenvoigt_2008}
\bibinfo{author}{\bibnamefont{Schlenvoigt}, \bibfnamefont{H.-P.}},
  \bibinfo{author}{\bibfnamefont{K.}~\bibnamefont{Haupt}},
  \bibinfo{author}{\bibfnamefont{A.}~\bibnamefont{Debus}},
  \bibinfo{author}{\bibfnamefont{F.}~\bibnamefont{Budde}},
  \bibinfo{author}{\bibfnamefont{O.}~\bibnamefont{J\"ackel}},
  \bibinfo{author}{\bibfnamefont{S.}~\bibnamefont{Pfotenhauer}},
  \bibinfo{author}{\bibfnamefont{H.}~\bibnamefont{Schwoerer}},
  \bibinfo{author}{\bibfnamefont{E.}~\bibnamefont{Rohwer}},
  \bibinfo{author}{\bibfnamefont{J.~G.} \bibnamefont{Gallacher}},
  \bibinfo{author}{\bibfnamefont{E.}~\bibnamefont{Brunetti}},
  \bibinfo{author}{\bibfnamefont{R.~P.} \bibnamefont{Shanks}},
  \bibinfo{author}{\bibfnamefont{S.~M.} \bibnamefont{Wiggins}}, \emph{et~al.},
  \bibinfo{year}{2008}, \bibinfo{journal}{Nature Phys.}
  \textbf{\bibinfo{volume}{4}}, \bibinfo{pages}{130}.

\bibitem[{\citenamefont{Schoenlein}
  \emph{et~al.}(1996)\citenamefont{Schoenlein, Leemans, Chin, Volfbeyn, Glover,
  Balling, Zolotorev, Kim, Chattopadhyay, and Shank}}]{Schoenlein_1996}
\bibinfo{author}{\bibnamefont{Schoenlein}, \bibfnamefont{R.~W.}},
  \bibinfo{author}{\bibfnamefont{W.~P.} \bibnamefont{Leemans}},
  \bibinfo{author}{\bibfnamefont{A.~H.} \bibnamefont{Chin}},
  \bibinfo{author}{\bibfnamefont{P.}~\bibnamefont{Volfbeyn}},
  \bibinfo{author}{\bibfnamefont{T.~E.} \bibnamefont{Glover}},
  \bibinfo{author}{\bibfnamefont{P.}~\bibnamefont{Balling}},
  \bibinfo{author}{\bibfnamefont{M.}~\bibnamefont{Zolotorev}},
  \bibinfo{author}{\bibfnamefont{K.-J.} \bibnamefont{Kim}},
  \bibinfo{author}{\bibfnamefont{S.}~\bibnamefont{Chattopadhyay}}, and
  \bibinfo{author}{\bibfnamefont{C.~V.} \bibnamefont{Shank}},
  \bibinfo{year}{1996}, \bibinfo{journal}{Science}
  \textbf{\bibinfo{volume}{274}}, \bibinfo{pages}{236}.

\bibitem[{\citenamefont{Schroeder} \emph{et~al.}(2010)\citenamefont{Schroeder,
  Esarey, Geddes, Benedetti, and Leemans}}]{Schroeder_2010}
\bibinfo{author}{\bibnamefont{Schroeder}, \bibfnamefont{C.~B.}},
  \bibinfo{author}{\bibfnamefont{E.}~\bibnamefont{Esarey}},
  \bibinfo{author}{\bibfnamefont{C.~G.~R.} \bibnamefont{Geddes}},
  \bibinfo{author}{\bibfnamefont{C.}~\bibnamefont{Benedetti}}, and
  \bibinfo{author}{\bibfnamefont{W.~P.} \bibnamefont{Leemans}},
  \bibinfo{year}{2010}, \bibinfo{journal}{Phys. Rev. ST Accel. Beams}
  \textbf{\bibinfo{volume}{13}}, \bibinfo{pages}{101301}.

\bibitem[{\citenamefont{Sch\"utzhold}
  \emph{et~al.}(2008)\citenamefont{Sch\"utzhold, Gies, and
  Dunne}}]{Schutzhold_2008}
\bibinfo{author}{\bibnamefont{Sch\"utzhold}, \bibfnamefont{R.}},
  \bibinfo{author}{\bibfnamefont{H.}~\bibnamefont{Gies}}, and
  \bibinfo{author}{\bibfnamefont{G.~V.} \bibnamefont{Dunne}},
  \bibinfo{year}{2008}, \bibinfo{journal}{Phys. Rev. Lett.}
  \textbf{\bibinfo{volume}{101}}, \bibinfo{pages}{130404}.

\bibitem[{\citenamefont{Schwinger}(1951)}]{Schwinger_1951}
\bibinfo{author}{\bibnamefont{Schwinger}, \bibfnamefont{J.}},
  \bibinfo{year}{1951}, \bibinfo{journal}{Phys. Rev.}
  \textbf{\bibinfo{volume}{82}}, \bibinfo{pages}{664}.

\bibitem[{\citenamefont{Schwoerer} \emph{et~al.}(2006)\citenamefont{Schwoerer,
  Pfotenhauer, J\"ackel, Amthor, Liesfeld, Ziegler, Sauerbrey, Ledingham, and
  Esirkepov}}]{Schwoerer_2006}
\bibinfo{author}{\bibnamefont{Schwoerer}, \bibfnamefont{H.}},
  \bibinfo{author}{\bibfnamefont{S.}~\bibnamefont{Pfotenhauer}},
  \bibinfo{author}{\bibfnamefont{O.}~\bibnamefont{J\"ackel}},
  \bibinfo{author}{\bibfnamefont{K.-U.} \bibnamefont{Amthor}},
  \bibinfo{author}{\bibfnamefont{B.}~\bibnamefont{Liesfeld}},
  \bibinfo{author}{\bibfnamefont{W.}~\bibnamefont{Ziegler}},
  \bibinfo{author}{\bibfnamefont{R.}~\bibnamefont{Sauerbrey}},
  \bibinfo{author}{\bibfnamefont{K.~W.~D.} \bibnamefont{Ledingham}}, and
  \bibinfo{author}{\bibfnamefont{T.}~\bibnamefont{Esirkepov}},
  \bibinfo{year}{2006}, \bibinfo{journal}{Nature (London)}
  \textbf{\bibinfo{volume}{439}}, \bibinfo{pages}{445}.

\bibitem[{\citenamefont{Seipt and K\"ampfer}(2011{\natexlab{a}})}]{Seipt_2011}
\bibinfo{author}{\bibnamefont{Seipt}, \bibfnamefont{D.}}, and
  \bibinfo{author}{\bibfnamefont{B.}~\bibnamefont{K\"ampfer}},
  \bibinfo{year}{2011}{\natexlab{a}}, \bibinfo{journal}{Phys. Rev. A}
  \textbf{\bibinfo{volume}{83}}, \bibinfo{pages}{022101}.

\bibitem[{\citenamefont{Seipt and K\"ampfer}(2011{\natexlab{b}})}]{Seipt_2011b}
\bibinfo{author}{\bibnamefont{Seipt}, \bibfnamefont{D.}}, and
  \bibinfo{author}{\bibfnamefont{B.}~\bibnamefont{K\"ampfer}},
  \bibinfo{year}{2011}{\natexlab{b}}, \bibinfo{journal}{Phys. Rev. ST Accel.
  Beams} \textbf{\bibinfo{volume}{14}}, \bibinfo{pages}{040704}.

\bibitem[{\citenamefont{Selst\o{}} \emph{et~al.}(2009)\citenamefont{Selst\o{},
  Lindroth, and Bengtsson}}]{Selsto_2009}
\bibinfo{author}{\bibnamefont{Selst\o{}}, \bibfnamefont{S.}},
  \bibinfo{author}{\bibfnamefont{E.}~\bibnamefont{Lindroth}}, and
  \bibinfo{author}{\bibfnamefont{J.}~\bibnamefont{Bengtsson}},
  \bibinfo{year}{2009}, \bibinfo{journal}{Phys. Rev. A}
  \textbf{\bibinfo{volume}{79}}, \bibinfo{pages}{043418}.

\bibitem[{\citenamefont{Sengupta}(1949)}]{Sengupta_1949}
\bibinfo{author}{\bibnamefont{Sengupta}, \bibfnamefont{N.~D.}},
  \bibinfo{year}{1949}, \bibinfo{journal}{Bull. Math. Soc. (Calcutta)}
  \textbf{\bibinfo{volume}{41}}, \bibinfo{pages}{187}.

\bibitem[{\citenamefont{Seto} \emph{et~al.}(2011)\citenamefont{Seto, Nagatomo,
  Koga, and Mima}}]{Seto_2011}
\bibinfo{author}{\bibnamefont{Seto}, \bibfnamefont{K.}},
  \bibinfo{author}{\bibfnamefont{H.}~\bibnamefont{Nagatomo}},
  \bibinfo{author}{\bibfnamefont{J.}~\bibnamefont{Koga}}, and
  \bibinfo{author}{\bibfnamefont{K.}~\bibnamefont{Mima}}, \bibinfo{year}{2011},
  \bibinfo{journal}{Phys. Plasmas} \textbf{\bibinfo{volume}{18}},
  \bibinfo{pages}{123101}.

\bibitem[{\citenamefont{Shahbaz} \emph{et~al.}(2010)\citenamefont{Shahbaz,
  B\"urvenich, and M\"uller}}]{Shahbaz_2010}
\bibinfo{author}{\bibnamefont{Shahbaz}, \bibfnamefont{A.}},
  \bibinfo{author}{\bibfnamefont{T.~J.} \bibnamefont{B\"urvenich}}, and
  \bibinfo{author}{\bibfnamefont{C.}~\bibnamefont{M\"uller}},
  \bibinfo{year}{2010}, \bibinfo{journal}{Phys. Rev. A}
  \textbf{\bibinfo{volume}{82}}, \bibinfo{pages}{013418}.

\bibitem[{\citenamefont{Shahbaz} \emph{et~al.}(2009)\citenamefont{Shahbaz,
  M\"uller, B\"urvenich, and Keitel}}]{Shahbaz_2009}
\bibinfo{author}{\bibnamefont{Shahbaz}, \bibfnamefont{A.}},
  \bibinfo{author}{\bibfnamefont{C.}~\bibnamefont{M\"uller}},
  \bibinfo{author}{\bibfnamefont{T.~J.} \bibnamefont{B\"urvenich}}, and
  \bibinfo{author}{\bibfnamefont{C.~H.} \bibnamefont{Keitel}},
  \bibinfo{year}{2009}, \bibinfo{journal}{Nucl. Phys. A}
  \textbf{\bibinfo{volume}{821}}, \bibinfo{pages}{106}.

\bibitem[{\citenamefont{Shahbaz} \emph{et~al.}(2007)\citenamefont{Shahbaz,
  M\"uller, Staudt, B\"urvenich, and Keitel}}]{Shahbaz_2007}
\bibinfo{author}{\bibnamefont{Shahbaz}, \bibfnamefont{A.}},
  \bibinfo{author}{\bibfnamefont{C.}~\bibnamefont{M\"uller}},
  \bibinfo{author}{\bibfnamefont{A.}~\bibnamefont{Staudt}},
  \bibinfo{author}{\bibfnamefont{T.~J.} \bibnamefont{B\"urvenich}}, and
  \bibinfo{author}{\bibfnamefont{C.~H.} \bibnamefont{Keitel}},
  \bibinfo{year}{2007}, \bibinfo{journal}{Phys. Rev. Lett.}
  \textbf{\bibinfo{volume}{98}}, \bibinfo{pages}{263901}.

\bibitem[{\citenamefont{Shen and {Meyer-ter-Vehn}}(2001)}]{Meyer-ter-Vehn_2001}
\bibinfo{author}{\bibnamefont{Shen}, \bibfnamefont{B.}}, and
  \bibinfo{author}{\bibfnamefont{J.}~\bibnamefont{{Meyer-ter-Vehn}}},
  \bibinfo{year}{2001}, \bibinfo{journal}{Phys. Rev. E}
  \textbf{\bibinfo{volume}{65}}, \bibinfo{pages}{016405}.

\bibitem[{\citenamefont{Shen}(1970)}]{Shen_1970}
\bibinfo{author}{\bibnamefont{Shen}, \bibfnamefont{C.~S.}},
  \bibinfo{year}{1970}, \bibinfo{journal}{Phys. Rev. Lett.}
  \textbf{\bibinfo{volume}{24}}, \bibinfo{pages}{410}.

\bibitem[{\citenamefont{Shvyd'ko} \emph{et~al.}(2011)\citenamefont{Shvyd'ko,
  Stoupin, Blank, and Terentyev}}]{Shvydko_2011}
\bibinfo{author}{\bibnamefont{Shvyd'ko}, \bibfnamefont{{\relax Yu}.}},
  \bibinfo{author}{\bibfnamefont{S.}~\bibnamefont{Stoupin}},
  \bibinfo{author}{\bibfnamefont{V.}~\bibnamefont{Blank}}, and
  \bibinfo{author}{\bibfnamefont{S.}~\bibnamefont{Terentyev}},
  \bibinfo{year}{2011}, \bibinfo{journal}{Nature Photon.}
  \textbf{\bibinfo{volume}{5}}, \bibinfo{pages}{539}.

\bibitem[{\citenamefont{Sieczka} \emph{et~al.}(2006)\citenamefont{Sieczka,
  Krajewska, Kaminski, Panek, and Ehlotzky}}]{Sieczka_2006}
\bibinfo{author}{\bibnamefont{Sieczka}, \bibfnamefont{P.}},
  \bibinfo{author}{\bibfnamefont{K.}~\bibnamefont{Krajewska}},
  \bibinfo{author}{\bibfnamefont{J.~Z.} \bibnamefont{Kaminski}},
  \bibinfo{author}{\bibfnamefont{P.}~\bibnamefont{Panek}}, and
  \bibinfo{author}{\bibfnamefont{F.}~\bibnamefont{Ehlotzky}},
  \bibinfo{year}{2006}, \bibinfo{journal}{Phys. Rev. A}
  \textbf{\bibinfo{volume}{73}}, \bibinfo{pages}{053409}.

\bibitem[{\citenamefont{Smeenk} \emph{et~al.}(2011)\citenamefont{Smeenk,
  Arissian, Zhou, Mysyrowicz, Villeneuve, Staudte, and Corkum}}]{Smeenk_2011}
\bibinfo{author}{\bibnamefont{Smeenk}, \bibfnamefont{C.~T.~L.}},
  \bibinfo{author}{\bibfnamefont{L.}~\bibnamefont{Arissian}},
  \bibinfo{author}{\bibfnamefont{B.}~\bibnamefont{Zhou}},
  \bibinfo{author}{\bibfnamefont{A.}~\bibnamefont{Mysyrowicz}},
  \bibinfo{author}{\bibfnamefont{D.~M.} \bibnamefont{Villeneuve}},
  \bibinfo{author}{\bibfnamefont{A.}~\bibnamefont{Staudte}}, and
  \bibinfo{author}{\bibfnamefont{P.~B.} \bibnamefont{Corkum}},
  \bibinfo{year}{2011}, \bibinfo{journal}{Phys. Rev. Lett.}
  \textbf{\bibinfo{volume}{106}}, \bibinfo{pages}{193002}.

\bibitem[{\citenamefont{Smorenburg}
  \emph{et~al.}(2010)\citenamefont{Smorenburg, Kamp, Geloni, and
  Luiten}}]{Smorenburg_2010}
\bibinfo{author}{\bibnamefont{Smorenburg}, \bibfnamefont{P.~W.}},
  \bibinfo{author}{\bibfnamefont{L.~P.~J.} \bibnamefont{Kamp}},
  \bibinfo{author}{\bibfnamefont{G.~A.} \bibnamefont{Geloni}}, and
  \bibinfo{author}{\bibfnamefont{O.~J.} \bibnamefont{Luiten}},
  \bibinfo{year}{2010}, \bibinfo{journal}{Laser Part. Beams}
  \textbf{\bibinfo{volume}{28}}, \bibinfo{pages}{553}.

\bibitem[{\citenamefont{Sokolov}
  \emph{et~al.}(2010{\natexlab{a}})\citenamefont{Sokolov, Naumova, Nees, and
  Mourou}}]{Sokolov_2010}
\bibinfo{author}{\bibnamefont{Sokolov}, \bibfnamefont{I.~V.}},
  \bibinfo{author}{\bibfnamefont{N.~M.} \bibnamefont{Naumova}},
  \bibinfo{author}{\bibfnamefont{J.~A.} \bibnamefont{Nees}}, and
  \bibinfo{author}{\bibfnamefont{G.~A.} \bibnamefont{Mourou}},
  \bibinfo{year}{2010}{\natexlab{a}}, \bibinfo{journal}{Phys. Rev. Lett.}
  \textbf{\bibinfo{volume}{105}}, \bibinfo{pages}{195005}.

\bibitem[{\citenamefont{Sokolov} \emph{et~al.}(2009)\citenamefont{Sokolov,
  Naumova, Nees, Mourou, and Yanovsky}}]{Sokolov_2009}
\bibinfo{author}{\bibnamefont{Sokolov}, \bibfnamefont{I.~V.}},
  \bibinfo{author}{\bibfnamefont{N.~M.} \bibnamefont{Naumova}},
  \bibinfo{author}{\bibfnamefont{J.~A.} \bibnamefont{Nees}},
  \bibinfo{author}{\bibfnamefont{G.~A.} \bibnamefont{Mourou}}, and
  \bibinfo{author}{\bibfnamefont{V.~P.} \bibnamefont{Yanovsky}},
  \bibinfo{year}{2009}, \bibinfo{journal}{Phys. Plasmas}
  \textbf{\bibinfo{volume}{16}}, \bibinfo{pages}{093115}.

\bibitem[{\citenamefont{Sokolov}
  \emph{et~al.}(2010{\natexlab{b}})\citenamefont{Sokolov, Nees, Yanovsky,
  Naumova, and Mourou}}]{Sokolov_2010b}
\bibinfo{author}{\bibnamefont{Sokolov}, \bibfnamefont{I.~V.}},
  \bibinfo{author}{\bibfnamefont{J.~A.} \bibnamefont{Nees}},
  \bibinfo{author}{\bibfnamefont{V.~P.} \bibnamefont{Yanovsky}},
  \bibinfo{author}{\bibfnamefont{N.~M.} \bibnamefont{Naumova}}, and
  \bibinfo{author}{\bibfnamefont{G.~A.} \bibnamefont{Mourou}},
  \bibinfo{year}{2010}{\natexlab{b}}, \bibinfo{journal}{Phys. Rev. E}
  \textbf{\bibinfo{volume}{81}}, \bibinfo{pages}{036412}.

\bibitem[{\citenamefont{Spohn}(2000)}]{Spohn_2000}
\bibinfo{author}{\bibnamefont{Spohn}, \bibfnamefont{H.}}, \bibinfo{year}{2000},
  \bibinfo{journal}{Europhys. Lett.} \textbf{\bibinfo{volume}{50}},
  \bibinfo{pages}{287}.

\bibitem[{\citenamefont{{SPring8}}(2011)}]{SPring8}
\bibinfo{author}{\bibnamefont{{SPring8}}} (\bibinfo{collaboration}{Super Photon
  Ring - 8 GeV}), \bibinfo{year}{2011},
  \urlprefix\url{http://www.spring8.or.jp/en/}.

\bibitem[{\citenamefont{Strickland and Mourou}(1985)}]{Strickland_1985}
\bibinfo{author}{\bibnamefont{Strickland}, \bibfnamefont{D.}}, and
  \bibinfo{author}{\bibfnamefont{G.}~\bibnamefont{Mourou}},
  \bibinfo{year}{1985}, \bibinfo{journal}{Opt. Commun.}
  \textbf{\bibinfo{volume}{55}}, \bibinfo{pages}{447}.

\bibitem[{\citenamefont{Suckewer and Jaegle}(2009)}]{Suckewer_2009}
\bibinfo{author}{\bibnamefont{Suckewer}, \bibfnamefont{S.}}, and
  \bibinfo{author}{\bibfnamefont{P.}~\bibnamefont{Jaegle}},
  \bibinfo{year}{2009}, \bibinfo{journal}{Laser Phys. Lett.}
  \textbf{\bibinfo{volume}{6}}, \bibinfo{pages}{411}.

\bibitem[{\citenamefont{Sun and Wu}(2011)}]{Sun_2011}
\bibinfo{author}{\bibnamefont{Sun}, \bibfnamefont{C.}}, and
  \bibinfo{author}{\bibfnamefont{Y.~K.} \bibnamefont{Wu}},
  \bibinfo{year}{2011}, \bibinfo{journal}{Phys. Rev. ST Accel. Beams}
  \textbf{\bibinfo{volume}{14}}, \bibinfo{pages}{044701}.

\bibitem[{\citenamefont{Sung} \emph{et~al.}(2010)\citenamefont{Sung, Lee, Yu,
  Jeong, and Lee}}]{Sung_2010}
\bibinfo{author}{\bibnamefont{Sung}, \bibfnamefont{J.~H.}},
  \bibinfo{author}{\bibfnamefont{S.~K.} \bibnamefont{Lee}},
  \bibinfo{author}{\bibfnamefont{T.~J.} \bibnamefont{Yu}},
  \bibinfo{author}{\bibfnamefont{T.~M.} \bibnamefont{Jeong}}, and
  \bibinfo{author}{\bibfnamefont{J.}~\bibnamefont{Lee}}, \bibinfo{year}{2010},
  \bibinfo{journal}{Opt. Lett.} \textbf{\bibinfo{volume}{35}},
  \bibinfo{pages}{3021}.

\bibitem[{\citenamefont{Tabak} \emph{et~al.}(1994)\citenamefont{Tabak, Hammer,
  Glinsky, Kruer, Wilks, Woodworth, Campbell, Perry, and Mason}}]{Tabak_1994}
\bibinfo{author}{\bibnamefont{Tabak}, \bibfnamefont{M.}},
  \bibinfo{author}{\bibfnamefont{J.}~\bibnamefont{Hammer}},
  \bibinfo{author}{\bibfnamefont{M.~E.} \bibnamefont{Glinsky}},
  \bibinfo{author}{\bibfnamefont{W.~L.} \bibnamefont{Kruer}},
  \bibinfo{author}{\bibfnamefont{S.~C.} \bibnamefont{Wilks}},
  \bibinfo{author}{\bibfnamefont{J.}~\bibnamefont{Woodworth}},
  \bibinfo{author}{\bibfnamefont{E.~M.} \bibnamefont{Campbell}},
  \bibinfo{author}{\bibfnamefont{M.~D.} \bibnamefont{Perry}}, and
  \bibinfo{author}{\bibfnamefont{R.~J.} \bibnamefont{Mason}},
  \bibinfo{year}{1994}, \bibinfo{journal}{Phys. Plasmas}
  \textbf{\bibinfo{volume}{1}}, \bibinfo{pages}{1626}.

\bibitem[{\citenamefont{Tajima and Dawson}(1979)}]{Tajima_1979}
\bibinfo{author}{\bibnamefont{Tajima}, \bibfnamefont{T.}}, and
  \bibinfo{author}{\bibfnamefont{J.~M.} \bibnamefont{Dawson}},
  \bibinfo{year}{1979}, \bibinfo{journal}{Phys. Rev. Lett.}
  \textbf{\bibinfo{volume}{43}}, \bibinfo{pages}{267}.

\bibitem[{\citenamefont{Tajima} \emph{et~al.}(2010)\citenamefont{Tajima, Habs,
  and Mourou}}]{Tajima_2010}
\bibinfo{author}{\bibnamefont{Tajima}, \bibfnamefont{T.}},
  \bibinfo{author}{\bibfnamefont{D.}~\bibnamefont{Habs}}, and
  \bibinfo{author}{\bibfnamefont{G.~A.} \bibnamefont{Mourou}},
  \bibinfo{year}{2010}, \bibinfo{journal}{Optik \& Photonik}
  \textbf{\bibinfo{volume}{5}}, \bibinfo{pages}{24}.

\bibitem[{\citenamefont{Tamburini} \emph{et~al.}(2012)\citenamefont{Tamburini,
  Liseykina, Pegoraro, and Macchi}}]{Tamburini_2011}
\bibinfo{author}{\bibnamefont{Tamburini}, \bibfnamefont{M.}},
  \bibinfo{author}{\bibfnamefont{T.~V.} \bibnamefont{Liseykina}},
  \bibinfo{author}{\bibfnamefont{F.}~\bibnamefont{Pegoraro}}, and
  \bibinfo{author}{\bibfnamefont{A.}~\bibnamefont{Macchi}},
  \bibinfo{year}{2012}, \bibinfo{journal}{Phys. Rev. E}
  \textbf{\bibinfo{volume}{85}}, \bibinfo{pages}{016407}.

\bibitem[{\citenamefont{Tamburini} \emph{et~al.}(2010)\citenamefont{Tamburini,
  Pegoraro, Di~Piazza, Keitel, and Macchi}}]{Tamburini_2010}
\bibinfo{author}{\bibnamefont{Tamburini}, \bibfnamefont{M.}},
  \bibinfo{author}{\bibfnamefont{F.}~\bibnamefont{Pegoraro}},
  \bibinfo{author}{\bibfnamefont{A.}~\bibnamefont{Di~Piazza}},
  \bibinfo{author}{\bibfnamefont{C.~H.} \bibnamefont{Keitel}}, and
  \bibinfo{author}{\bibfnamefont{A.}~\bibnamefont{Macchi}},
  \bibinfo{year}{2010}, \bibinfo{journal}{New J. Phys.}
  \textbf{\bibinfo{volume}{12}}, \bibinfo{pages}{123005}.

\bibitem[{\citenamefont{Taranukhin}(2000)}]{Taranukhin_2000}
\bibinfo{author}{\bibnamefont{Taranukhin}, \bibfnamefont{V.~D.}},
  \bibinfo{year}{2000}, \bibinfo{journal}{Laser Phys.}
  \textbf{\bibinfo{volume}{10}}, \bibinfo{pages}{330}.

\bibitem[{\citenamefont{Taranukhin and Shubin}(2001)}]{Taranukhin_2001}
\bibinfo{author}{\bibnamefont{Taranukhin}, \bibfnamefont{V.~D.}}, and
  \bibinfo{author}{\bibfnamefont{N.~{\relax Yu}.} \bibnamefont{Shubin}},
  \bibinfo{year}{2001}, \bibinfo{journal}{Quantum Electronics}
  \textbf{\bibinfo{volume}{31}}, \bibinfo{pages}{179}.

\bibitem[{\citenamefont{Taranukhin and Shubin}(2002)}]{Taranukhin_2002}
\bibinfo{author}{\bibnamefont{Taranukhin}, \bibfnamefont{V.~D.}}, and
  \bibinfo{author}{\bibfnamefont{N.~{\relax Yu}.} \bibnamefont{Shubin}},
  \bibinfo{year}{2002}, \bibinfo{journal}{J. Opt. Soc. Am. B}
  \textbf{\bibinfo{volume}{19}}, \bibinfo{pages}{1132}.

\bibitem[{\citenamefont{Teitelboim}(1971)}]{Teitelboim_1971}
\bibinfo{author}{\bibnamefont{Teitelboim}, \bibfnamefont{C.}},
  \bibinfo{year}{1971}, \bibinfo{journal}{Phys. Rev. D}
  \textbf{\bibinfo{volume}{4}}, \bibinfo{pages}{345}.

\bibitem[{\citenamefont{Telnov}(1990)}]{Telnov_1990}
\bibinfo{author}{\bibnamefont{Telnov}, \bibfnamefont{V.}},
  \bibinfo{year}{1990}, \bibinfo{journal}{Nucl. Instr. Meth. Phys. Res. A}
  \textbf{\bibinfo{volume}{294}}, \bibinfo{pages}{72}.

\bibitem[{\citenamefont{Teubner and Gibbon}(2009)}]{Teubner_2009}
\bibinfo{author}{\bibnamefont{Teubner}, \bibfnamefont{U.}}, and
  \bibinfo{author}{\bibfnamefont{P.}~\bibnamefont{Gibbon}},
  \bibinfo{year}{2009}, \bibinfo{journal}{Rev. Mod. Phys.}
  \textbf{\bibinfo{volume}{81}}, \bibinfo{pages}{445}.

\bibitem[{\citenamefont{Thoma}(2009{\natexlab{a}})}]{Thoma_RMP}
\bibinfo{author}{\bibnamefont{Thoma}, \bibfnamefont{M.~H.}},
  \bibinfo{year}{2009}{\natexlab{a}}, \bibinfo{journal}{Rev. Mod. Phys.}
  \textbf{\bibinfo{volume}{81}}, \bibinfo{pages}{959}.

\bibitem[{\citenamefont{Thoma}(2009{\natexlab{b}})}]{Thoma_EPJD}
\bibinfo{author}{\bibnamefont{Thoma}, \bibfnamefont{M.~H.}},
  \bibinfo{year}{2009}{\natexlab{b}}, \bibinfo{journal}{Eur. Phys. J. D}
  \textbf{\bibinfo{volume}{55}}, \bibinfo{pages}{271}.

\bibitem[{\citenamefont{Ting} \emph{et~al.}(1995)\citenamefont{Ting, Fischer,
  Fisher, Evans, Burris, Krall, Esarey, and Sprangle}}]{Ting_1995}
\bibinfo{author}{\bibnamefont{Ting}, \bibfnamefont{A.}},
  \bibinfo{author}{\bibfnamefont{R.}~\bibnamefont{Fischer}},
  \bibinfo{author}{\bibfnamefont{A.}~\bibnamefont{Fisher}},
  \bibinfo{author}{\bibfnamefont{K.}~\bibnamefont{Evans}},
  \bibinfo{author}{\bibfnamefont{R.}~\bibnamefont{Burris}},
  \bibinfo{author}{\bibfnamefont{J.}~\bibnamefont{Krall}},
  \bibinfo{author}{\bibfnamefont{E.}~\bibnamefont{Esarey}}, and
  \bibinfo{author}{\bibfnamefont{P.}~\bibnamefont{Sprangle}},
  \bibinfo{year}{1995}, \bibinfo{journal}{J. Appl. Phys.}
  \textbf{\bibinfo{volume}{78}}, \bibinfo{pages}{575}.

\bibitem[{\citenamefont{Ting} \emph{et~al.}(1996)\citenamefont{Ting, Fischer,
  Fisher, Moore, Hafizi, Elton, Krushelnick, Burris, Jackel, Evans, Weaver,
  Sprangle} \emph{et~al.}}]{Ting_1996}
\bibinfo{author}{\bibnamefont{Ting}, \bibfnamefont{A.}},
  \bibinfo{author}{\bibfnamefont{R.}~\bibnamefont{Fischer}},
  \bibinfo{author}{\bibfnamefont{A.}~\bibnamefont{Fisher}},
  \bibinfo{author}{\bibfnamefont{C.~I.} \bibnamefont{Moore}},
  \bibinfo{author}{\bibfnamefont{B.}~\bibnamefont{Hafizi}},
  \bibinfo{author}{\bibfnamefont{R.}~\bibnamefont{Elton}},
  \bibinfo{author}{\bibfnamefont{K.}~\bibnamefont{Krushelnick}},
  \bibinfo{author}{\bibfnamefont{R.}~\bibnamefont{Burris}},
  \bibinfo{author}{\bibfnamefont{S.}~\bibnamefont{Jackel}},
  \bibinfo{author}{\bibfnamefont{K.}~\bibnamefont{Evans}},
  \bibinfo{author}{\bibfnamefont{J.~N.} \bibnamefont{Weaver}},
  \bibinfo{author}{\bibfnamefont{P.}~\bibnamefont{Sprangle}}, \emph{et~al.},
  \bibinfo{year}{1996}, \bibinfo{journal}{Nucl. Instr. Meth. Phys. Res. A}
  \textbf{\bibinfo{volume}{375}}, \bibinfo{pages}{ABS68}.

\bibitem[{\citenamefont{Tinsley}(2005)}]{Tinsley_2005}
\bibinfo{author}{\bibnamefont{Tinsley}, \bibfnamefont{T.~M.}},
  \bibinfo{year}{2005}, \bibinfo{journal}{Phys. Rev. D}
  \textbf{\bibinfo{volume}{71}}, \bibinfo{pages}{073010}.

\bibitem[{\citenamefont{Titov} \emph{et~al.}(2009)\citenamefont{Titov,
  K\"ampfer, and Takabe}}]{Kampfer_2009}
\bibinfo{author}{\bibnamefont{Titov}, \bibfnamefont{A.~I.}},
  \bibinfo{author}{\bibfnamefont{B.}~\bibnamefont{K\"ampfer}}, and
  \bibinfo{author}{\bibfnamefont{H.}~\bibnamefont{Takabe}},
  \bibinfo{year}{2009}, \bibinfo{journal}{Phys. Rev. ST Accel. Beams}
  \textbf{\bibinfo{volume}{12}}, \bibinfo{pages}{111301}.

\bibitem[{\citenamefont{Tommasini} \emph{et~al.}(2008)\citenamefont{Tommasini,
  Ferrando, Michinel, and Seco}}]{Tommasini_2008}
\bibinfo{author}{\bibnamefont{Tommasini}, \bibfnamefont{D.}},
  \bibinfo{author}{\bibfnamefont{A.}~\bibnamefont{Ferrando}},
  \bibinfo{author}{\bibfnamefont{H.}~\bibnamefont{Michinel}}, and
  \bibinfo{author}{\bibfnamefont{M.}~\bibnamefont{Seco}}, \bibinfo{year}{2008},
  \bibinfo{journal}{Phys. Rev. A} \textbf{\bibinfo{volume}{77}},
  \bibinfo{pages}{042101}.

\bibitem[{\citenamefont{Tommasini and Michinel}(2010)}]{Tommasini_2010}
\bibinfo{author}{\bibnamefont{Tommasini}, \bibfnamefont{D.}}, and
  \bibinfo{author}{\bibfnamefont{H.}~\bibnamefont{Michinel}},
  \bibinfo{year}{2010}, \bibinfo{journal}{Phys. Rev. A}
  \textbf{\bibinfo{volume}{82}}, \bibinfo{pages}{011803}.

\bibitem[{\citenamefont{Toncian} \emph{et~al.}(2006)\citenamefont{Toncian,
  Borghesi, Fuchs, d'Humi\`{e}res, Antici, Audebert, Brambrink, Cecchetti,
  Pipahl, Romagnani, and Willi}}]{Toncian_2006}
\bibinfo{author}{\bibnamefont{Toncian}, \bibfnamefont{T.}},
  \bibinfo{author}{\bibfnamefont{M.}~\bibnamefont{Borghesi}},
  \bibinfo{author}{\bibfnamefont{J.}~\bibnamefont{Fuchs}},
  \bibinfo{author}{\bibfnamefont{E.}~\bibnamefont{d'Humi\`{e}res}},
  \bibinfo{author}{\bibfnamefont{P.}~\bibnamefont{Antici}},
  \bibinfo{author}{\bibfnamefont{P.}~\bibnamefont{Audebert}},
  \bibinfo{author}{\bibfnamefont{E.}~\bibnamefont{Brambrink}},
  \bibinfo{author}{\bibfnamefont{C.~A.} \bibnamefont{Cecchetti}},
  \bibinfo{author}{\bibfnamefont{A.}~\bibnamefont{Pipahl}},
  \bibinfo{author}{\bibfnamefont{L.}~\bibnamefont{Romagnani}}, and
  \bibinfo{author}{\bibfnamefont{O.}~\bibnamefont{Willi}},
  \bibinfo{year}{2006}, \bibinfo{journal}{Science}
  \textbf{\bibinfo{volume}{312}}, \bibinfo{pages}{410}.

\bibitem[{\citenamefont{{TPL}}(2011)}]{TPL}
\bibinfo{author}{\bibnamefont{{TPL}}} (\bibinfo{collaboration}{Texas Petawatt
  Laser}), \bibinfo{year}{2011},
  \urlprefix\url{http://www.ph.utexas.edu/~utlasers/}.

\bibitem[{\citenamefont{Tsai}(1993)}]{Tsai_1993}
\bibinfo{author}{\bibnamefont{Tsai}, \bibfnamefont{Y.~S.}},
  \bibinfo{year}{1993}, \bibinfo{journal}{Phys. Rev. D}
  \textbf{\bibinfo{volume}{48}}, \bibinfo{pages}{96}.

\bibitem[{\citenamefont{Tuchin}(2010)}]{Tuchin_2010}
\bibinfo{author}{\bibnamefont{Tuchin}, \bibfnamefont{K.}},
  \bibinfo{year}{2010}, \bibinfo{journal}{Phys. Lett. B}
  \textbf{\bibinfo{volume}{686}}, \bibinfo{pages}{29}.

\bibitem[{\citenamefont{Varfolomeev}(1966)}]{Varfolomeev_1966}
\bibinfo{author}{\bibnamefont{Varfolomeev}, \bibfnamefont{A.~A.}},
  \bibinfo{year}{1966}, \bibinfo{journal}{Sov. Phys. JETP}
  \textbf{\bibinfo{volume}{23}}, \bibinfo{pages}{681}.

\bibitem[{\citenamefont{Verschl and
  Keitel}(2007{\natexlab{a}})}]{Verschl_2007a}
\bibinfo{author}{\bibnamefont{Verschl}, \bibfnamefont{M.}}, and
  \bibinfo{author}{\bibfnamefont{C.~H.} \bibnamefont{Keitel}},
  \bibinfo{year}{2007}{\natexlab{a}}, \bibinfo{journal}{J. Phys. B}
  \textbf{\bibinfo{volume}{40}}, \bibinfo{pages}{F69}.

\bibitem[{\citenamefont{Verschl and
  Keitel}(2007{\natexlab{b}})}]{Verschl_2007b}
\bibinfo{author}{\bibnamefont{Verschl}, \bibfnamefont{M.}}, and
  \bibinfo{author}{\bibfnamefont{C.~H.} \bibnamefont{Keitel}},
  \bibinfo{year}{2007}{\natexlab{b}}, \bibinfo{journal}{Europhys. Lett.}
  \textbf{\bibinfo{volume}{77}}, \bibinfo{pages}{64004}.

\bibitem[{\citenamefont{Volkov}(1935)}]{Volkov_1935}
\bibinfo{author}{\bibnamefont{Volkov}, \bibfnamefont{D.~M.}},
  \bibinfo{year}{1935}, \bibinfo{journal}{Z. Phys.}
  \textbf{\bibinfo{volume}{94}}, \bibinfo{pages}{250}.

\bibitem[{\citenamefont{{Vulcan}}(2011)}]{Vulcan}
\bibinfo{author}{\bibnamefont{{Vulcan}}}, \bibinfo{year}{2011},
  \urlprefix\url{http://www.clf.rl.ac.uk/Facilities/Vulcan/12248.aspx}.

\bibitem[{\citenamefont{{Vulcan 10PW}}(2011)}]{Vulcan_10PW}
\bibinfo{author}{\bibnamefont{{Vulcan 10PW}}}, \bibinfo{year}{2011},
  \urlprefix\url{http://www.clf.rl.ac.uk/New+Initiatives/The+Vulcan+10+Petawatt+Project/14684.aspx}.

\bibitem[{\citenamefont{Wagner}
  \emph{et~al.}(2010{\natexlab{a}})\citenamefont{Wagner, Ware, Su, and
  Grobe}}]{Grobe_Spin2010}
\bibinfo{author}{\bibnamefont{Wagner}, \bibfnamefont{R.~E.}},
  \bibinfo{author}{\bibfnamefont{M.~R.} \bibnamefont{Ware}},
  \bibinfo{author}{\bibfnamefont{Q.}~\bibnamefont{Su}}, and
  \bibinfo{author}{\bibfnamefont{R.}~\bibnamefont{Grobe}},
  \bibinfo{year}{2010}{\natexlab{a}}, \bibinfo{journal}{Phys. Rev. A}
  \textbf{\bibinfo{volume}{81}}, \bibinfo{pages}{024101}.

\bibitem[{\citenamefont{Wagner}
  \emph{et~al.}(2010{\natexlab{b}})\citenamefont{Wagner, Ware, Su, and
  Grobe}}]{Wagner_Spin2010}
\bibinfo{author}{\bibnamefont{Wagner}, \bibfnamefont{R.~E.}},
  \bibinfo{author}{\bibfnamefont{M.~R.} \bibnamefont{Ware}},
  \bibinfo{author}{\bibfnamefont{Q.}~\bibnamefont{Su}}, and
  \bibinfo{author}{\bibfnamefont{R.}~\bibnamefont{Grobe}},
  \bibinfo{year}{2010}{\natexlab{b}}, \bibinfo{journal}{Phys. Rev. A}
  \textbf{\bibinfo{volume}{81}}, \bibinfo{pages}{052104}.

\bibitem[{\citenamefont{Walker and Dracoulis}(1999)}]{Walker_1999}
\bibinfo{author}{\bibnamefont{Walker}, \bibfnamefont{P.}}, and
  \bibinfo{author}{\bibfnamefont{G.}~\bibnamefont{Dracoulis}},
  \bibinfo{year}{1999}, \bibinfo{journal}{Nature (London)}
  \textbf{\bibinfo{volume}{399}}, \bibinfo{pages}{35}.

\bibitem[{\citenamefont{Walser} \emph{et~al.}(2002)\citenamefont{Walser,
  Urbach, Hatsagortsyan, Hu, and Keitel}}]{Walser_2002}
\bibinfo{author}{\bibnamefont{Walser}, \bibfnamefont{M.~W.}},
  \bibinfo{author}{\bibfnamefont{D.~J.} \bibnamefont{Urbach}},
  \bibinfo{author}{\bibfnamefont{K.~Z.} \bibnamefont{Hatsagortsyan}},
  \bibinfo{author}{\bibfnamefont{S.~X.} \bibnamefont{Hu}}, and
  \bibinfo{author}{\bibfnamefont{C.~H.} \bibnamefont{Keitel}},
  \bibinfo{year}{2002}, \bibinfo{journal}{Phys. Rev. A}
  \textbf{\bibinfo{volume}{65}}, \bibinfo{pages}{043410}.

\bibitem[{\citenamefont{Wang} \emph{et~al.}(2008)\citenamefont{Wang, Granados,
  Pedaci, Alessi, Luther, Berrill, and Rocca}}]{Wang_2008}
\bibinfo{author}{\bibnamefont{Wang}, \bibfnamefont{Y.}},
  \bibinfo{author}{\bibfnamefont{E.}~\bibnamefont{Granados}},
  \bibinfo{author}{\bibfnamefont{F.}~\bibnamefont{Pedaci}},
  \bibinfo{author}{\bibfnamefont{D.}~\bibnamefont{Alessi}},
  \bibinfo{author}{\bibfnamefont{B.}~\bibnamefont{Luther}},
  \bibinfo{author}{\bibfnamefont{M.}~\bibnamefont{Berrill}}, and
  \bibinfo{author}{\bibfnamefont{J.~J.} \bibnamefont{Rocca}},
  \bibinfo{year}{2008}, \bibinfo{journal}{Nature Photon.}
  \textbf{\bibinfo{volume}{2}}, \bibinfo{pages}{94}.

\bibitem[{\citenamefont{Wang} \emph{et~al.}(2011)\citenamefont{Wang, Liu, Shen,
  Zhang, Teng, and Wei}}]{Wang_2011}
\bibinfo{author}{\bibnamefont{Wang}, \bibfnamefont{Z.}},
  \bibinfo{author}{\bibfnamefont{C.}~\bibnamefont{Liu}},
  \bibinfo{author}{\bibfnamefont{Z.}~\bibnamefont{Shen}},
  \bibinfo{author}{\bibfnamefont{Q.}~\bibnamefont{Zhang}},
  \bibinfo{author}{\bibfnamefont{H.}~\bibnamefont{Teng}}, and
  \bibinfo{author}{\bibfnamefont{Z.}~\bibnamefont{Wei}}, \bibinfo{year}{2011},
  \bibinfo{journal}{Opt. Lett.} \textbf{\bibinfo{volume}{36}},
  \bibinfo{pages}{3194}.

\bibitem[{\citenamefont{Weidenm{\"u}ller}(2011)}]{Weidenmueller_2011}
\bibinfo{author}{\bibnamefont{Weidenm{\"u}ller}, \bibfnamefont{H.~A.}},
  \bibinfo{year}{2011}, \bibinfo{journal}{Phys. Rev. Lett.}
  \textbf{\bibinfo{volume}{106}}, \bibinfo{pages}{122502}.

\bibitem[{\citenamefont{Weisskopf}(1936)}]{Weisskopf_1936}
\bibinfo{author}{\bibnamefont{Weisskopf}, \bibfnamefont{V.}},
  \bibinfo{year}{1936}, \bibinfo{journal}{K. Dan. Vidensk. Selsk. Mat. Fys.
  Medd.} \textbf{\bibinfo{volume}{14}}, \bibinfo{pages}{1}.

\bibitem[{\citenamefont{Wilson}(2001)}]{Wilson_b_2001}
\bibinfo{author}{\bibnamefont{Wilson}, \bibfnamefont{E.~J.~N.}},
  \bibinfo{year}{2001}, \emph{\bibinfo{title}{An Introduction to Particle
  Accelerators}} (\bibinfo{publisher}{Oxford University Press, Oxford}).

\bibitem[{\citenamefont{Winterfeldt}
  \emph{et~al.}(2008)\citenamefont{Winterfeldt, Spielmann, and
  Gerber}}]{Winterfeldt_2008}
\bibinfo{author}{\bibnamefont{Winterfeldt}, \bibfnamefont{C.}},
  \bibinfo{author}{\bibfnamefont{C.}~\bibnamefont{Spielmann}}, and
  \bibinfo{author}{\bibfnamefont{G.}~\bibnamefont{Gerber}},
  \bibinfo{year}{2008}, \bibinfo{journal}{Rev. Mod. Phys.}
  \textbf{\bibinfo{volume}{80}}, \bibinfo{pages}{117}.

\bibitem[{\citenamefont{Wong} \emph{et~al.}(2011)\citenamefont{Wong, Grigoriu,
  Roslund, Ho, and Rabitz}}]{Wong_2011}
\bibinfo{author}{\bibnamefont{Wong}, \bibfnamefont{I.}},
  \bibinfo{author}{\bibfnamefont{A.}~\bibnamefont{Grigoriu}},
  \bibinfo{author}{\bibfnamefont{J.}~\bibnamefont{Roslund}},
  \bibinfo{author}{\bibfnamefont{T.-S.} \bibnamefont{Ho}}, and
  \bibinfo{author}{\bibfnamefont{H.}~\bibnamefont{Rabitz}},
  \bibinfo{year}{2011}, \bibinfo{journal}{Phys. Rev. A}
  \textbf{\bibinfo{volume}{84}}, \bibinfo{pages}{053429}.

\bibitem[{\citenamefont{Woodward}(1947)}]{Woodward_1947}
\bibinfo{author}{\bibnamefont{Woodward}, \bibfnamefont{P.~M.}},
  \bibinfo{year}{1947}, \bibinfo{journal}{J. Inst. Electr. Eng.}
  \textbf{\bibinfo{volume}{93}}, \bibinfo{pages}{1554}.

\bibitem[{\citenamefont{{XCELS}}(2012)}]{XCELS}
\bibinfo{author}{\bibnamefont{{XCELS}}} (\bibinfo{collaboration}{Exawatt Center
  for Extreme Light Studies}), \bibinfo{year}{2012},
  \urlprefix\url{http://www.xcels.iapras.ru/}.

\bibitem[{\citenamefont{Xiang} \emph{et~al.}(2010)\citenamefont{Xiang, Niu, Qi,
  Li, and Gong}}]{Xiang_2010}
\bibinfo{author}{\bibnamefont{Xiang}, \bibfnamefont{Y.}},
  \bibinfo{author}{\bibfnamefont{Y.}~\bibnamefont{Niu}},
  \bibinfo{author}{\bibfnamefont{Y.}~\bibnamefont{Qi}},
  \bibinfo{author}{\bibfnamefont{R.}~\bibnamefont{Li}}, and
  \bibinfo{author}{\bibfnamefont{S.}~\bibnamefont{Gong}}, \bibinfo{year}{2010},
  \bibinfo{journal}{J. Mod. Opt.} \textbf{\bibinfo{volume}{57}},
  \bibinfo{pages}{385}.

\bibitem[{\citenamefont{Yakovlev}(1965)}]{Yakovlev_1966}
\bibinfo{author}{\bibnamefont{Yakovlev}, \bibfnamefont{V.~P.}},
  \bibinfo{year}{1965}, \bibinfo{journal}{Sov. Phys. JETP}
  \textbf{\bibinfo{volume}{22}}, \bibinfo{pages}{223}.

\bibitem[{\citenamefont{Yamakawa} \emph{et~al.}(2003)\citenamefont{Yamakawa,
  Akahane, Fukuda, Aoyama, Inoue, and Ueda}}]{Yamakawa_2003}
\bibinfo{author}{\bibnamefont{Yamakawa}, \bibfnamefont{K.}},
  \bibinfo{author}{\bibfnamefont{Y.}~\bibnamefont{Akahane}},
  \bibinfo{author}{\bibfnamefont{Y.}~\bibnamefont{Fukuda}},
  \bibinfo{author}{\bibfnamefont{M.}~\bibnamefont{Aoyama}},
  \bibinfo{author}{\bibfnamefont{N.}~\bibnamefont{Inoue}}, and
  \bibinfo{author}{\bibfnamefont{H.}~\bibnamefont{Ueda}}, \bibinfo{year}{2003},
  \bibinfo{journal}{Phys. Rev. A} \textbf{\bibinfo{volume}{68}},
  \bibinfo{pages}{065403}.

\bibitem[{\citenamefont{Yamakawa} \emph{et~al.}(2004)\citenamefont{Yamakawa,
  Akahane, Fukuda, Aoyama, Inoue, Ueda, and Utsumi}}]{Yamakawa_2004}
\bibinfo{author}{\bibnamefont{Yamakawa}, \bibfnamefont{K.}},
  \bibinfo{author}{\bibfnamefont{Y.}~\bibnamefont{Akahane}},
  \bibinfo{author}{\bibfnamefont{Y.}~\bibnamefont{Fukuda}},
  \bibinfo{author}{\bibfnamefont{M.}~\bibnamefont{Aoyama}},
  \bibinfo{author}{\bibfnamefont{N.}~\bibnamefont{Inoue}},
  \bibinfo{author}{\bibfnamefont{H.}~\bibnamefont{Ueda}}, and
  \bibinfo{author}{\bibfnamefont{T.}~\bibnamefont{Utsumi}},
  \bibinfo{year}{2004}, \bibinfo{journal}{Phys. Rev. Lett.}
  \textbf{\bibinfo{volume}{92}}, \bibinfo{pages}{123001}.

\bibitem[{\citenamefont{Yanovsky} \emph{et~al.}(2008)\citenamefont{Yanovsky,
  Chvykov, Kalinchenko, Rousseau, Planchon, Matsuoka, Maksimchuk, Nees,
  Cheriaux, Mourou, and Krushelnick}}]{Yanovsky_2008}
\bibinfo{author}{\bibnamefont{Yanovsky}, \bibfnamefont{V.}},
  \bibinfo{author}{\bibfnamefont{V.}~\bibnamefont{Chvykov}},
  \bibinfo{author}{\bibfnamefont{G.}~\bibnamefont{Kalinchenko}},
  \bibinfo{author}{\bibfnamefont{P.}~\bibnamefont{Rousseau}},
  \bibinfo{author}{\bibfnamefont{T.}~\bibnamefont{Planchon}},
  \bibinfo{author}{\bibfnamefont{T.}~\bibnamefont{Matsuoka}},
  \bibinfo{author}{\bibfnamefont{A.}~\bibnamefont{Maksimchuk}},
  \bibinfo{author}{\bibfnamefont{J.}~\bibnamefont{Nees}},
  \bibinfo{author}{\bibfnamefont{G.}~\bibnamefont{Cheriaux}},
  \bibinfo{author}{\bibfnamefont{G.}~\bibnamefont{Mourou}}, and
  \bibinfo{author}{\bibfnamefont{K.}~\bibnamefont{Krushelnick}},
  \bibinfo{year}{2008}, \bibinfo{journal}{Opt. Express}
  \textbf{\bibinfo{volume}{16}}, \bibinfo{pages}{2109}.

\bibitem[{\citenamefont{Zakowicz}(2005)}]{Zakowicz_2005}
\bibinfo{author}{\bibnamefont{Zakowicz}, \bibfnamefont{S.}},
  \bibinfo{year}{2005}, \bibinfo{journal}{J. Math. Phys.}
  \textbf{\bibinfo{volume}{46}}, \bibinfo{pages}{032304}.

\bibitem[{\citenamefont{Zeitoun} \emph{et~al.}(2004)\citenamefont{Zeitoun,
  Faivre, Sebban, Mocek, Hallou, Fajardo, Aubert, Balcou, Burgy, Douillet,
  Kazamias, de~Lach\`{e}ze-Murel} \emph{et~al.}}]{Zeitoun_2004}
\bibinfo{author}{\bibnamefont{Zeitoun}, \bibfnamefont{P.}},
  \bibinfo{author}{\bibfnamefont{G.}~\bibnamefont{Faivre}},
  \bibinfo{author}{\bibfnamefont{S.}~\bibnamefont{Sebban}},
  \bibinfo{author}{\bibfnamefont{T.}~\bibnamefont{Mocek}},
  \bibinfo{author}{\bibfnamefont{A.}~\bibnamefont{Hallou}},
  \bibinfo{author}{\bibfnamefont{M.}~\bibnamefont{Fajardo}},
  \bibinfo{author}{\bibfnamefont{D.}~\bibnamefont{Aubert}},
  \bibinfo{author}{\bibfnamefont{P.}~\bibnamefont{Balcou}},
  \bibinfo{author}{\bibfnamefont{F.}~\bibnamefont{Burgy}},
  \bibinfo{author}{\bibfnamefont{D.}~\bibnamefont{Douillet}},
  \bibinfo{author}{\bibfnamefont{S.}~\bibnamefont{Kazamias}},
  \bibinfo{author}{\bibfnamefont{G.}~\bibnamefont{de~Lach\`{e}ze-Murel}},
  \emph{et~al.}, \bibinfo{year}{2004}, \bibinfo{journal}{Nature}
  \textbf{\bibinfo{volume}{431}}, \bibinfo{pages}{426}.

\bibitem[{\citenamefont{Zeldovich and Popov}(1972)}]{Zeldovich_1972}
\bibinfo{author}{\bibnamefont{Zeldovich}, \bibfnamefont{Y.}}, and
  \bibinfo{author}{\bibfnamefont{V.~S.} \bibnamefont{Popov}},
  \bibinfo{year}{1972}, \bibinfo{journal}{Sov. Phys. Usp.}
  \textbf{\bibinfo{volume}{14}}, \bibinfo{pages}{673}.

\bibitem[{\citenamefont{Zeldovich and Sunyaev}(1969)}]{Zeldovich_1969}
\bibinfo{author}{\bibnamefont{Zeldovich}, \bibfnamefont{{\relax Ya}.~B.}}, and
  \bibinfo{author}{\bibfnamefont{R.~A.} \bibnamefont{Sunyaev}},
  \bibinfo{year}{1969}, \bibinfo{journal}{Astrophys. Space Sci.}
  \textbf{\bibinfo{volume}{4}}, \bibinfo{pages}{301}.

\bibitem[{\citenamefont{Zhang} \emph{et~al.}(2008)\citenamefont{Zhang, Song,
  and Zhang}}]{Zhang_2008}
\bibinfo{author}{\bibnamefont{Zhang}, \bibfnamefont{P.}},
  \bibinfo{author}{\bibfnamefont{Y.}~\bibnamefont{Song}}, and
  \bibinfo{author}{\bibfnamefont{Z.}~\bibnamefont{Zhang}},
  \bibinfo{year}{2008}, \bibinfo{journal}{Phys. Rev. A}
  \textbf{\bibinfo{volume}{78}}, \bibinfo{pages}{013811}.

\bibitem[{\citenamefont{Zhi and Sokolov}(2009)}]{Zhi_2009}
\bibinfo{author}{\bibnamefont{Zhi}, \bibfnamefont{M.~C.}}, and
  \bibinfo{author}{\bibfnamefont{A.~V.} \bibnamefont{Sokolov}},
  \bibinfo{year}{2009}, \bibinfo{journal}{Phys. Rev. A}
  \textbf{\bibinfo{volume}{80}}, \bibinfo{pages}{023415}.

\bibitem[{\citenamefont{Zhidkov} \emph{et~al.}(2002)\citenamefont{Zhidkov,
  Koga, Sasaki, and Uesaka}}]{Zhidkov_2002}
\bibinfo{author}{\bibnamefont{Zhidkov}, \bibfnamefont{A.}},
  \bibinfo{author}{\bibfnamefont{J.}~\bibnamefont{Koga}},
  \bibinfo{author}{\bibfnamefont{A.}~\bibnamefont{Sasaki}}, and
  \bibinfo{author}{\bibfnamefont{M.}~\bibnamefont{Uesaka}},
  \bibinfo{year}{2002}, \bibinfo{journal}{Phys. Rev. Lett.}
  \textbf{\bibinfo{volume}{88}}, \bibinfo{pages}{185002}.

\bibitem[{\citenamefont{Zimmer} \emph{et~al.}(2012)\citenamefont{Zimmer,
  Dominguez, Falomir, and Loewe}}]{Zimmer_2012}
\bibinfo{author}{\bibnamefont{Zimmer}, \bibfnamefont{O.}},
  \bibinfo{author}{\bibfnamefont{C.~A.} \bibnamefont{Dominguez}},
  \bibinfo{author}{\bibfnamefont{H.}~\bibnamefont{Falomir}}, and
  \bibinfo{author}{\bibfnamefont{M.}~\bibnamefont{Loewe}},
  \bibinfo{year}{2012}, \bibinfo{journal}{Phys. Rev. D}
  \textbf{\bibinfo{volume}{85}}, \bibinfo{pages}{013004}.

\bibitem[{\citenamefont{Zon}(1999)}]{Zon_1999}
\bibinfo{author}{\bibnamefont{Zon}, \bibfnamefont{B.~A.}},
  \bibinfo{year}{1999}, \bibinfo{journal}{J. Exp. Theor. Phys.}
  \textbf{\bibinfo{volume}{89}}, \bibinfo{pages}{219}.

\bibitem[{\citenamefont{Zon}(2000)}]{Zon_2000}
\bibinfo{author}{\bibnamefont{Zon}, \bibfnamefont{B.~A.}},
  \bibinfo{year}{2000}, \bibinfo{journal}{J. Exp. Theor. Phys.}
  \textbf{\bibinfo{volume}{91}}, \bibinfo{pages}{899}.

\end{thebibliography}

\end{document}